\documentclass[acmtog, authorversion]{acmart}

\newcommand{\final}{1}

\usepackage{algorithmicx}
\usepackage{algorithm}
\usepackage{algpseudocode}
\usepackage{graphicx}
\usepackage{acro}
\usepackage{xspace}
\usepackage{multirow}
\usepackage{caption}
\usepackage{subcaption}
\usepackage{listings}
 \definecolor{Maroon}{HTML}{800000}
\definecolor{Gray}{gray}{0.9}
\usepackage{lipsum,booktabs}
\usepackage{siunitx}
\usepackage{amsmath,amsfonts}
\usepackage{enumitem}
\usepackage{supertabular}
\usepackage{placeins}
\usepackage{float}
\usepackage{stfloats}
\usepackage{amsmath}
\usepackage{bbm}

\usepackage{pgfplots}
\usepackage{pbox}
\usepackage{nowidow}
\usepackage{mathtools}
\usepackage{microtype}
\usepackage{nicefrac}

\usepackage{tabularx}
\newcolumntype{Y}{>{\raggedright\arraybackslash}X}
\usepackage{array}       %
\usepackage{multirow}    %
\usepackage{booktabs}    %

\newtheorem*{remark}{Remark}

\usepackage{subcaption}
\usepackage{array}

\usepackage{enumitem,enumerate}
\setlist{noitemsep, nolistsep, leftmargin=*}

\definecolor{ahmedColor}{rgb}{1,0.43,0.43}
\newcommand{\ahmed}[1]{{\color{ahmedColor} ahmed: #1}}

\definecolor{todoColor}{rgb}{0, 0, 1}
\newcommand{\todo}[1]{{\color{todoColor} todo: #1}}

\definecolor{changeColor}{rgb}{0,0,0}
\newcommand{\change}[1]{{\color{changeColor} #1}}

\newcommand{\warning}[1]{{\emph{\color{red} #1}}}
\newcommand{\note}[1]{{\emph{\color{blue} #1}}}

\newcommand{\nothing}[1]{}

\ifthenelse{\equal{\final}{1}}
{
\renewcommand{\ahmed}[1]{}
\renewcommand{\todo}[1]{}
\renewcommand{\warning}[1]{}
\renewcommand{\note}[1]{}

}
{}

\definecolor{grayColor}{rgb}{0.5,0.5,0.5}

\copyrightyear{2025}
\acmYear{2025}
\acmDOI{10.1145/3757377.3763882}
\acmConference[SA Conference Papers '25]{SIGGRAPH Asia 2025 Conference Papers}{December 15--18, 2025}{Hong Kong, Hong Kong}

\acmBooktitle{SIGGRAPH Asia 2025 Conference Papers (SA Conference Papers '25), December 15--18, 2025, Hong Kong, Hong Kong}
\acmISBN{979-8-4007-2137-3/2025/12}

\acmVolume{37}
\acmNumber{4}
\acmMonth{8}

\begin{document}

\title{Low-Rank Adaptation of Neural Fields}

\author{Anh Truong}
\affiliation{%
  \department{Department of Electrical Engineering and Computer Science}
  \institution{Massachusetts Institute of Technology}
  \streetaddress{32 Vassar St}
  \city{Cambridge}
  \state{MA}
  \postcode{02139}
  \country{USA}}
\email{anh_t@mit.edu}

\author{Ahmed H. Mahmoud}
\affiliation{%
  \department{Computer Science \& Artificial Intelligence Laboratory}
  \institution{Massachusetts Institute of Technology}
  \streetaddress{32 Vassar St}
  \city{Cambridge}
  \state{MA}
  \postcode{02139}
  \country{USA}}
\email{ahdhn@mit.edu}

\author{Mina Konakovi\'{c} Lukovi\'{c}}
\affiliation{%
  \department{Department of Electrical Engineering and Computer Science}
  \institution{Massachusetts Institute of Technology}
  \streetaddress{32 Vassar St}
  \city{Cambridge}
  \state{MA}
  \postcode{02139}
  \country{USA}}
\email{minakl@mit.edu}

\author{Justin Solomon}
\affiliation{%
  \department{Department of Electrical Engineering and Computer Science}
  \institution{Massachusetts Institute of Technology}
  \streetaddress{32 Vassar St}
  \city{Cambridge}
  \state{MA}
  \postcode{02139}
  \country{USA}}
\email{jsolomon@mit.edu}

\renewcommand{\shortauthors}{Truong, Mahmoud, Konakovi\'{c} Lukovi\'{c}, Solomon}



\begin{abstract}
  Processing visual data often involves small adjustments or sequences of changes, e.g., image filtering, surface smoothing, and animation.
  While established graphics techniques like normal mapping and video compression exploit redundancy to encode such small changes efficiently,
  the problem of encoding small changes to neural fields---neural network parameterizations of visual or physical functions---has received less attention.
  We propose a parameter-efficient strategy for updating neural fields using low-rank adaptations (LoRA)\@.
  LoRA, a method from the parameter-efficient fine-tuning LLM community, encodes small updates to pre-trained models with minimal computational overhead.
  We adapt LoRA for instance-specific neural fields, avoiding the need for large pre-trained models and yielding lightweight updates.
  We validate our approach with experiments in image filtering, geometry editing, video compression, and energy-based editing,
  demonstrating its effectiveness and versatility for representing neural field updates.
\end{abstract}

\maketitle

\begin{figure}
  \centering
  \includegraphics[width=0.98\linewidth]{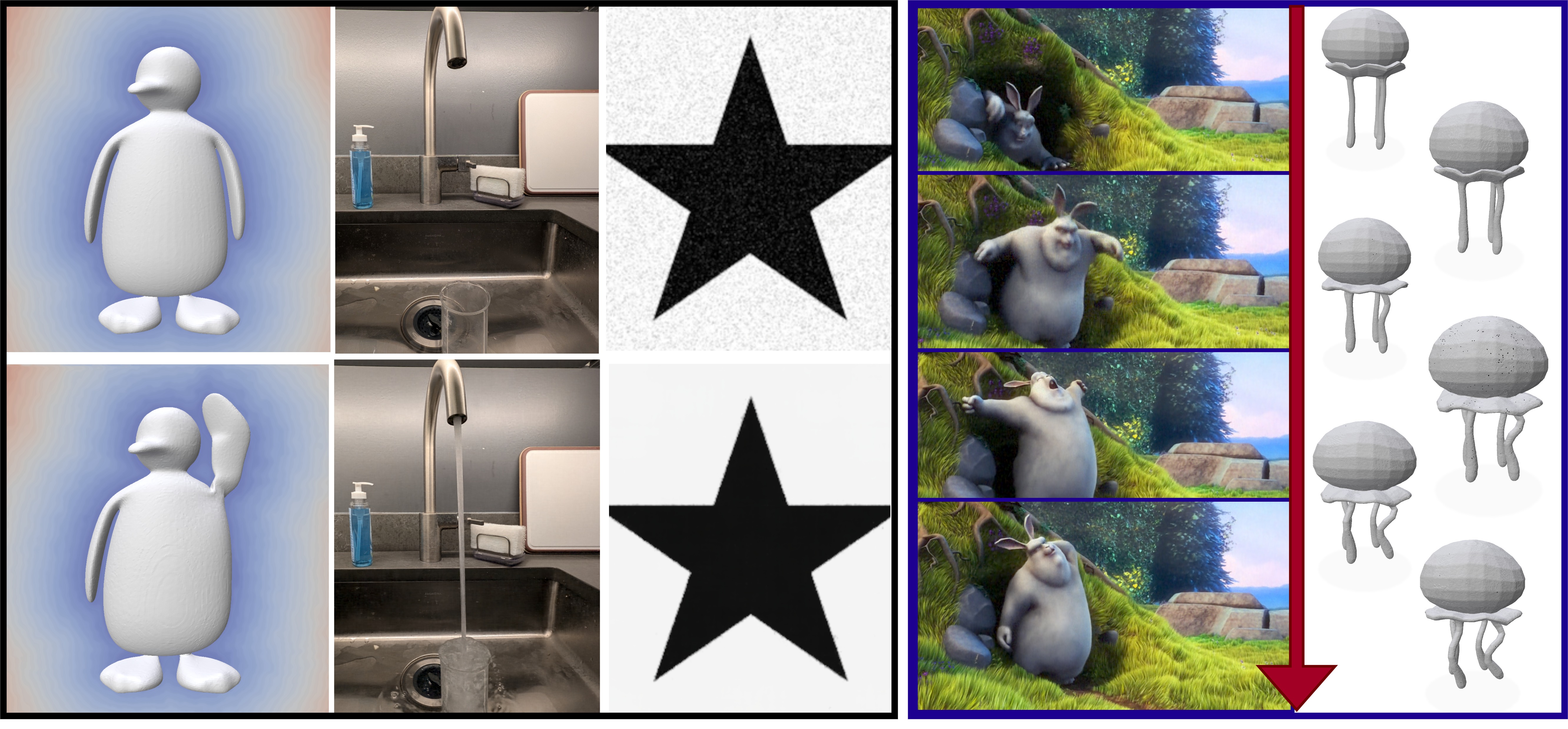}
  \caption{We encode diverse types of variations---including (from left to right) surface deformations, image changes, energy-based denoising, videos, and physical simulation---as compact low-rank updates to pre-trained neural fields. Despite operating with $7-8\times$ fewer parameters than for full fine-tuning, our approach achieves visually faithful reconstructions across a range of tasks.}
  \label{fig:method}
\end{figure}

\section{Introduction}
\label{sec:intro}

Processing visual data often involves small adjustments or sequences of changes. For example, image filtering usually involves local operations that alter the appearance but preserve the overall structure of an input image. Similarly, local surface edits---such as smoothing or sharpening---produce perturbative displacements from the original surface.  Finally, videos consist of image frames where the difference between any two consecutive frames is likely small. 

It should be possible to store small changes compactly. Built on this observation, classical methods in graphics reduce redundancy for different representations by only storing what is necessary to realize a change. For instance, normal maps store surface displacements as compact textures encoding bumps and dents~\cite{Cook:1984:ST}. Similarly, video codecs reduce temporal redundancy by using intermediate P-frames that store offsets from previous frames~\cite{Beach:2018:VCH}. In contrast, methods for compactly perturbing neural fields---an emerging representation in graphics and vision---have received little attention.

Neural fields represent physical or visual quantities (e.g., distances, density, color) in the weights of neural networks~\cite{Xie:2022:neuralfieldsvisualcomputing}.
Neural fields have desirable properties (e.g., continuity, differentiability, compact size, and ease of querying) that make them attractive for visual computing applications like deformation~\cite{Mehta:2022:ALS}, elastic simulation~\cite{Modi:2024:SMF}, and image processing~\cite{Luzi:2024:titandeepimage}.  
Like the classical examples above, many of these tasks yield small updates to a pre-trained model. 

Editing a neural field, however, is far from straightforward, as there is a highly non-linear relationship between changes in its weights and the resulting changes in the output. Many specialized methods have been proposed to edit specific classes of neural fields (e.g., NeRFs~\cite{Liu:2021:ECR} and neural SDFs~\cite{Wang:2021:SDFDisplacementField}) but require large memory footprints and/or elaborate training pipelines, e.g., involving full fine-tuning or auxiliary networks.

In this paper, we present a parameter-efficient strategy for updating neural fields to capture small changes in an input function using low-rank adaptations (LoRA)~\cite{Hu:2022:LLR}. LoRA is a popular method in the large language model (LLM) community for parameter-efficient fine-tuning of pre-trained networks to adapt to minor distributional shifts. 

\begin{figure}
	\captionsetup[subfloat]{labelformat=empty}

  \includegraphics[width=0.49\textwidth]{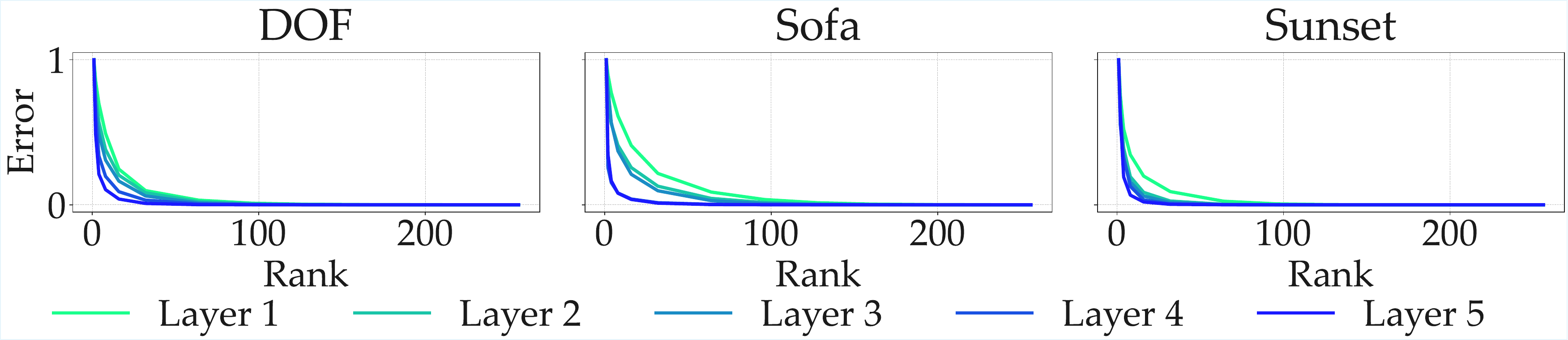}

	\caption{Low-rank weight approximation error across layers of three fine-tuned image neural fields. A minor edit is performed to the target image of each neural field, and each neural field is fine-tuned to overfit to its edited image. The vertical axis measures the (normalized) approximation error of a rank-constrained factorization of the weight difference, accounting for the input distribution of the layer. A low rank approximation is sufficient to recover the fully fine-tuned weight with minimal error.}
  \Description{Low-rank structure of weight difference obtained from full-finetuning of an image neural field} 
	\label{fig:low-rank motivating experiment}
\end{figure}

Our key insight is that typical updates to the function represented by a neural field---e.g., filtering or other local modifications---correspond to small shifts in the underlying data distribution. This makes strategies from parameter-efficient fine-tuning relevant. In particular, we observe that the \emph{low-rank} structure imposed by LoRA provides a natural parameterization for such updates. As a motivating experiment, we fine-tuned image neural fields with weights $W_{\text{base}}$ to match minor variations in their target images, resulting in updated weights $W_{\text{fine-tuned}}$. We found that a \textbf{low-rank factorization of the weight difference $W_\textrm{fine-tuned} -W_\textrm{base}$ approximates the true update with little error} when accounting for the layer's input distribution (see Figure~\ref{fig:low-rank motivating experiment} and the Supplemental Material). Rather than computing this low-rank difference post hoc via full fine-tuning, LoRA allows us to \emph{directly} learn the low-rank update. This makes it a principled and lightweight tool for neural field editing.

Unlike prior LoRA applications in visual computing (e.g., style transfer~\cite{Shah:2025:ZAS, Chenxi:2024:ALI}) which fine-tune large, pre-trained models on datasets containing numerous examples, we apply LoRA to \emph{instance-specific} neural fields. These are small networks overfit to individual signals (e.g., an image or an SDF)\@. This setting avoids reliance on large pre-trained models and better suits common editing tasks in graphics.

We demonstrate the versatility of LoRAs for neural field updates with experiments in image, SDF, video, and energy minimization applications.
Our experiments show that our method can effectively adapt pre-trained neural fields to faithfully encode edits while using \textbf{7-8$\mathbf{\times}$ fewer parameters than conventional fine-tuning} (see Figure~\ref{fig:whiteboard lora vs FT highlight}).

In summary, our contributions are:\change{\footnote{\change{Code is available at~\url{https://github.com/dinhanhtruong/LoRA-NF}.}}}
\begin{itemize}
    \item A parameter-efficient strategy for perturbing a neural field using LoRA.
    \item Experimental validation demonstrating the effectiveness and versatility of LoRA for neural field updates across a range of applications, including image filtering, geometry editing, video compression, and energy minimization (Figure~\ref{fig:method}).
\end{itemize}
\change{While time-efficient model adaptation is a growing research area, the empirical benefits presented in this work focus on storage efficiency.}

We begin by reviewing related work in~\S\ref{sec:related} and introducing the preliminaries of neural fields in~\S\ref{sec:prelim}. We then present our LoRA-based strategy for updating generic neural fields in~\S\ref{sec:lora_nf}, describe its implementation for specific field types in~\S\ref{sec:lora_for_edits}, and conclude with our experimental findings in~\S\ref{sec:experiments}.

\section{Related Work}
\label{sec:related}

\subsection{Neural Fields} \label{neural fields overview}
Neural fields are coordinate-based neural networks that serve as a flexible representation for data in visual computing (see survey by~\citet{Xie:2022:neuralfieldsvisualcomputing}). By parameterizing spatially- and time-varying physical properties of scenes or objects which lack closed-form expressions, neural fields support a wide range of applications, e.g., surface modeling~\cite{park2019deepsdf}, scene reconstruction~\cite{mescheder2019occupancy}, inverse rendering~\cite{Mildenhall:2021:NRS}, signal processing~\cite{sitzmann2020implicit}, and physics-informed simulation~\cite{Raissi:2019:PINN}. Their desirable properties---continuity, full differentiability, and independence from spatial discretization---make them attractive for these tasks.

Neural fields are typically trained either to overfit a single instance (e.g., an image or SDF) or to generalize across a collection of instances using shared weights and instance-specific latent codes~\cite{park2019deepsdf}. The former---instance-specific neural fields---are increasingly used as a primary representation of graphics data~\cite{Davies:2021:ONN}, as they permit dedicating the full capacity of the network to a single signal. Once trained, these models can serve as drop-in replacements for the data they encode---e.g., a neural SDF can be used for downstream geometry processing tasks such as fast closest-point queries~\cite{Sharp:2022:STD} or Constructive Solid Geometry operations upon minor refinement ~\cite{Marschner:2023:CSG}. Compared to traditional graphics representations, instance-specific neural fields are often more memory-efficient.

To further compress neural fields, various methods target the network weights themselves. Techniques such as weight quantization, model pruning~\cite{Chen:2021:NNR}, and low-rank tensor factorizations~\cite{Chen:2022:TTR} have been proposed for lossy compression of network parameters. We focus on this instance-specific setting.

Most neural fields are parameterized using multi-layer perceptrons (MLPs), leveraging the universal approximation theorem~\cite{Kim:2003:ABF}. To accelerate training and improve fidelity, MLPs are frequently augmented with auxiliary data structures, e.g., grids of latent vectors~\cite{Muller:2022:ING,martel2021acorn} or sparse hierarchical encodings~\cite{takikawa2021neural}. While these hybrid architectures often improve convergence and quality~\cite{Takikawa:2023:CNG}, they can introduce implementation complexity and increase memory usage at inference time, especially at high resolutions. For simplicity and broad compatibility, we restrict our method to standard MLP-based neural fields without auxiliary parameters.

\subsection{Neural Field Editing}
Editing a neural field to fit observed data \change{or minimize energy functionals} is not straightforward, as there is a highly non-linear relationship between changes in \change{the weights of the network} and the resulting changes in the network output. Recent works have proposed neural field editing techniques which fall under three broad categories~\cite{Xie:2022:neuralfieldsvisualcomputing}: (1) network fine-tuning where parameters of a pre-trained neural field are further optimized to fit edited data observations~\cite{Liu:2021:ECR} \change{or conform to energy terms ~\cite{Marschner:2023:CSG}}; (2) using hypernetworks trained to map data distributions to neural field parameters~\cite{Chiang:2022:S3S}; (3) latent code fine-tuning/interpolation for networks conditioned on latent codes~\cite{hao2020dualsdf}. Our investigation focuses on network fine-tuning, because hypernetworks require access to a data distribution which may be unavailable and unwanted for performing simple edits, and latent-code models are outside the scope of this work (see \S\ref{neural fields overview}). 

Closest to our approach, prior works in neural field fine-tuning have explored applying updates to a subset of a pre-trained neural field's parameters. \citet{Liu:2021:ECR} propose fine-tuning later layers of a pre-trained NeRF jointly with their latent codes. Their hybrid approach allows them to avoid the cost of fully tuning the network, but earlier layers remain fixed, limiting the expressivity of the network update.  \citet{irene2024} observe that neurons in the final layer of a NeRF's color MLP encode either view-dependent or diffuse appearance. By selectively fine-tuning only the neurons associated with diffuse appearance, their method is able to quickly re-color NeRF scenes. Similarly, \citet{Lee:2023:icenerf} analyze neuron activations to select and fine-tune only those that contribute the most to the color of specific spatial regions. Recently, \citet{kang:2025:codecnerf} apply low-rank updates to factorized tri-plane features, which are aggregated and used as input to a fine-tuned MLP\@. These methods, however, assume that the pre-trained model has a specific NeRF architecture and/or that only a specific attribute (e.g. color) is to be edited. %

\subsection{Low-Rank Adaptations}
LoRA~\cite{Hu:2022:LLR} is a widely-used strategy for parameter-efficient fine-tuning of foundation models such as LLMs. By imposing a low-rank constraint on model weight updates, LoRA is effective at adapting pre-trained models to downstream tasks with few additional parameters. Beyond their original setting in LLM fine-tuning, LoRAs have been applied to other foundation models, predominantly diffusion models for image, video, and 3D data~\cite{Lu:2024:LLR,Dagli:2024:NRU, ouyang:2024:i2vedit} in tasks such as stylization~\cite{Chenxi:2024:ALI, wang:2022:rewritingGAN}, text-based editing~\cite{qi2024tailor3d, zhao:2024:motiondirector}, and generative modeling~\cite{wang2024prolificdreamer}. 

While prior works demonstrate impressive task adaptation using LoRA, they typically operate in settings where large-scale pre-trained models are available, e.g., LLMs or diffusion models. These models contain billions of parameters and are often the result of costly and resource-intensive training pipelines. In contrast, our setting focuses on direct adaptation of instance-specific neural fields where no such large-scale pre-training is involved.

\begin{figure}
    \captionsetup[subfloat]{labelformat=empty}
	\subfloat[Input ($\mathcal{D}$)]{\includegraphics[width=0.12\textwidth]{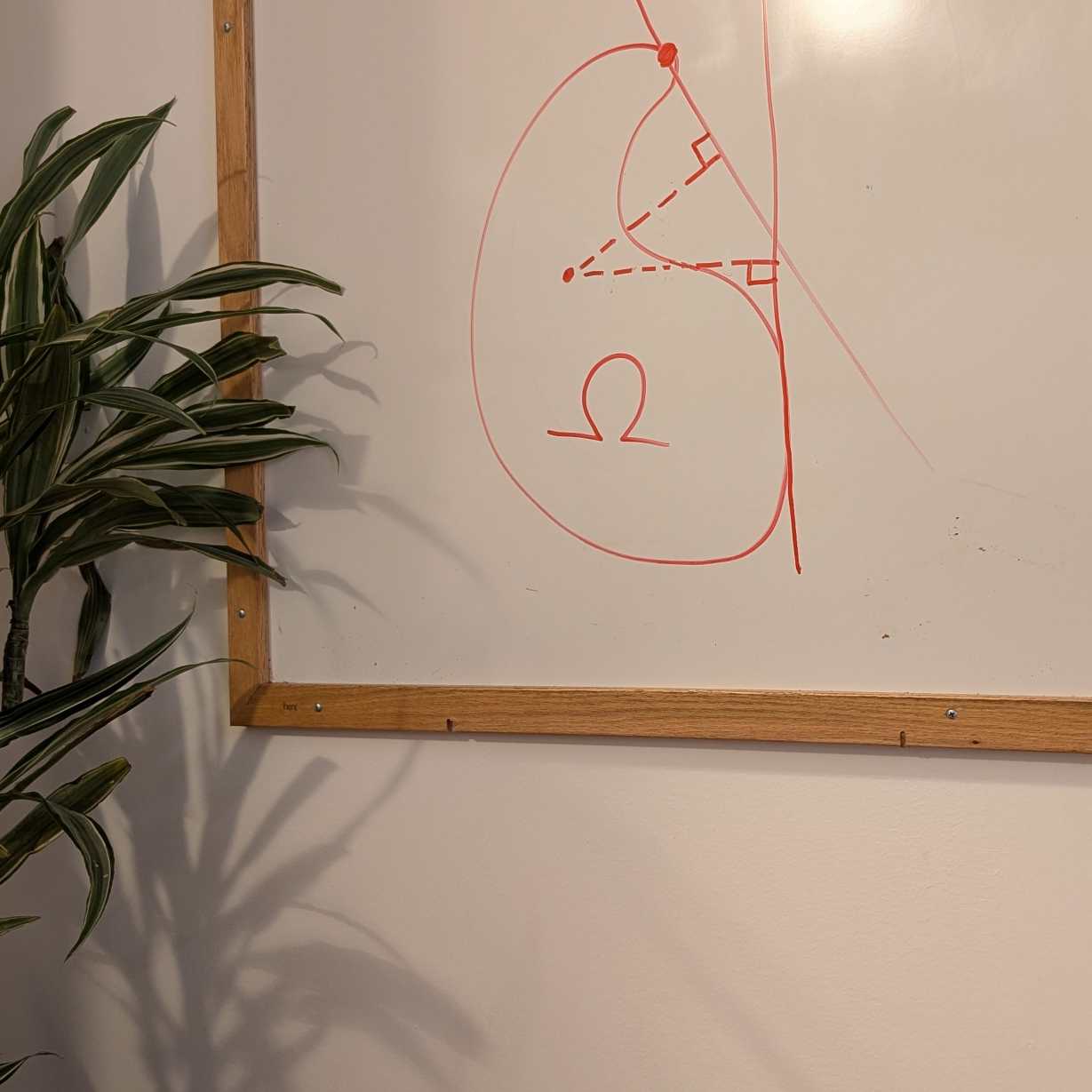}}
	\subfloat[Target ($\mathcal{D}'$)]{\includegraphics[width=0.12\textwidth]{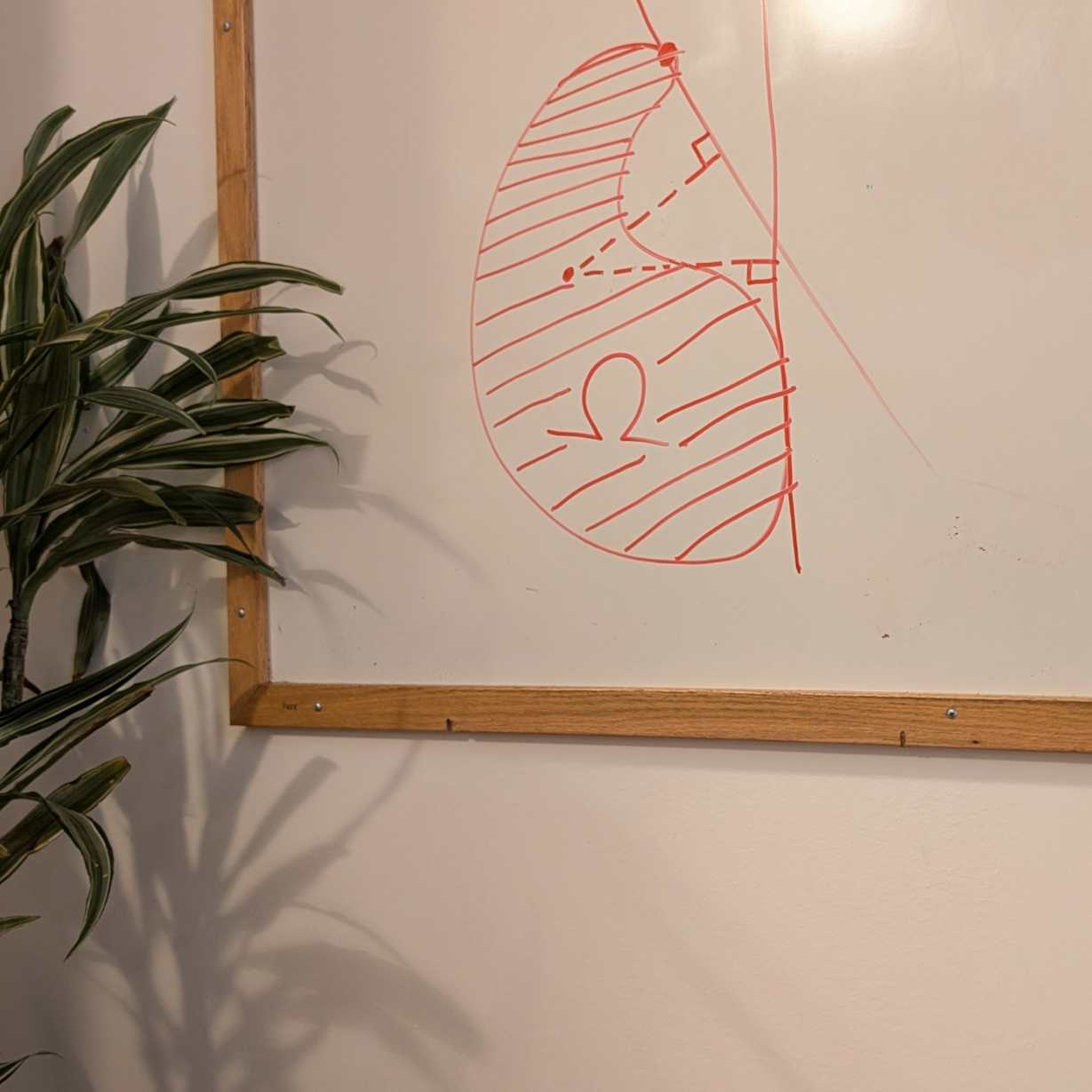}}
    \subfloat[FT: PSNR=42.12 \\ \phantom{FT:} \# Params=340k]{\includegraphics[width=0.12\textwidth]{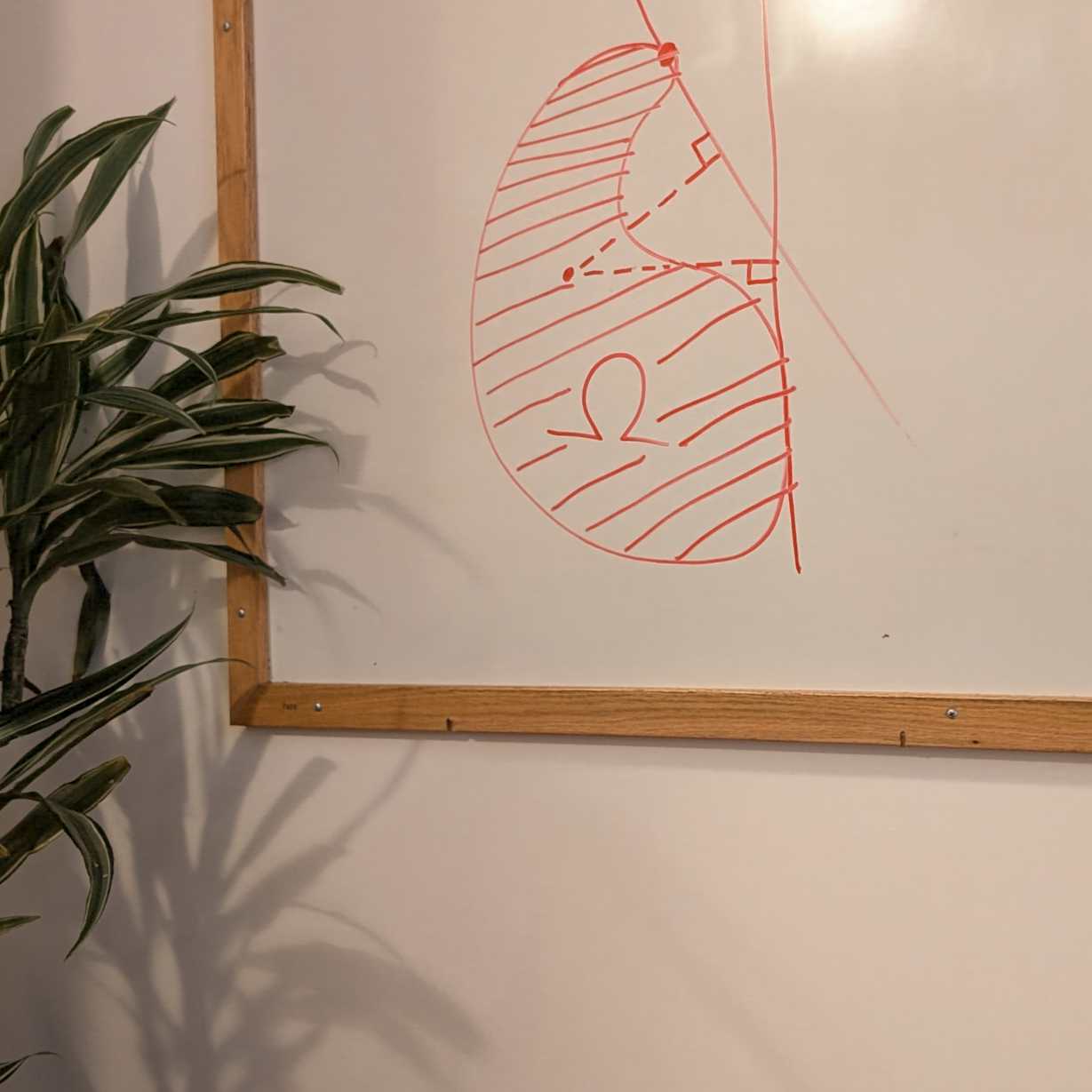}}
	\subfloat[LoRA: PSNR=41.31 \\ \phantom{LoRA:} \# Params=46k]{
        \includegraphics[width=0.12\textwidth]{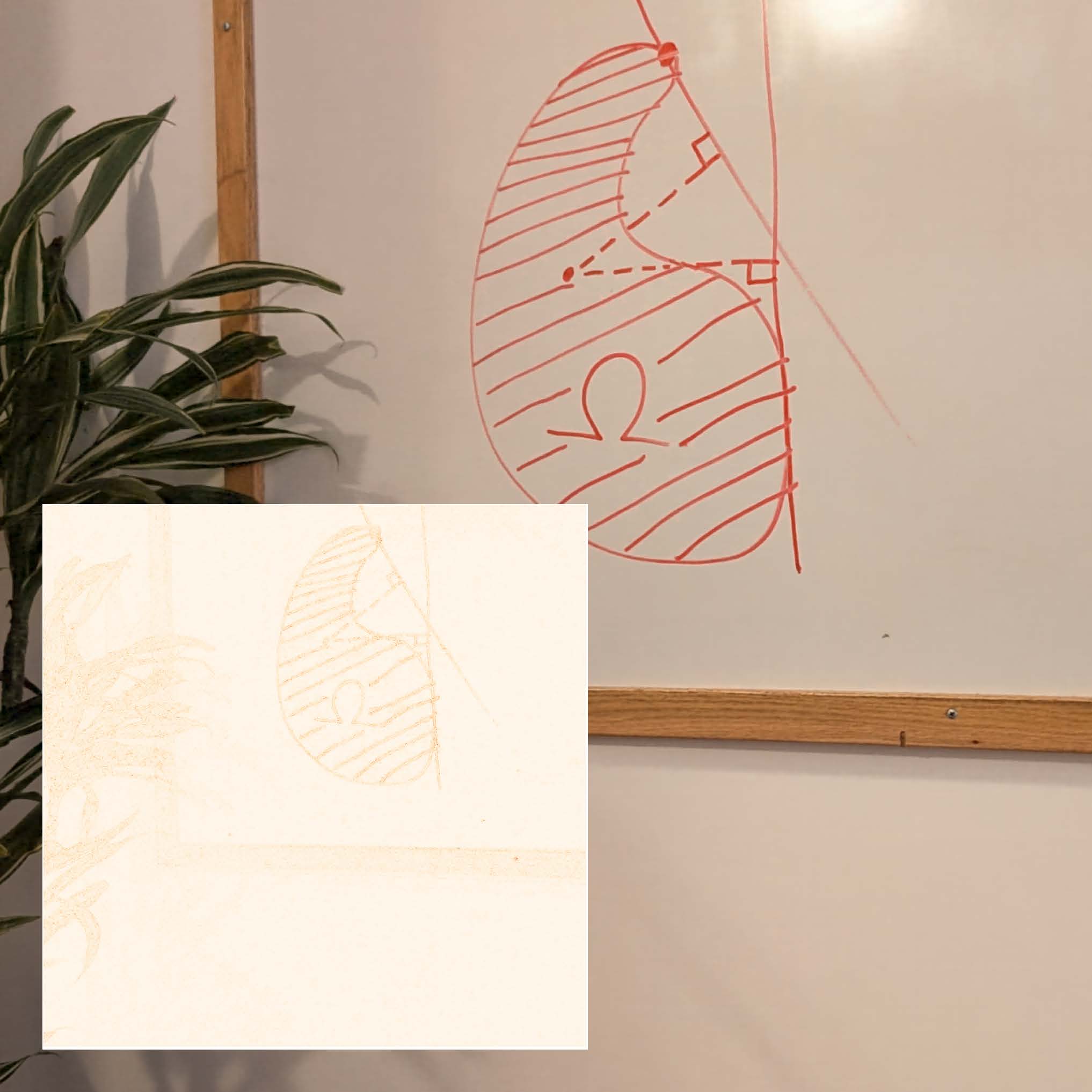}
    }
	\caption{Comparison between full fine-tuning (FT) and our LoRA-based approach for adapting a neural field to an image variation. The original image $\mathcal{D}$ is edited to produce $\mathcal{D}'$ by adding red marker strokes. LoRA reconstructs the edited image with high fidelity achieving a PSNR of 41.31 vs.\ 42.12 for FT while using $\approx$7.4$\times$ fewer parameters (46k vs.\ 340k). The inset shows the difference image between LoRA and FT outputs.}
	\label{fig:whiteboard lora vs FT highlight}
\end{figure}

\section{Preliminaries}
\label{sec:prelim}

Here, we establish notation used in the rest of the paper. We define a neural field as a continuous function $f_\theta: \mathbb{R}^m \rightarrow \mathbb{R}^n$ parameterized by the weights $\theta$ of a neural network, mapping spatial coordinates to values. The input/output dimensions $m$ and $n$ depend on the target signal. For example, a SDF has $m=3$, $n=1$ and maps 3D coordinates to signed distances, while an RGB image has $m=2$, $n=3$ and maps 2D coordinates to RGB values. Throughout this work, we assume $f_\theta$ is overfit to a \emph{single} instance of graphics data (e.g., one image or shape) by minimizing the reconstruction error between the neural field output and target signal. %

We adopt the most basic neural field architecture: a multi-layer perceptron (MLP) that maps spatial coordinates to field values. An MLP consists of alternating linear layers and elementwise non-linearities, e.g., ReLU. Following prior work~\cite{Muller:2022:ING, Mildenhall:2021:NRS}, we first apply a frequency (positional) encoding $\gamma$ to the input coordinates before feeding them to the MLP:
$$\gamma(x) \coloneqq (\sin(2^0x), \cos(2^0x), \ldots,\sin(2^{L-1}x),\cos(2^{L-1}x)),$$ 
where $L \in \mathbb{N}$.
Frequency encodings have proven effective in helping MLPs to regress high-frequency content. Composing the steps of our construction, we therefore consider neural fields of the form $f_\theta(x)\coloneqq \text{MLP}_\theta(\gamma(x)).$

\section{Low-rank Adaptation of Neural Fields} 
\label{sec:lora_nf}

The input to our method consists of (1)~a neural field $f_\theta$ representing a single graphics instance $\mathcal{D}$ (e.g., an image or a SDF) and (2)~an edited variant $\mathcal{D}'$ of $\mathcal{D}$\change{, which can either be explicitly provided (\S\ref{sec:imagevariations}, \S\ref{sec:geometricdeformations}) or implicitly characterized, e.g., as the solution of an energy minimization problem (\S\ref{sec: energy minimization}). In the latter case, the corresponding energy functional characterizing $\mathcal{D'}$ is taken as input.} %

Our method outputs a set of additive weight updates to $f_\theta$ as low-rank adapters, defined below. When these updates are applied to $f_\theta$, they approximate the edited instance $\mathcal{D}'$. This approach encodes the edit made to $\mathcal{D}$ as a low-rank update to the base neural field $f_\theta$. %
\change{We also explore a extension of this setup for \emph{sequences} of edits (\S\ref{sec:video}).}

As discussed in \S\ref{sec:related}, existing methods primarily update neural fields by directly fine-tuning the network on \change{new training objectives}. This approach typically involves either storing a full copy of the base model weights or discarding the original model entirely. Neither of these options is ideal: storing a complete copy is often redundant, especially for minor edits, while losing access to the base model eliminates the ability to encode multiple edits from a shared starting point. Drawing inspiration from recent advances in parameter-efficient fine-tuning, we address this challenge by leveraging \textbf{low-rank adaptations} applied to neural fields.

A LoRA is a rank-constrained additive update to the weight matrices of a neural network. To define this update, recall that an MLP with $h+1$ layers can be written in the following form:
\begin{equation}
\textrm{MLP}_\theta(x) = W_{h}\sigma(W_{h-1}\sigma(W_{h-2}\sigma(\cdots(W_1\sigma(W_0x))\cdots))),
\end{equation}
where the matrices $W_i$ contain weights of the neural network (and the bias terms implicitly) and $\sigma(\cdot)$ is an activation. $\theta$ contains the elements of the matrices $W_i$. 

Let $W_i \in \mathbb{R}^{d^{\mathrm{out}}_i \times d^{\mathrm{in}}_i}$ be one of the weight matrices in $\{W_0,\ldots,W_{h}\}$ of a pre-trained network, which performs a linear transformation from $\mathbb{R}^{d^{\mathrm{in}}_i}$ to $\mathbb{R}^{d^{\mathrm{out}}_i}$. Rather than fine-tuning $W_i$ directly, the network is fine-tuned by updating $W_i\mapsto W_i + \Delta W_i$, where $W_i$ remains fixed and its adapter $\Delta W_i$ satisfies
$\text{rank}(\Delta W_i) < \min \{d^{\mathrm{in}}_i, d^{\mathrm{out}}_i\}.$

In particular, following \citet{Hu:2022:LLR}, we factorize%
\begin{equation}
\Delta W_i\coloneqq B_iA_i,
\end{equation}
where $A_i \in \mathbb{R}^{r \times d^{\mathrm{in}}_i}$, $B_i \in \mathbb{R}^{d^{\mathrm{out}}_i \times r}$, and $r$ is an adjustable parameter. By construction, the rank of the weight update $\Delta W_i$ is at most $r$. %

While prior methods predominantly apply LoRA to large foundation models trained on extensive datasets, \emph{we apply LoRA directly to the weights of a neural field to encode edits to the field compactly}. That is, given a neural field $f_\theta$ representing a graphics instance $\mathcal{D}$ as well as an edited version $\mathcal{D}'$, we encode the change from $\mathcal{D}$ to $\mathcal{D}'$ as a LoRA applied to $f_\theta$. %

By encoding edits to a neural field as LoRAs, we inherit many of the advantages LoRAs offer in their original context of LLMs\@. Since LoRAs generalize full fine-tuning (by setting $r = \min \{d^{\mathrm{in}}_i, d^{\mathrm{out}}_i\}$), they enable finer control over the tradeoff between memory footprint and the expressiveness of the network update. For a pre-trained weight matrix $W_0 \in \mathbb{R}^{d^{\mathrm{out}}_i \times d^{\mathrm{in}}_i}$, a rank-$r$ LoRA requires only $r (d^{\mathrm{in}}_i + d^{\mathrm{out}}_i)$ parameters, which is significantly fewer than the $d^{\mathrm{out}}_i \times d^{\mathrm{in}}_i$ parameters needed for full fine-tuning, especially when $r \ll \nicefrac{d^{\mathrm{in}}_i d^{\mathrm{out}}_i}{d^{\mathrm{in}}_i + d^{\mathrm{out}}_i}$.

In this paper, we focus on minor edits, where the information contained in the pre-trained neural field remains relevant. As pointed out in the motivating experiment (\S\ref{sec:intro}), LoRAs are a natural fit under this assumption, as they are designed to capture incremental changes to the pre-trained network. Furthermore, as edits are translated into weight updates for a neural field, downstream tasks can be performed by directly querying the LoRA-updated neural field without reference to the original data $\mathcal{D}, \mathcal{D}'$. We also empirically explore the relationship between the magnitude of the edit between $\mathcal{D}, \mathcal{D}'$ and the performance of LoRAs (\S\ref{sec:variation magnitude results}).

\section{Encoding Visual Data Variations with LoRA}
\label{sec:lora_for_edits}
We explore four graphics applications where our LoRA-based representation can capture changes. \change{For each one, using LoRA to adapt a neural field involves defining a minimization problem with respect to LoRA parameters whose general form is
\begin{equation}
    \min_{\{A_i\}_{i=0}^h, \{B_i\}_{i=0}^h} \,\mathcal{J}\left(f_{\theta+\mathrm{LoRA}}, \mathcal{D}, \mathcal{D}' \right), \label{eq:1}
\end{equation}
where $\mathcal{J}$ is an application-specific objective functional and $f_{\theta+\mathrm{LoRA}}$ denotes $f_\theta$ with the pre-trained parameters $\theta=(W_0,\ldots,W_h)$ frozen and a trainable LoRA update $((A_0,B_0),\ldots, (A_h,B_h))$ applied to every weight matrix. In the following subsections, we describe a few possible realizations of $\mathcal{J}$, including loss functions for regression and energy functionals. 
}

\subsection{Image Variations}\label{sec:imagevariations}

We represent changes to an input image $\mathcal{D}\in\mathbb{R}^{H\times W\times3}$ as updates to a base MLP $f_\theta$ that maps (sub-)pixel locations to RGB values. \change{To encode the change, we optimize for LoRA parameters following problem~\eqref{eq:1} by defining
\begin{equation} \label{eq:image objective fn}
    \mathcal{J}\left(f_{\theta+\mathrm{LoRA}}, \mathcal{D}' \right) \coloneqq \mathop{\mathbb{E}}_{x \sim \text{unif}(\mathcal{X})} \left[\mathcal{L}\left(f_{\theta+\mathrm{LoRA}}(x), \mathcal{D}'_x\right) \right]
\end{equation}
}
where $\mathcal{L}$ is the loss function used to obtain the base model $f_\theta$, $\mathcal{X}$ is a dense set of 2D point samples in the domain of the image, and $\mathcal{D}'_x$ denotes the target field value sampled at $x \in \mathcal{X}$ evaluated using $\mathcal{D}'$. We use the relative $L2$ loss for $\mathcal{L}$.

\subsection{Geometric Deformations}\label{sec:geometricdeformations}

Many geometry processing tasks involve small changes that must be stored or transmitted efficiently, e.g., consecutive frames of animated surface or local edits from smoothing~\cite{Desbrun:1999:IFO} and stylization~\cite{Liu:2019:CS}. 
Here, we represent an input surface $\mathcal{D}$ implicitly as a level set of a neural SDF ~\cite{park2019deepsdf} $f_\theta$. Surface edits are generated using a modeling tool via, e.g., cage-based or ARAP deformations~\cite{sorkine2007rigid} applied to a meshed version of $\mathcal{D}$.

\change{Here we define the objective $\mathcal{J}$ similarly to equation~\eqref{eq:image objective fn}, where $\mathcal{D}_x'$ now represents SDF values at 3D point samples $x$ instead of RGB colors.} Following~\citet{Muller:2022:ING}, our reconstruction loss $\mathcal L$ for both base model training and LoRA fine-tuning is the mean average percentage error (MAPE), defined as 
$\frac{\mid \text{prediction} - \text{target} \mid}{\mid \text{target} \mid + 0.01}$.

\subsection{Energy Minimization } \label{sec: energy minimization}
Many visual computing tasks (e.g., smoothing, deformation, or denoising) are framed as energy minimization problems~\cite{Taubin:2012:ITG,sorkine2007rigid,Crane:2013:RFC,chambolle:2010:TV_intro}. These tasks modify data by seeking critical points of energy functionals via numerical optimization. Recent work extends this paradigm to neural fields~\cite{xu:2022:signalprocINR, Marschner:2023:CSG}, and we show that LoRA remains compatible.

In this setting, $\mathcal{D'}$ is unknown but characterized as a solution to an energy minimization problem. \change{A general formulation of the objective functional in problem~\eqref{eq:1} is 
\begin{equation} \label{general energy min problem}
    \!\!\mathcal{J}\!\left(f_{\theta+\mathrm{LoRA}}, \mathcal{D} \right) \!\coloneqq\!\!\!\!\! \mathop{\mathbb{E}}_{x \sim \text{unif}(\mathcal{X})} \!\!\!\!\!\!\!\left[\mathcal{L}\!\left(f_{\theta+\mathrm{LoRA}}(x), \mathcal{D}_x\right) \! + \!\lambda E(f_{\theta+\mathrm{LoRA}}, x)\right]
\end{equation}
where $E$ is a user-defined energy functional}, $\mathcal{L}$ is a data fidelity loss encouraging similarity to the original data $\mathcal{D}$, and $\lambda \in \mathbb{R}_{>0}$ controls the relative importance of the energy and data fidelity terms.

\subsection{Encoding Sequential Changes} \label{sec:video}

Whereas the previous applications focus on small edits to a single instance, real-world data often exhibits large or sequential variations, such as temporal changes in videos or animated character motion. To explore this setting, we decompose large variations into a \emph{sequence} of smaller frames and encode each using LoRA\@. 

In doing so, two strategies naturally emerge: (1)~train an \emph{independent} LoRA for each frame, adapting a shared base model as in \S\ref{sec:imagevariations}; or (2)~apply LoRA updates \emph{sequentially}, where each update builds on its predecessor. The latter tests our method's ability to handle long sequences while controlling accumulated error. We evaluate both approaches and describe the sequential version below.

We treat videos and character animations as temporal sequences of static variations, extending the setups in \S\ref{sec:imagevariations} and \S\ref{sec:geometricdeformations}. To model these frame-to-frame changes, we optimize for LoRA parameters solving problem ~\eqref{eq:1}. 

In more detail, given a sequence of $n$ frames, we first overfit a base network $f_{\theta,1}$ to frame 1. Then, for each subsequent frame $i \in \{2, \ldots, n\}$, we train a LoRA to capture the update from the preceding frame. After applying the LoRA to obtain $f_{\theta,i}$, we freeze its parameters and repeat the process for the next frame. Each LoRA induces a rank-$r$ update, so frame $k$ corresponds to a rank-$r(k-1)$ update to the original base model.

\begin{remark}
Our proposed pipeline for processing frames sequentially simply optimizes LoRAs one-at-a-time.  
This algorithm has identical efficiency per frame as the static setting in \S\ref{sec:imagevariations} and \S\ref{sec:geometricdeformations}.  
We emphasize that this experiment mainly serves as a stress test of our approach for temporal data.
\end{remark}

\newcommand{\sdfimgwidth}{0.09\textwidth}

\begin{figure*} 
    
    \centering
     \begin{tabular}{c | c c  c  c  c  c  c c }
        Shape & Input/\textcolor{Maroon}{Target} & $r=1$ & $r=4$ & $r=8$ & $r=16$ & $r=32$ & $r=64$ & FT \\
        \# Params &  & 2.6 k & 9.6 k & 19.0 k & 37.7 k & 75.1 k & 141.8 k & 272.9 k \\
         \change {\% Params} & & \change{0.9} & \change{3.5} & \change{7.0}  & \change{13.8} & \change{27.6} & \change{51.9} & \change{100}  \\
         \hline

         & IoU & 0.990 & 0.994 & 0.996 & 0.997 & 0.998 & 0.998 & 0.997 \\
         \rotatebox{90}{Penguin} &
         \includegraphics[width=\sdfimgwidth]{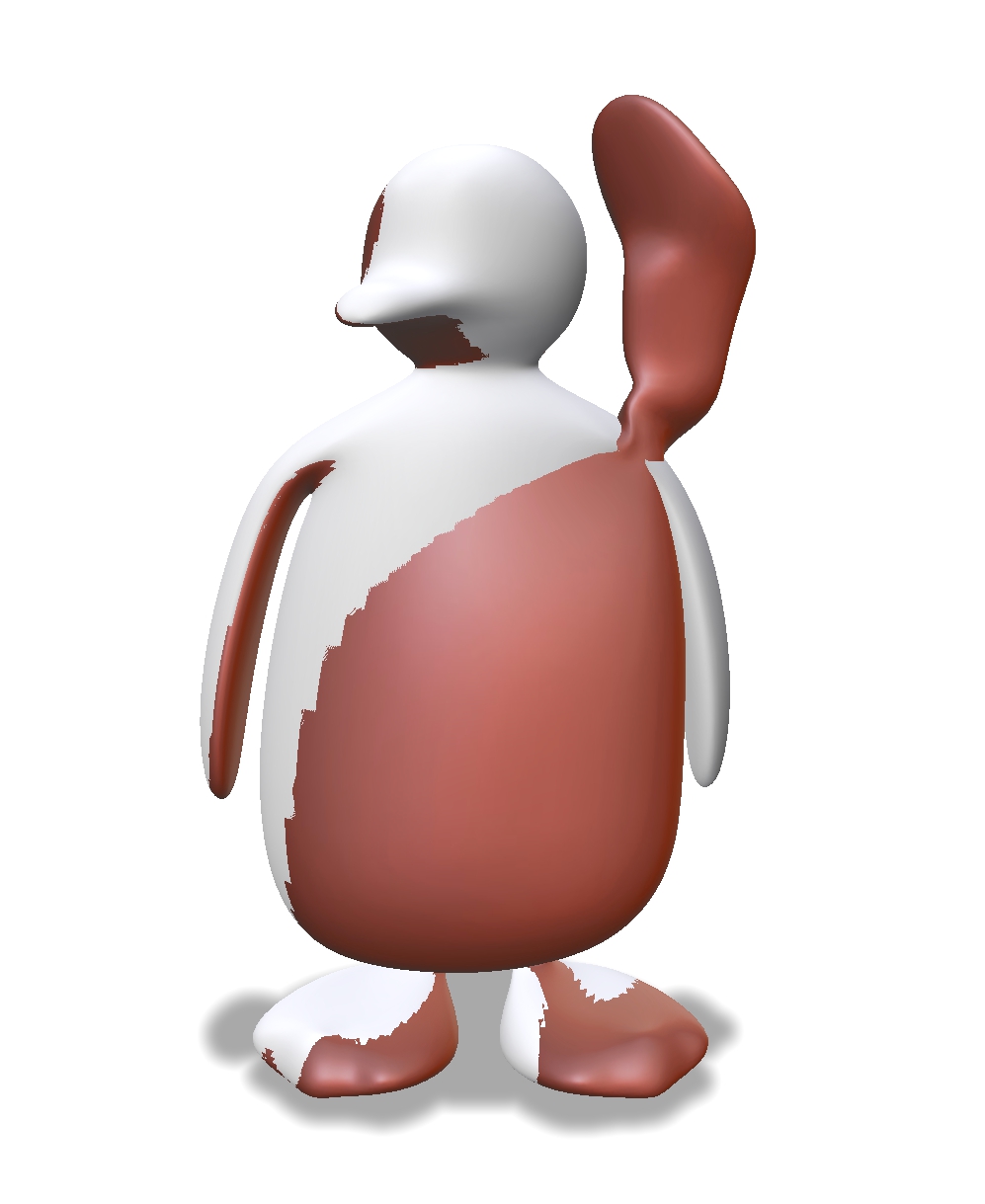} & %
         \includegraphics[width=\sdfimgwidth]{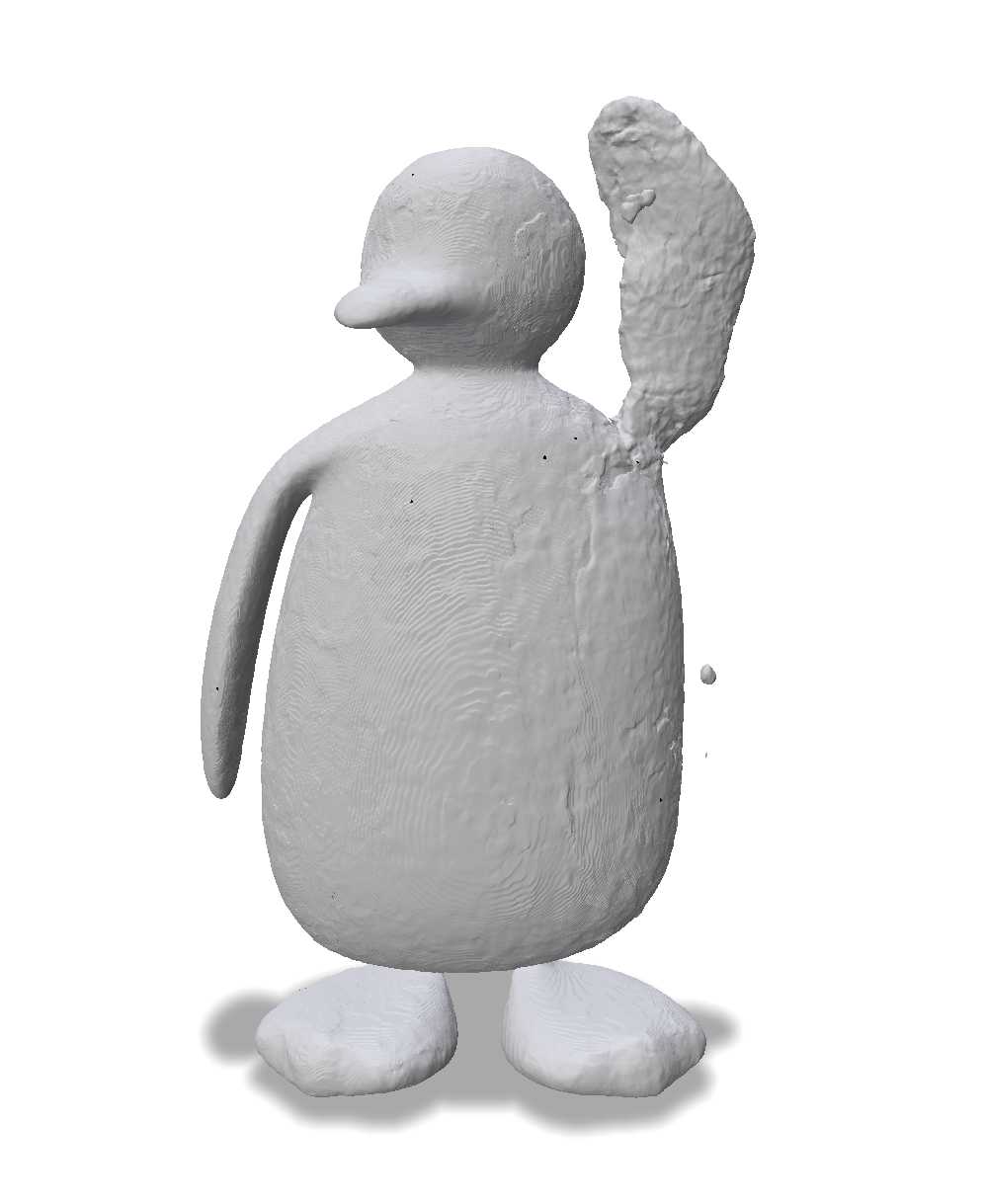} & %
         \includegraphics[width=\sdfimgwidth]{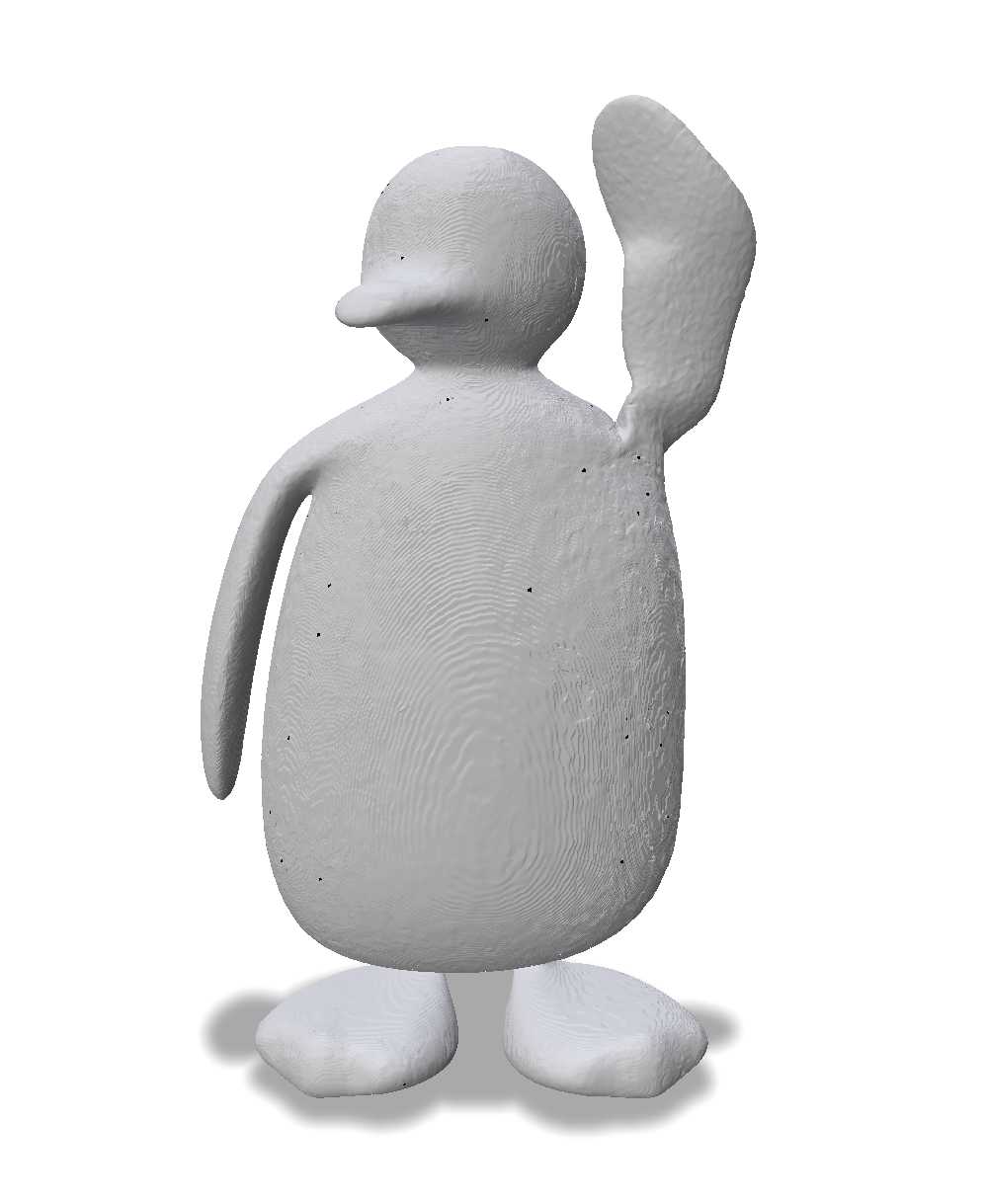} & %
         \includegraphics[width=\sdfimgwidth]{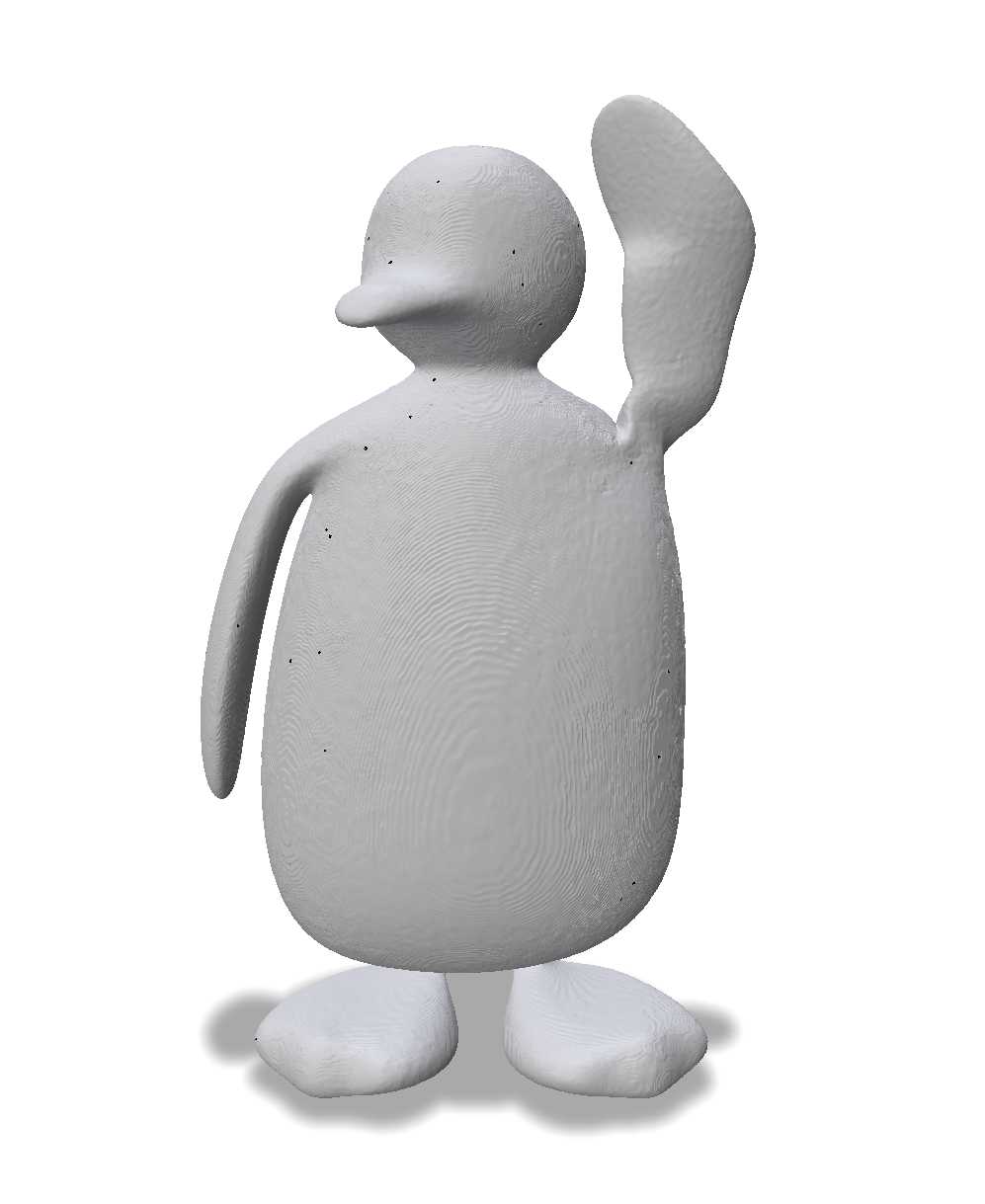} & %
         \includegraphics[width=\sdfimgwidth]{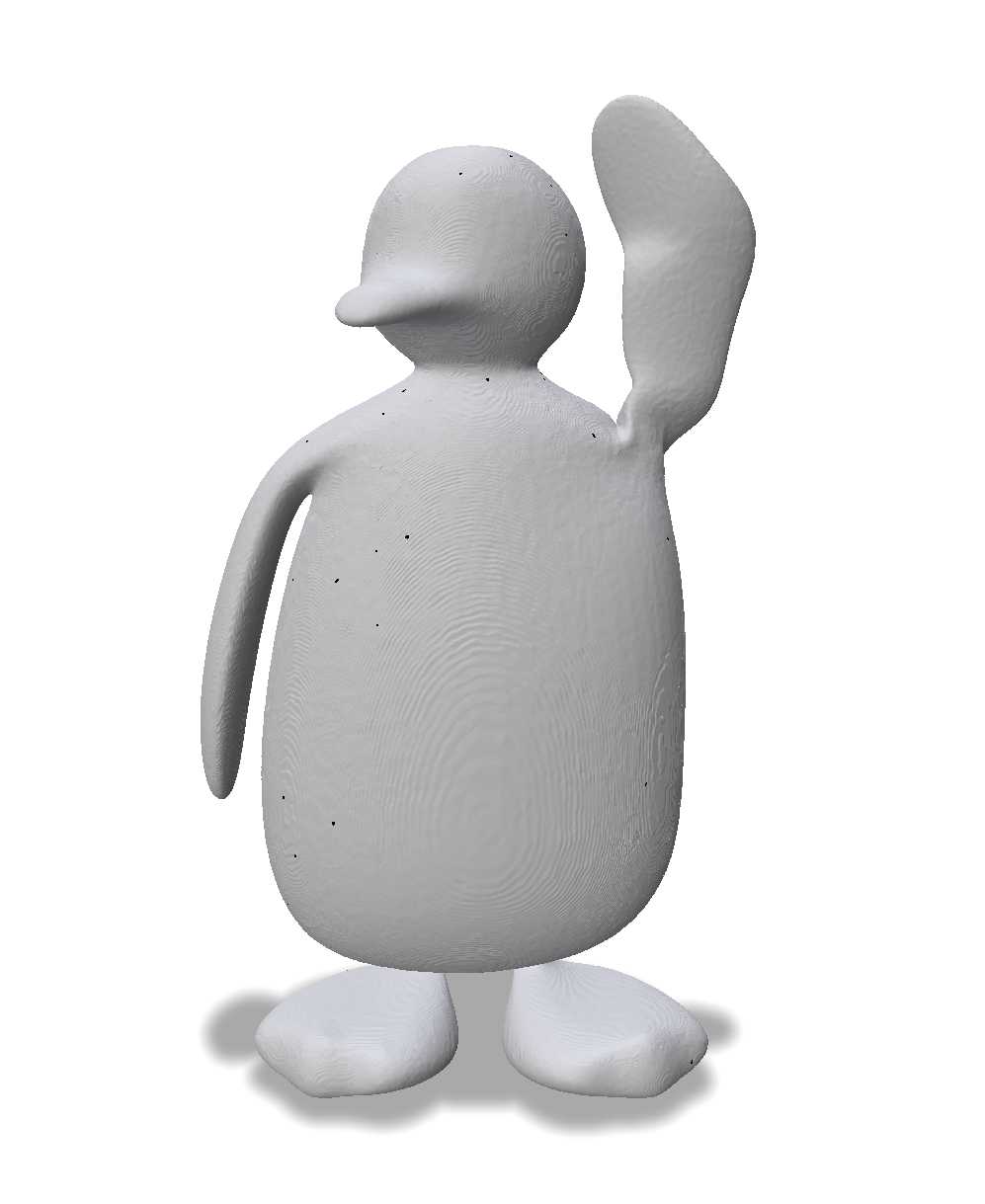} & %
         \includegraphics[width=\sdfimgwidth]{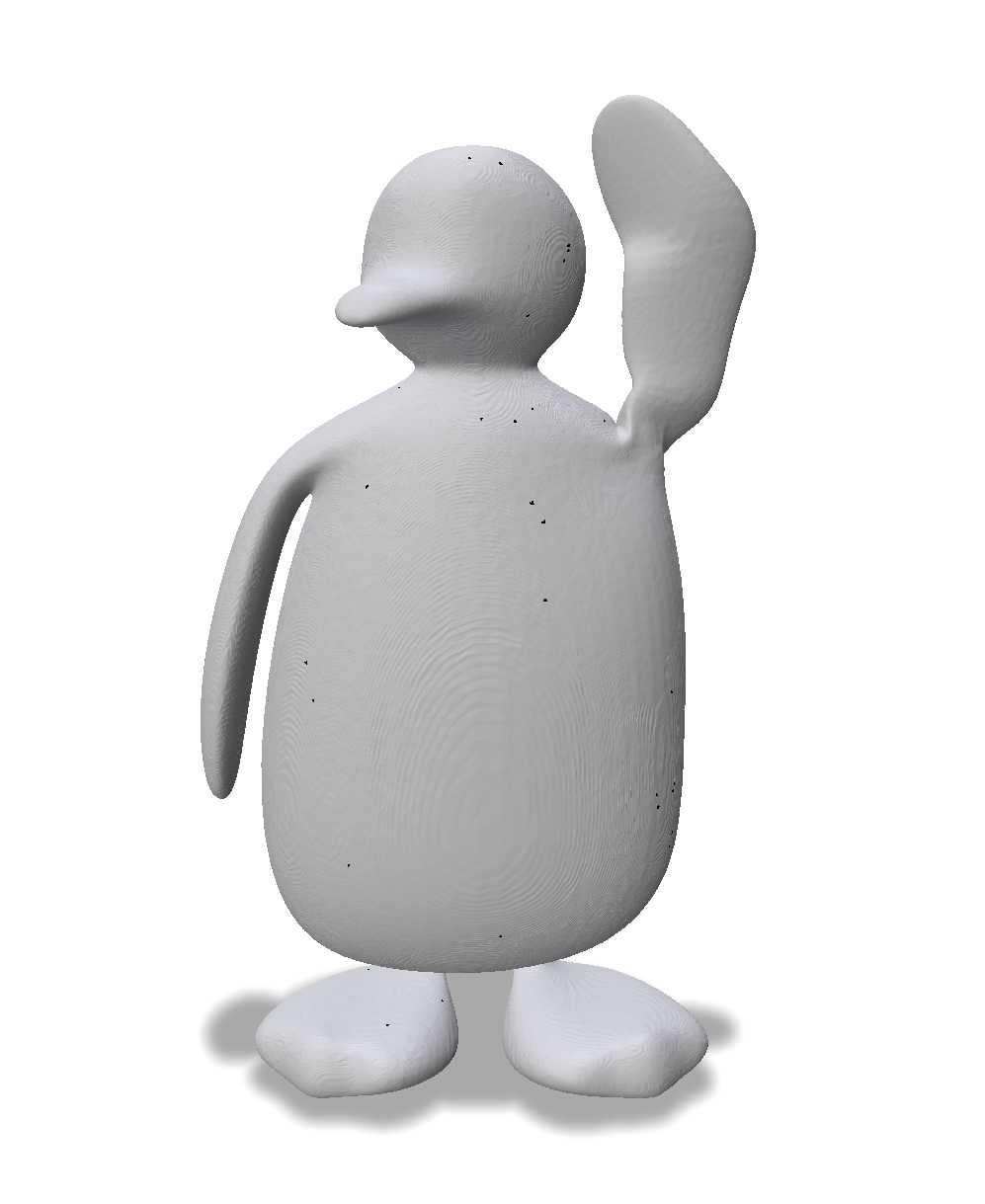} & %
         \includegraphics[width=\sdfimgwidth]{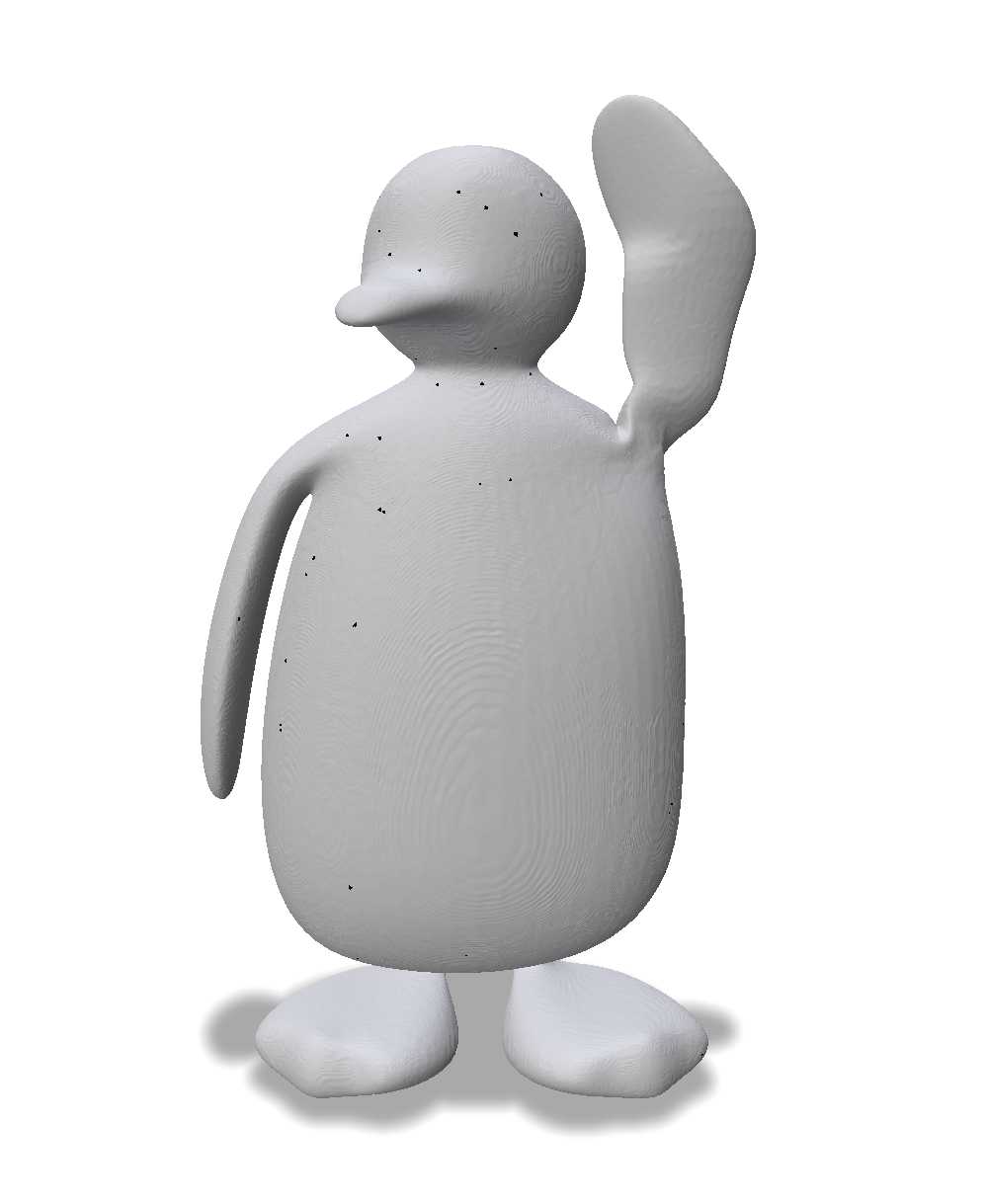} & %
         \includegraphics[width=\sdfimgwidth]{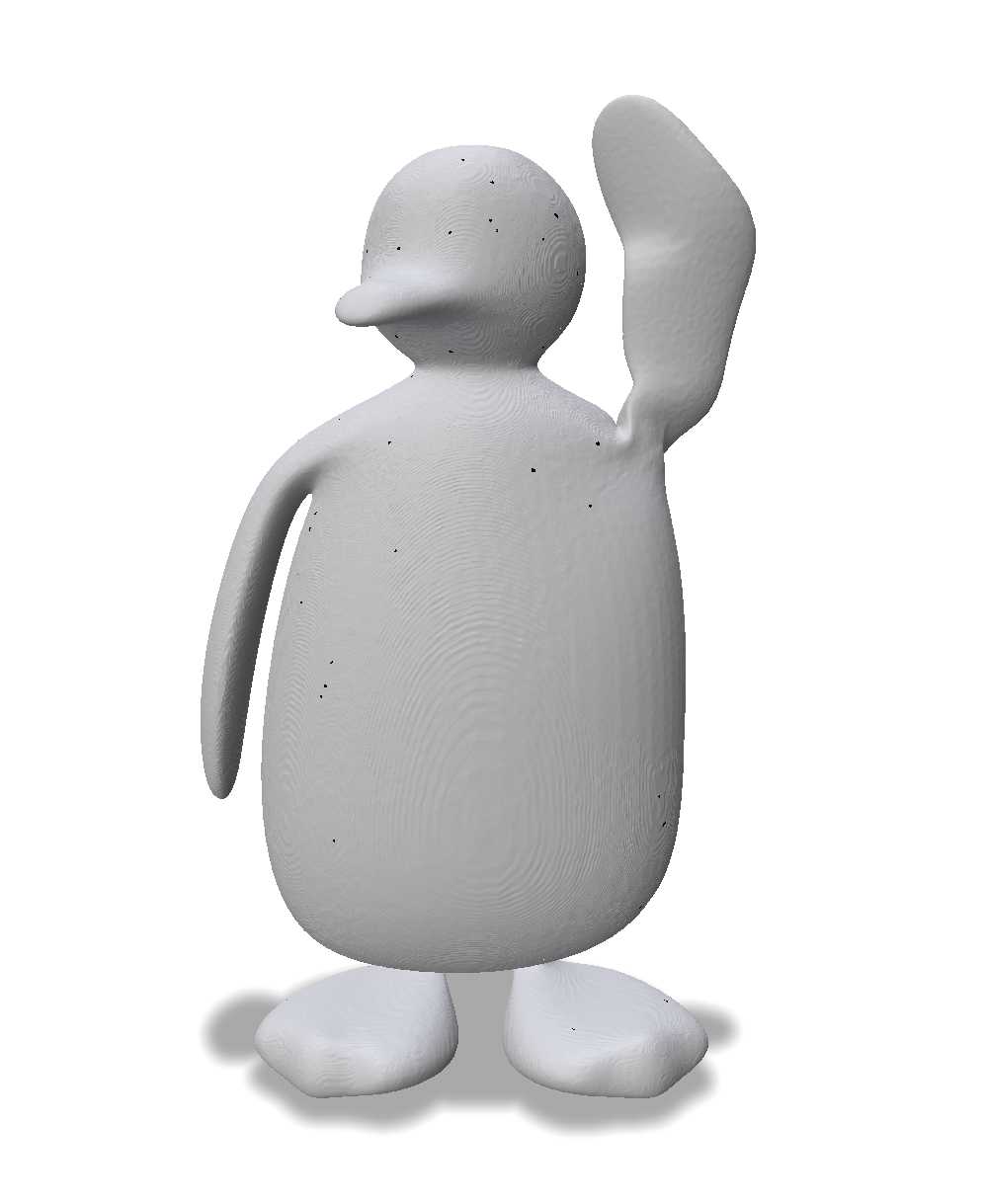} \\ %
        \hline  

         & IoU & 0.922 & 0.984 & 0.993 & 0.996 & 0.997 & 0.998 & 0.998 \\
         \rotatebox{90}{Horse} &
         \includegraphics[width=\sdfimgwidth]{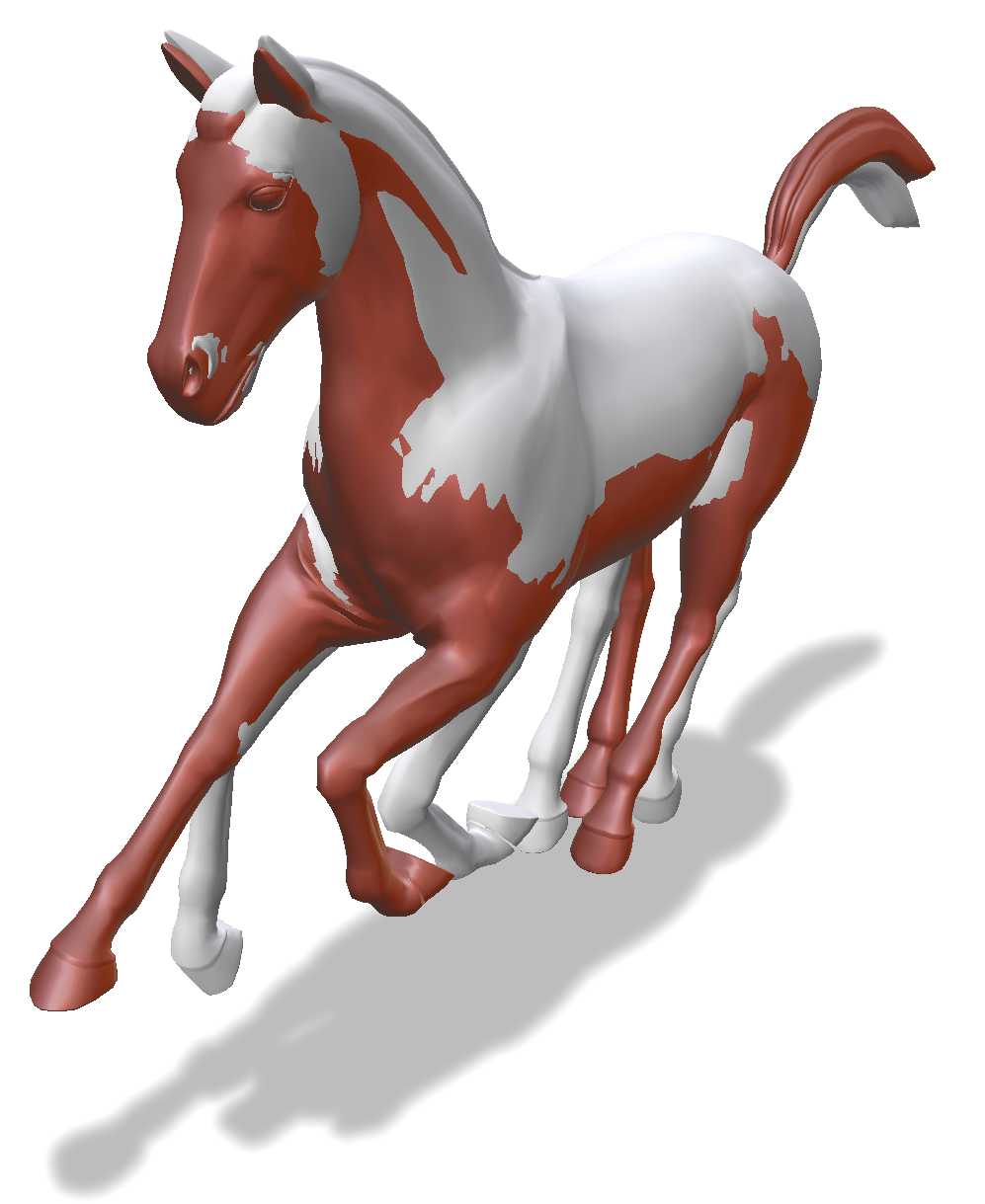} & %
         \includegraphics[width=\sdfimgwidth]{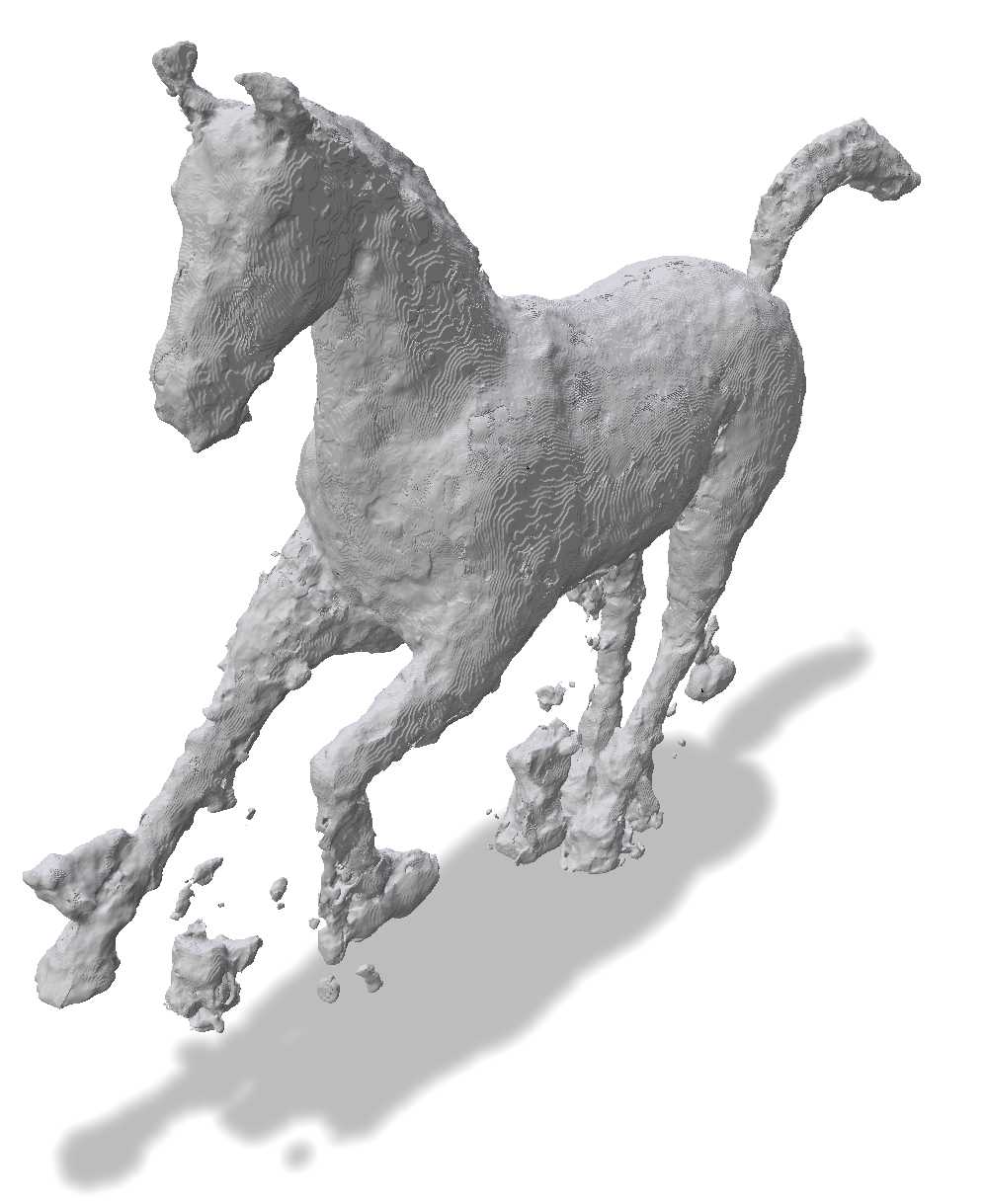} & %
         \includegraphics[width=\sdfimgwidth]{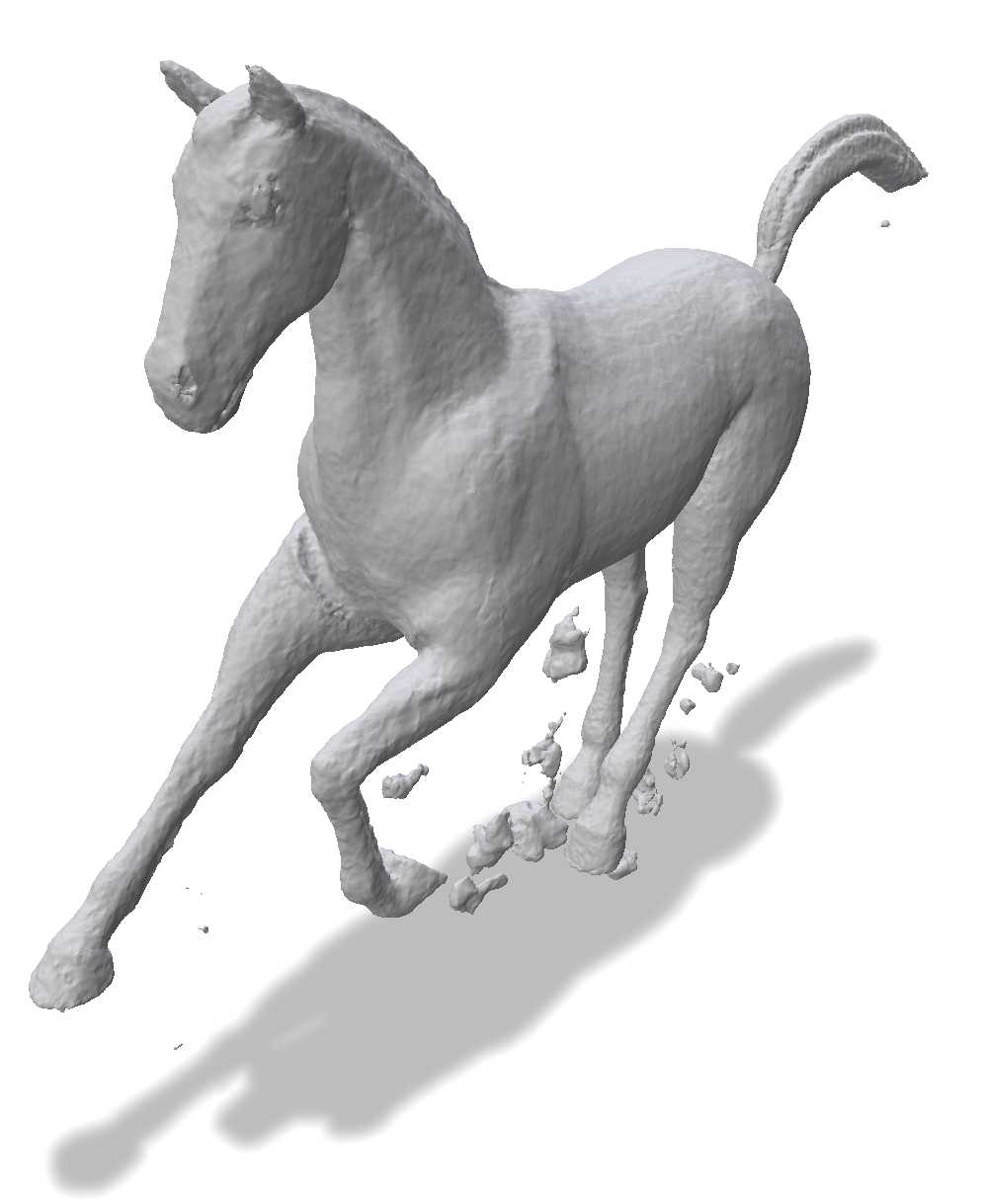} & %
         \includegraphics[width=\sdfimgwidth]{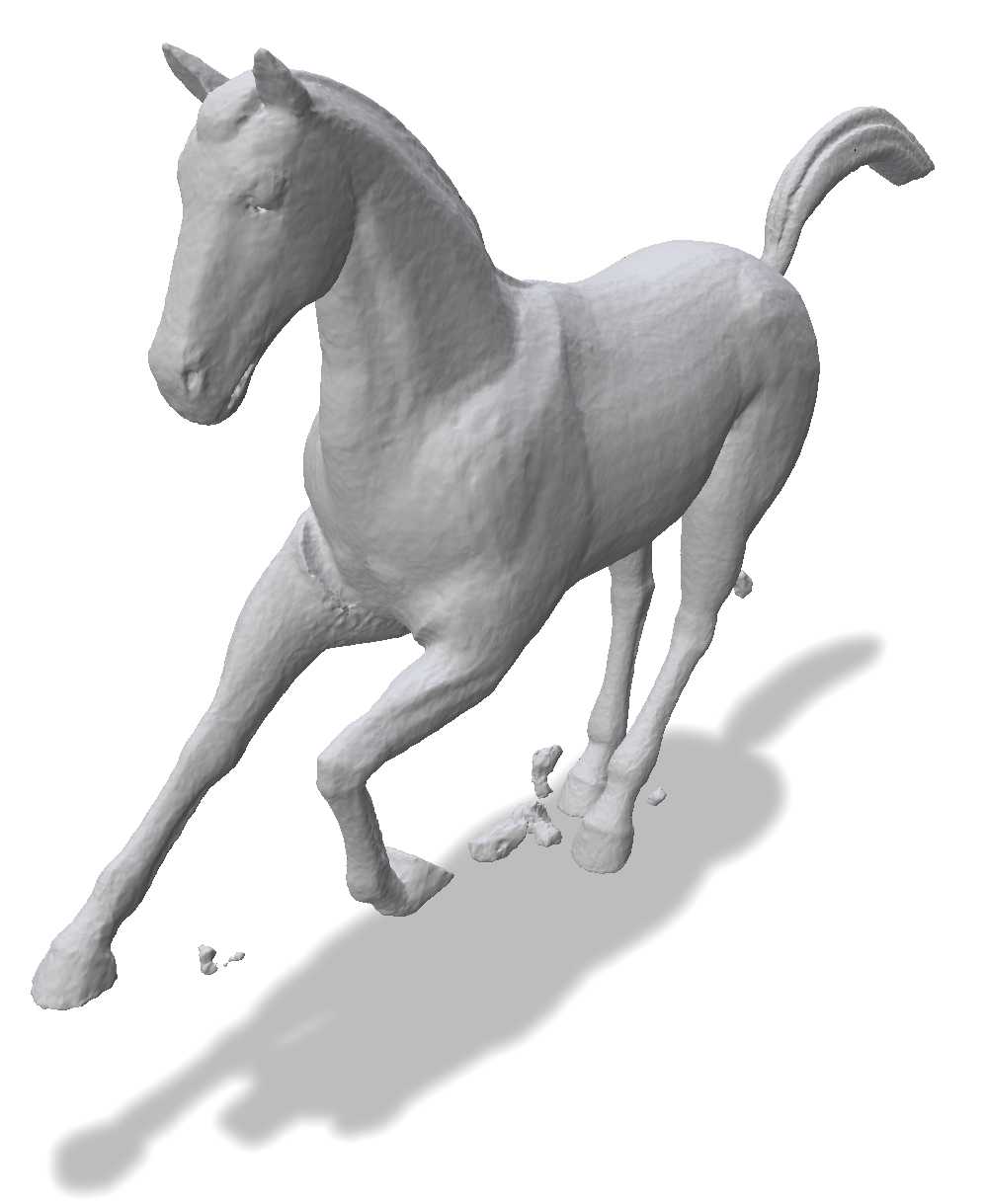} & %
         \includegraphics[width=\sdfimgwidth]{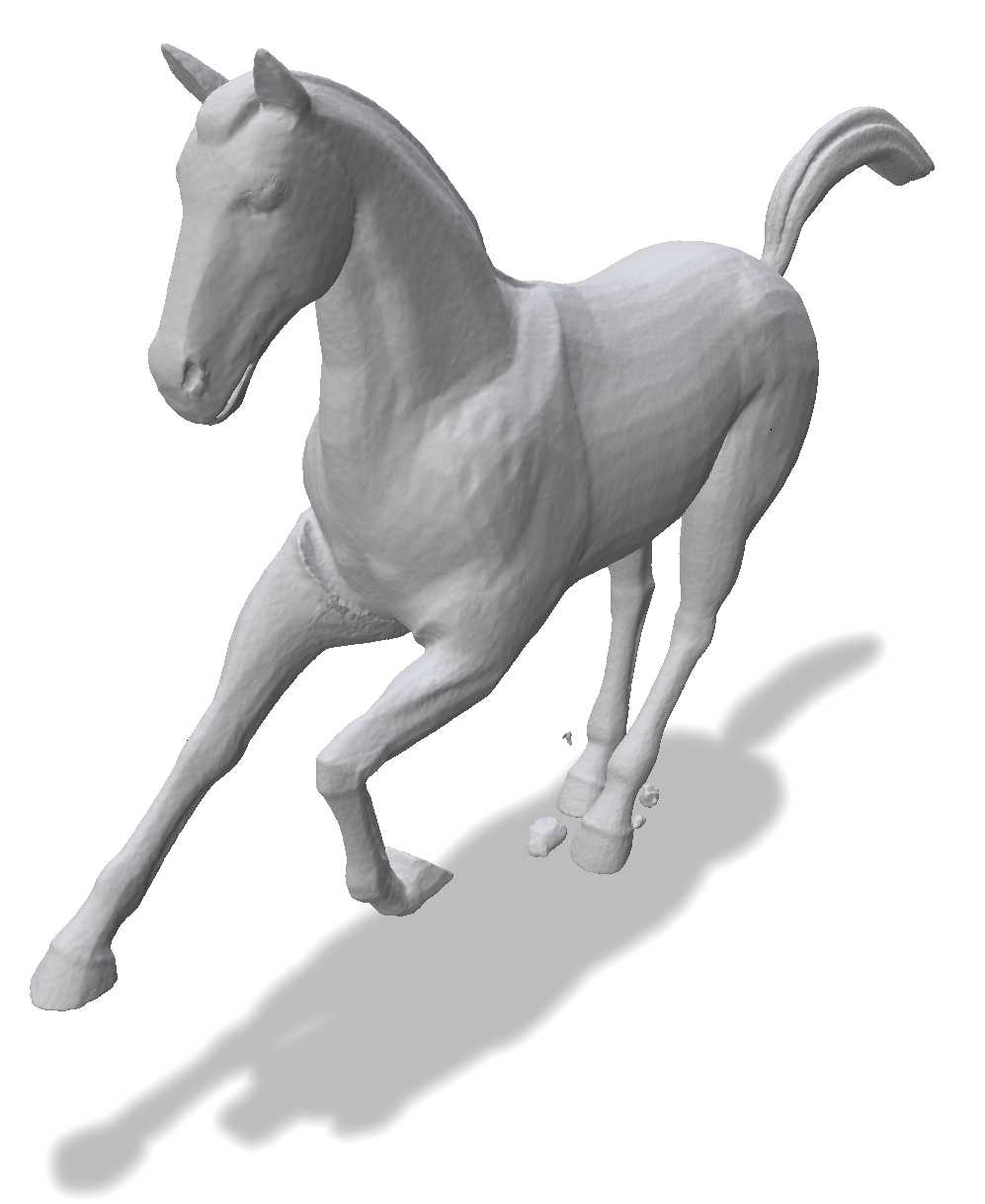} & %
         \includegraphics[width=\sdfimgwidth]{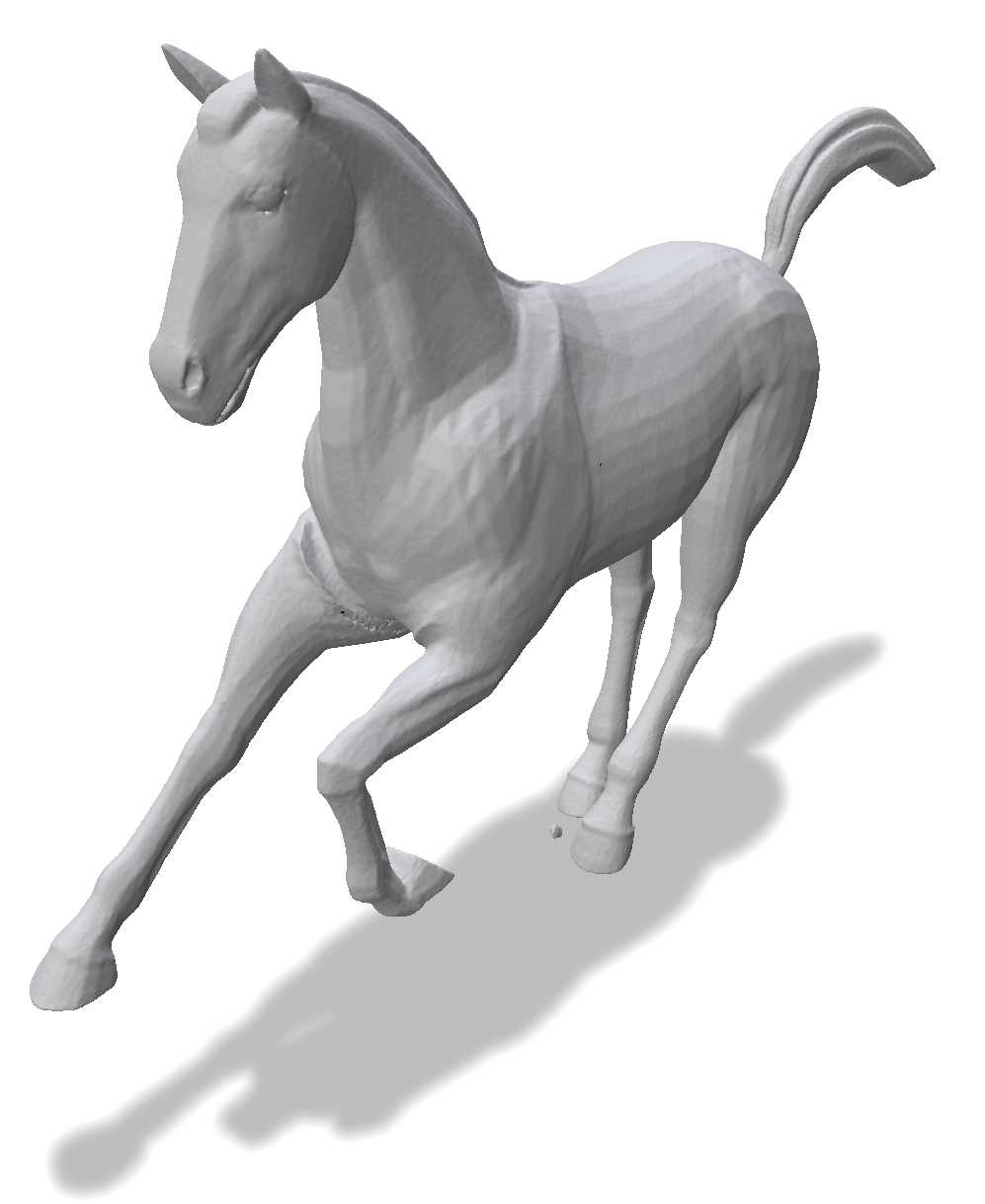} & %
         \includegraphics[width=\sdfimgwidth]{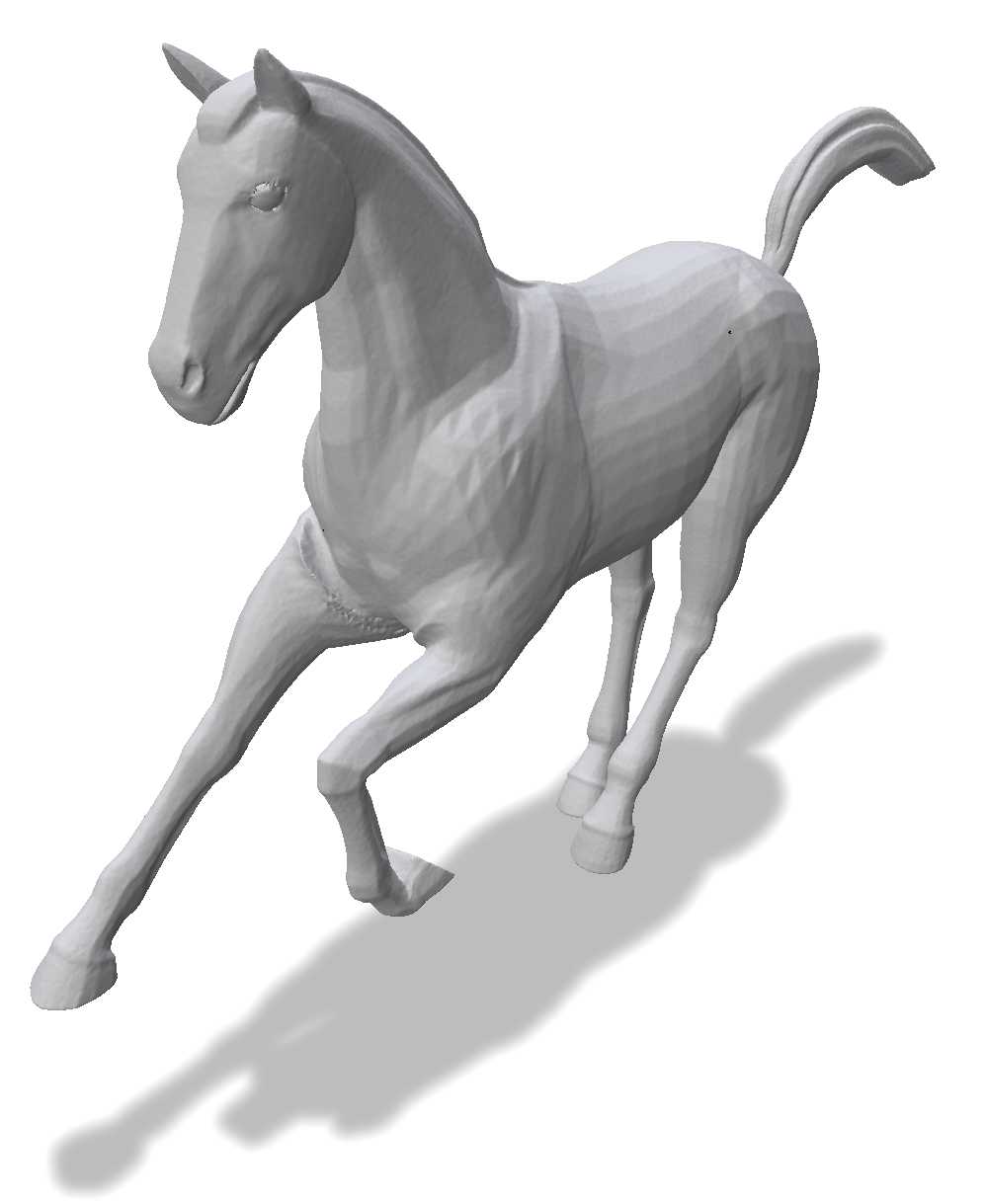} & %
         \includegraphics[width=\sdfimgwidth]{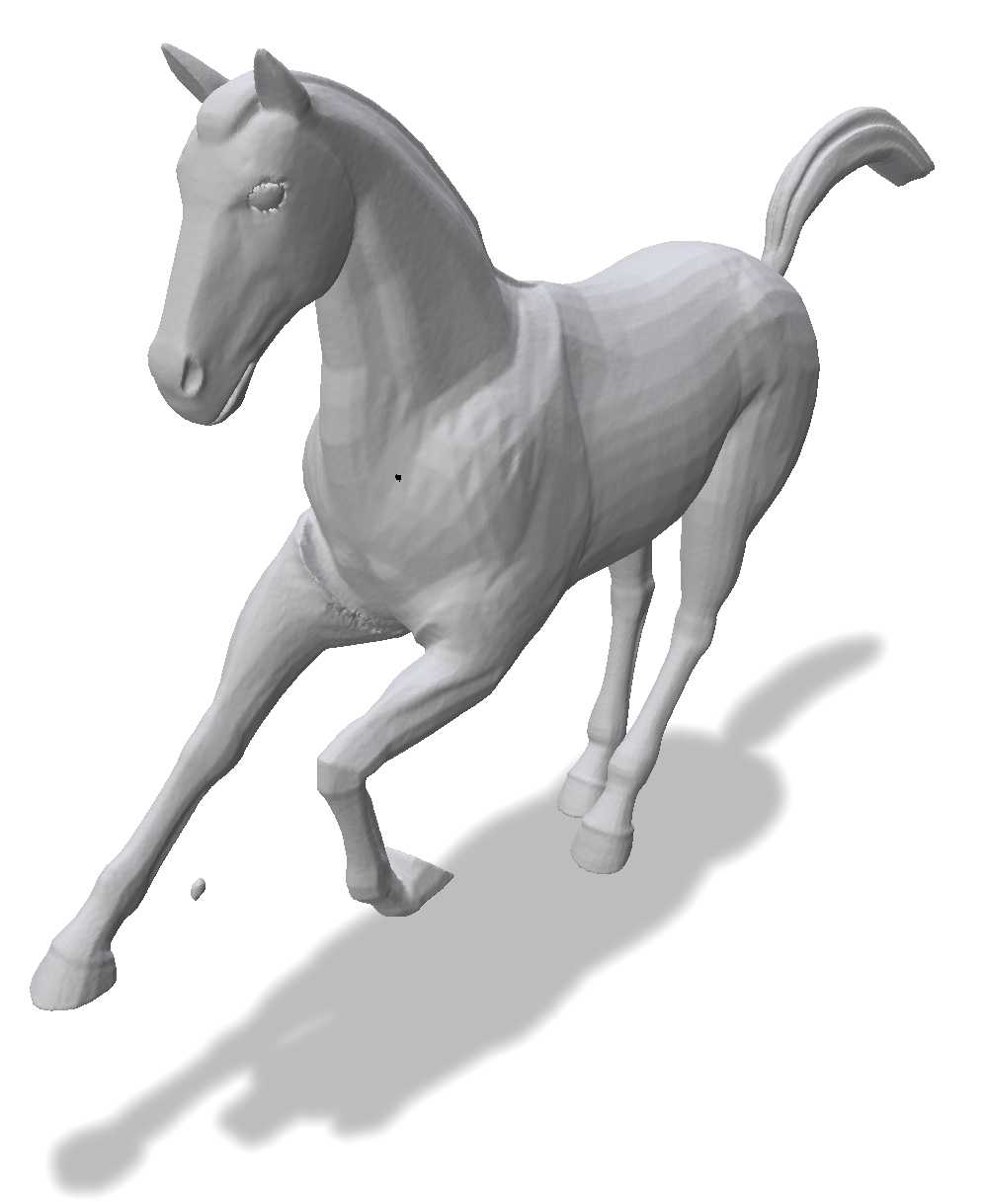} \\ %
        \hline    
        
          & IoU & 0.870 & 0.965 & 0.976 & 0.989 & 0.992 & 0.992 & 0.997 \\
         \rotatebox{90}{Armadillo} &
         \includegraphics[width=\sdfimgwidth]{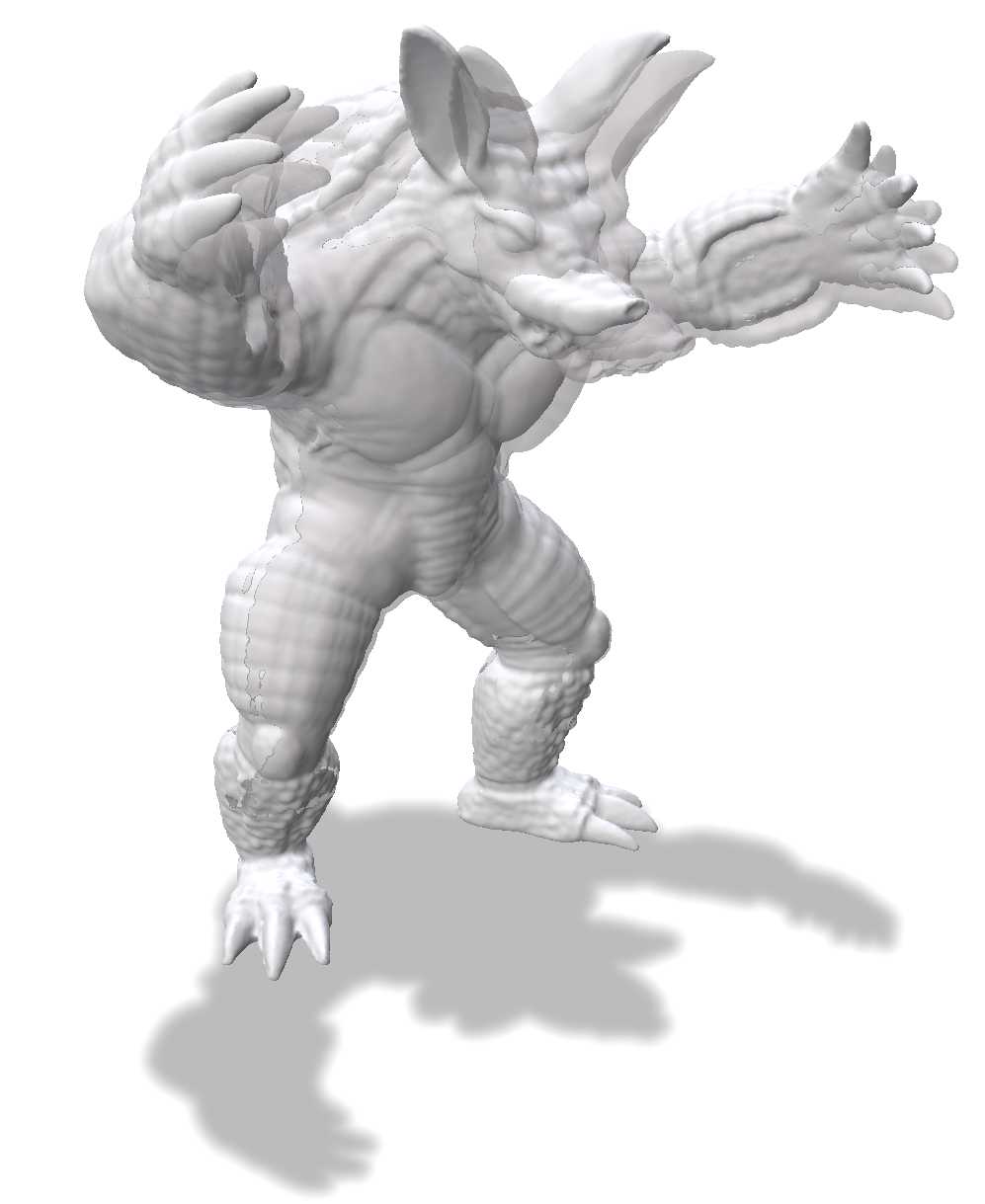} & %
         \includegraphics[width=\sdfimgwidth]{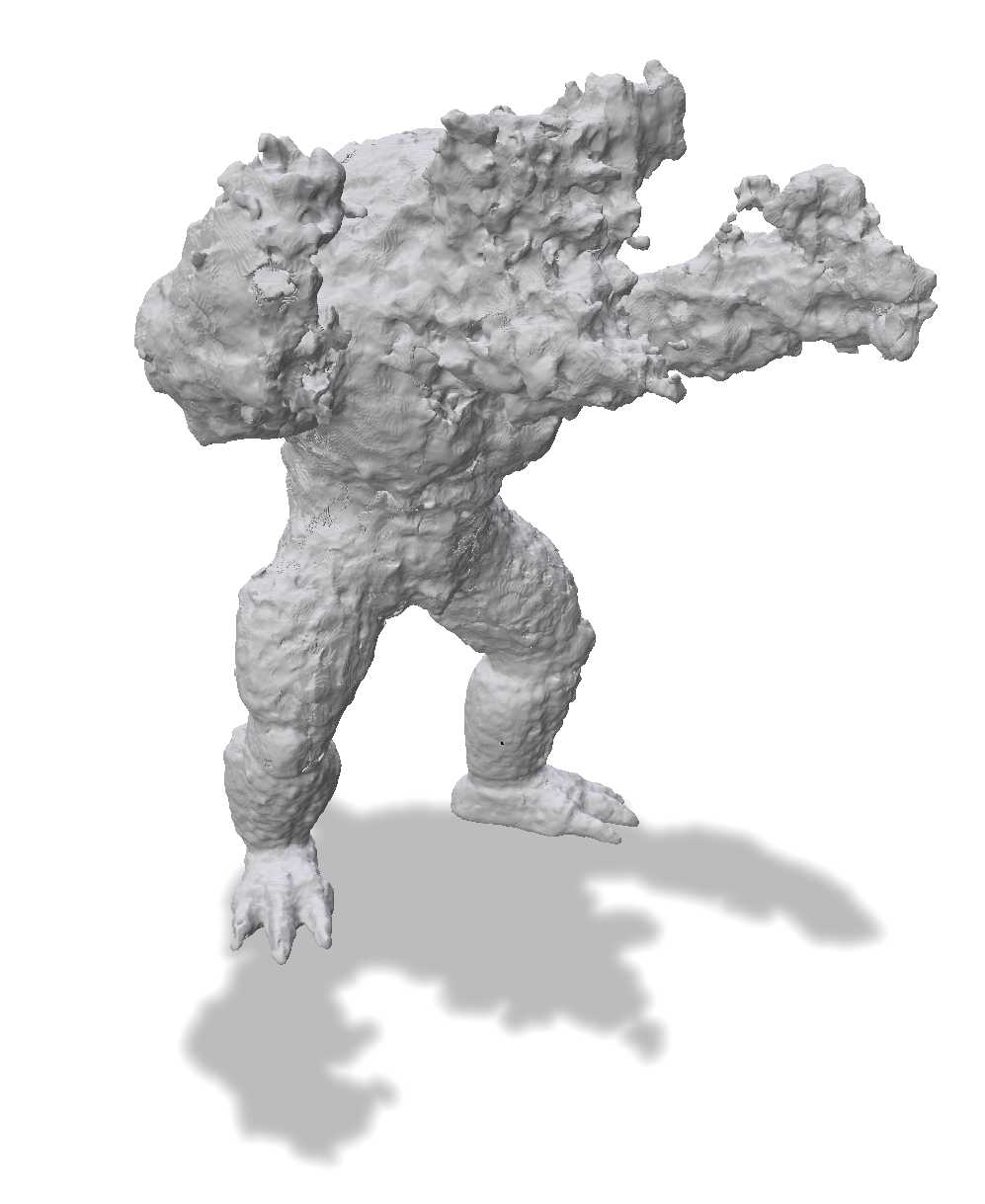} & %
         \includegraphics[width=\sdfimgwidth]{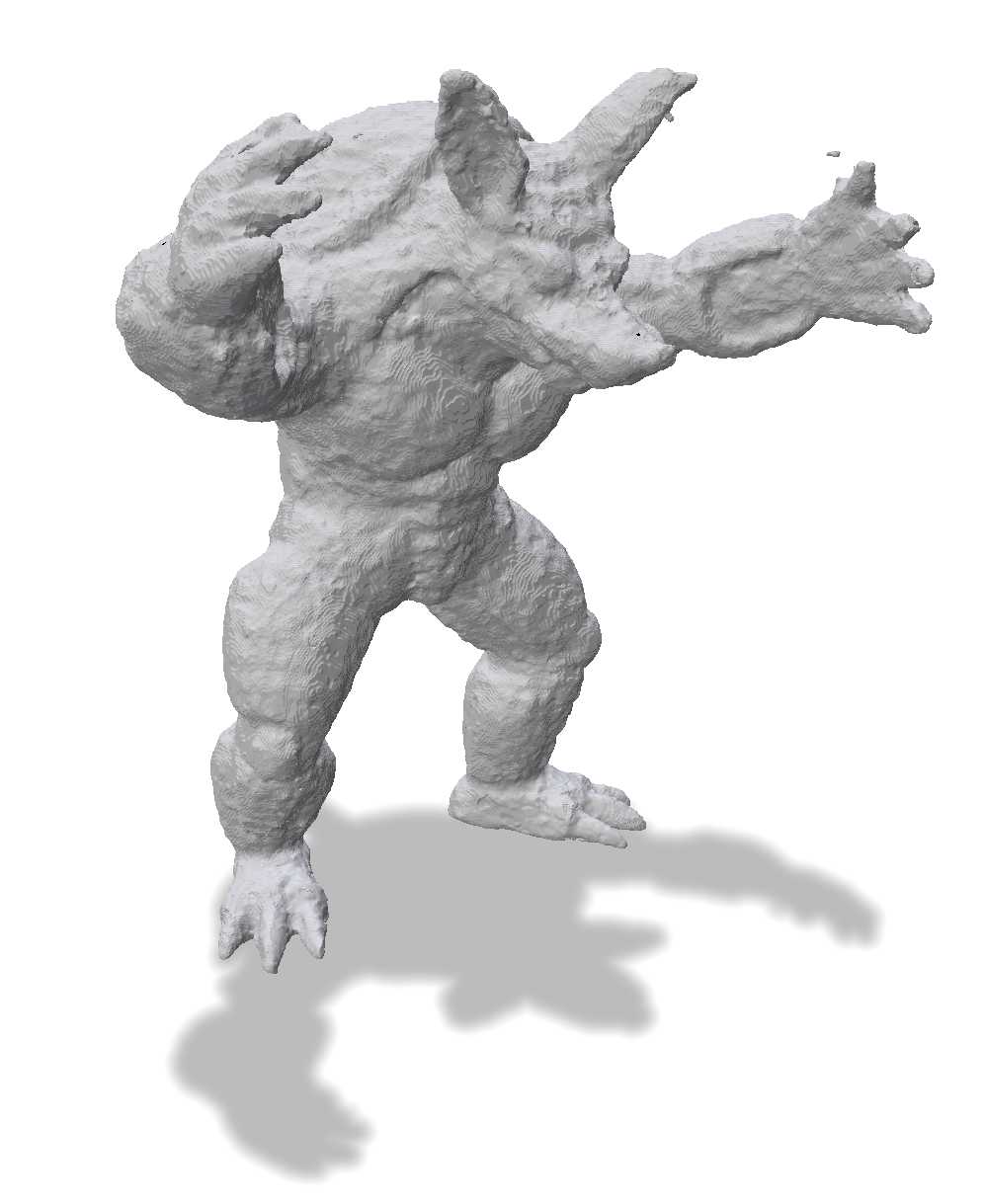} & %
         \includegraphics[width=\sdfimgwidth]{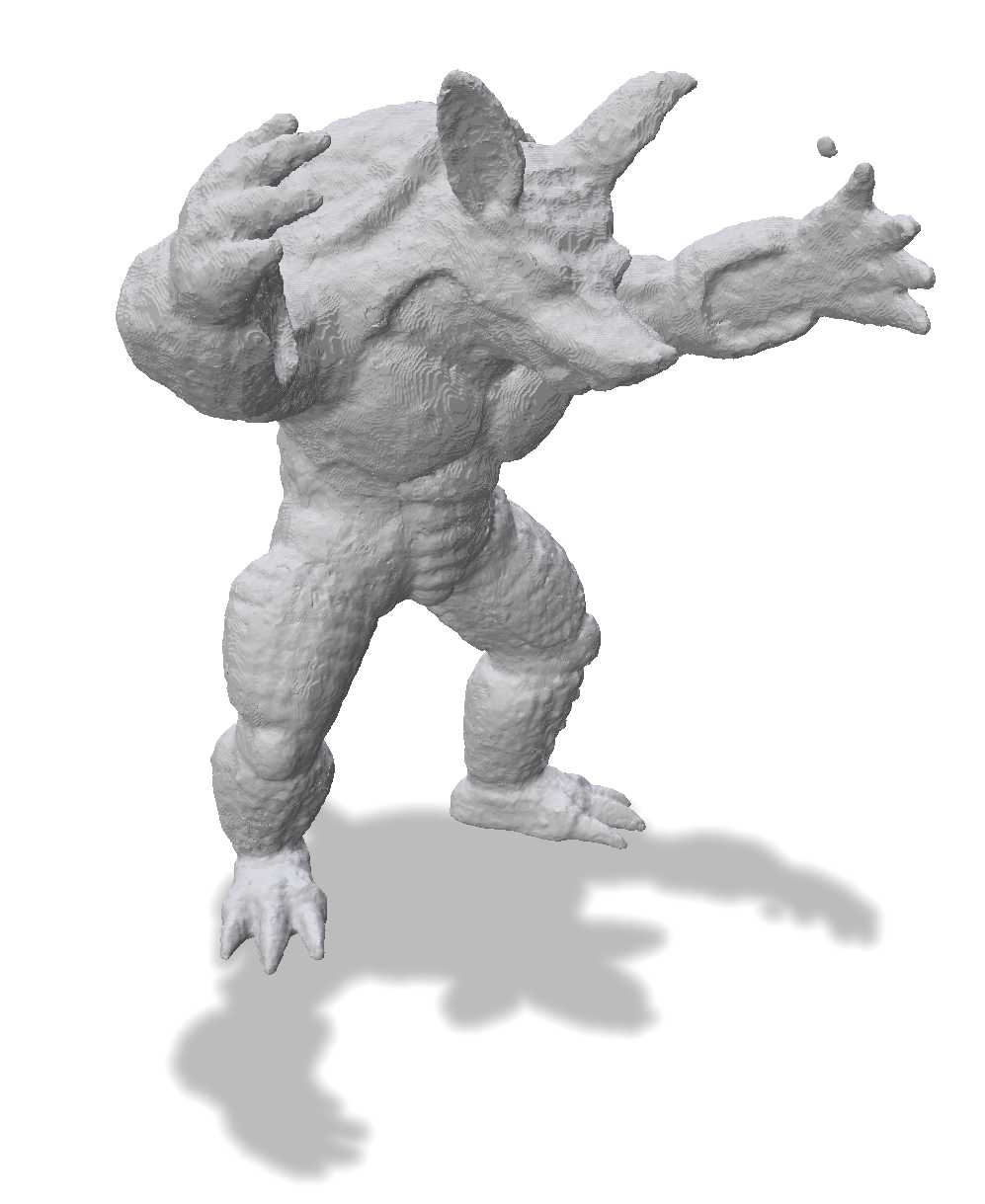} & %
         \includegraphics[width=\sdfimgwidth]{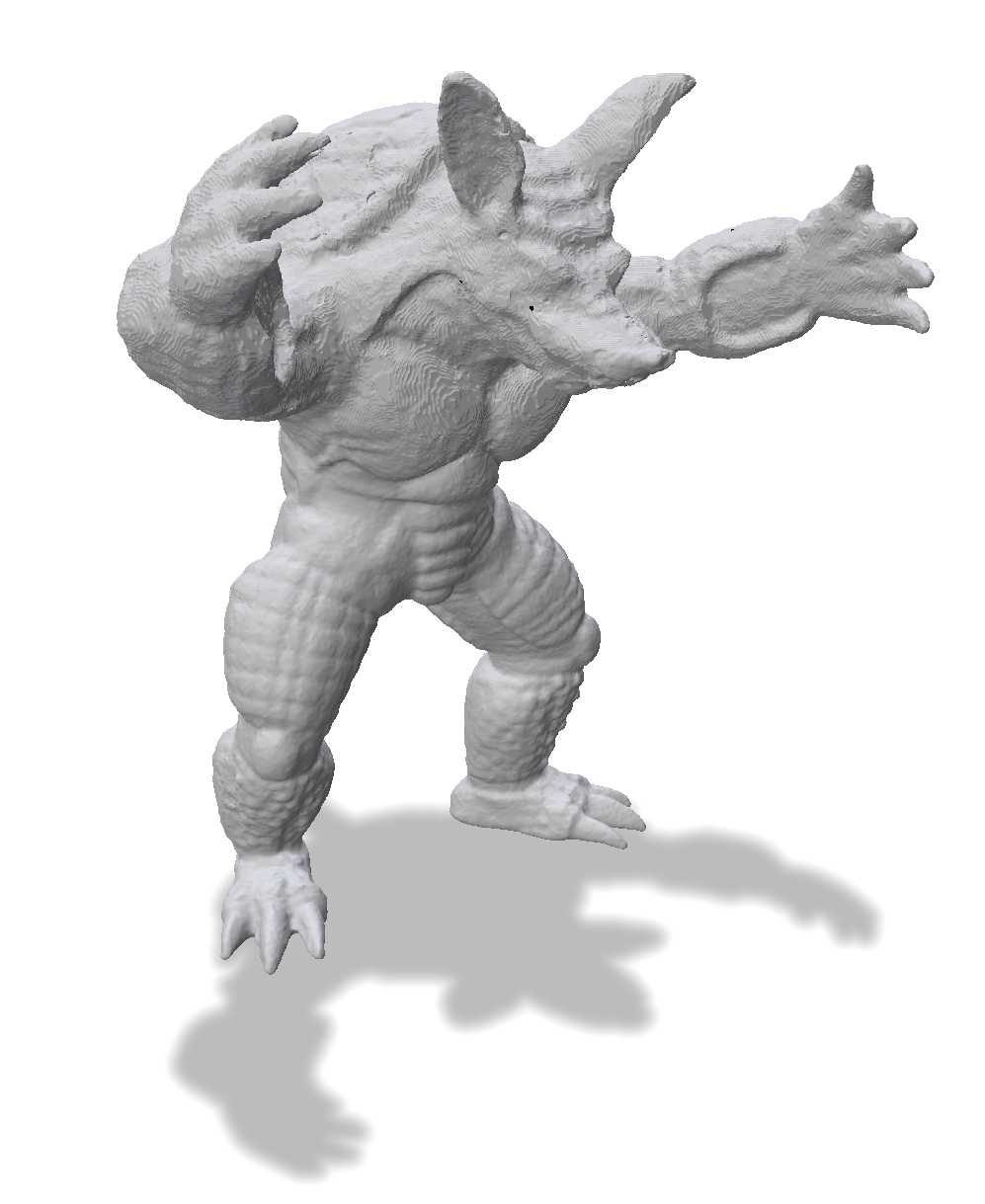} & %
         \includegraphics[width=\sdfimgwidth]{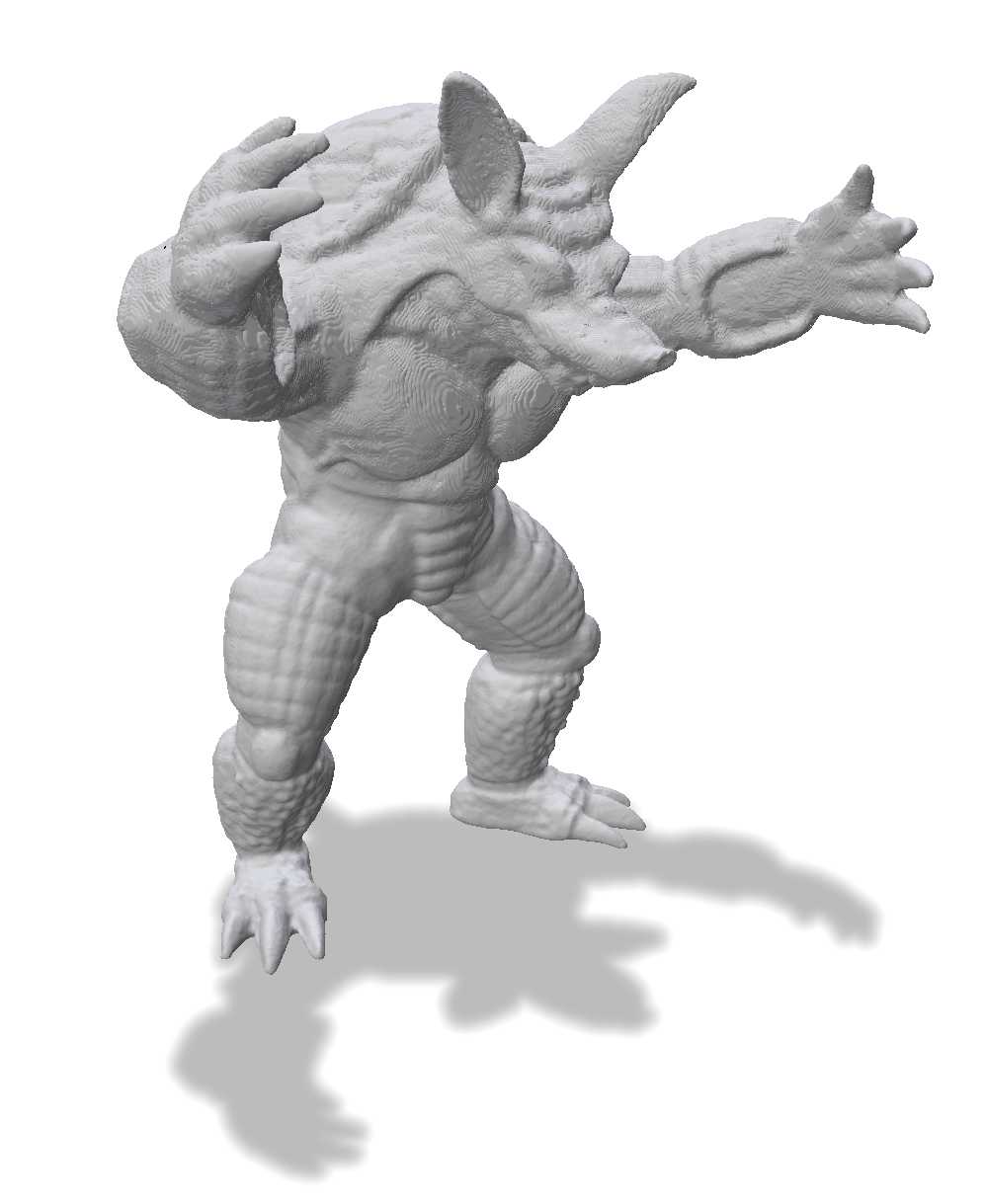} & %
         \includegraphics[width=\sdfimgwidth]{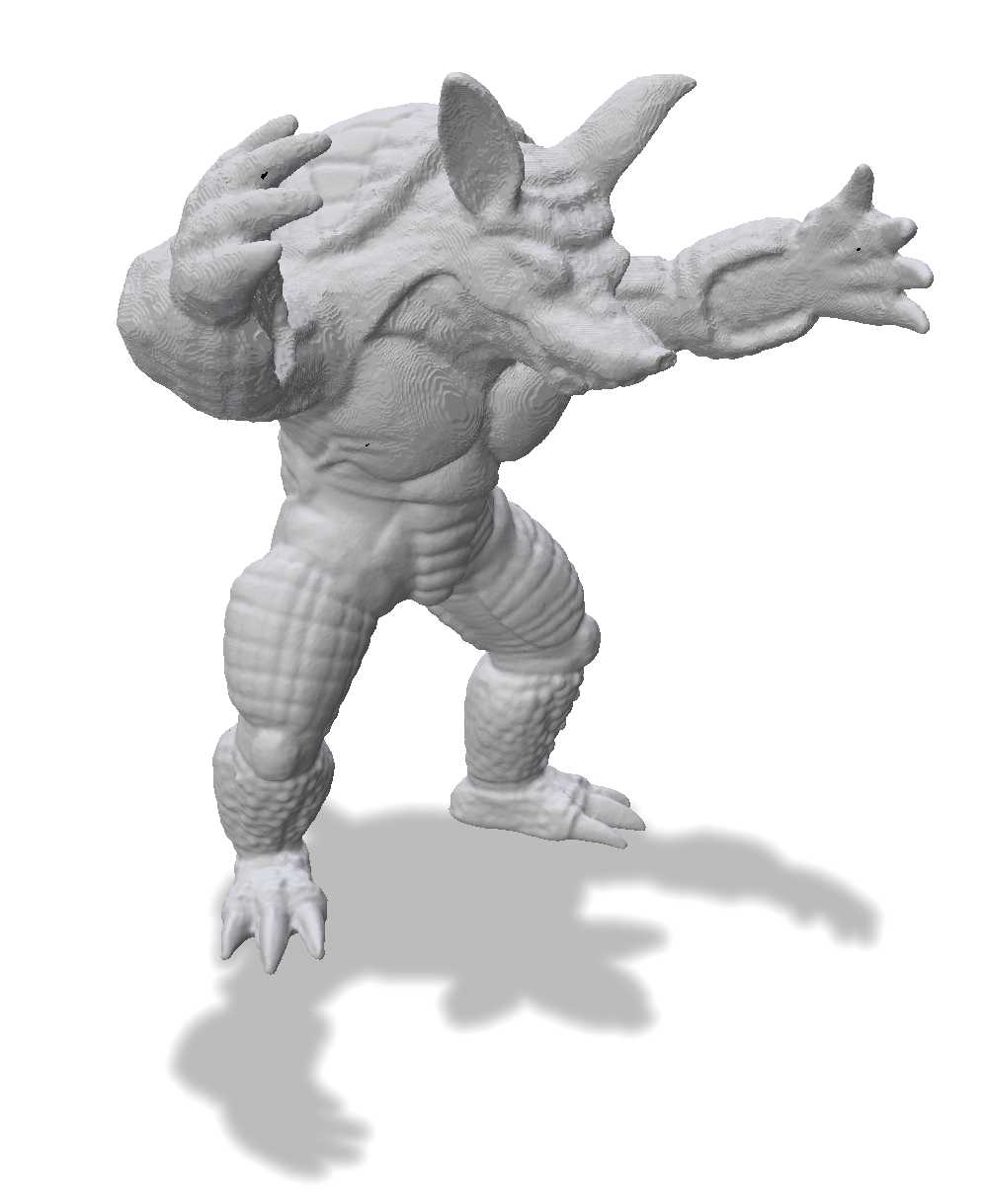} & %
         \includegraphics[width=\sdfimgwidth]{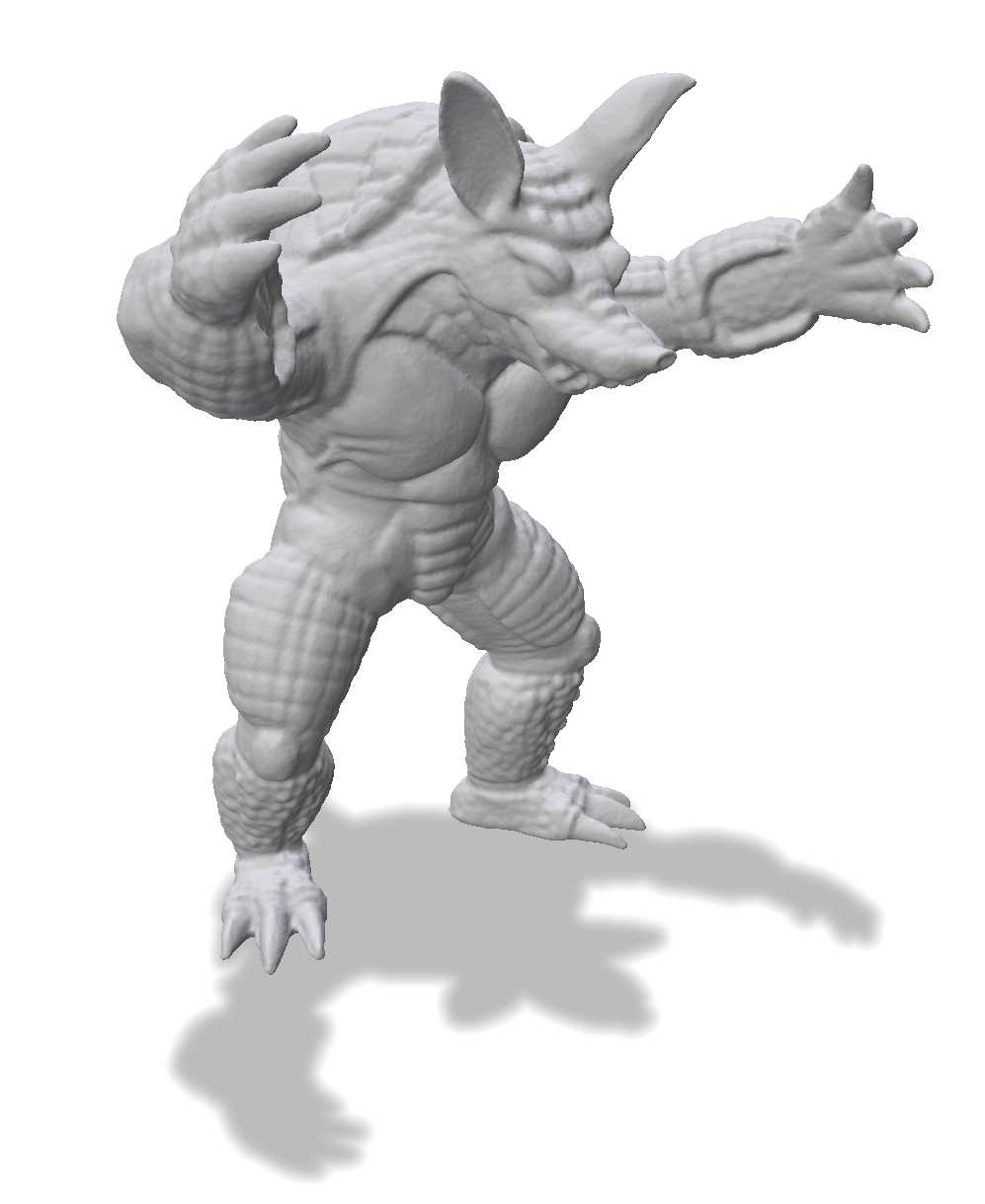} \\ %
         \hline

     \end{tabular}
     \caption{Comparison of LoRA-based SDF reconstruction quality (IoU) using increasing rank $r$ against full fine-tuning (FT) for encoding surface deformations. \change{The number of fine-tuning parameters as a percentage of the base model size is shown in the top row.}}
     \label{fig:sdf_ablation_reconstructions}
 \end{figure*}

\section{Implementation Details}
\label{sec:details}

\paragraph*{Architecture.} In all experiments, we use a standard MLP as our base model $f_\theta$ with a frequency encoding (\ref{sec:prelim}) and ReLU activations on the hidden layers. We use 4, 5, and 6 hidden layers for SDF, image, and video, respectively. All hidden layers have width 256. For the frequency encoding, we use $L=6$ for SDF experiments and $L=10$ elsewhere. In practice, one could select base model hyperparameters according to the complexity of the input data. Crucially, the maximum rank hyperparameter $r$ controls the size of the low-rank adapters, and we investigate the impact of the choice of $r$ in \S\ref{sec:lora rank ablation}.

\paragraph*{Initialization.} Following~\citet{Hayou:2024:lorainit}, we initialize each $A_i$ matrix of our LoRAs using a normal distribution $\mathcal{N}\left(0,\nicefrac{1}{d_i^{in}}\right)$ where $d_i^{in}$ is the input dimension of the corresponding base weight $W_i \in \mathbb{R}^{d_i^{out} \times d_i^{in}}$; every $B_i$ matrix is initialized to 0. The update $\Delta W_i=B_iA_i$ is scaled by $\nicefrac{1}{r_i}$ where $r_i$ is the maximum rank of the current adapter.

\paragraph*{Setup.} We implement our base neural field and LoRA updates in PyTorch\@. We conduct our experiments on a NVIDIA RTX 3090 GPU, with 24 GB of memory and 1395 MHz clock frequency. We compute SDF samples from meshes using the PySDF Python library~\cite{Yu:2023:PySDF}. We train and fine-tune using the Adam algorithm~\cite{Kingma:2014:AAM}. We observed reasonably fast convergence with a learning rate of $10^{-4}$ for the base neural field $f_\theta$. For fair comparison between LoRA and full fine-tuning, we empirically searched for the largest learning rate such that the loss does not diverge: $5\times 10^{-3}$ for LoRA and $10^{-4}$ for full fine-tuning. An improvement-based convergence threshold is used to terminate fine-tuning, and a maximum step budget ($3\times 10^{4}$) is set across all experiments. We rescale coordinates to fit in $[-1,1]^n$.

\section{Results}
\label{sec:experiments}

Here, we discuss the results of each application presented in $\S\ref{sec:lora_for_edits}$, comparing the performance of encoding variations using LoRA against full fine-tuning of base neural field $f_\theta$. We also conduct experiments investigating the impact of the LoRA rank hyperparameter (\S\ref{sec:lora rank ablation}), the relationship between rank, and the magnitude of variation between $\mathcal{D}, \mathcal{D'}$ (\S\ref{sec:variation magnitude results}). \change{Finally, we compare against post hoc low-rank factorization (\S\ref{sec:svd baseline results})} and a network overfitting baseline under a fixed parameter count (\S\ref{sec:mlp baseline results}).

\paragraph*{Metrics.} For surface fitting, we evaluate the zero level set of LoRA-augmented neural SDFs against the target surface $\mathcal{D}'$ using intersection-over-union (IoU), computed using 134 million points sampled uniformly in the bounding box, as in~\citet{Muller:2022:ING}. IoU ranges from 0 (no overlap) to 1 (perfect match). For images and videos, we report peak signal-to-noise ratio (PSNR) and mean squared error (MSE) against the reference frames.

\begin{figure}
	\includegraphics[width=0.22\textwidth]{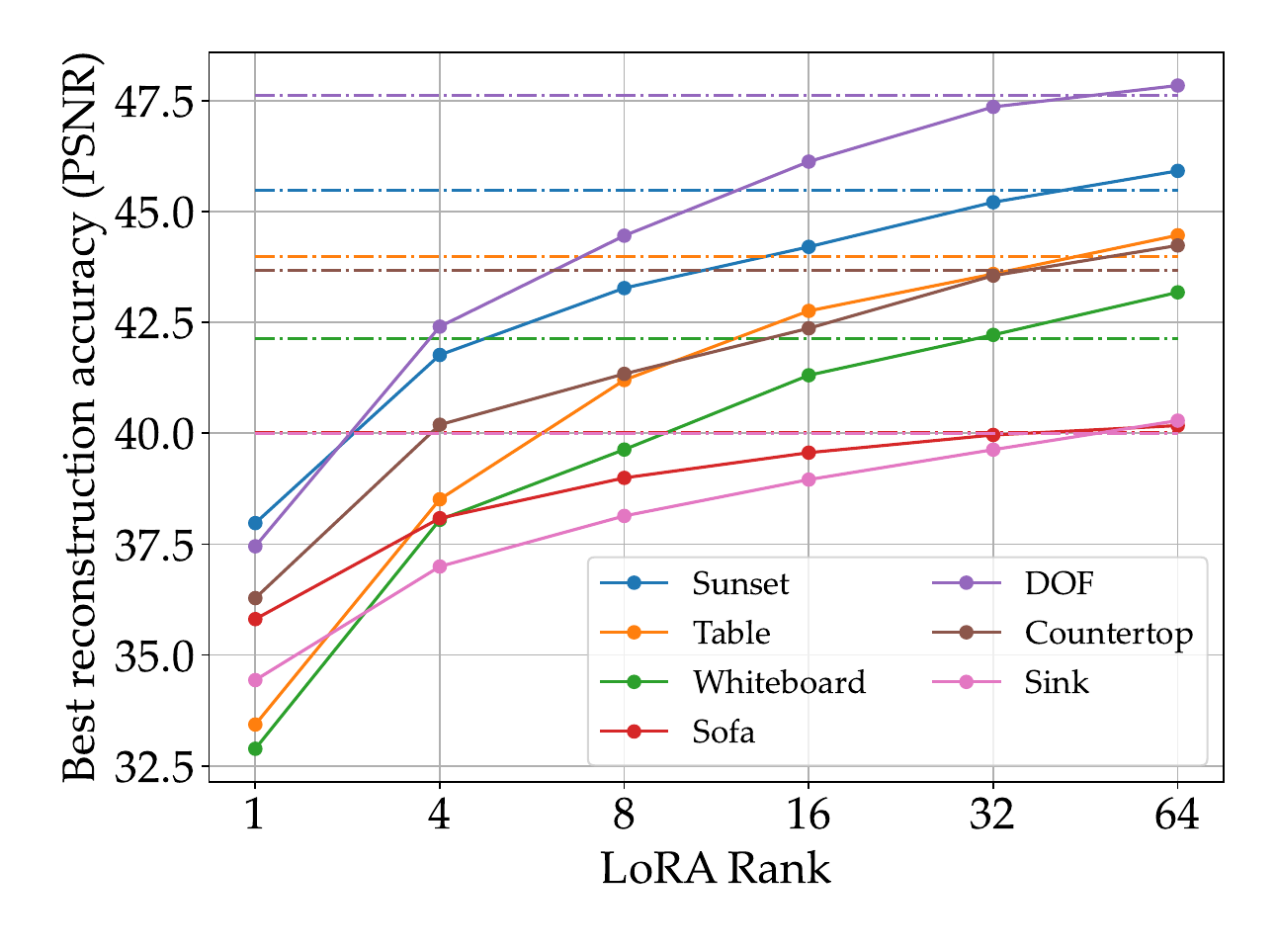}
    \includegraphics[width=0.22\textwidth]{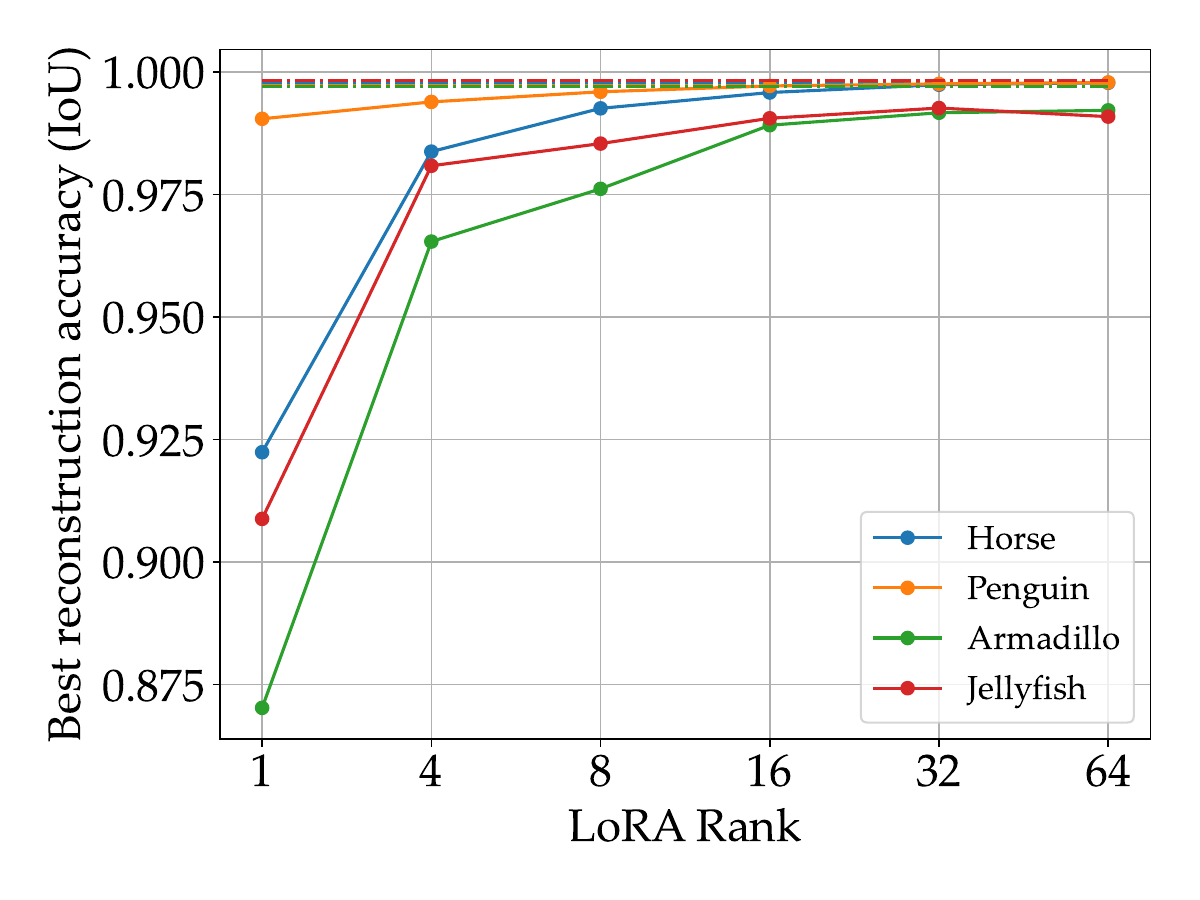}
	\caption{Image (left) and SDF (right) reconstruction quality of LoRA vs.\ full-finetuning for image variations and surface deformations, respectively. Dotted line denotes full fine-tuning. See Figure~\ref{fig:main image result table} and Figure~\ref{fig:sdf_ablation_reconstructions} for the corresponding reconstructions.}
	\label{fig:image rank ablation psnr plot}
\end{figure}

\begin{figure}    
    \captionsetup[subfloat]{labelformat=empty}
    \subfloat[]{\includegraphics[width=0.09\textwidth]{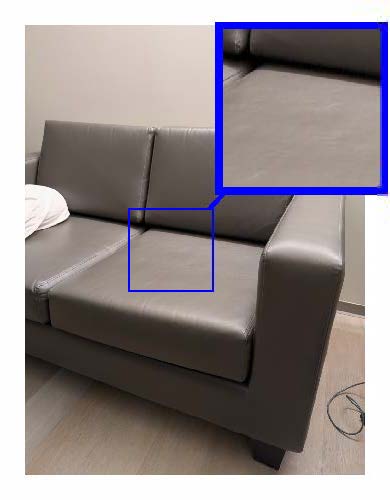}}
    \subfloat[]{\includegraphics[width=0.09\textwidth]{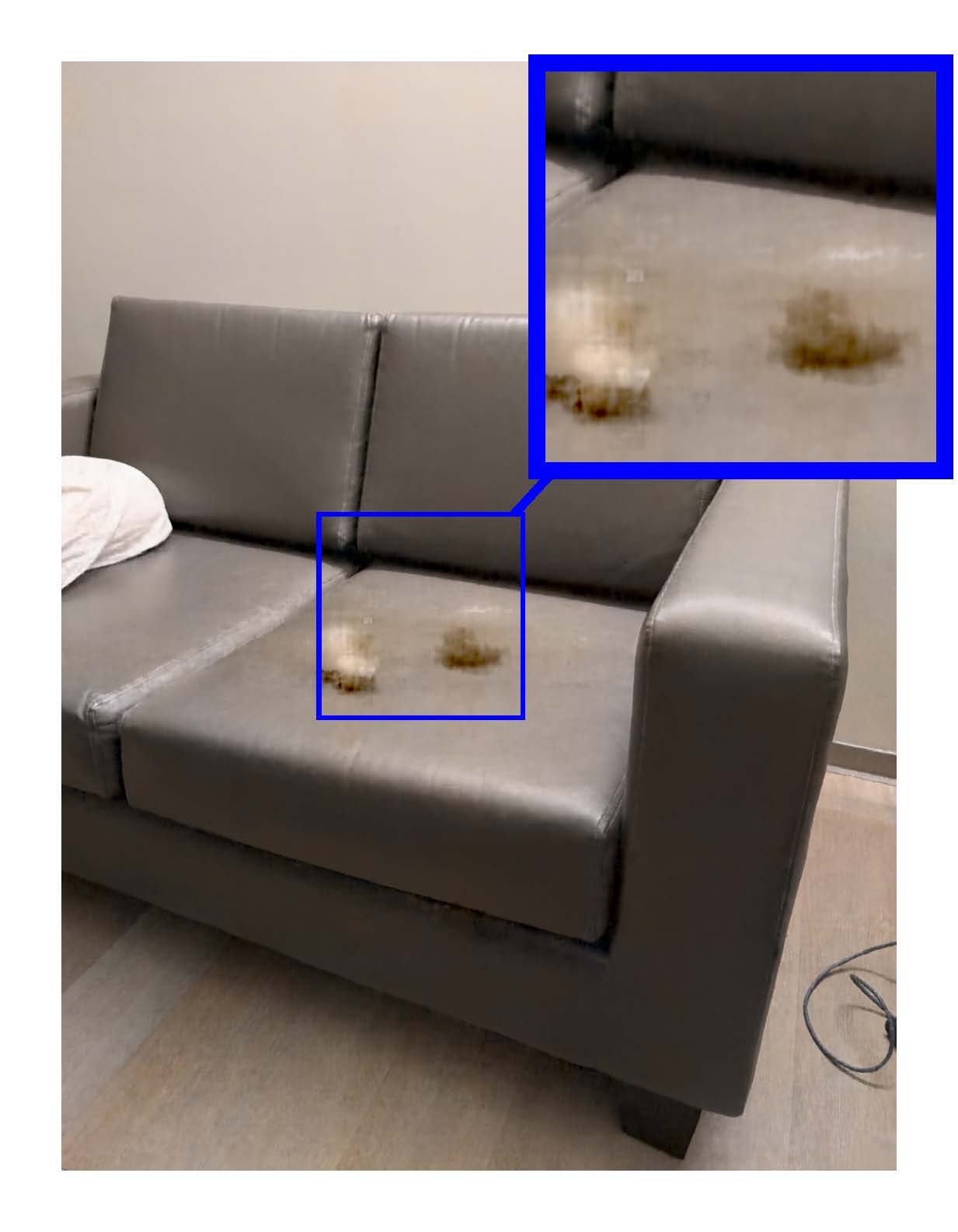}}
    \subfloat[]{\includegraphics[width=0.09\textwidth]{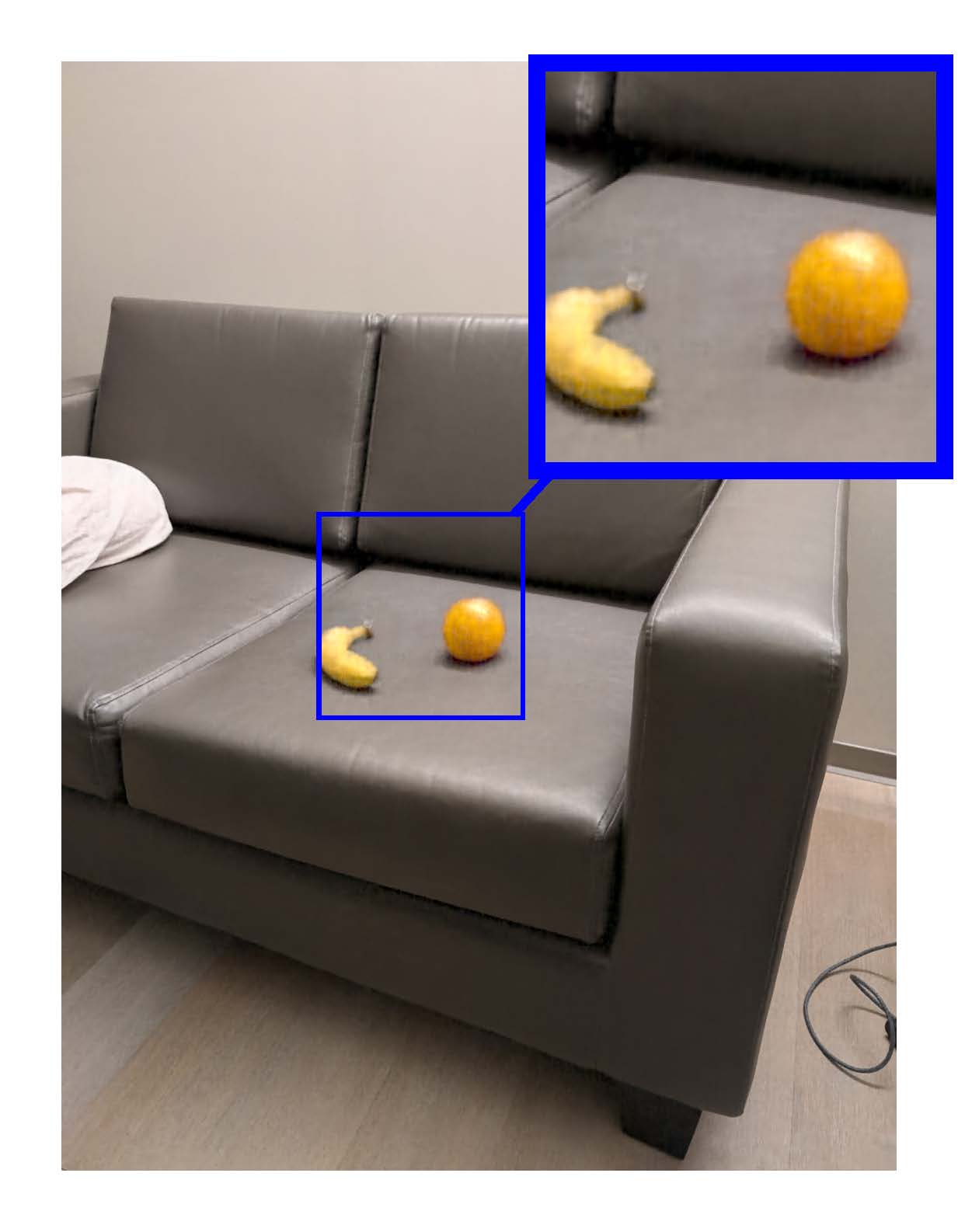}}
    \subfloat[]{\includegraphics[width=0.09\textwidth]{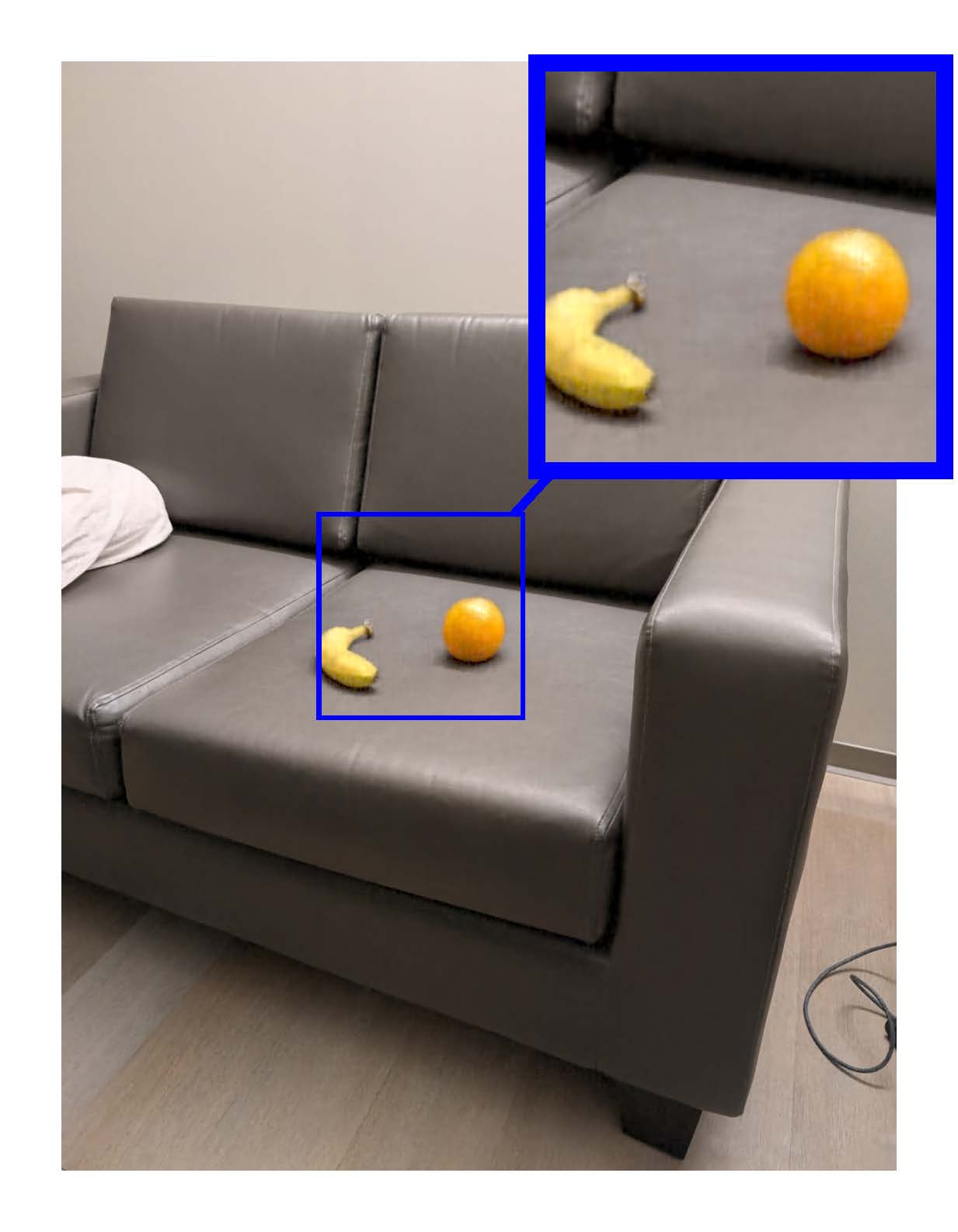}}
    \subfloat[]{\includegraphics[width=0.09\textwidth]{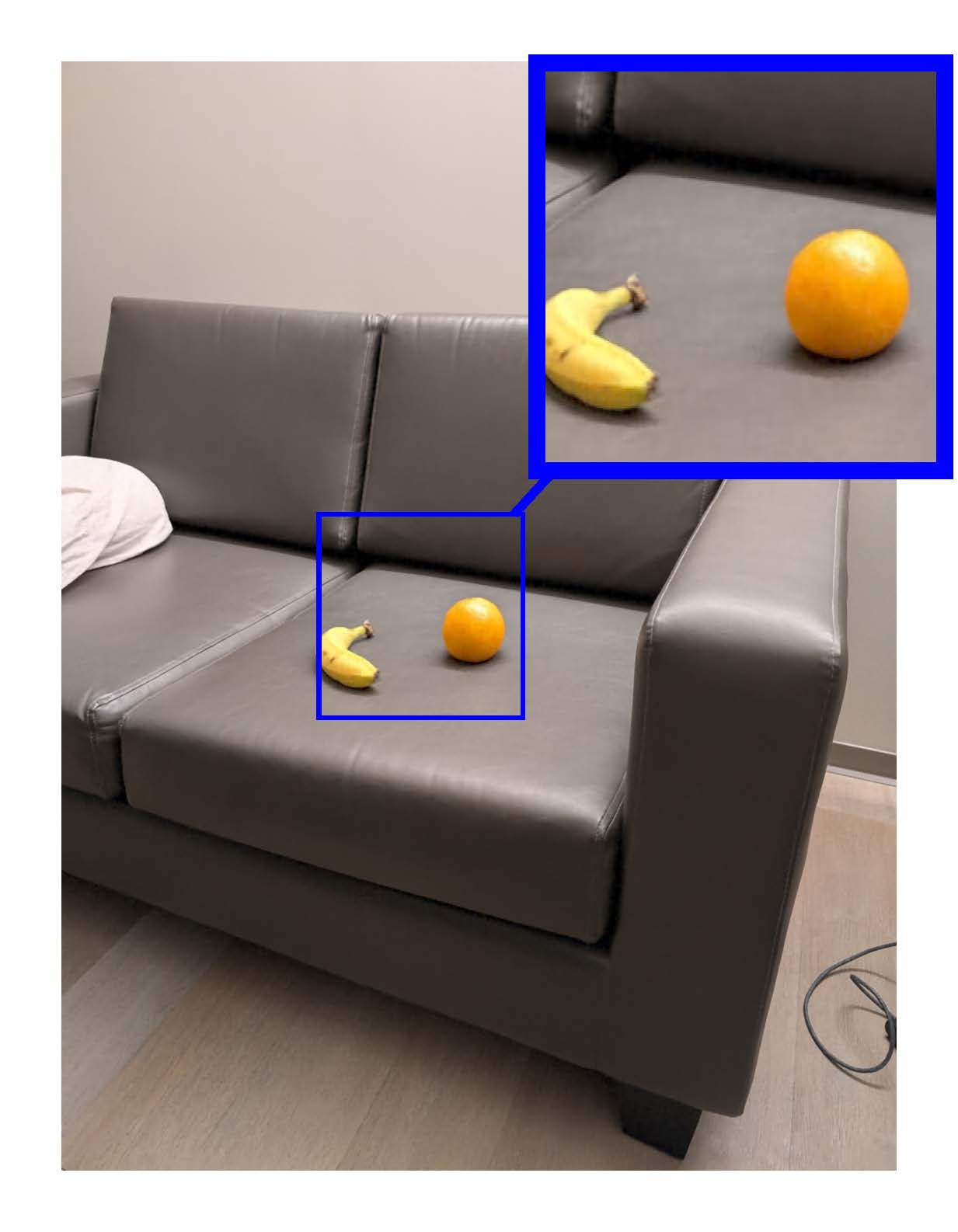}}
    
    \vspace{-6mm}
    \subfloat[0]{\includegraphics[    width=0.09\textwidth]{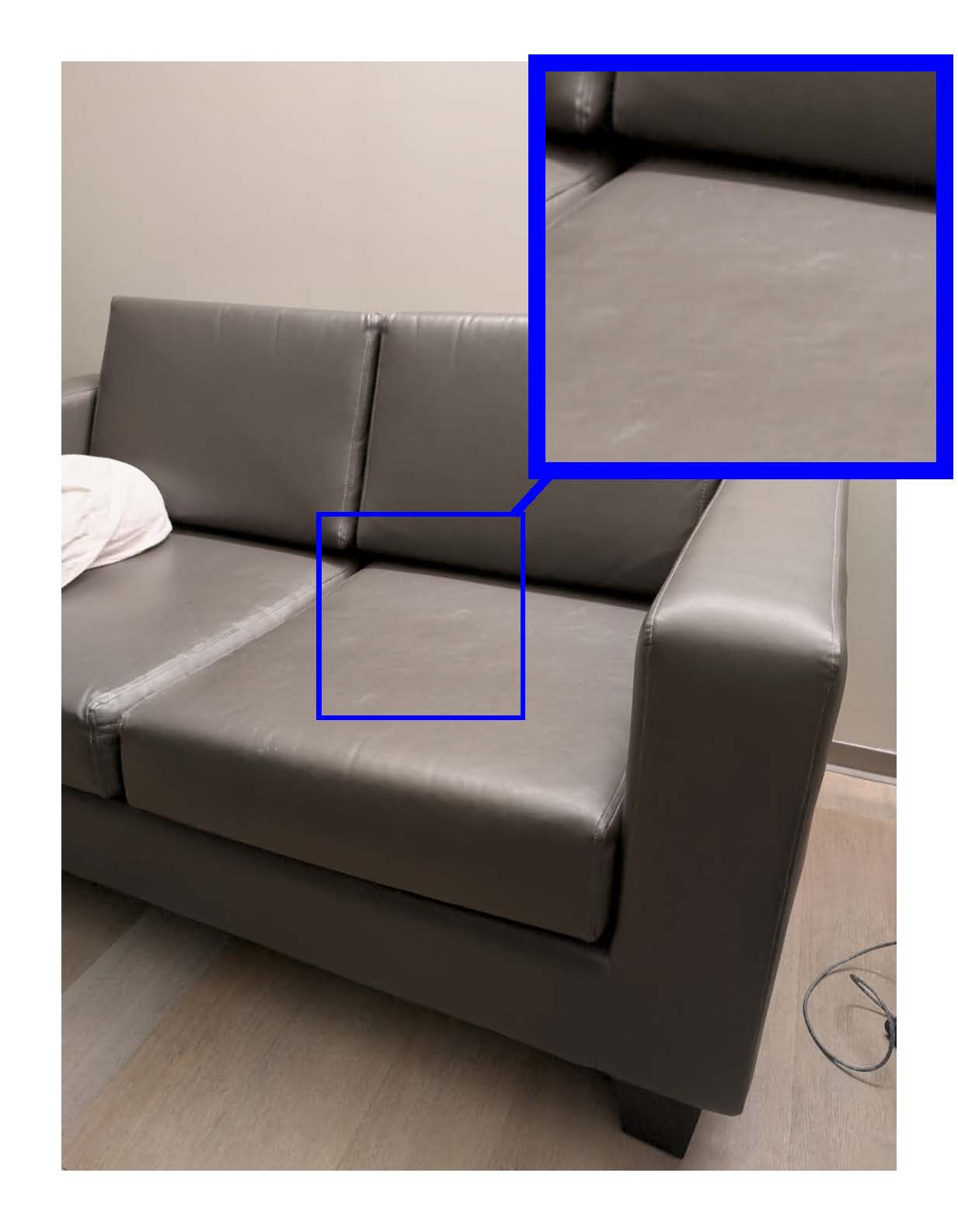}}
    \subfloat[100]{\includegraphics[  width=0.09\textwidth]{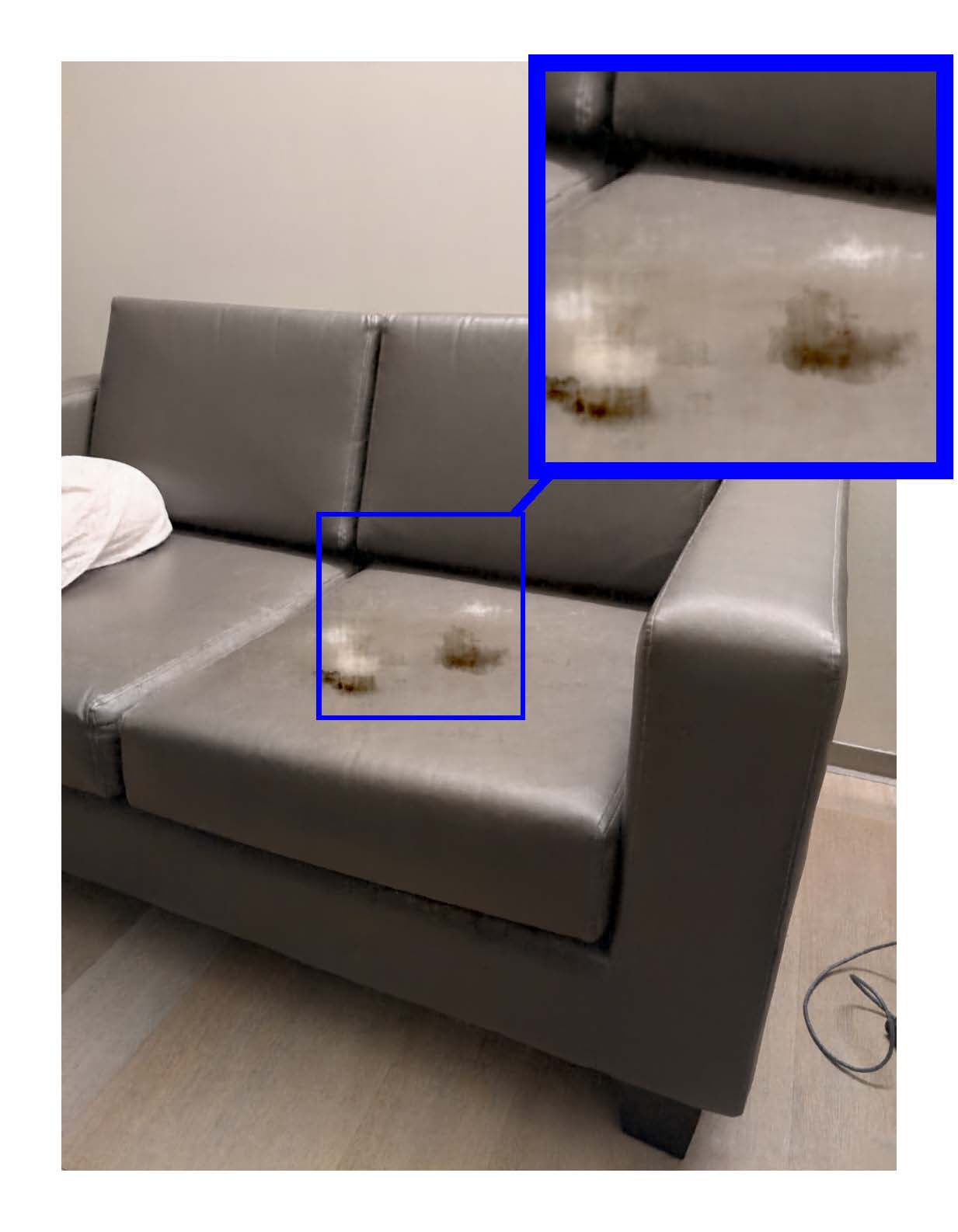}}
    \subfloat[1000]{\includegraphics[ width=0.09\textwidth]{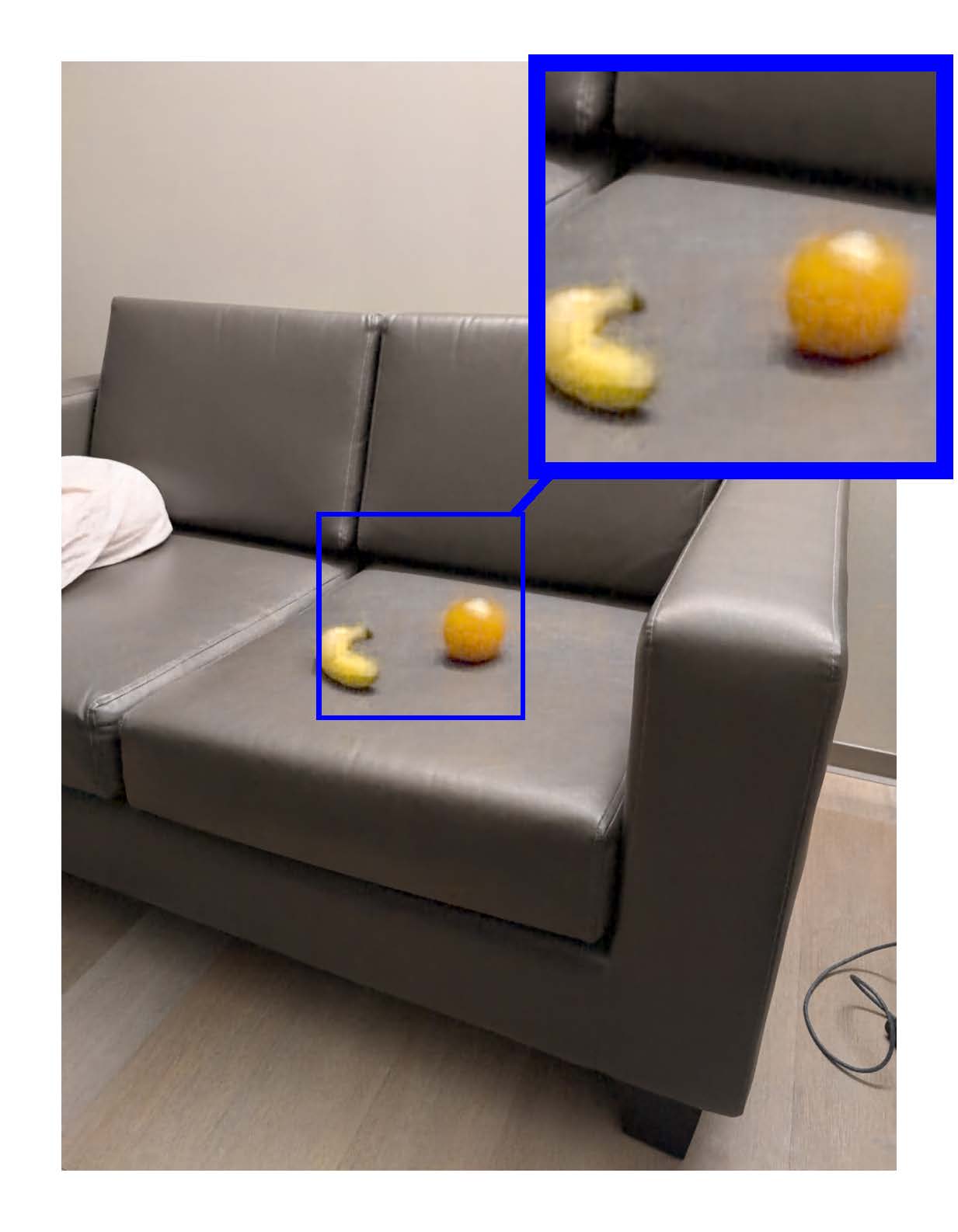}}
    \subfloat[2000]{\includegraphics[ width=0.09\textwidth]{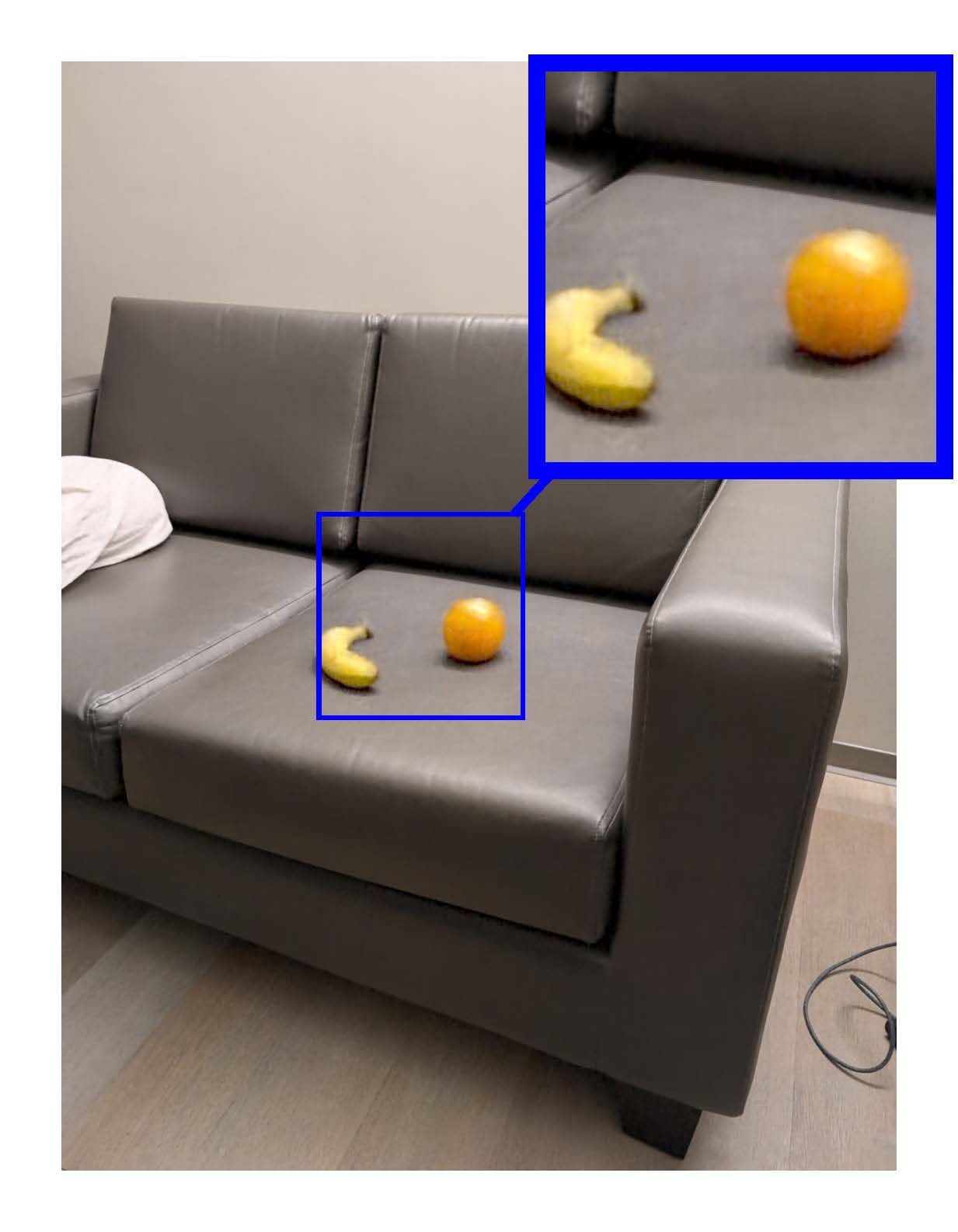}}
    \subfloat[Final]{\includegraphics[width=0.09\textwidth]{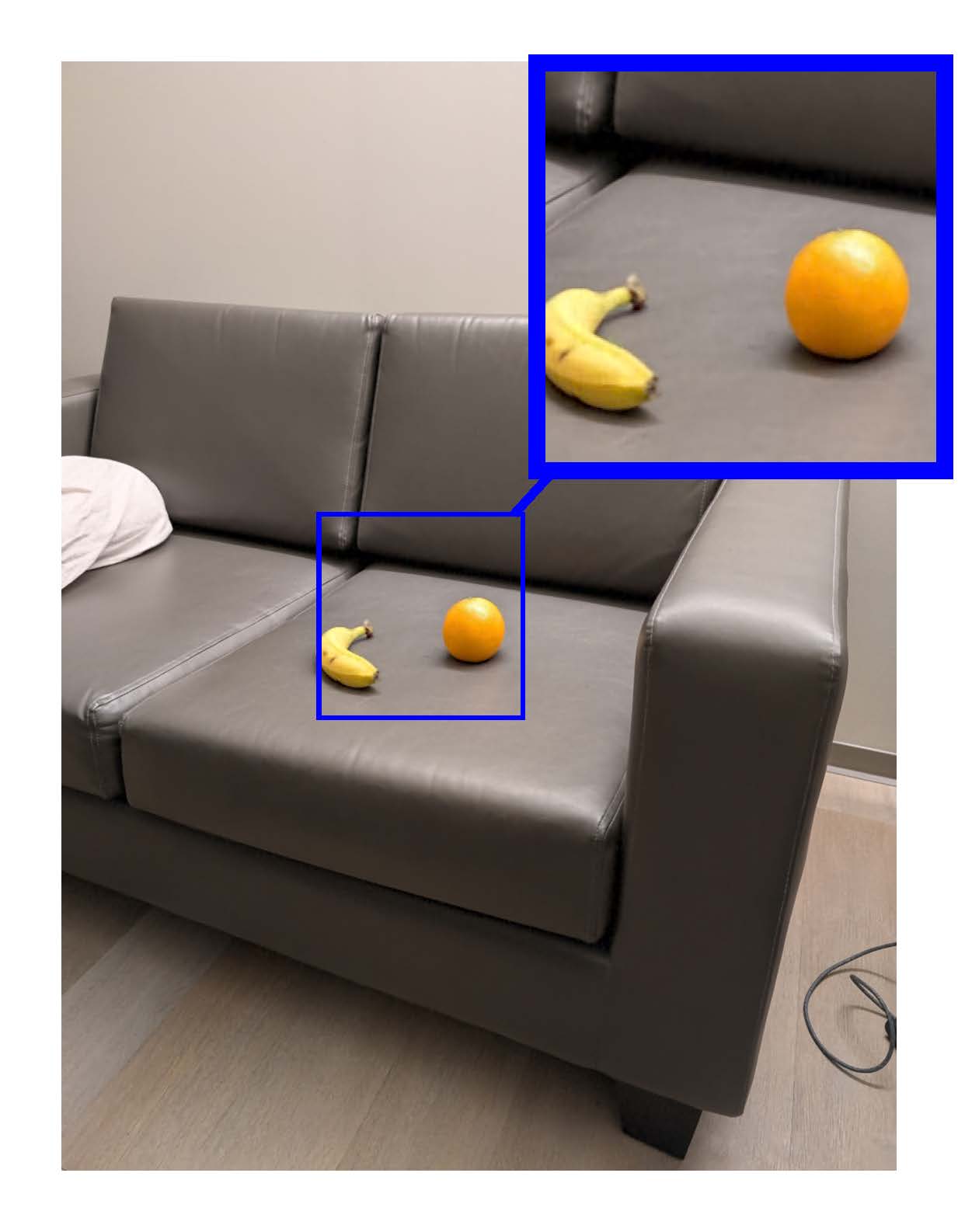}}
    
    \caption{Intermediate image reconstructions using LoRA (top) and full fine-tuning (bottom). Each column corresponds to a training step. LoRA reconstructions mirror full fine-tuning at convergence and throughout training.}
    \label{fig:image lora vs FT during training}
\end{figure}

\subsection{Image Variations}
\label{sec:core image results}

Figures~\ref{fig:whiteboard lora vs FT highlight} and~\ref{fig:main image result table} present results on eight image pairs $\mathcal{D}, \mathcal{D}'$ exhibiting diverse variations, including object insertions/removals, lighting changes, temporal changes, depth-of-field adjustments, and post-processing filters. Our LoRA-based representation achieves PSNR values comparable to full fine-tuning—while using \textbf{7--8$\mathbf{\times}$ fewer parameters}. Visually, LoRA reconstructions are nearly indistinguishable from fine-tuned ones (Figure~\ref{fig:main image result table}) even under challenging edits like the inserted marker strokes in Figure~\ref{fig:whiteboard lora vs FT highlight}. LoRA faithfully reconstructs all tested variations, with only minor degradation for post-processing effects (e.g., stylized filters) which globally alter high-frequency detail in $\mathcal{D}$.

Additionally, LoRA preserves unedited regions of $\mathcal{D}$ during training, similar to full fine-tuning, and produces competitive intermediate results after only a few iterations (Figure~\ref{fig:image lora vs FT during training}). This suggests that the low-rank constraint imposed by LoRA does not adversely affect the training trajectory relative to full fine-tuning, despite allowing a higher stable learning rate.

\subsection{Geometric Deformations}
\label{sec:core sdf results}
Deformations in this experiment are generated by applying skeleton-based transformations to $\mathcal{D}$. Figure~\ref{fig:sdf_ablation_reconstructions} shows that LoRA and full fine-tuning reconstructions differ in IoU by under 0.02 at the default rank ($r=16$). Visually, both methods preserve fine geometric features---e.g., the ridges on the armadillo---and are nearly indistinguishable. LoRA achieves this while using \emph{85--90\% fewer parameters} than full fine-tuning. When present, error typically appears as high-frequency noise in the isosurface reconstruction. As with image variations, LoRA's intermediate reconstructions closely track those of full fine-tuning: artifacts resolve at a similar rate, concentrate in deformed regions, and are minimal in unmodified areas (see Supplemental Material). These results suggest that LoRA can significantly compress neural SDF updates while maintaining stable and predictable training dynamics.

\subsection{Impact of LoRA Rank}
\label{sec:lora rank ablation} 
We train LoRAs with varying ranks ($r=1,4,8,16,32,64$) for images and SDFs to examine the tradeoff between LoRA size and update capacity. As shown in Figure~\ref{fig:image rank ablation psnr plot}, reconstruction quality increases monotonically with $r$, and ranks $r \geq 16$ consistently match full fine-tuning for both modalities. Beyond $r=16$ (i.e., $>12\%$ of full fine-tuning parameters), quality continues to improve gradually. At $r = 64$, LoRA slightly outperforms full fine-tuning for images and closely approaches it for SDFs (within $0.003$ IoU on average).

Performance degrades gracefully for very small $r$: noise artifacts grow as expressivity decreases, but many variations remain recognizable even with $r=1$, a $99\%$ parameter reduction (see Figures~\ref{fig:main image result table} and~\ref{fig:sdf_ablation_reconstructions}). This aligns with our motivating experiment (Figure~\ref{fig:low-rank motivating experiment}) which revealed a rank threshold below which low-rank factorizations struggle to approximate the full update. When error is visible, it is spatially localized to regions with the largest edits and minimal elsewhere (see Figure~\ref{fig:whiteboard lora vs FT highlight}, right inset; Figure~\ref{fig:sdf_ablation_reconstructions}, $r=1,4$). Overall, LoRA offers a predictable and controllable tradeoff between parameter efficiency and reconstruction fidelity.

\subsection{Effect of Variation Magnitude}
\label{sec:variation magnitude results}
We investigate the relationship between the magnitude of the variation between $\mathcal{D},\mathcal{D}'$ and the reconstruction quality using LoRA compared to full fine-tuning.

For images, we controllably edit $\mathcal{D}$ in two ways. Firstly, we apply low-frequency Gaussian noise to the RGB values of $\mathcal{D}$
(by computing per-pixel noise $\Delta x \in \mathbb{R}^{H\times W\times 3}$ where $\Delta x_{ijc} \sim \mathcal{N}(0,\sigma^2)$, applying a low-pass filter to $\Delta x$, scaling by $k \geq 1$, and setting $x'\coloneqq x+k\Delta x$) to obtain $\mathcal{D}'$. The variation magnitude is controlled by the noise magnitude $k$. Secondly, we experiment with $\mathcal{D'}$ obtained from applying non-linear distortion filters to $\mathcal{D}$ in image processing software; here, the magnitude is controlled by the software-defined distortion intensity parameter. Figure~\ref{fig:image variation magnitude, gaussian noise} suggests that LoRA is most competitive with full fine-tuning when the variation is minor. The error gap between successive LoRA ranks and full fine-tuning grows with the difference between $\mathcal{D}, \mathcal{D'}$. For significant variations where $f_\theta$ no longer provides a meaningful initialization for fine-tuning, the LoRAs ineffectively adapt the base model. This is unsurprising, as in practice LoRAs are almost universally applied to capture \emph{small} distributional shifts.

For geometry, we take $\mathcal{D}$ to be a source mesh and take intermediate meshes from a physical simulation ~\cite{leticia:2025:elastodynamic} of $\mathcal{D}$ to use as $\mathcal{D}'$. As the mesh continually deforms away from its rest state, we use the simulation time step as a proxy for the variation magnitude. Figure~\ref{fig:jellyfish} reveals the same pattern previously observed with images.

\begin{figure}
	\captionsetup[subfloat]{labelformat=empty,justification=centering}
	\subfloat[{Noisy $\mathcal{D}$ \\ \SI{22.27}{dB}}]{
		\includegraphics[width=0.10\textwidth]{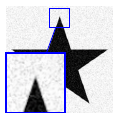}
	}
	\subfloat[{LoRA \\ \SI{23.43}{dB}}]{
		\includegraphics[width=0.10\textwidth]{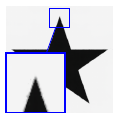}
	}
	\subfloat[Chambolle\\ \SI{23.25}{dB}]{
		\includegraphics[width=0.10\textwidth]{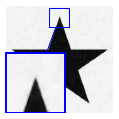}
	}
	\subfloat[Original]{
		\includegraphics[width=0.10\textwidth]{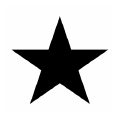}
    }    

	\subfloat[Noisy $\mathcal{D}$\\ \SI{32.26}{dB}]{
		\includegraphics[width=0.10\textwidth]{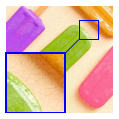}
	}
	\subfloat[{LoRA\\ \SI{33.09}{dB}}]{
		\includegraphics[width=0.10\textwidth]{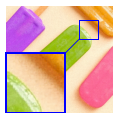}
	}
	\subfloat[Chambolle\\ \SI{33.40}{dB}]{
		\includegraphics[width=0.10\textwidth]{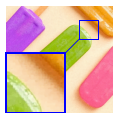}
	}	
    \subfloat[Original]{
		\includegraphics[width=0.10\textwidth]{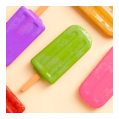}
    }

	\caption{Total Variation image denoising. Noisy input image regressed by $f_\theta$ (left), LoRA TV energy minimization output (center left), Chambolle's projection algorithm applied to the noisy image (center right), and the original image. PSNR against the original noise-free image is shown. Image by Edy HG\@.}
	\label{fig:TV_denoising_building_star}
\end{figure}

\subsection{Energy Minimization}
\label{sec:energy minimization results}

We experiment with the \emph{Total Variation} energy defined as $E_{TV}(f, x) \coloneqq  {\left\lVert \nabla_x f(x)\right\rVert_2}$,
where $\nabla_x f(x)$ denotes the spatial gradient of an input function $f$. Solving problem ~\eqref{general energy min problem} with $E_{TV}$ is a neural field adaptation of the well-studied Total Variation denoising (TVD) problem~\cite{ROF:1992:TV_paper}. TVD is known to remove noise while preserving sharp edges in the underlying signal. Unlike conventional low-pass filters, the result of TVD is obtained through numerical optimization.

In this setting, $f_\theta$ parameterizes a noisy input image (obtained by adding Gaussian noise), and $f_{\theta+\text{LoRA}}$ seeks to denoise. We estimate the gradient via a finite difference. We compare our denoised result with that of applying Chambolle's projection algorithm---a standard approach for TVD on images---applied to the output of $f_\theta$. Figure \ref{fig:TV_denoising_building_star} shows that minimizing $E_{TV}$ over LoRA parameters yields an edge-preserving denoising effect, as desired.

\begin{figure}
  \centering
  \begin{subfigure}{0.49\columnwidth}
    \includegraphics[width=\textwidth]{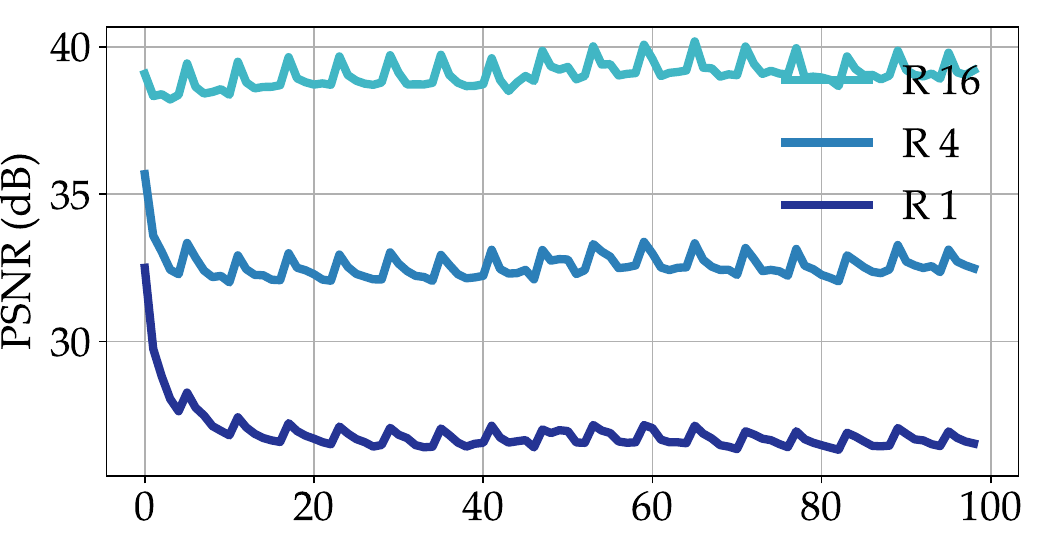}
    \caption{Our video PSNR for different ranks.}
    \label{fig:videoranks}
  \end{subfigure}  
  \begin{subfigure}{0.46\columnwidth}
    \includegraphics[width=\textwidth]{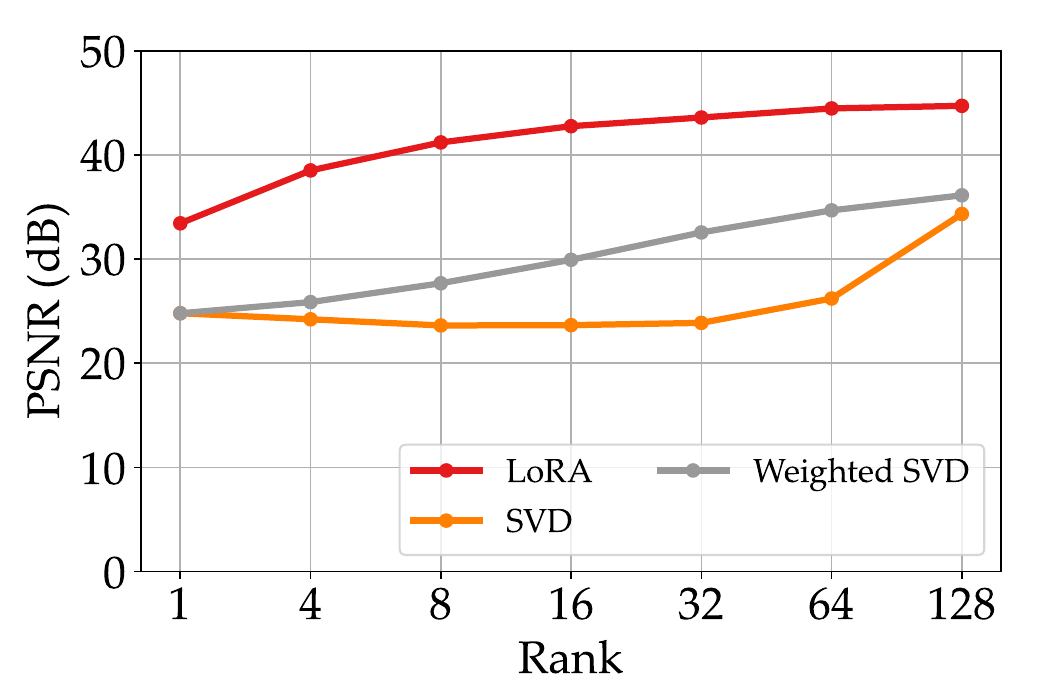}
    \caption{LoRA vs. SVD methods.}
    \label{fig:svdbaseline}
  \end{subfigure}
  \caption{Comparison of our neural field update methods across different ranks. (a) Per-frame video reconstruction quality (PSNR) for different LoRA ranks ($R=1, 4, 16$) showing stable performance over 100 frames. (b)~Reconstruction quality comparison between our LoRA method, truncated SVD, and weighted SVD across varying ranks. LoRA consistently outperforms both SVD baselines, with the performance gap increasing at higher ranks.}
  \label{fig:video_and_svd}
\end{figure}

\subsection{Sequential LoRAs}
\label{sec:video results}
We apply the sequential LoRA training procedure outlined in \S\ref{sec:video} to encode real 1080p video footage (Figure~\ref{fig:trafficvideo}) and an animation (see Supplemental Material) with 100 and 130 frames, respectively. Even with basic sequential optimization, our method is able to encode long sequences of frame-to-frame changes with no apparent long-term error accumulation. Figure~\ref{fig:videoranks} shows per-frame reconstruction accuracy for the video footage using sequential LoRAs with varying ranks. While there is local fluctuation between small subsequencies of frames, the PSNR globally remains within $39.2 \pm 1 \text{ dB}$ for the captured video ($r=16$) and  $37.1 \pm 1.8 \text{ dB}$ for the animation. Even at very low ranks ($r \leq 4$), error drift is minimal past the initial frames. See the Supplemental Material for video reconstructions.

We also evaluate the parallel encoding strategy described in \S\ref{sec:video} on animated surfaces, where each frame $\mathcal{D}'$ is obtained from physical simulation and is \emph{independently} encoded as a LoRA update to the rest-state model ($t=0$). This setup is trivially parallelizable, taking no more time than training a single LoRA. Figure~\ref{fig:jellyfish} shows that this approach performs well for a small number of frames ($\leq$5) across all ranks $r\geq 4$ but struggles with $r=1$, especially at larger time steps. This behavior is consistent with the variation magnitude observations in \S\ref{sec:variation magnitude results}.

\subsection{Post-hoc Low-Rank Factorization Baseline}
\label{sec:svd baseline results}
To assess parameter efficiency, we compare against low-rank factorization of the weight difference matrices performed \emph{after} full fine-tuning as a baseline. As mentioned in the motivating experiment in \S\ref{sec:intro}, a low-rank factorization of each layer's weight difference $\Delta W \coloneqq W_\textrm{fine-tuned} -W_\textrm{base}$ approximates the true difference with relatively low error (with respect to a matrix norm). We now compare the output reconstruction quality using LoRA against (1)~the rank-$r$ truncated singular value decomposition (SVD) of $\Delta W$, and (2)~a weighted variant of truncated SVD that accounts for each network layer's input distribution (see Supplemental Materials). 

\begin{figure*}
	\centering
	\captionsetup[subfloat]{labelformat=empty}

	\subfloat[Input ($\mathcal{D}$)]{ \includegraphics[width=0.17\textwidth]{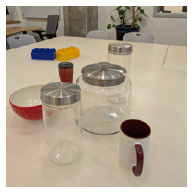}}
	\subfloat[Target ($\mathcal{D}'$)]{ \includegraphics[width=0.17\textwidth]{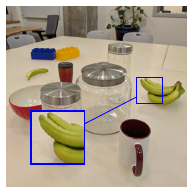}}
	\subfloat[SVD \\PSNR=\SI{23.66}{dB}]{\includegraphics[width=0.17\textwidth]{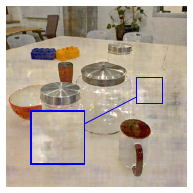}}
	\subfloat[W-SVD \\PSNR=\SI{29.93}{dB}]{\includegraphics[width=0.17\textwidth]{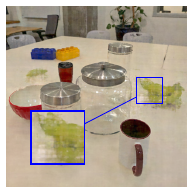}}
	\subfloat[LoRA \\PSNR=\SI{42.76}{dB}]{\includegraphics[width=0.17\textwidth]{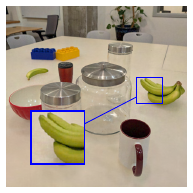}}

	\caption{\change{Comparison of image reconstructions using our LoRA vs.\ the truncated singular value decomposition of the full network weight differences (SVD) and a weighted variant of truncated SVD (W-SVD). All approaches use $r=16$.}}
	\label{fig:svd_comparison outputs}
\end{figure*}

Figure~\ref{fig:svdbaseline} shows that LoRA image reconstruction PSNR consistently exceeds both SVD and weighted SVD by \SI{10}{dB} across all ranks. In Figure~\ref{fig:svd_comparison outputs}, we show an examine where vanilla SVD struggles to encode the desired change, while weighted SVD only captures a blurry change. Furthermore, both SVD approaches produce significant global artifacts which---unlike LoRA---are pronounced in unmodified regions of the target image. Although truncated SVD provides the theoretically optimal low-rank factorization for each individual layer's $\Delta W$ (w.r.t. Frobenius norm), this experiment suggests that the online and reconstruction-loss-aware factorization using LoRA significantly benefits the reconstruction accuracy of the neural field's final output. 

\begin{figure}%
	\captionsetup[subfloat]{labelformat=empty}
	\subfloat[Input ($\mathcal{D}$)]{\includegraphics[width=0.12\textwidth]{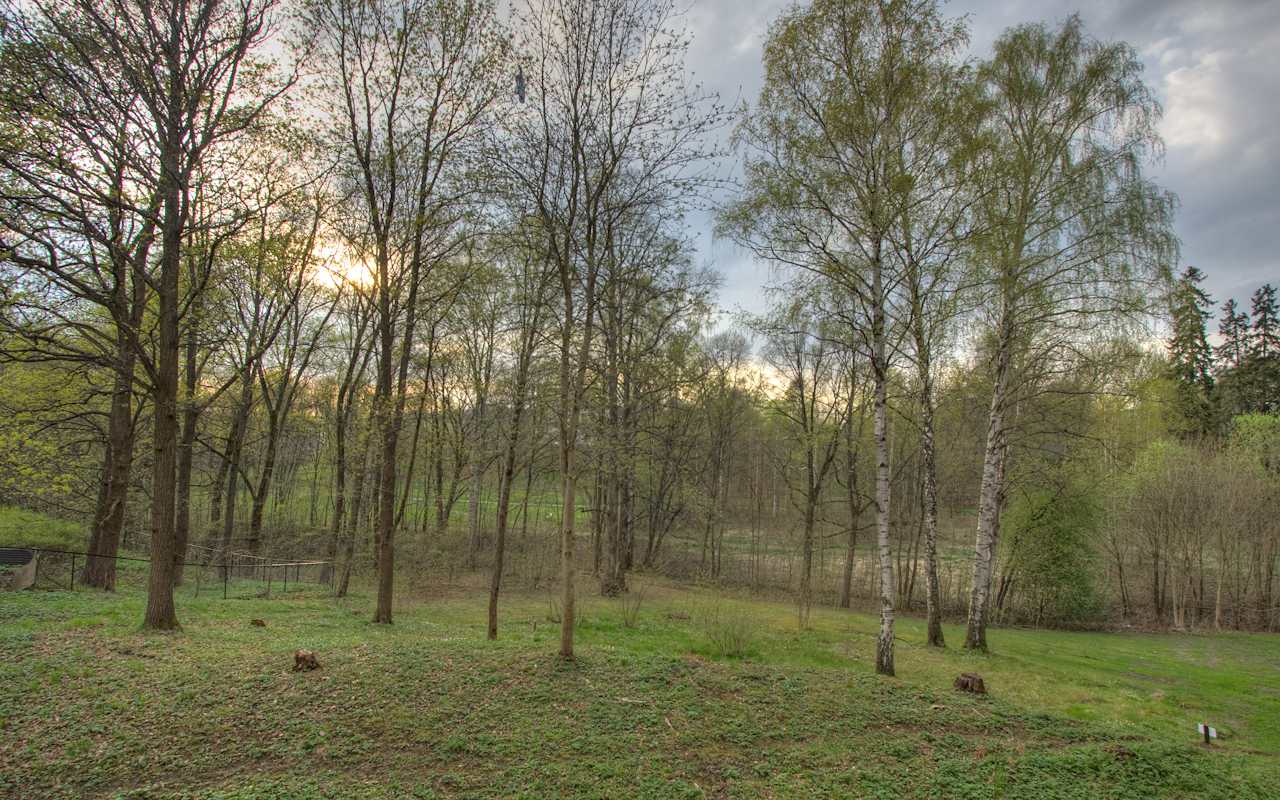}}
    \subfloat[Target ($\mathcal{D}'$)]{\includegraphics[width=0.12\textwidth]{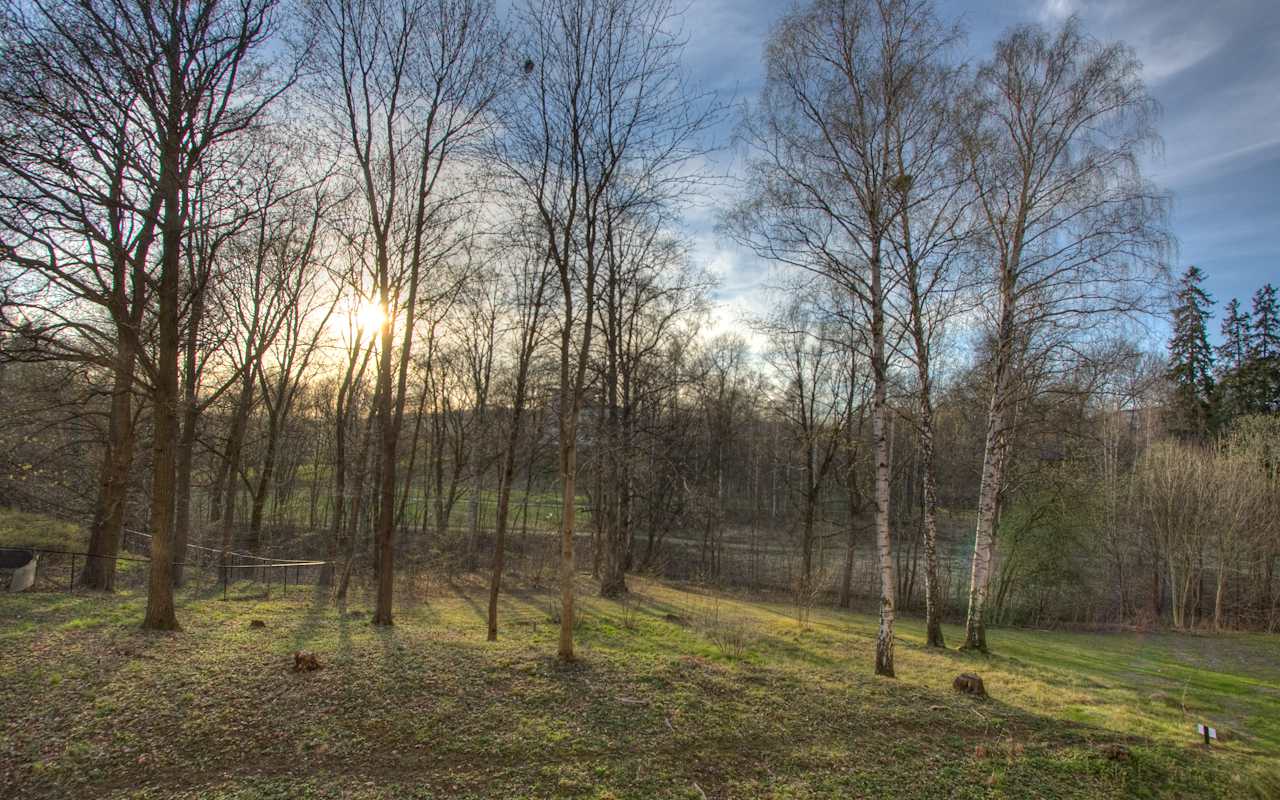}}    
	\subfloat[Small MLP ]{\includegraphics[width=0.12\textwidth]{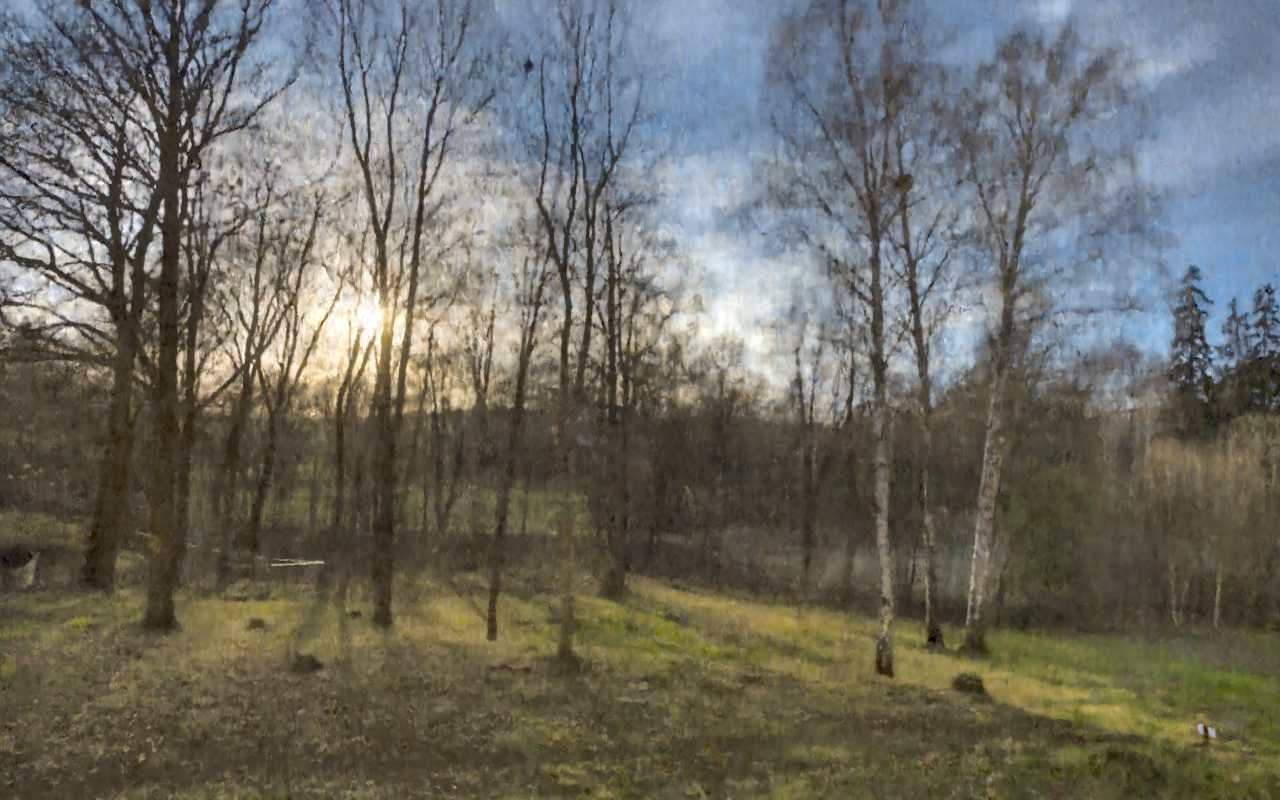}}
    \subfloat[LoRA]{\includegraphics[width=0.12\textwidth]{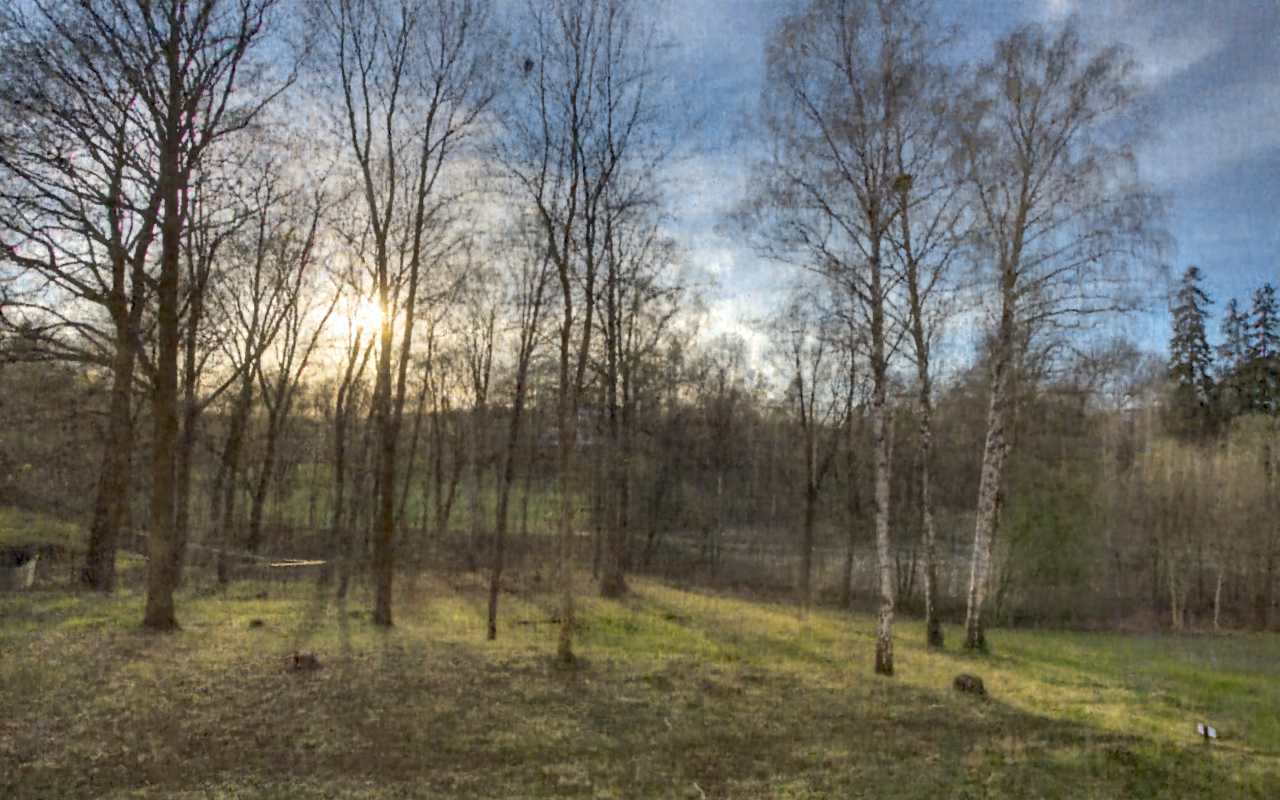}}
	\caption{We compare reconstructions of image variations using LoRA vs. with a small MLP containing the same number of parameters (46.9k). From left to right: input image $\mathcal{D}$, target variation $\mathcal{D'}$, reconstruction from a small MLP trained from scratch, and the result from our LoRA method. The small MLP struggles to capture as many fine details as LoRA (PSNR: \SI{26.3}{dB} vs.\ \SI{28.2}{dB} respectively). Input photos by Eirik Solheim.}
	\label{fig:small_mlp}
\end{figure}

\subsection{Fixed-parameter-count Baseline}
\label{sec:mlp baseline results}
To determine whether the low-rank constraint on the weight updates encourages them to encode a minor \emph{offset} from $\mathcal{D}$ rather than simply parameterizing the \emph{entire} edited instance $\mathcal{D}'$, we perform a fixed-parameter-count comparison. We compare the reconstruction quality of $\mathcal{D}'$ using $f_{\theta+ \text{LoRA}}$ against that using a new neural field $f_{\theta'}$ trained from scratch with the same number of parameters as the LoRA, i.e., \texttt{size}(LoRA) = \texttt{size}($\theta'$). $f_{\theta'}$ has the same architecture as $f_\theta$ but a uniformly lower hidden layer size chosen so that $f_{\theta'}$ has parameter count as close to \texttt{size}(LoRA) as possible. We carry out this comparison on a difficult input image which the base model can barely regress. Figure~\ref{fig:small_mlp} shows a decrease in reconstruction quality for both our LoRAs and the small MLP baseline. However, with the same number of parameters, our LoRAs visibly preserve details that the baseline cannot resolve, outperforming the baseline by $1.9 \text{ dB}$.

\section{Conclusion} 
\label{sec:conclusion}
Redundancy is pervasive in graphics and has long been leveraged for efficient representations across different signal modalities---such as in video compression and mesh simplification---where redundancy can be understood in lower-dimensional domains. Neural fields, however, pose a unique challenge: they lift low-dimensional signals into high-dimensional representations, making it less obvious how to think about redundancy. In this work, we demonstrate that low-rank adaptation offers a principled and effective lens for understanding and leveraging redundancy in neural fields.

We show LoRA updates enable versatile neural field editing for images and surfaces, plus extensions to energy minimization and dynamic data. Our method provides fine-grained control over the expressivity-storage trade-off. We empirically demonstrate this trade-off is predictable where reconstruction quality degrades gracefully at low ranks and high variations, with error artifacts appearing in expected output regions.

Our method's training speed is limited by the base neural field's size and architecture. Since LoRA optimization requires backpropagating through the entire base model each iteration, with the backward pass dominating tuning time, our unoptimized implementation runs no faster per iteration than full fine-tuning despite matching intermediate outputs. Both methods achieve adequate results in under 5 minutes but need 20-30 minutes for full convergence. Thus, exploring more time-efficient LoRA updates is a valuable direction for future work.

\begin{acks}
\label{sec:ack}
The MIT Geometric Data Processing Group acknowledges the generous support of Army Research Office grant W911NF2110293, of National Science Foundation grants IIS2335492 and OAC2403239, from the CSAIL Future of Data and FinTechAI programs, from the MIT--IBM Watson AI Laboratory, from the Wistron Corporation, from the MIT Generative AI Impact Consortium, from the Toyota--CSAIL Joint Research Center, and from Schmidt Sciences. We thank Leticia Mattos da Silva for providing the Jellyfish animation, Silvia Sell\'{a}n for assistance with manuscript preparation, and Ana Dodik for the image of Bella the dog and for valuable discussions. Big Buck Bunny (2008 Blender Foundation) is licensed under the Creative Commons Attribution 3.0 (CC BY 3.0) license.
\end{acks}

\balance
\bibliographystyle{ACM-Reference-Format}
\bibliography{lorafields}

\begin{figure*}
    \centering
     \begin{tabular}{c | c c c c}
         & Frame \#1 & Frame \#25 & Frame \#50 & Frame \#99 \\
         \hline
         \rotatebox{90}{Ground Truth} &
         \includegraphics[width=0.18\textwidth]{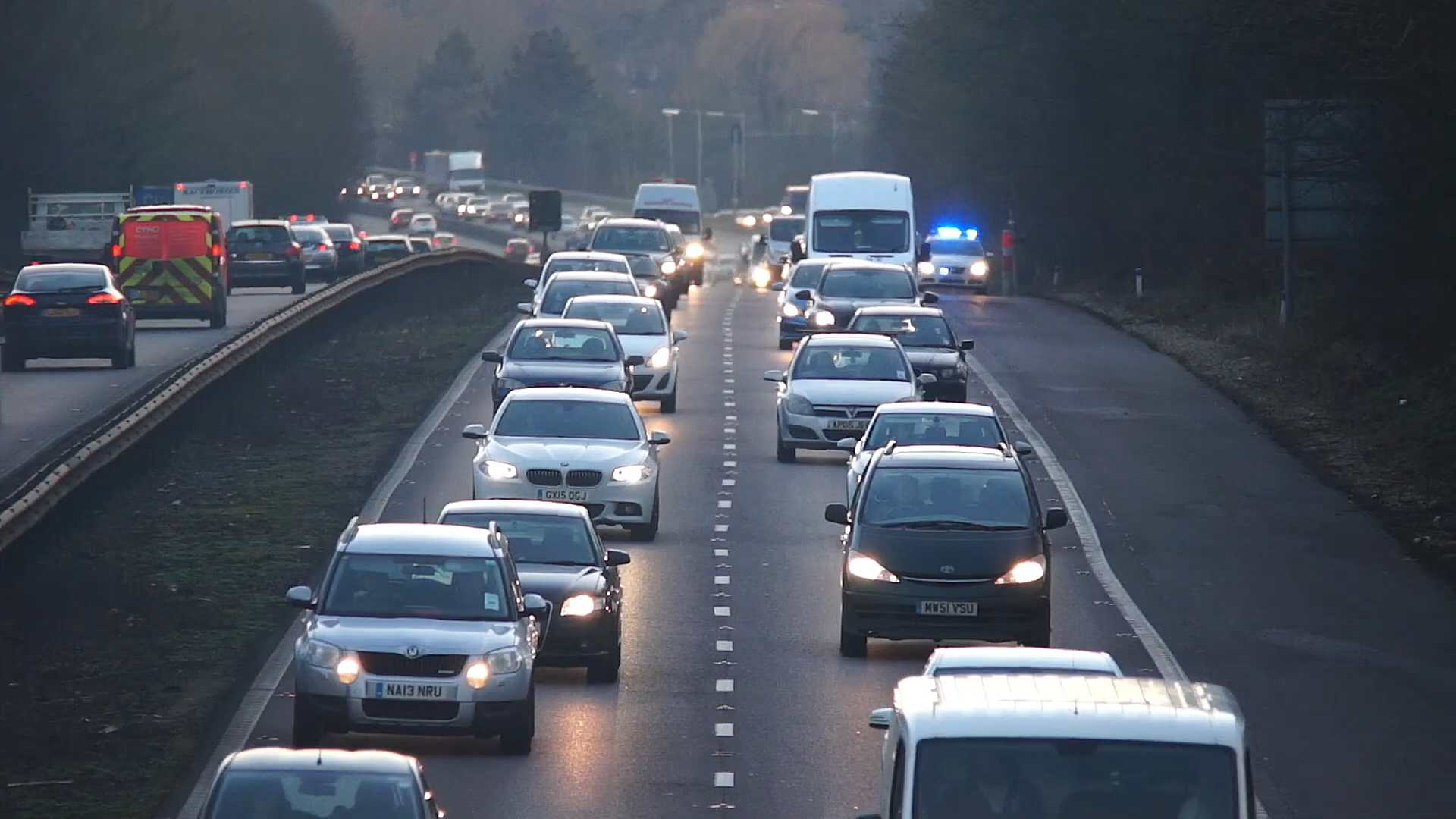} &
         \includegraphics[width=0.18\textwidth]{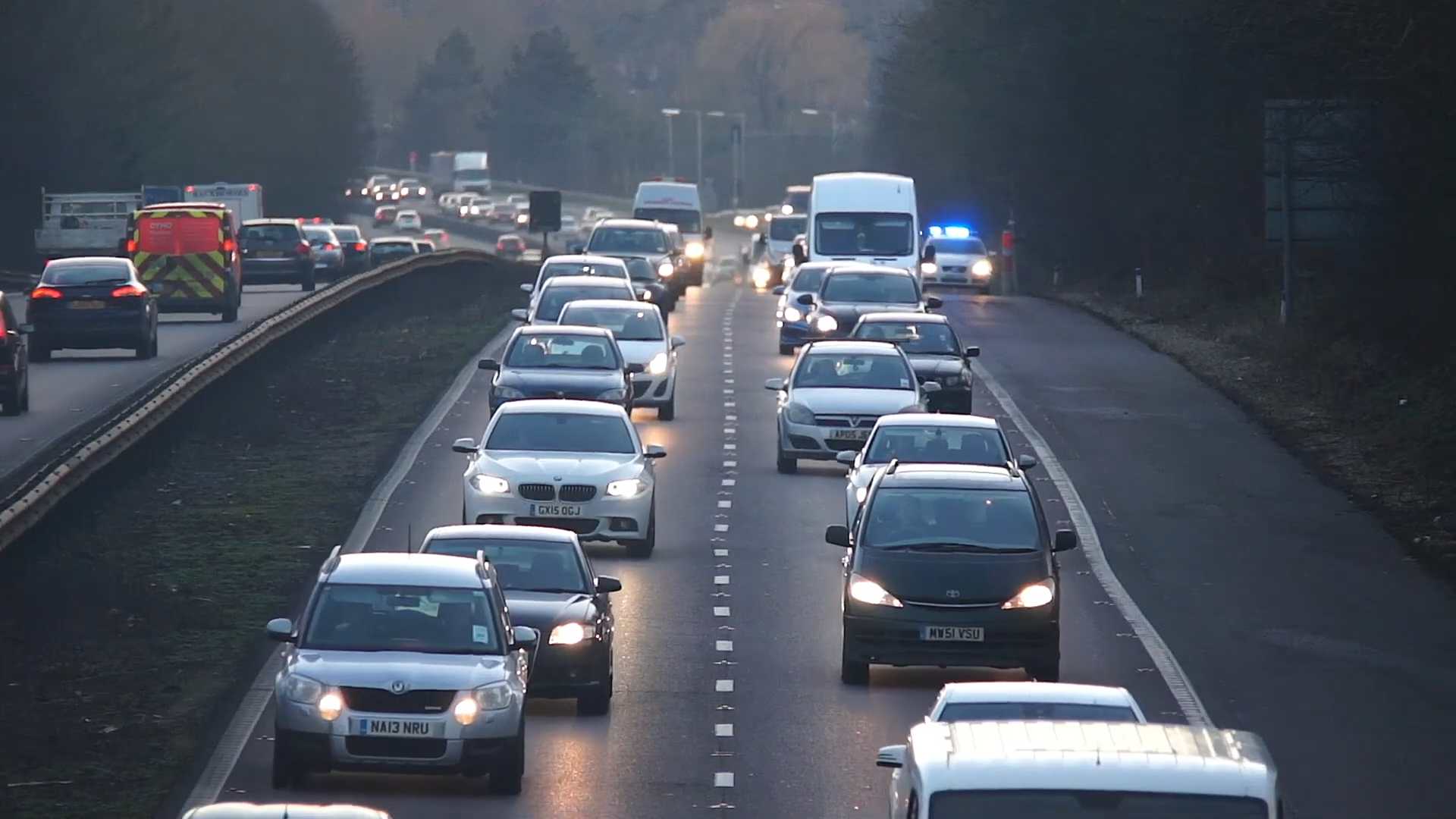} &
        \includegraphics[width=0.18\textwidth]{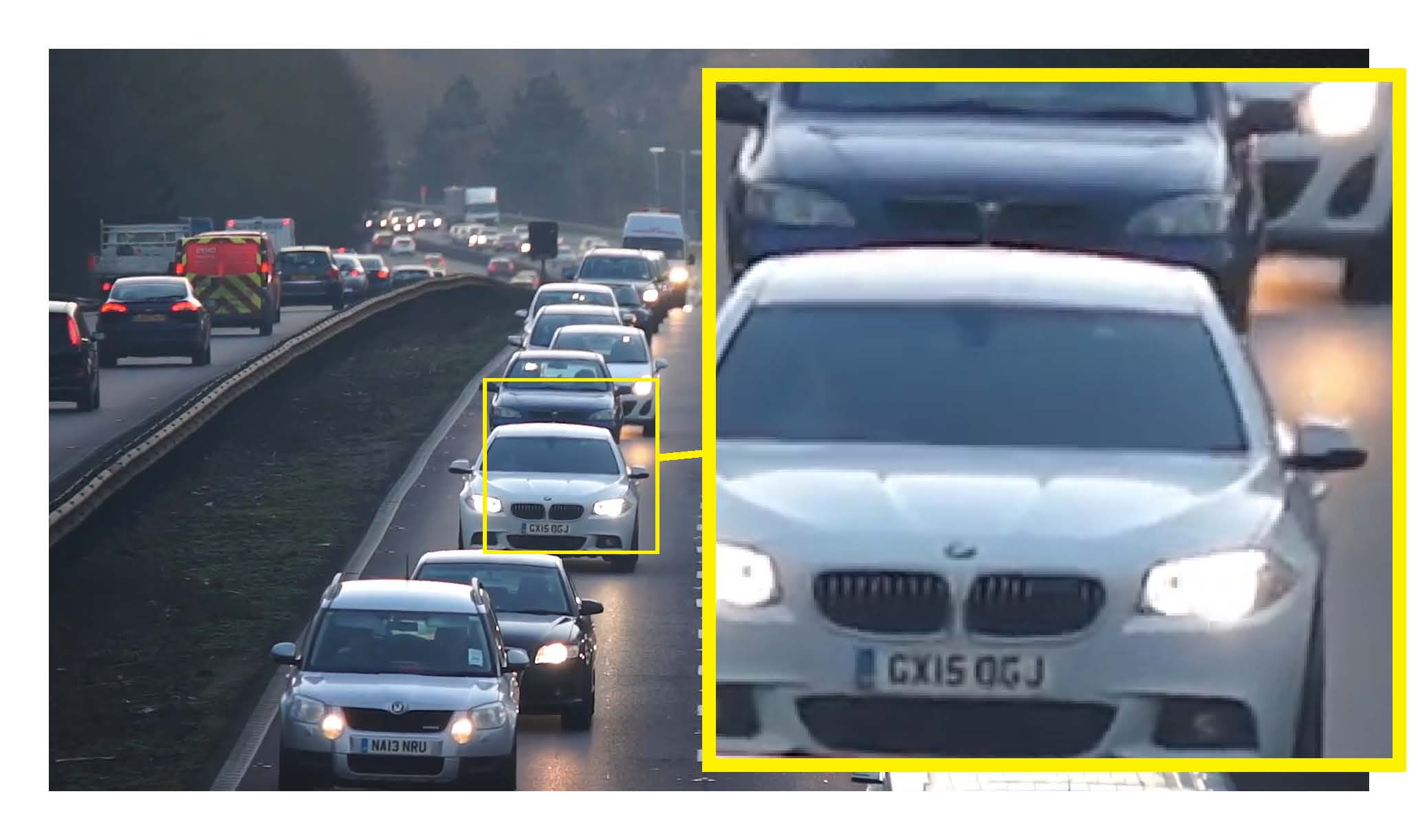} &
         \includegraphics[width=0.18\textwidth]{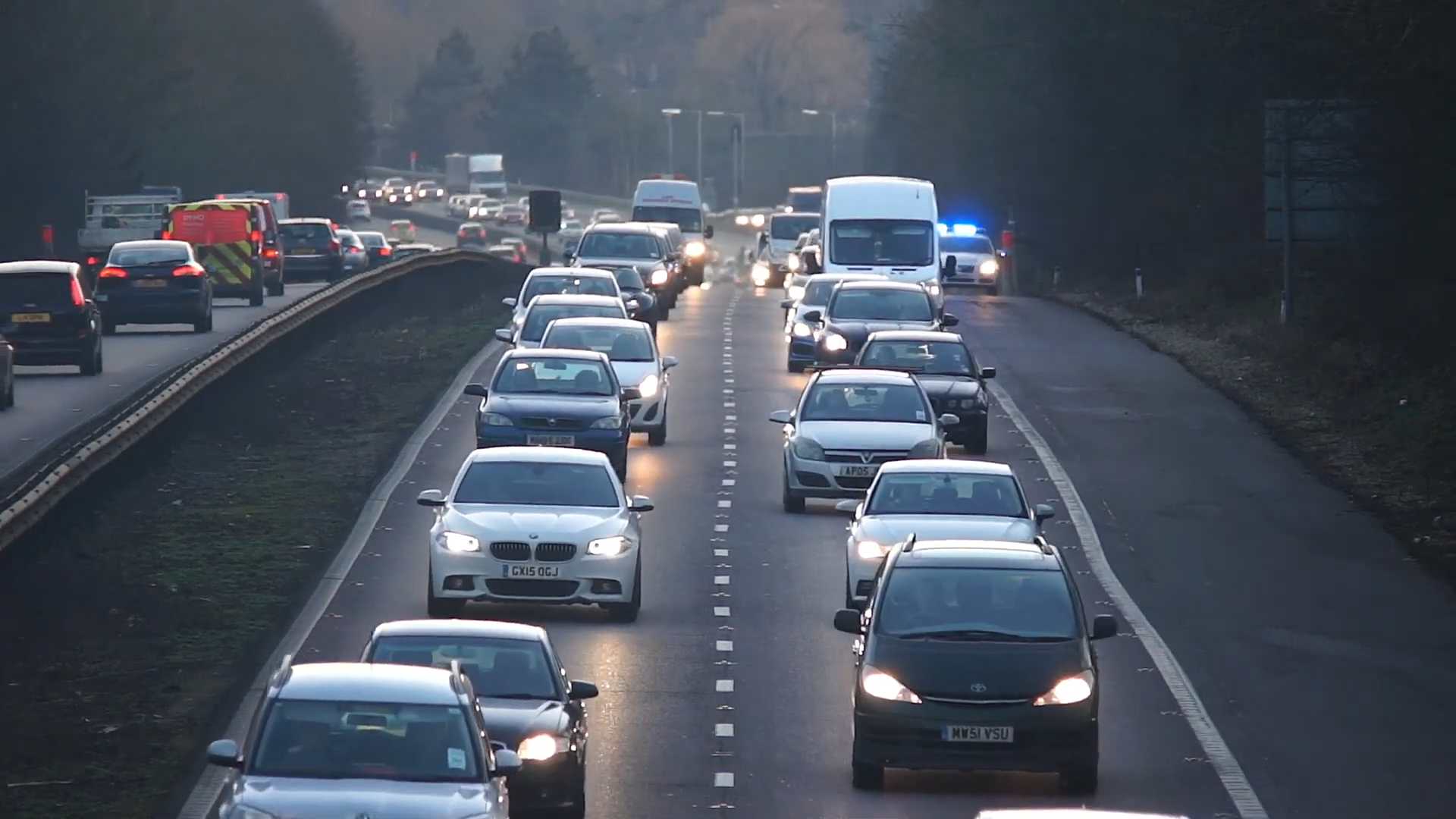} \\        
         
         \rotatebox{90}{$f_\theta$ fine-tuned} &
         \includegraphics[width=0.18\textwidth]{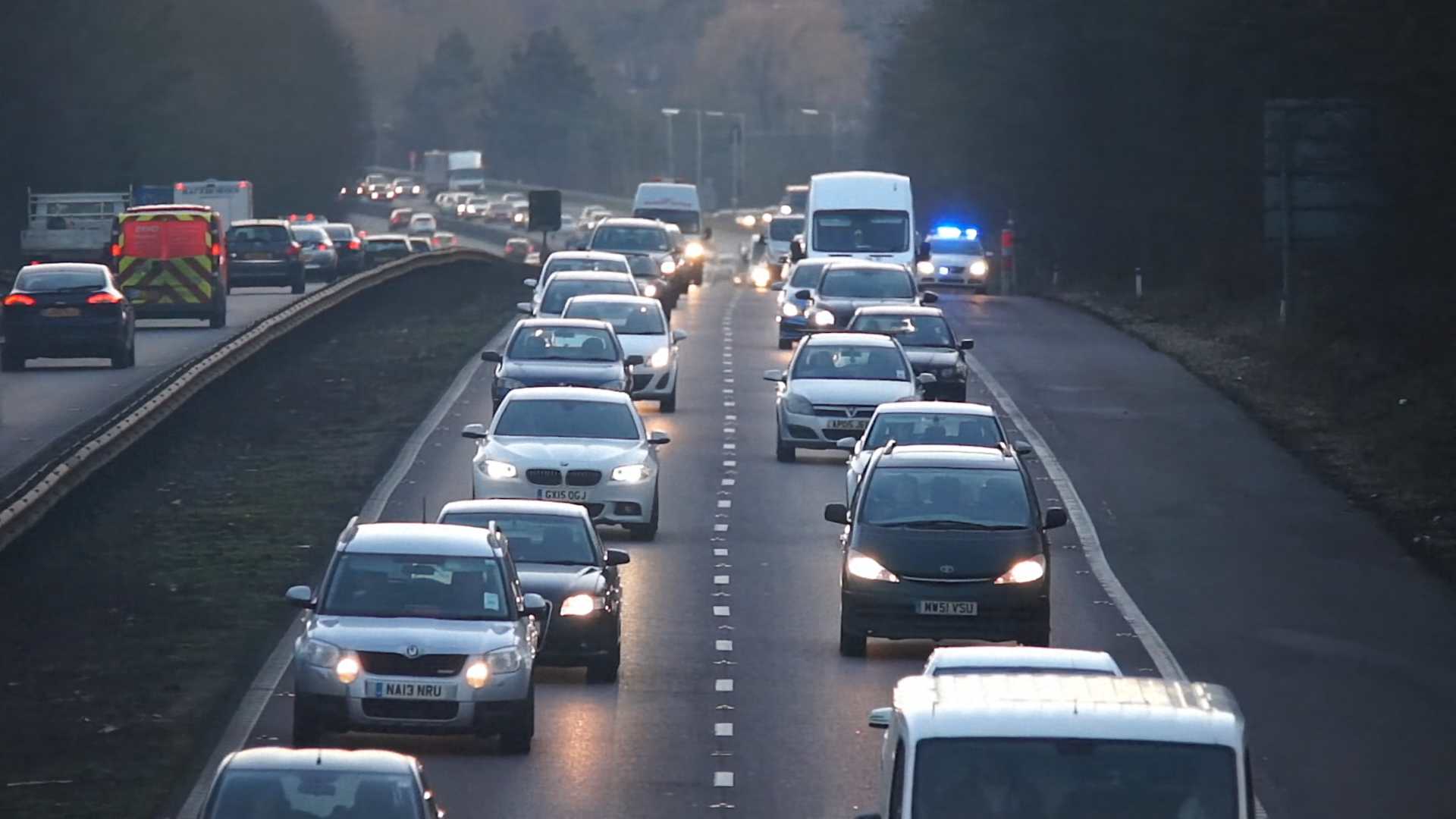} &
         \includegraphics[width=0.18\textwidth]{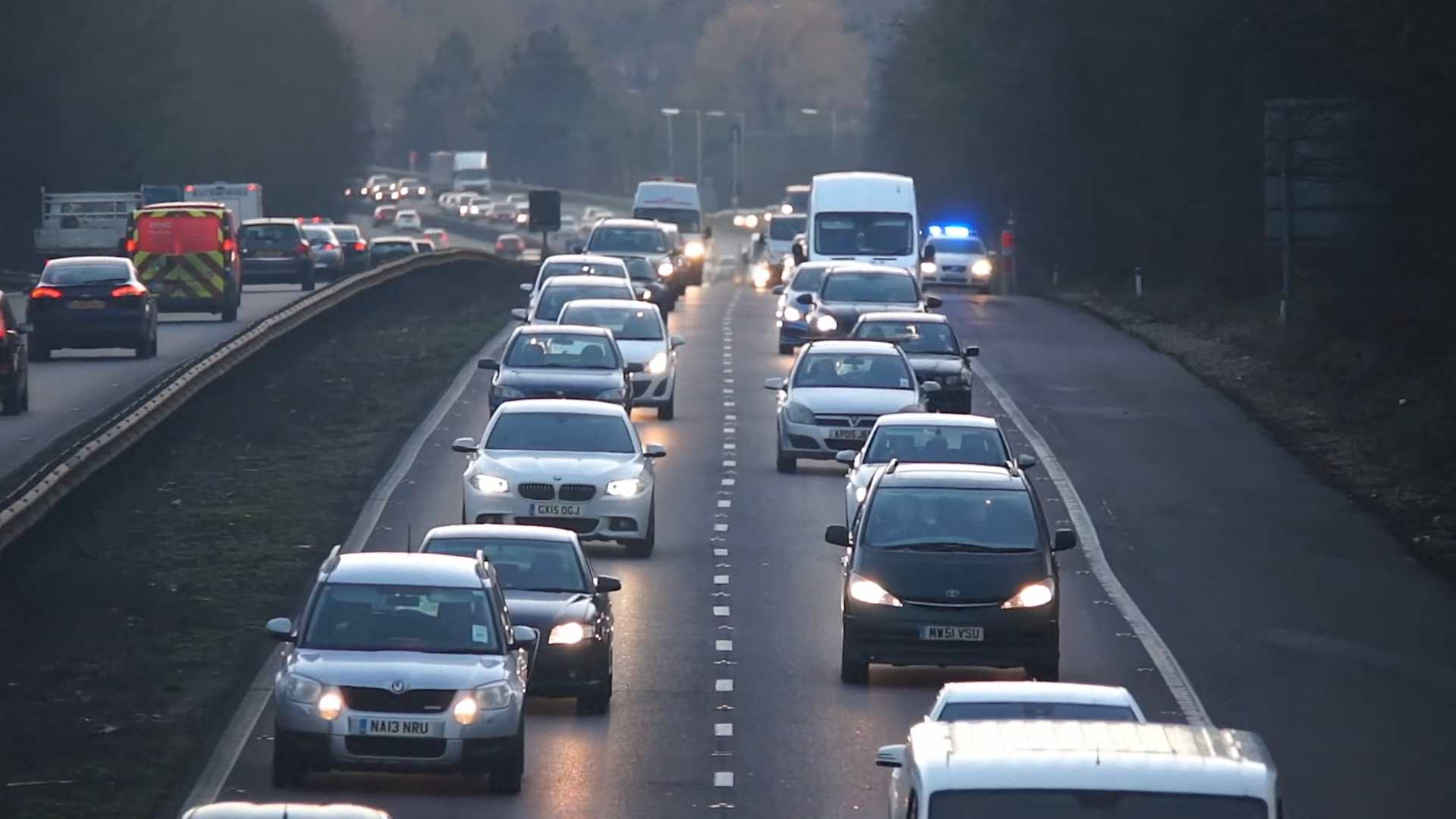} &
        \includegraphics[width=0.18\textwidth]{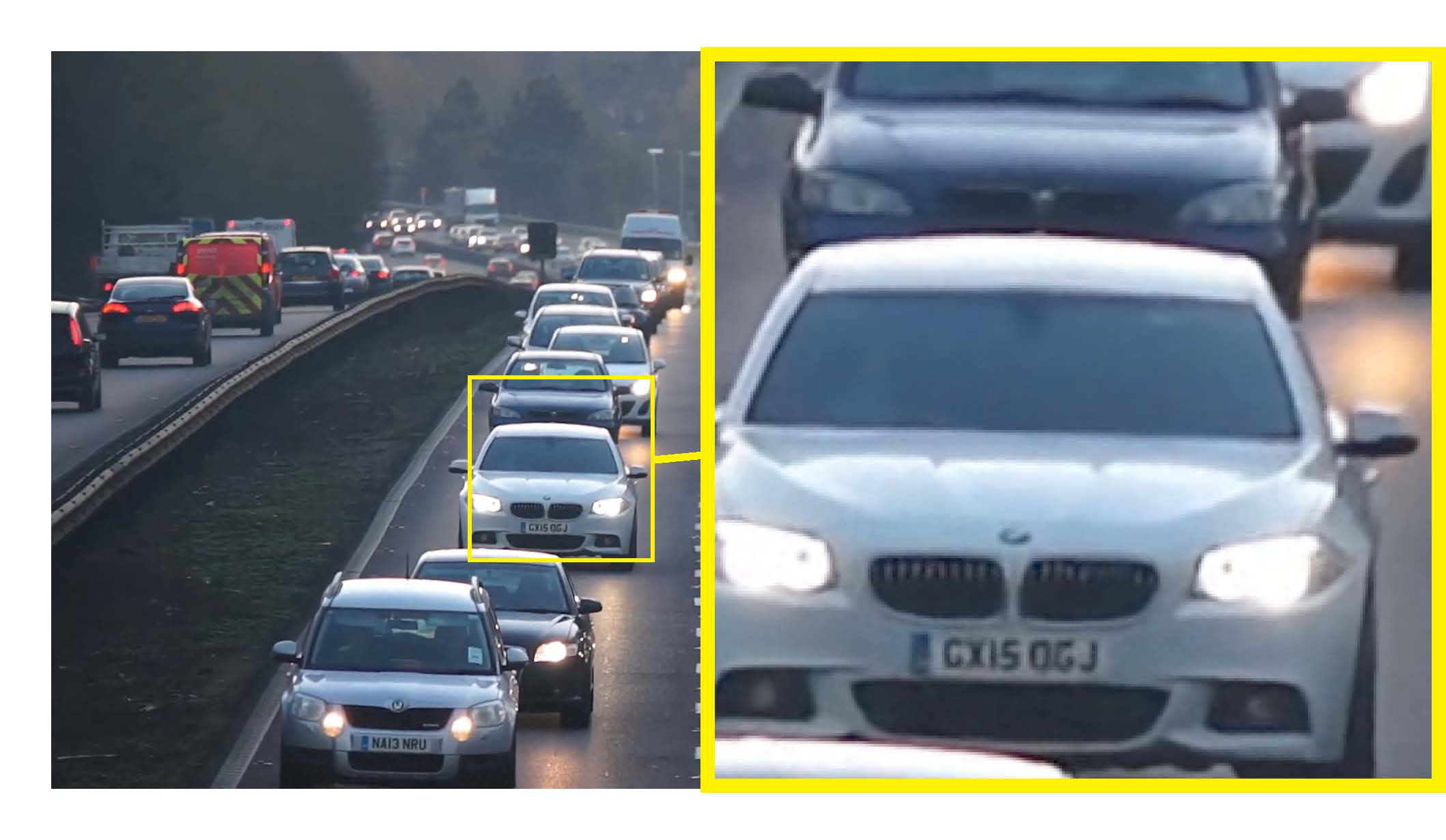} &
         \includegraphics[width=0.18\textwidth]{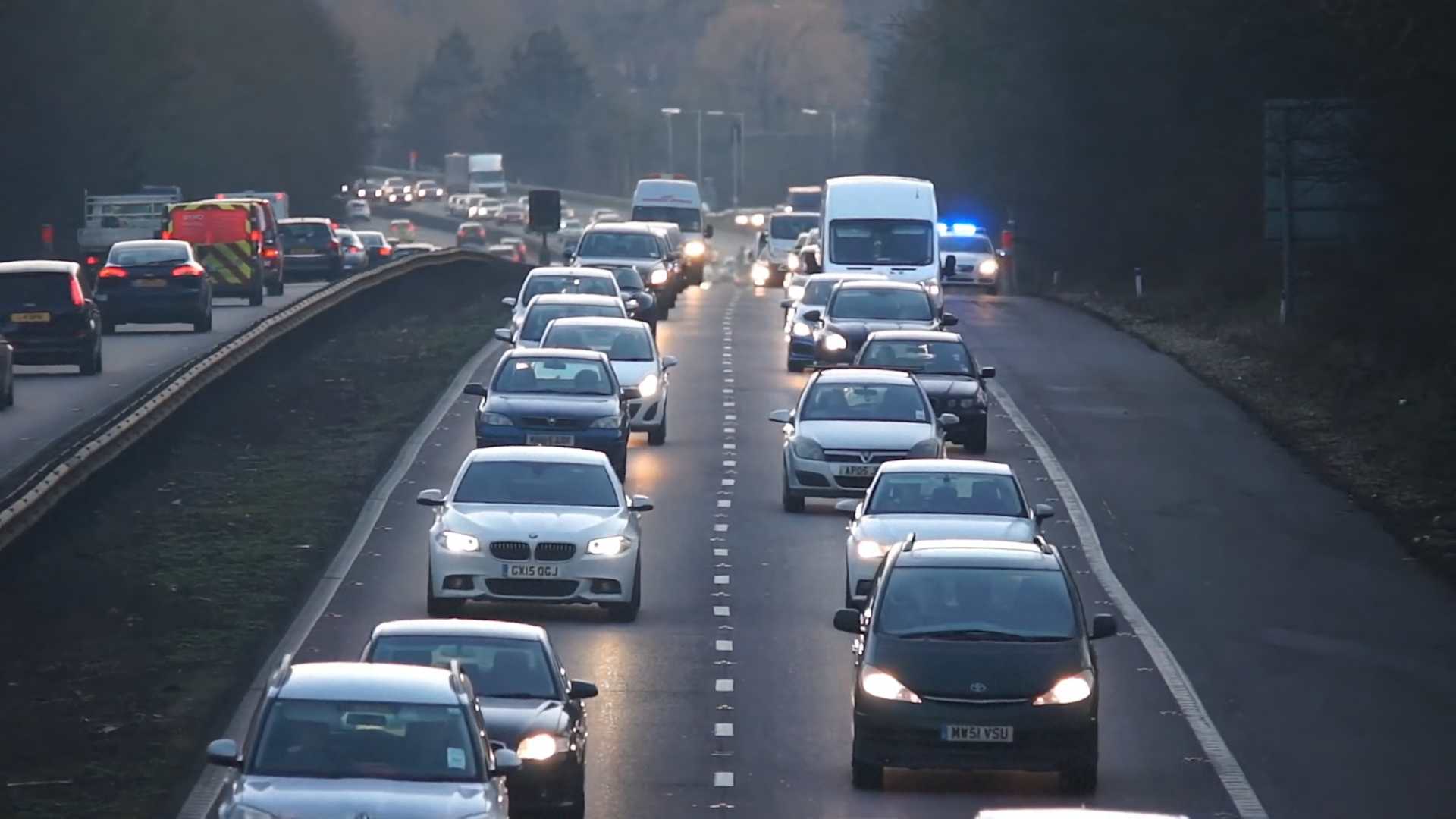} \\
         & PSNR: \SI{41.5}{dB} & PSNR: \SI{42.6}{dB} & PSNR: \SI{43.4}{dB} & PSNR: \SI{43.9}{dB} \\
         
         \rotatebox{90}{$f_\theta$ with LoRAs} &
         \includegraphics[width=0.18\textwidth]{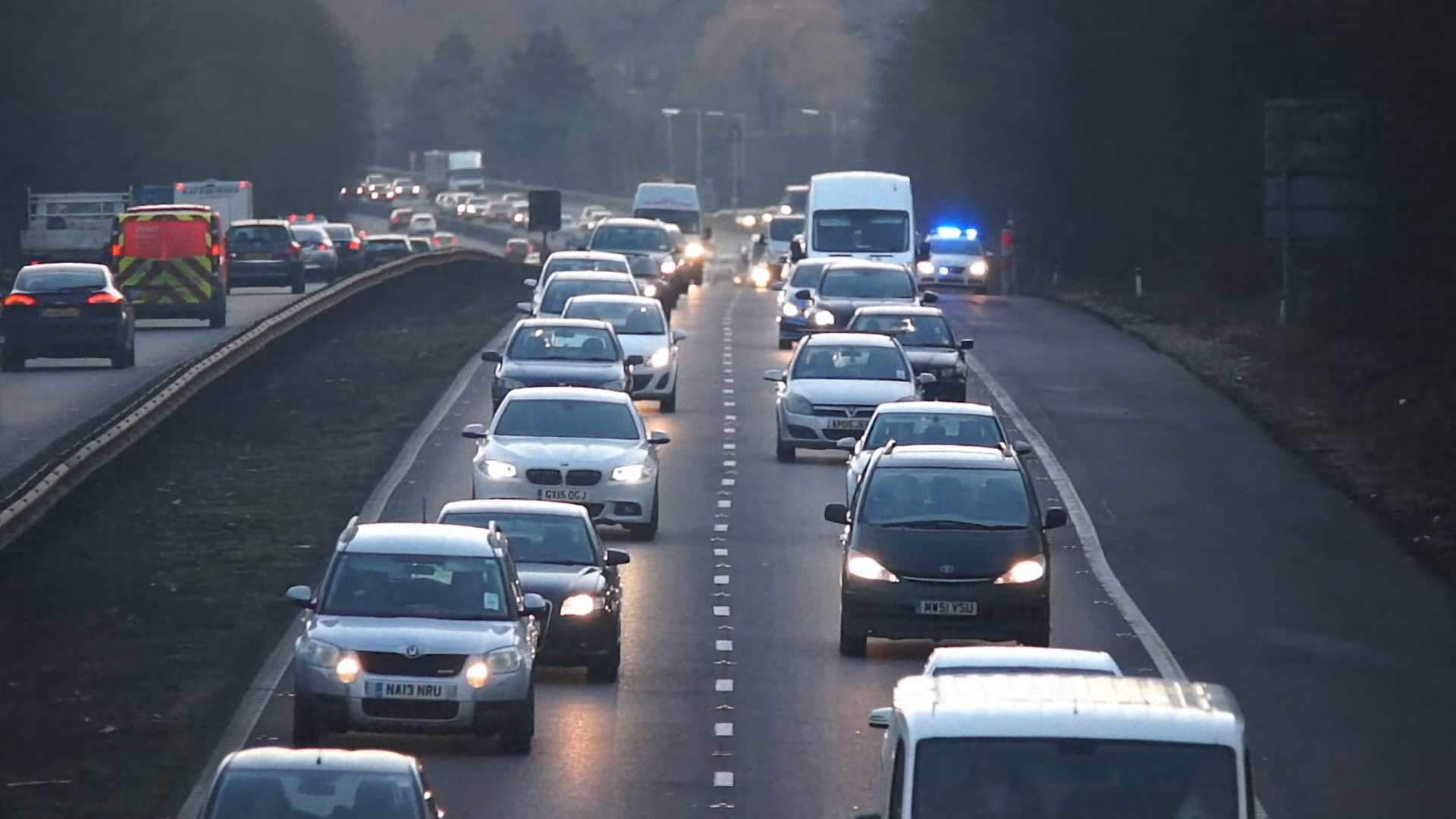} &
         \includegraphics[width=0.18\textwidth]{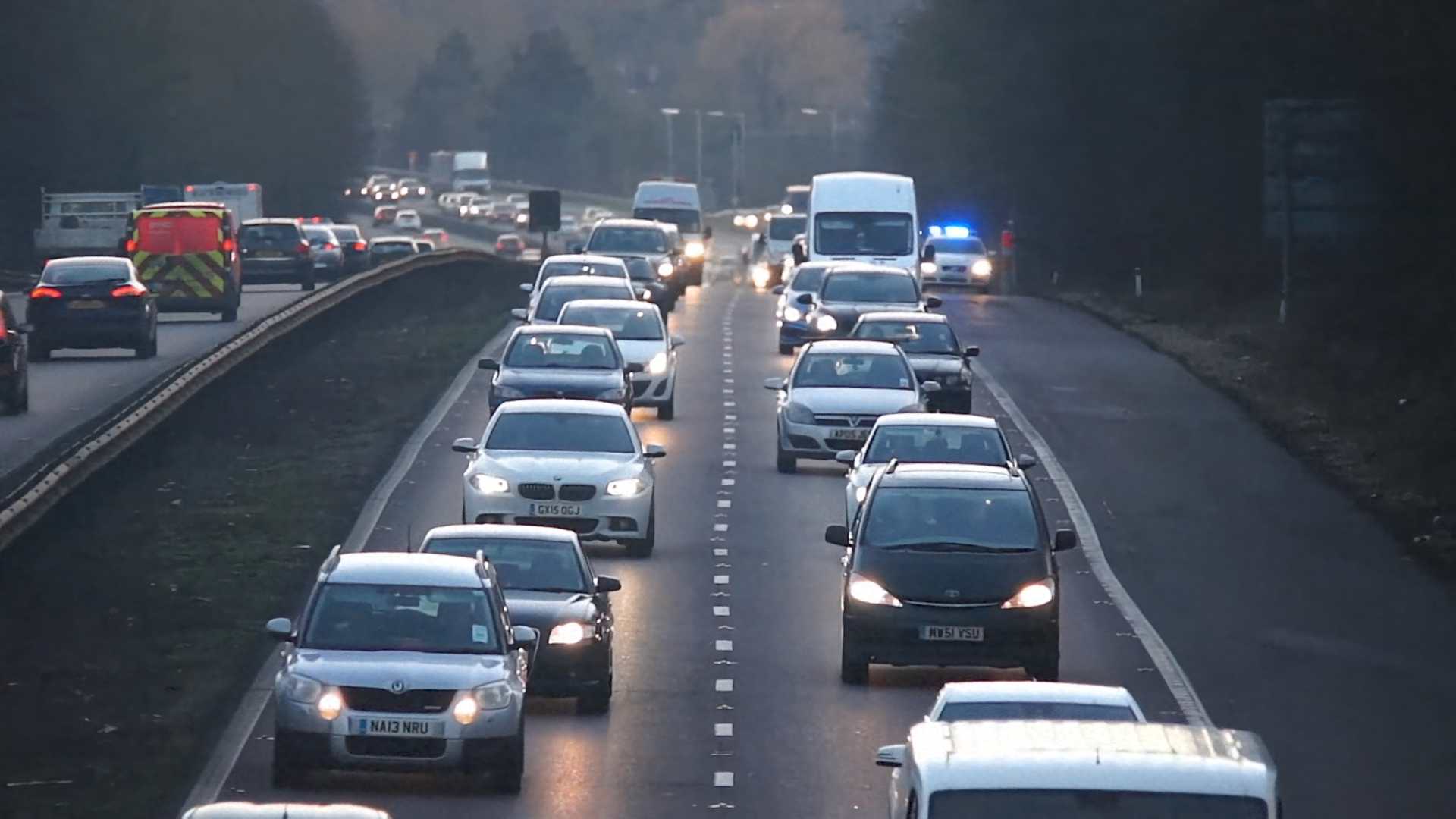} &
        \includegraphics[width=0.18\textwidth]{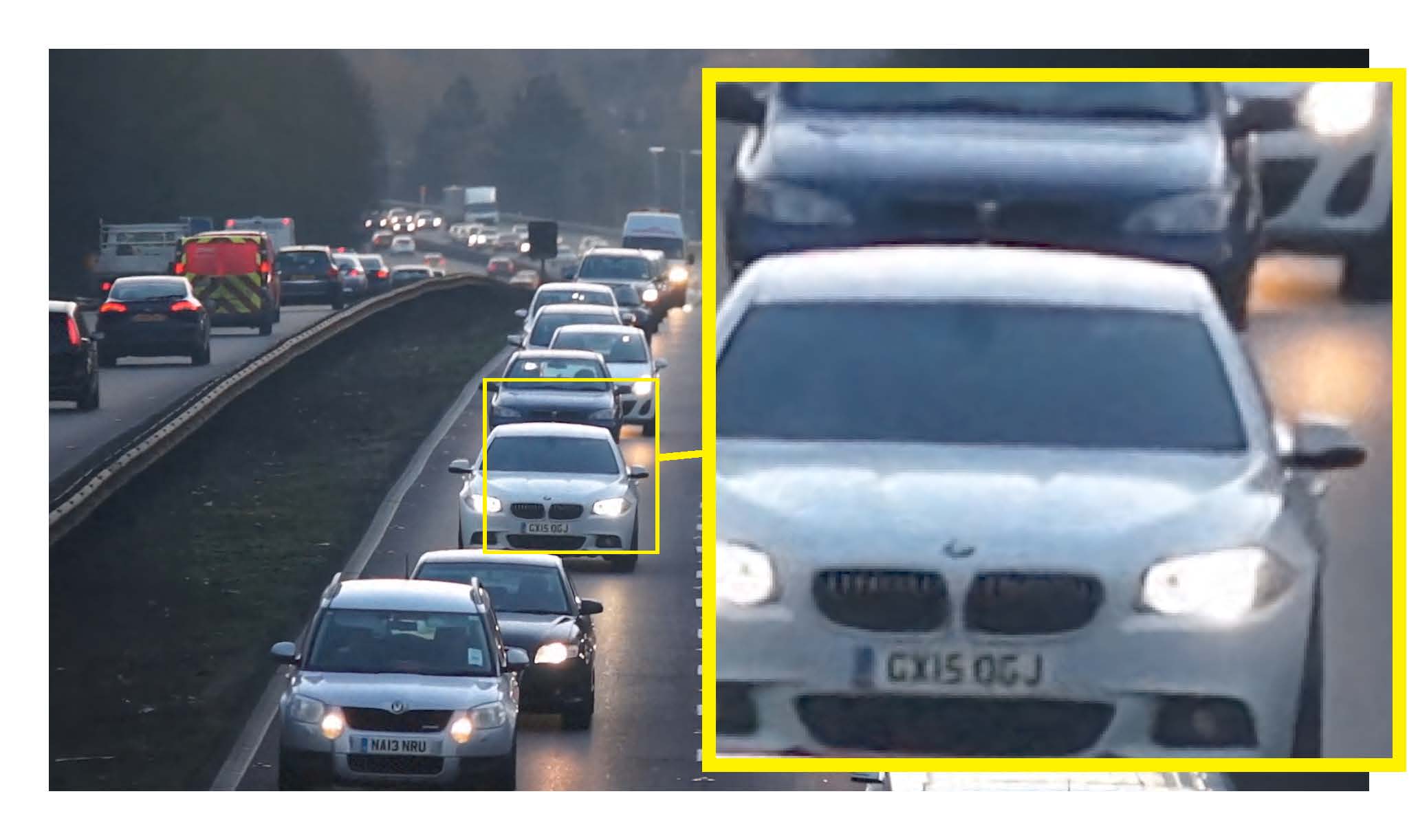} &
         \includegraphics[width=0.18\textwidth]{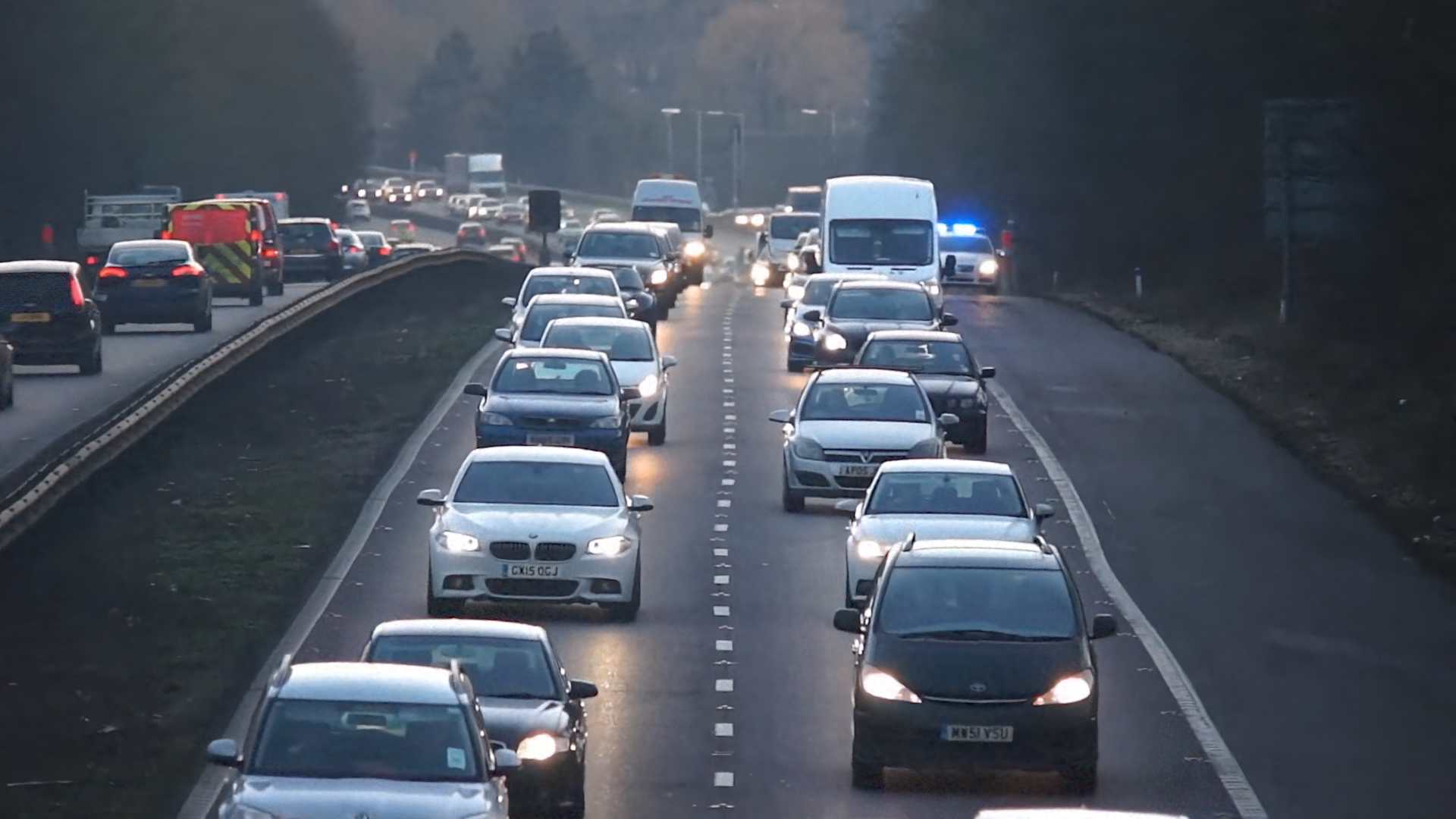} \\
         & PSNR: \SI{39.1}{dB}& PSNR: \SI{39.0}{dB} & PSNR:\SI{39.2}{dB} & PSNR:\SI{39.2}{dB} \\

     \end{tabular}
     \caption{We encode a high-resolution video as sequential LoRA updates from an initial frame (see \S\ref{sec:video}). Here, we visualize reconstructions of four frames using full fine-tuning (406.1k parameters) vs.\ LoRA (54.6k parameters) and report the reconstruction accuracy.}
     \label{fig:trafficvideo}
 \end{figure*}

\begin{figure}

	\includegraphics[width=0.45\textwidth]{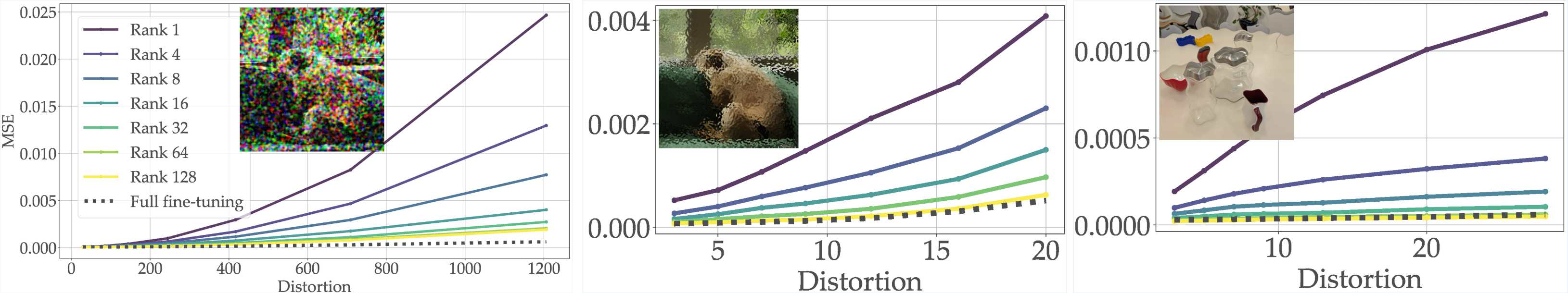}

	\caption{We synthetically vary images by adding low-frequency Gaussian noise (left), applying a glass style filter (center), and applying a wave-style distortion filter (right). The graphs in the top row show reconstruction error for increasing variation magnitudes across different LoRA ranks and full fine-tuning.}
	\label{fig:image variation magnitude, gaussian noise}
\end{figure}

\newcommand{\jellyimgwidth}{0.04\textwidth}

\begin{figure}[H]
    
    \centering
     \begin{tabular}{ c  c  c  c  c  c c }
        $r=1$ & $r=4$ & $r=8$ & $r=16$ & $r=32$ & $r=64$  & FT \\
         \hline         

          0.940 & 0.981 & 0.985 & 0.989 & 0.991 & 0.992 & 0.997 \\
         \includegraphics[width=\jellyimgwidth]{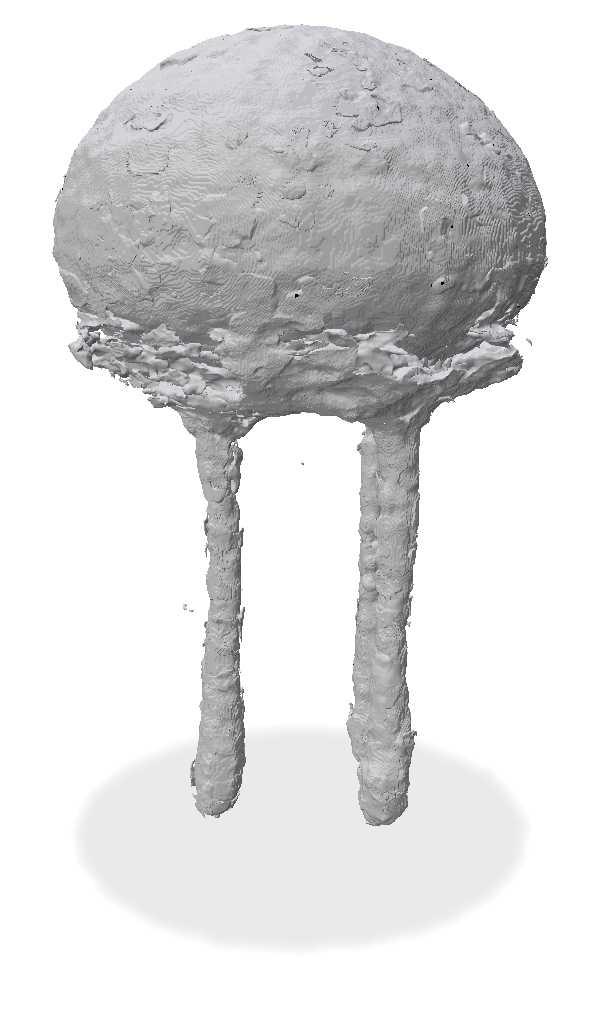} & %
         \includegraphics[width=\jellyimgwidth]{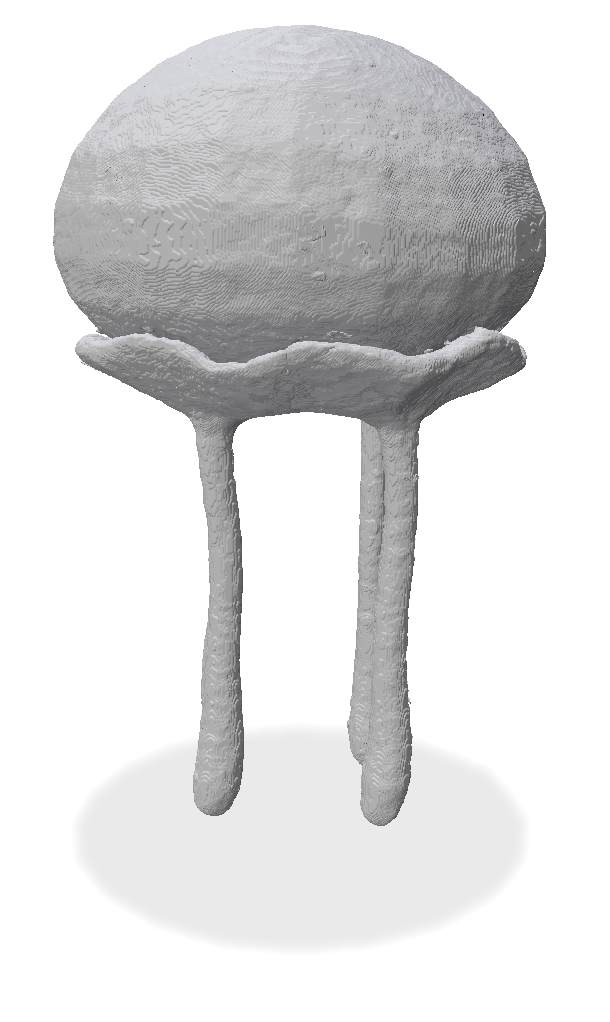} & %
         \includegraphics[width=\jellyimgwidth]{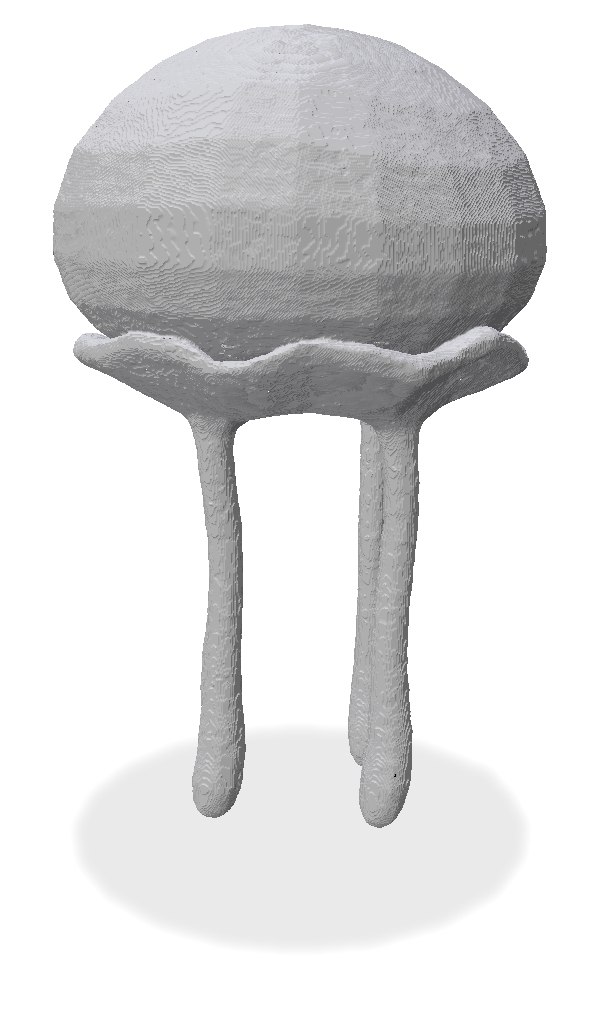} & %
         \includegraphics[width=\jellyimgwidth]{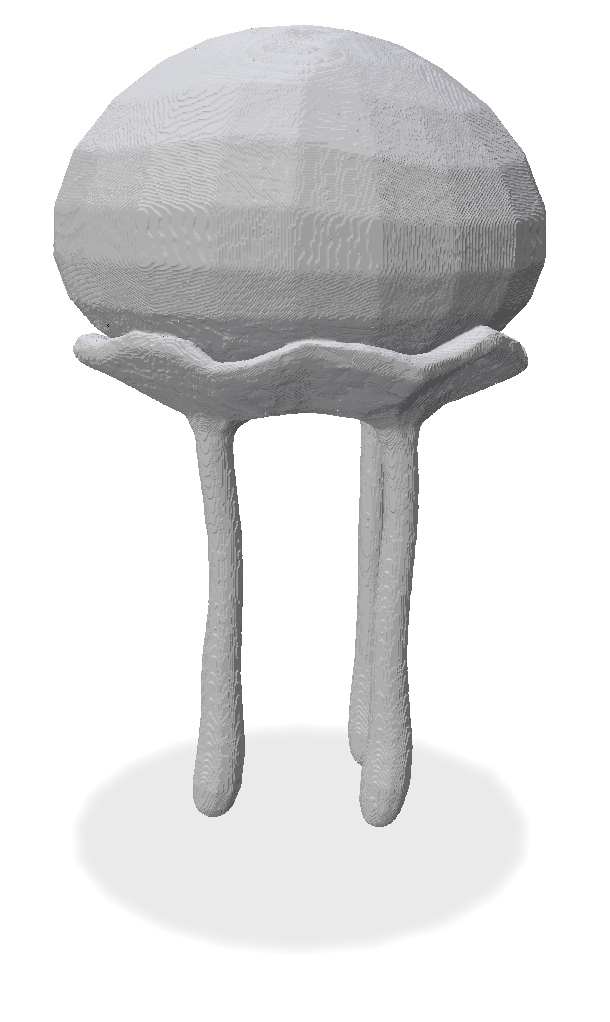} & %
         \includegraphics[width=\jellyimgwidth]{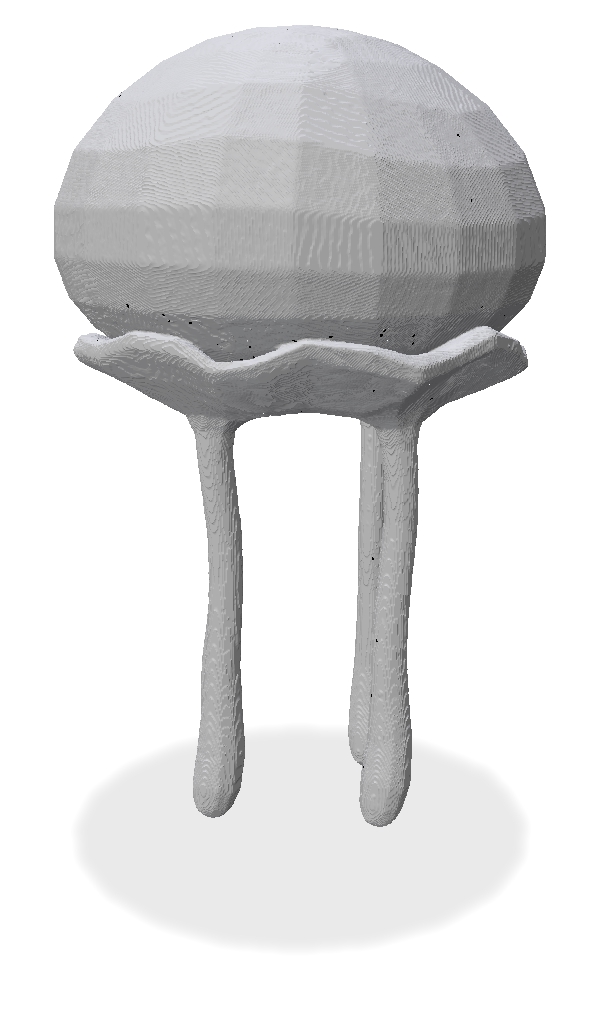} & %
         \includegraphics[width=\jellyimgwidth]{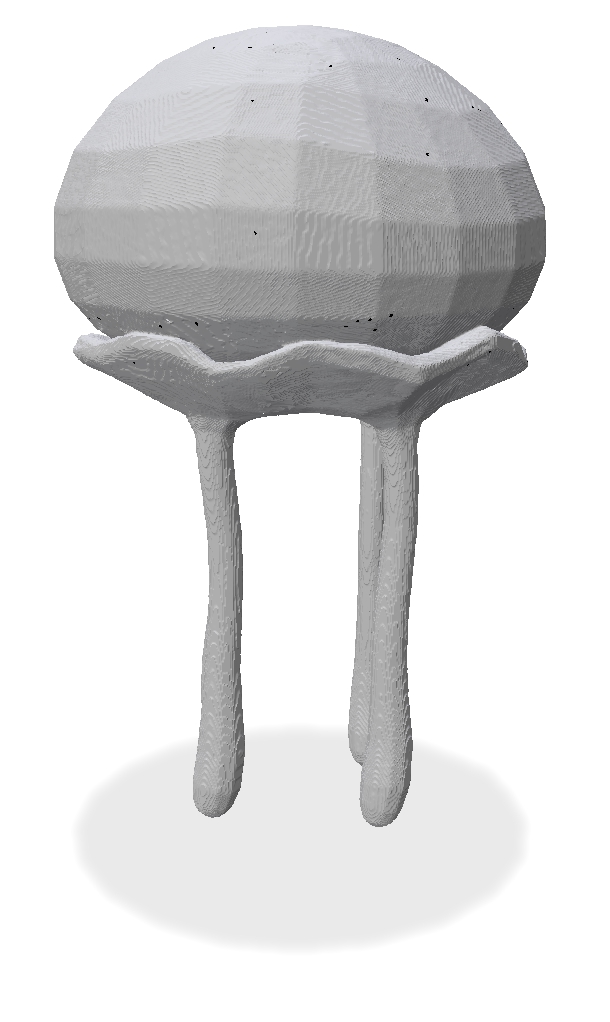} & %
         \includegraphics[width=\jellyimgwidth]{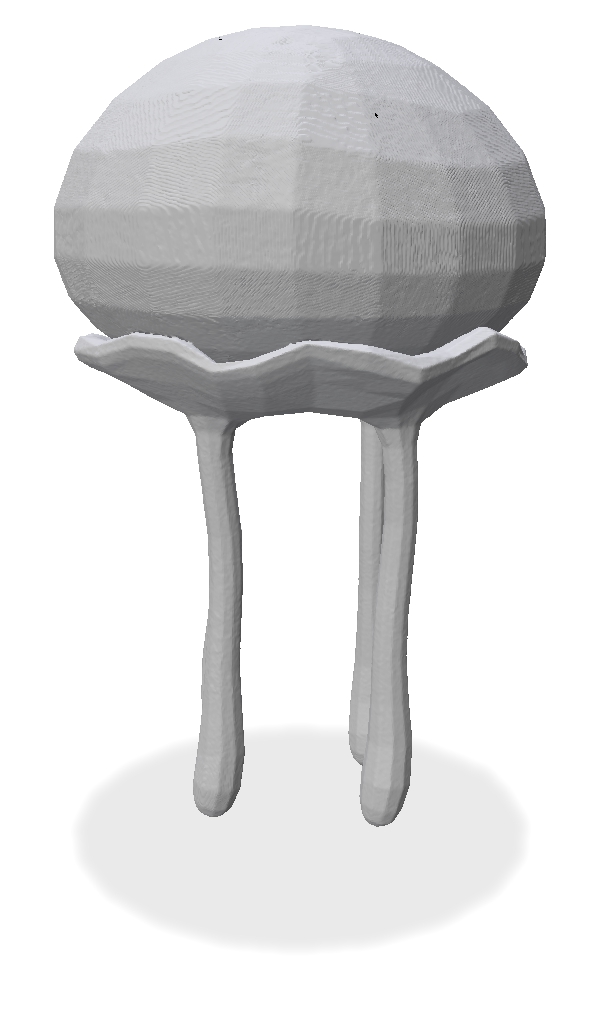} \\ %
         \hline         

          0.934 & 0.978 & 0.988 & 0.991 & 0.992 & 0.992 & 0.998 \\
         \includegraphics[width=\jellyimgwidth]{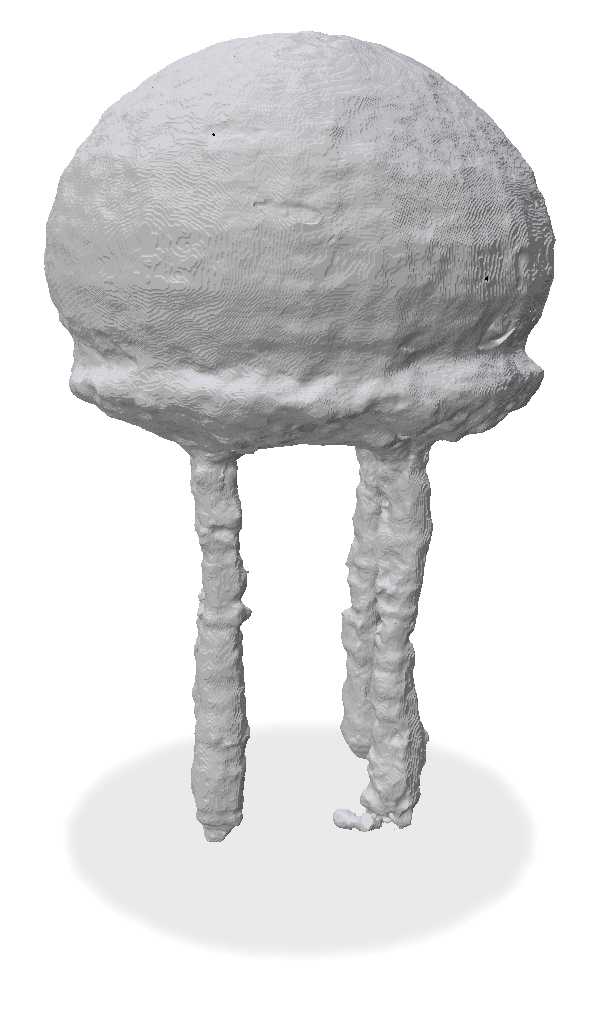} & %
         \includegraphics[width=\jellyimgwidth]{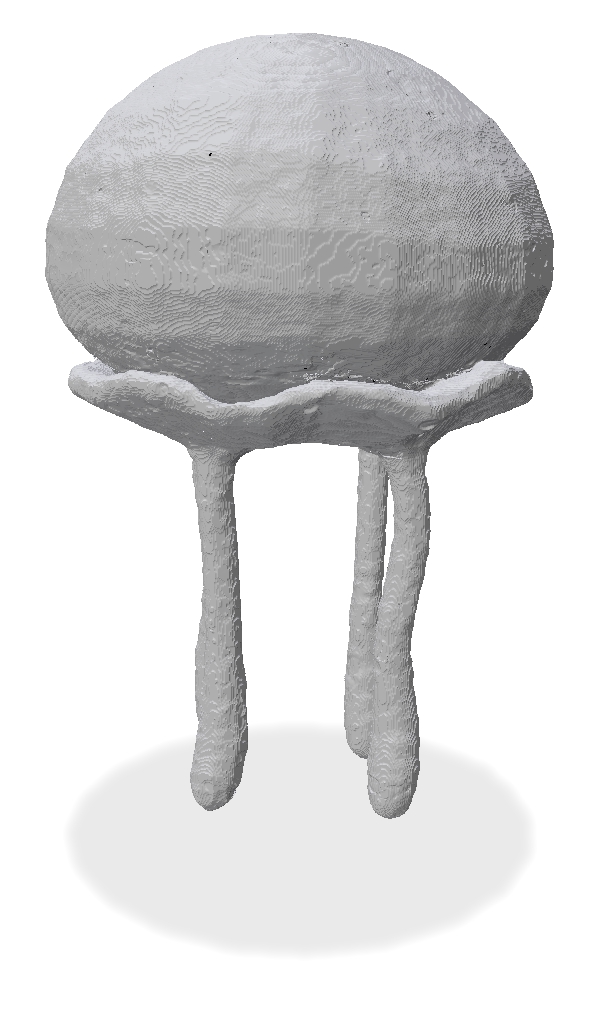} & %
         \includegraphics[width=\jellyimgwidth]{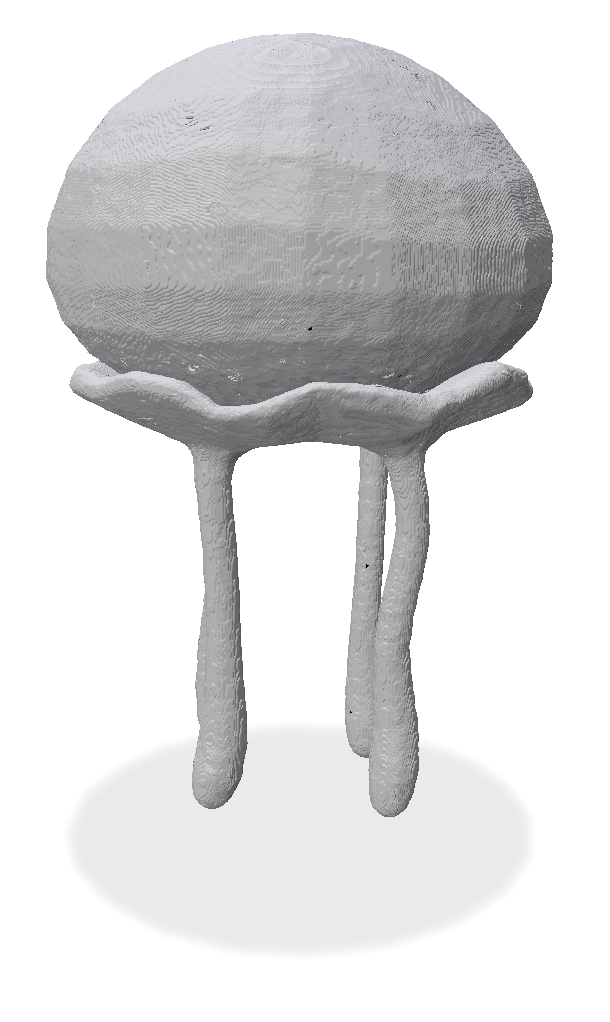} & %
         \includegraphics[width=\jellyimgwidth]{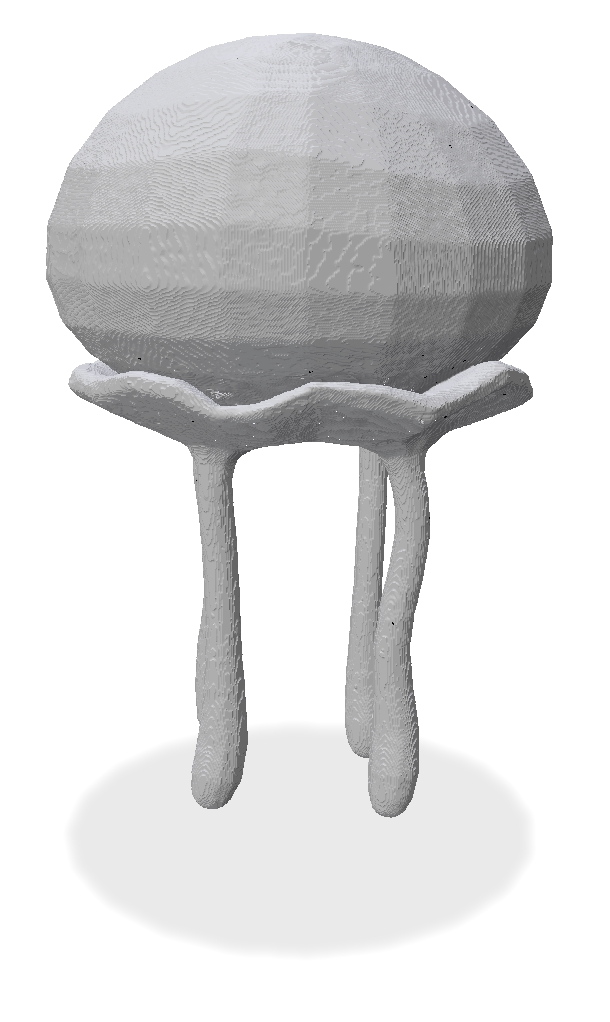} & %
         \includegraphics[width=\jellyimgwidth]{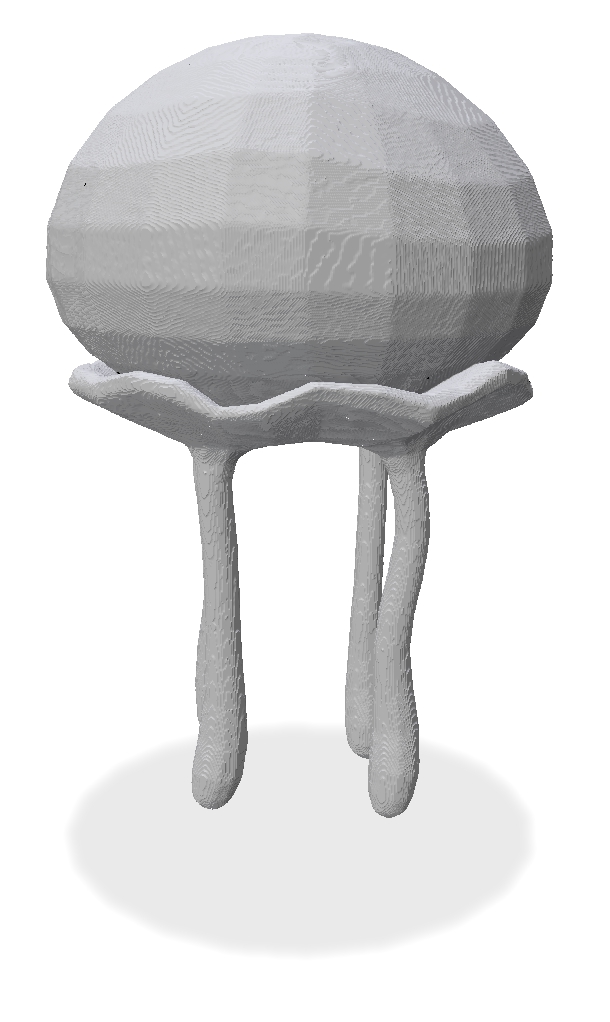} & %
         \includegraphics[width=\jellyimgwidth]{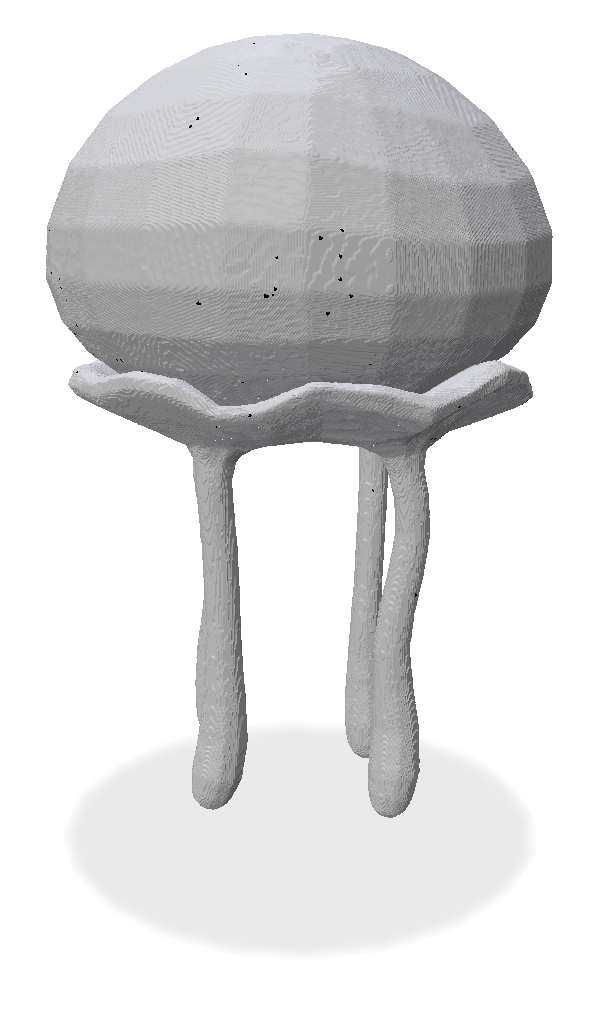} & %
         \includegraphics[width=\jellyimgwidth]{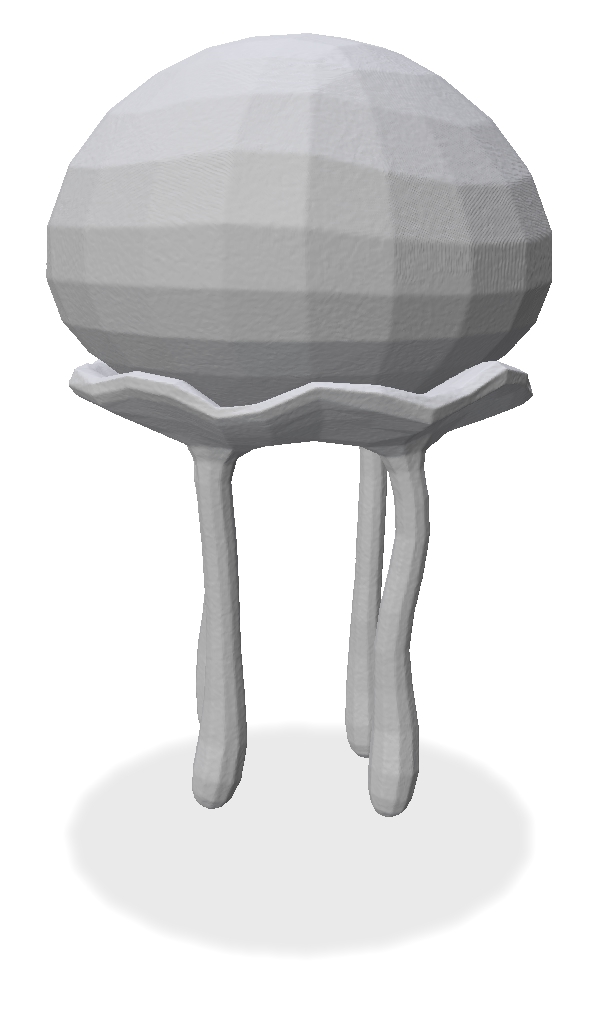} \\ %
         \hline         

          0.909 & 0.980 & 0.985 & 0.991 & 0.993 & 0.991 & 0.998 \\
         \includegraphics[width=\jellyimgwidth]{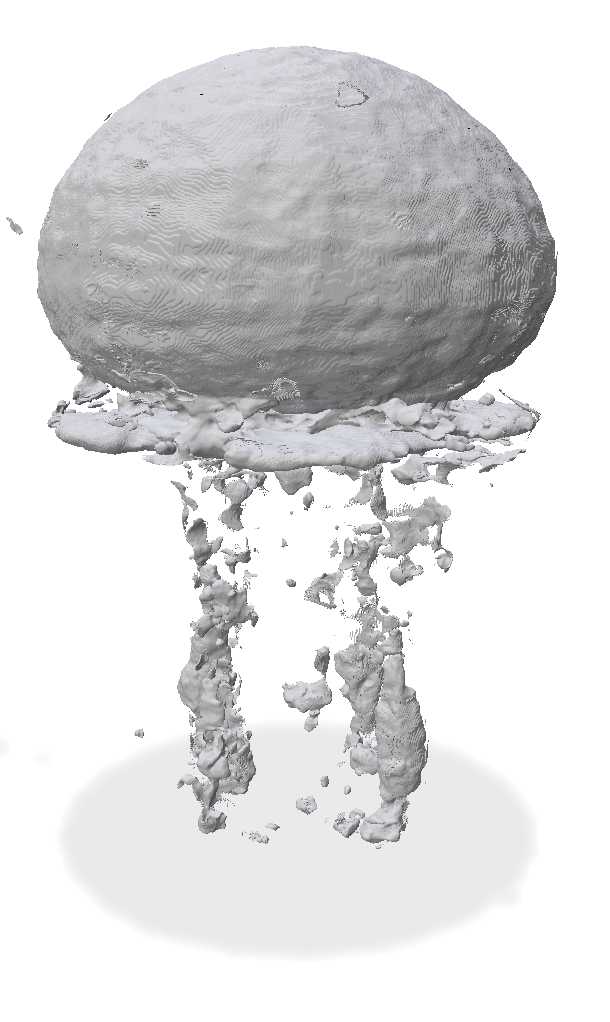} & %
         \includegraphics[width=\jellyimgwidth]{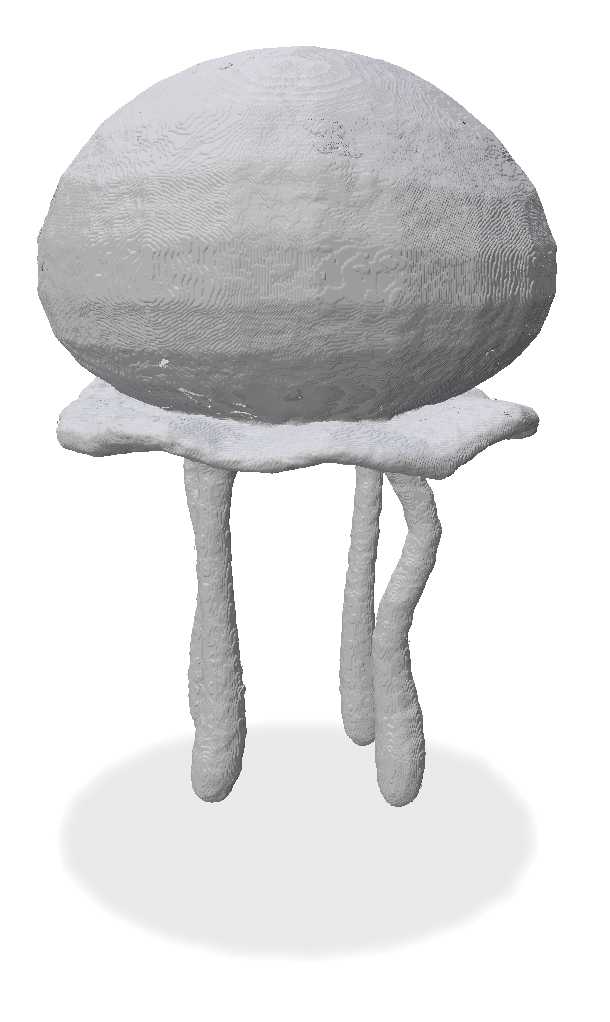} & %
         \includegraphics[width=\jellyimgwidth]{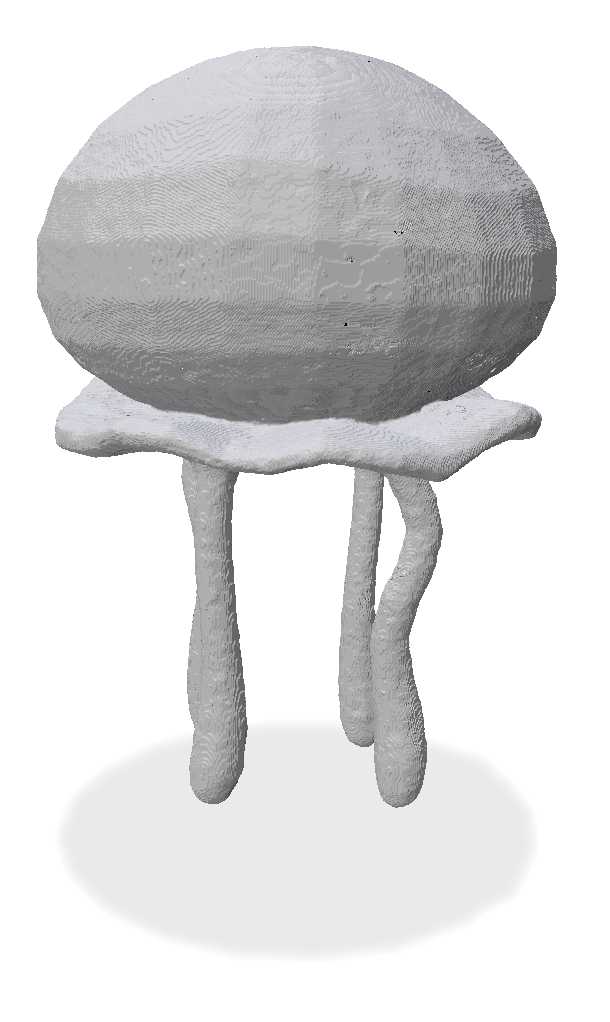} & %
         \includegraphics[width=\jellyimgwidth]{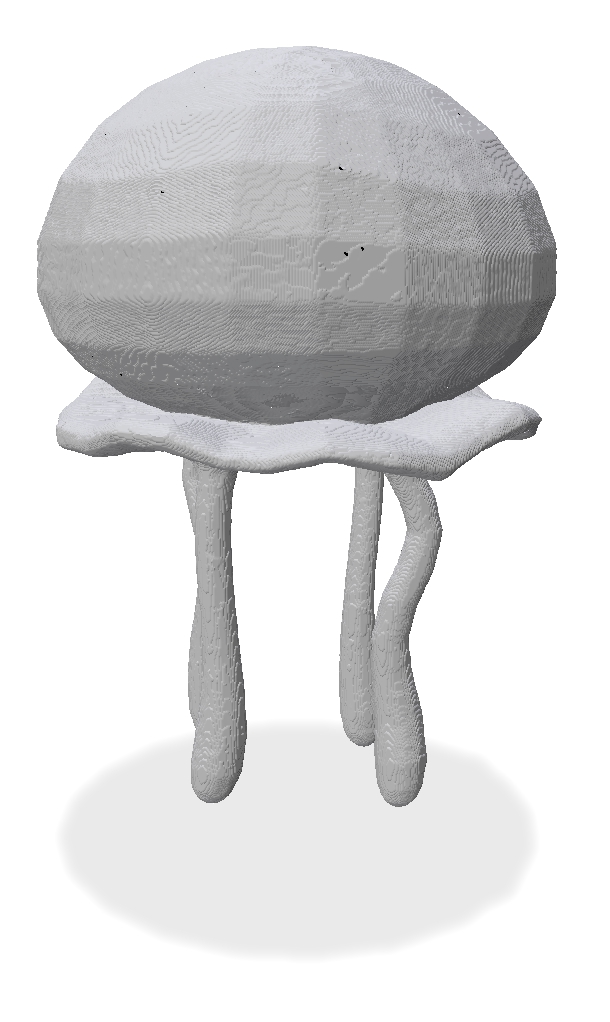} & %
         \includegraphics[width=\jellyimgwidth]{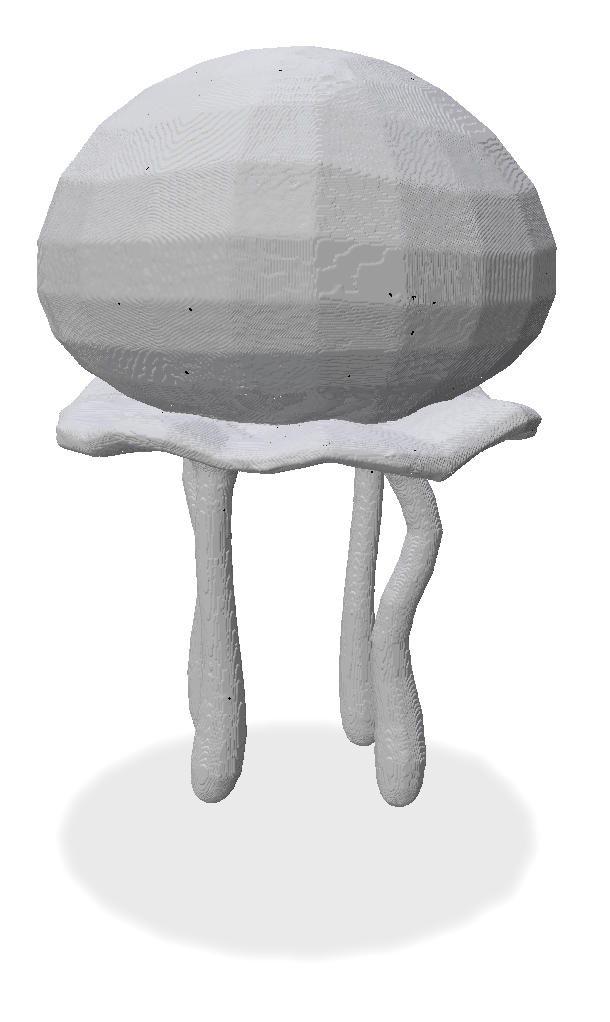} & %
         \includegraphics[width=\jellyimgwidth]{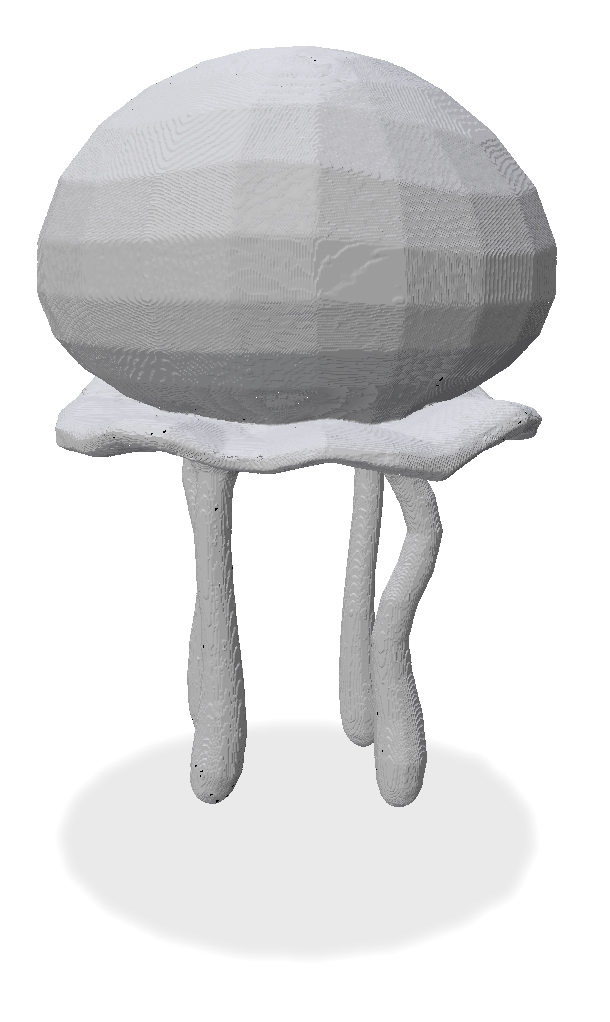} & %
         \includegraphics[width=\jellyimgwidth]{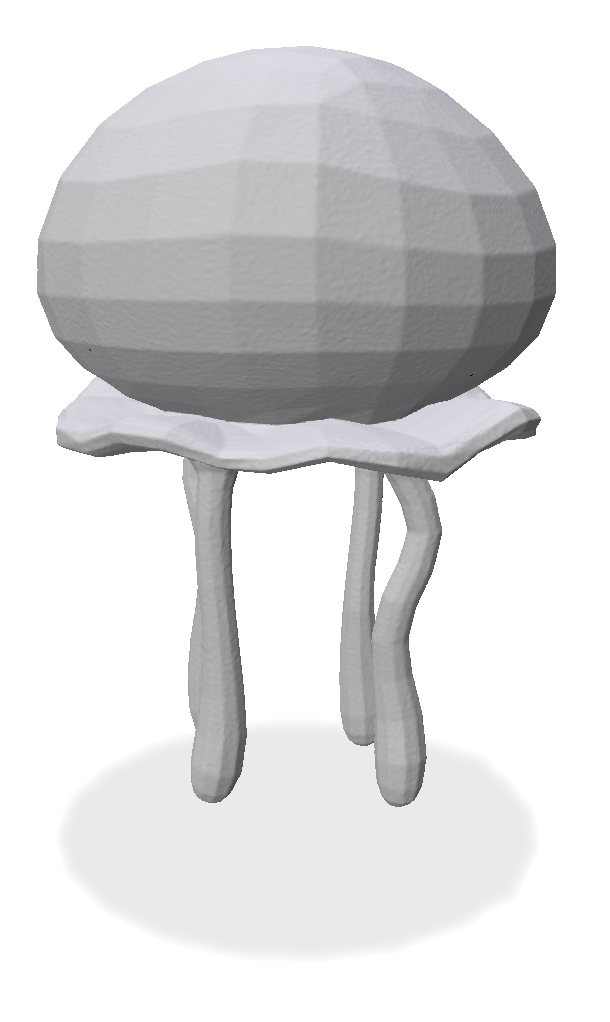} \\ %
         \hline         

          0.891 & 0.987 & 0.988 & 0.992 & 0.993 & 0.993 & 0.998 \\
         \includegraphics[width=\jellyimgwidth]{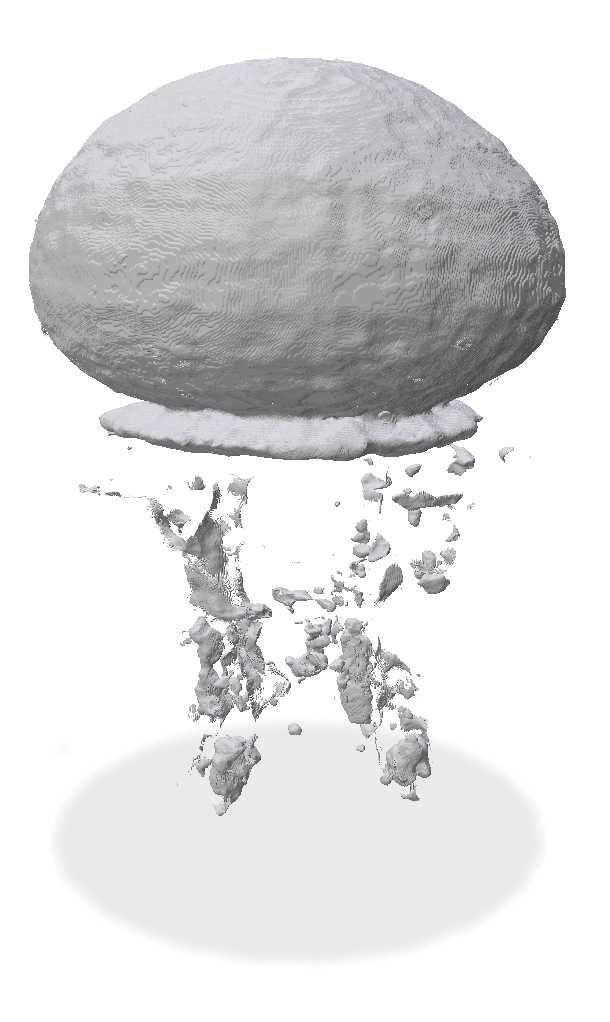} & %
         \includegraphics[width=\jellyimgwidth]{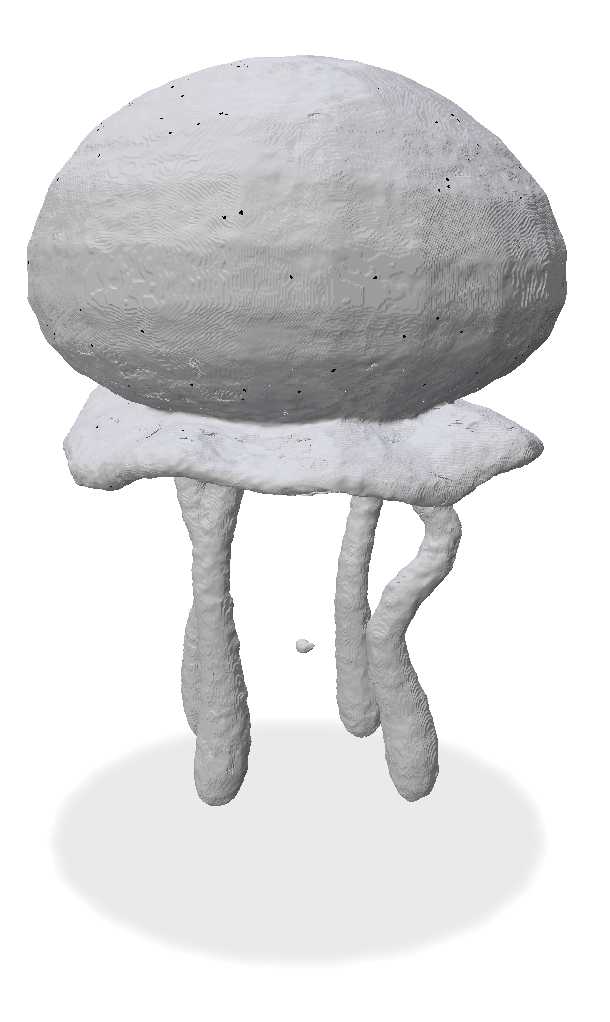} & %
         \includegraphics[width=\jellyimgwidth]{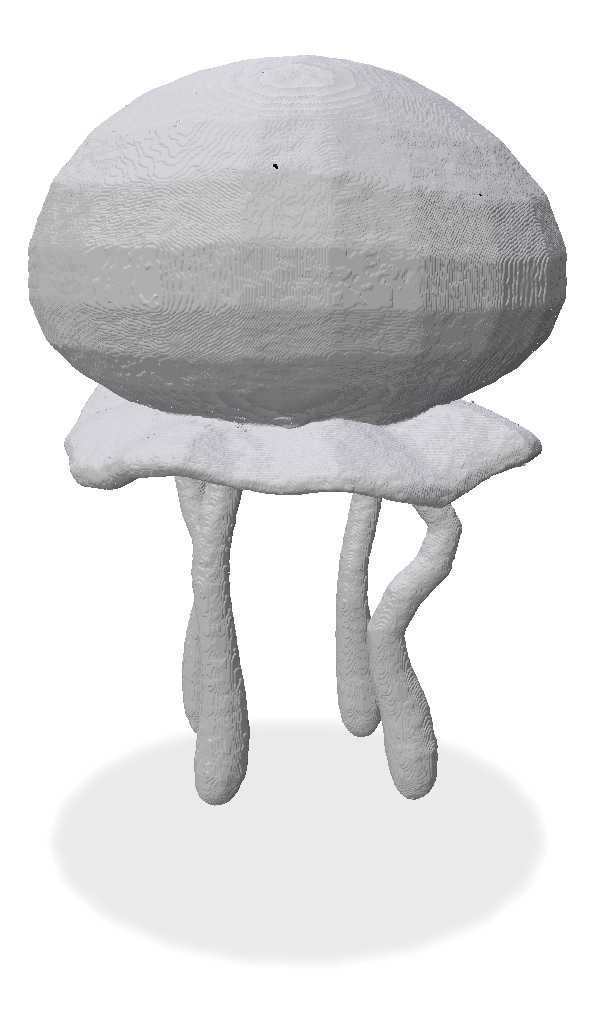} & %
         \includegraphics[width=\jellyimgwidth]{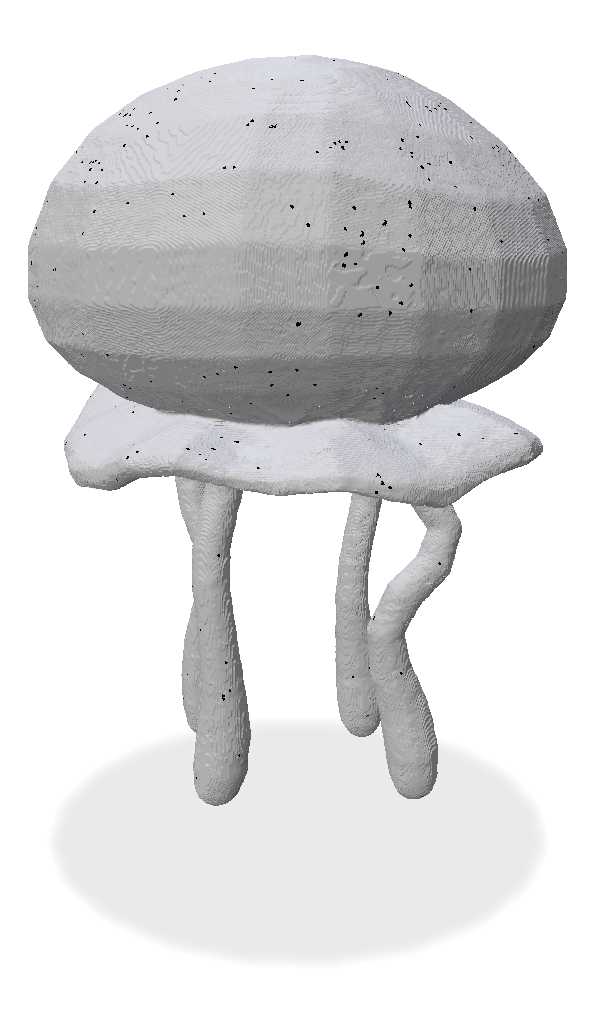} & %
         \includegraphics[width=\jellyimgwidth]{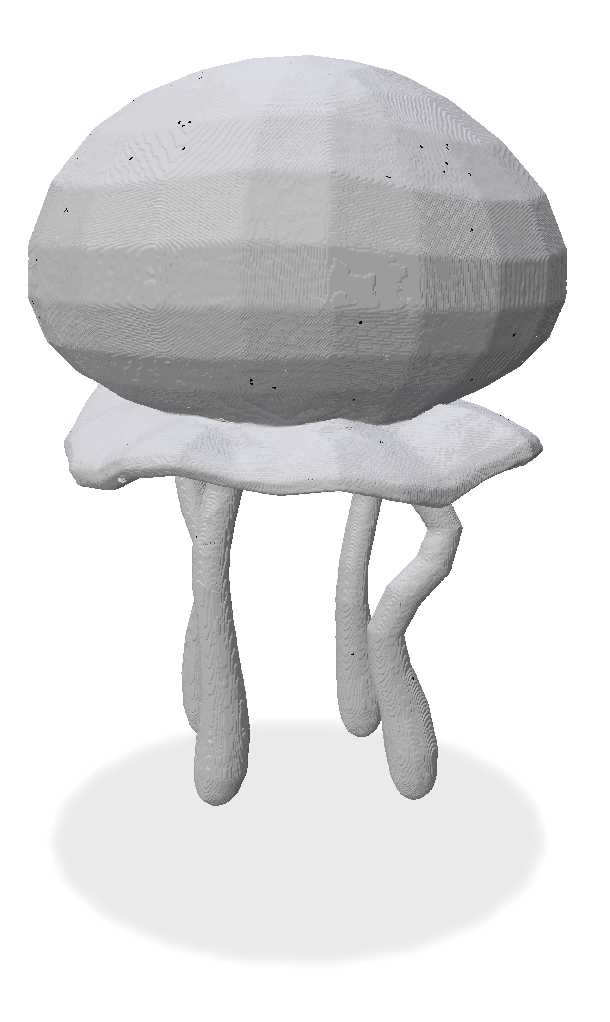} & %
         \includegraphics[width=\jellyimgwidth]{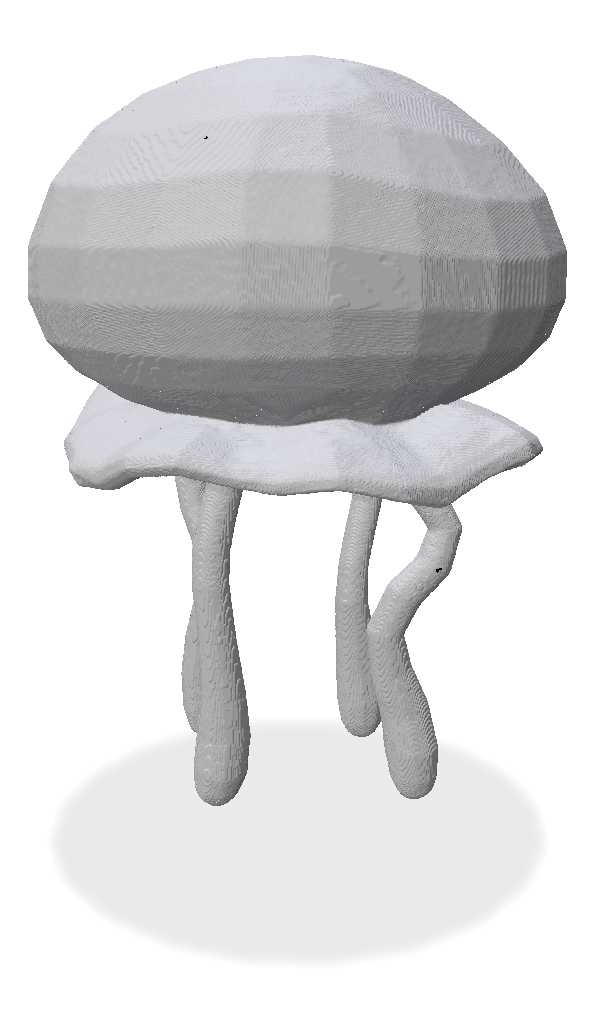} & %
         \includegraphics[width=\jellyimgwidth]{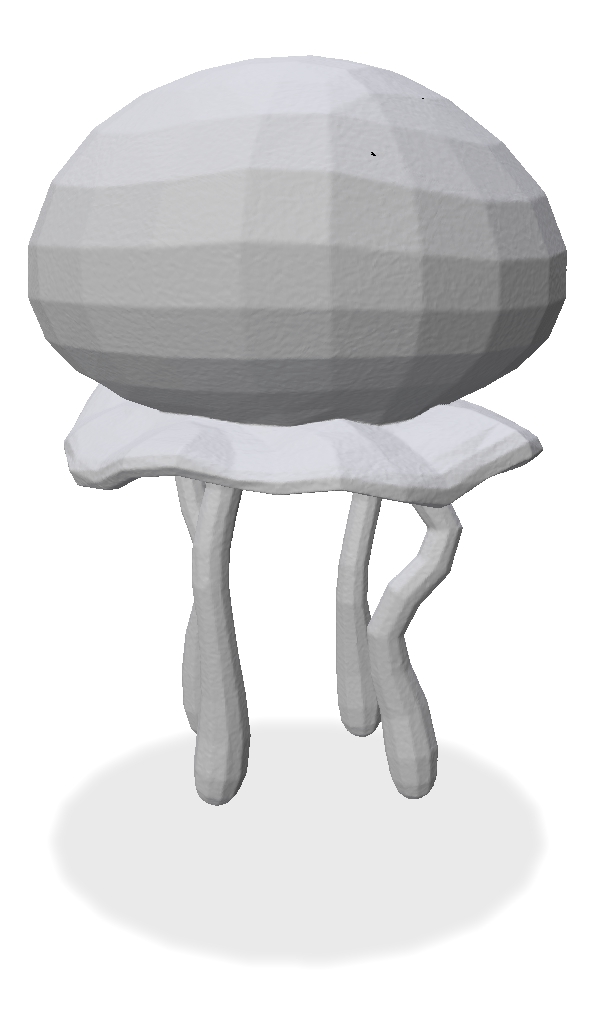} \\ %
         \hline         

          0.870 & 0.979 & 0.987 & 0.990 & 0.992 & 0.994 & 0.998 \\
         \includegraphics[width=\jellyimgwidth]{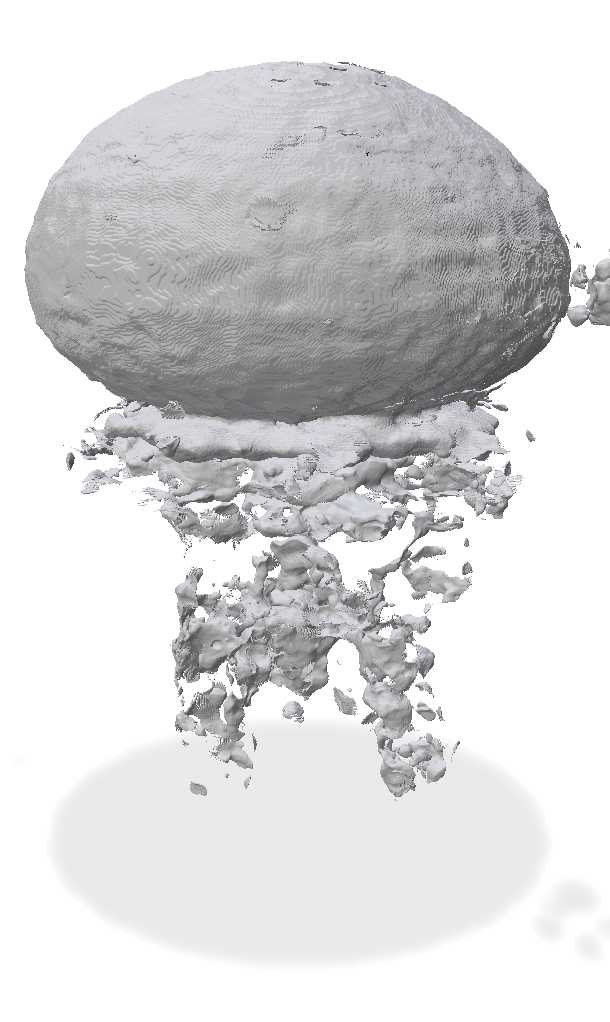} & %
         \includegraphics[width=\jellyimgwidth]{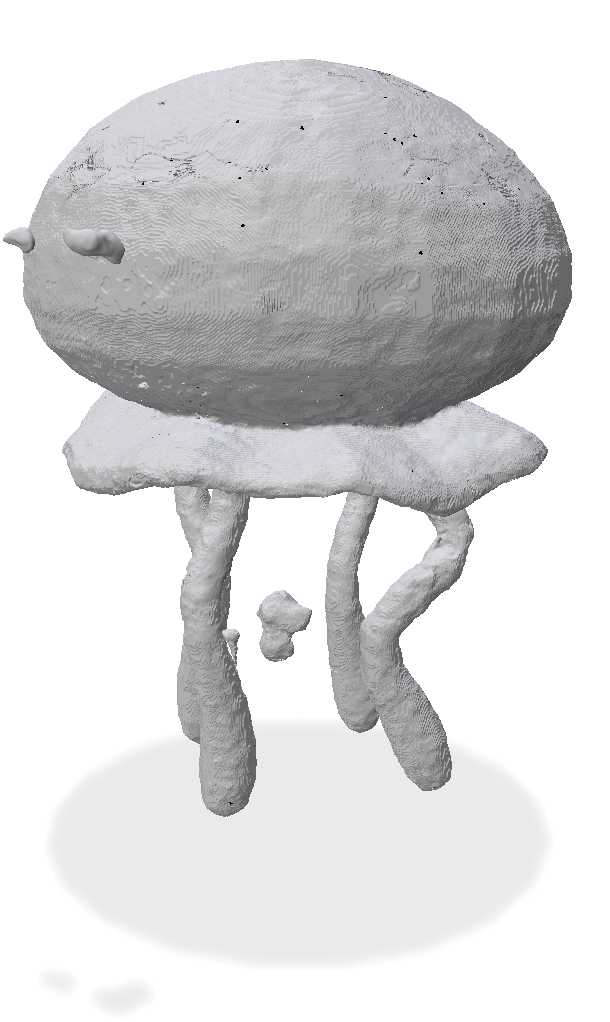} & %
         \includegraphics[width=\jellyimgwidth]{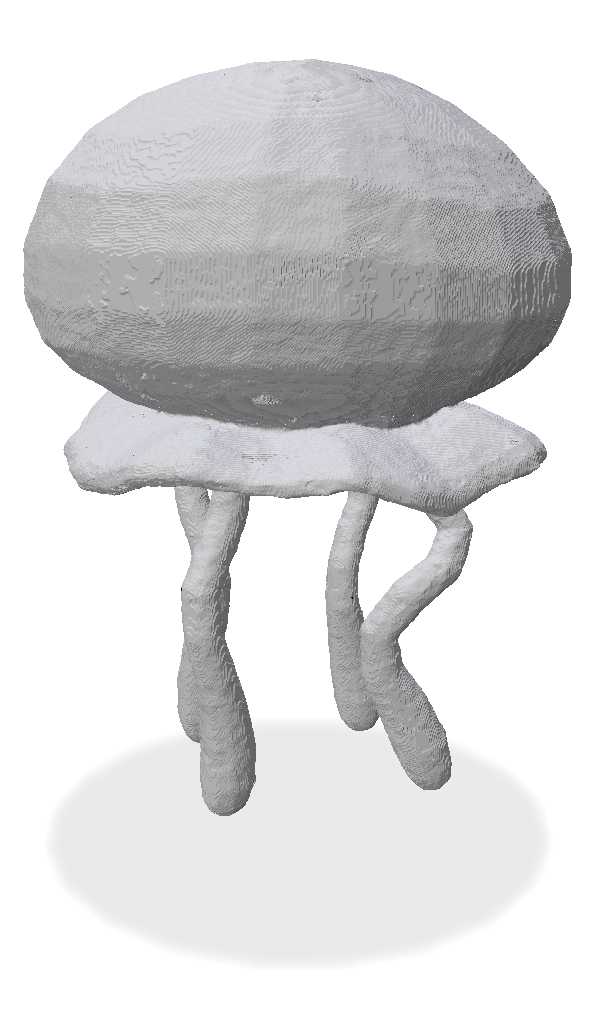} & %
         \includegraphics[width=\jellyimgwidth]{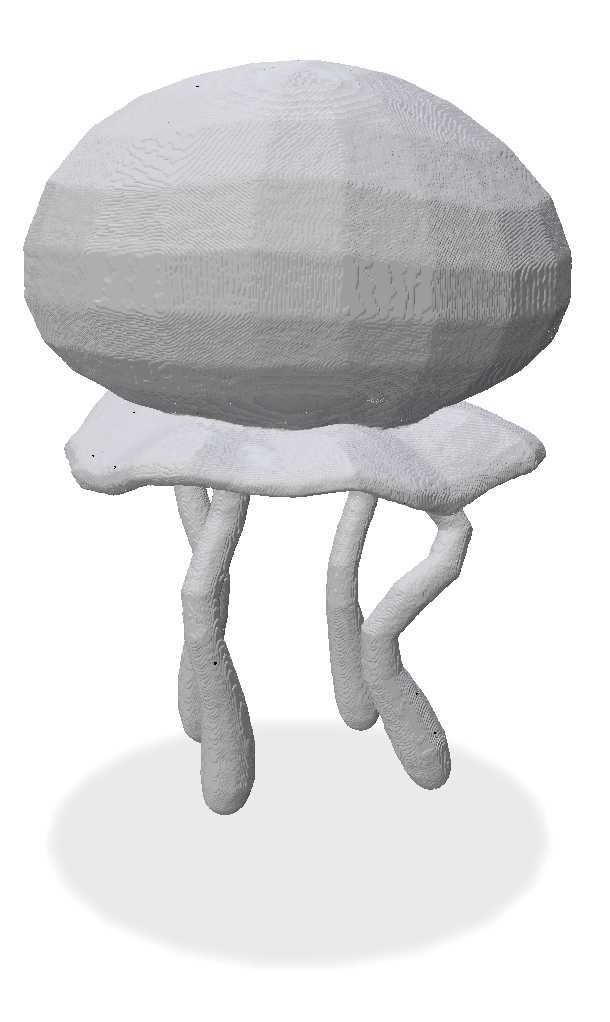} & %
         \includegraphics[width=\jellyimgwidth]{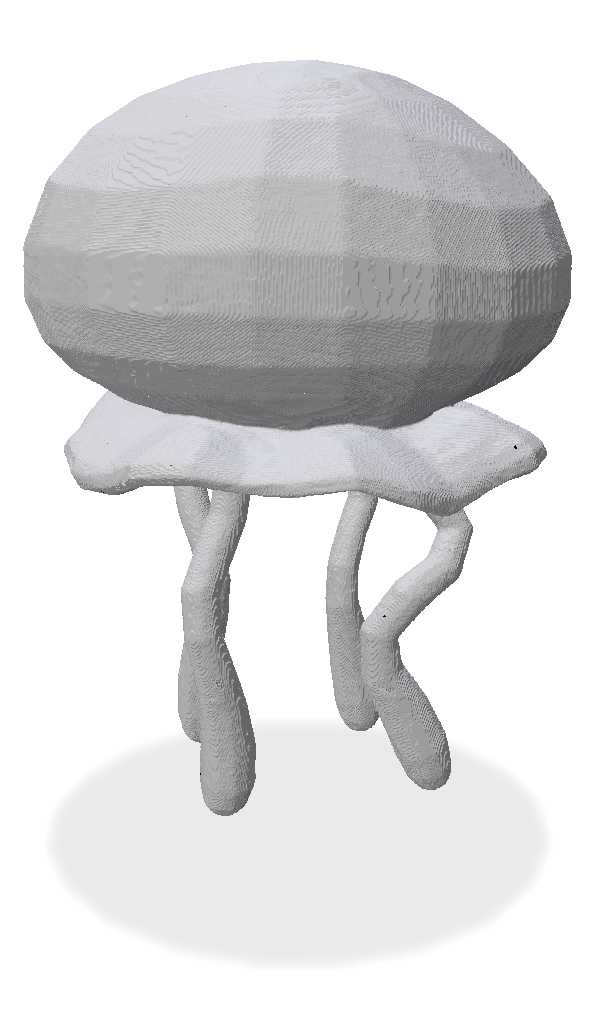} & %
         \includegraphics[width=\jellyimgwidth]{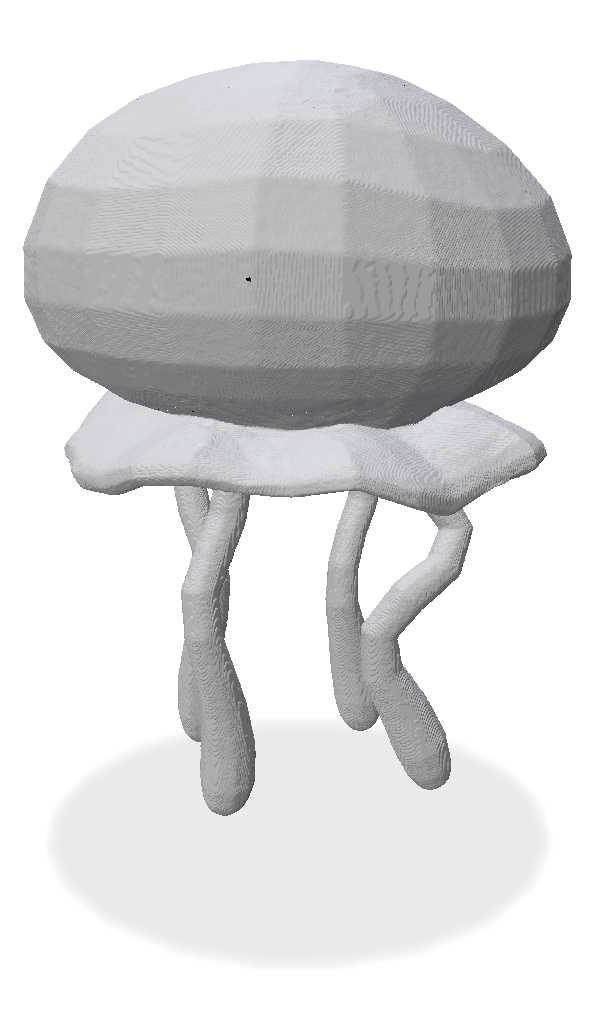} & %
         \includegraphics[width=\jellyimgwidth]{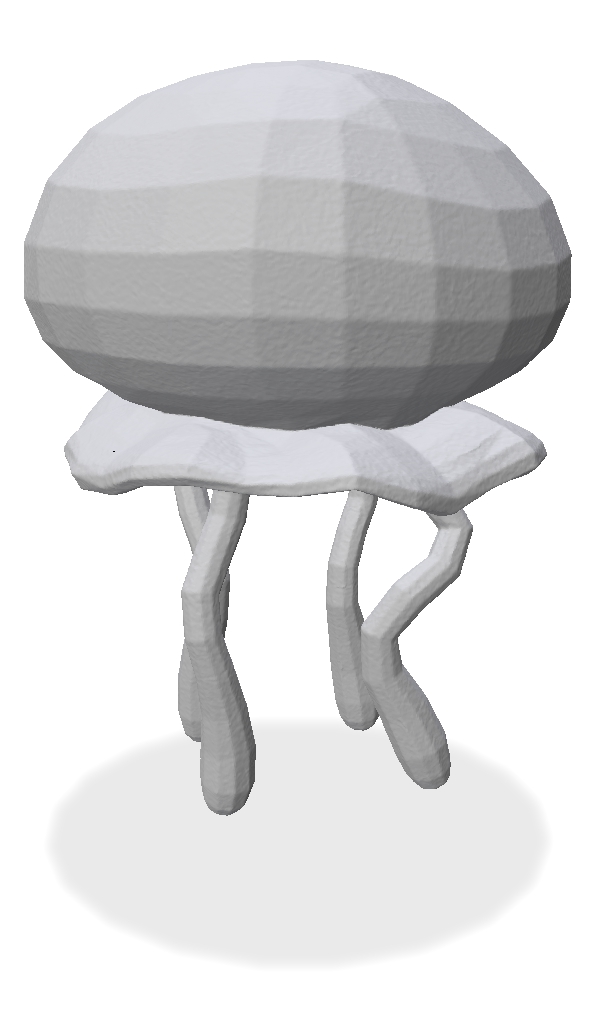} \\ %
         \hline         

          0.796 & 0.983 & 0.988 & 0.989 & 0.993 & 0.993 & 0.998 \\
         \includegraphics[width=\jellyimgwidth]{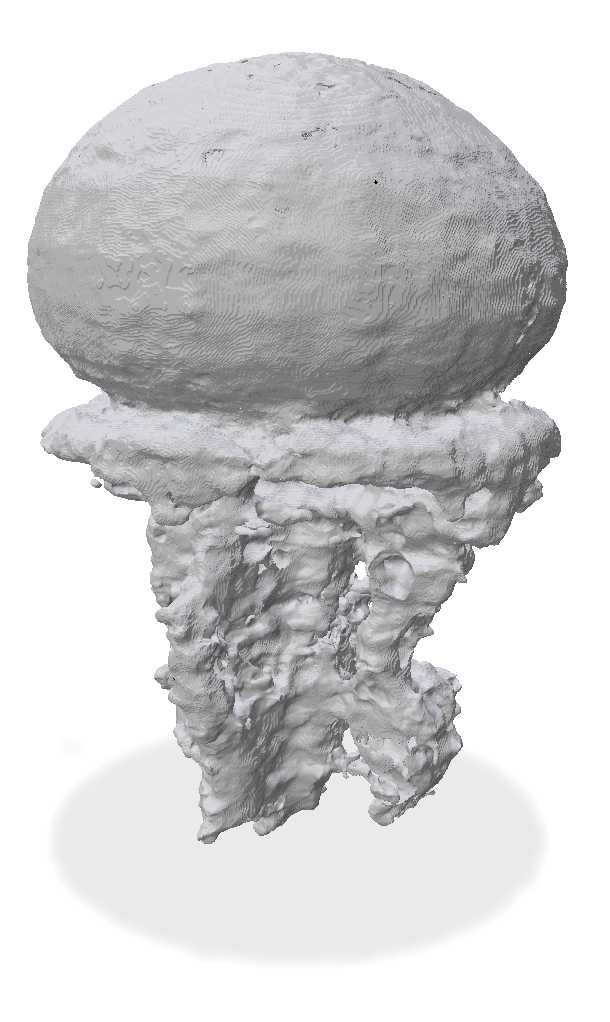} & %
         \includegraphics[width=\jellyimgwidth]{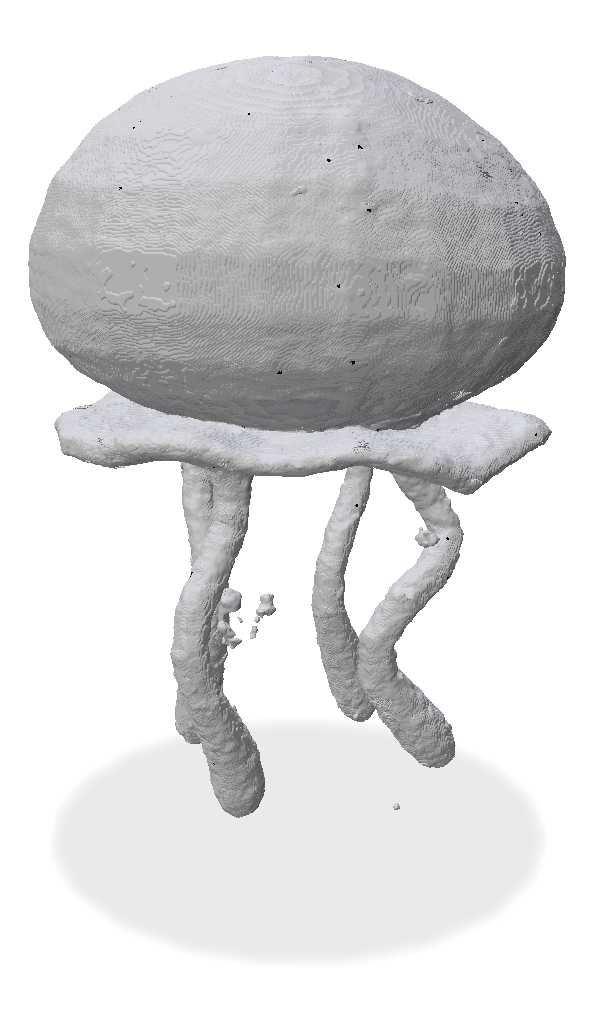} & %
         \includegraphics[width=\jellyimgwidth]{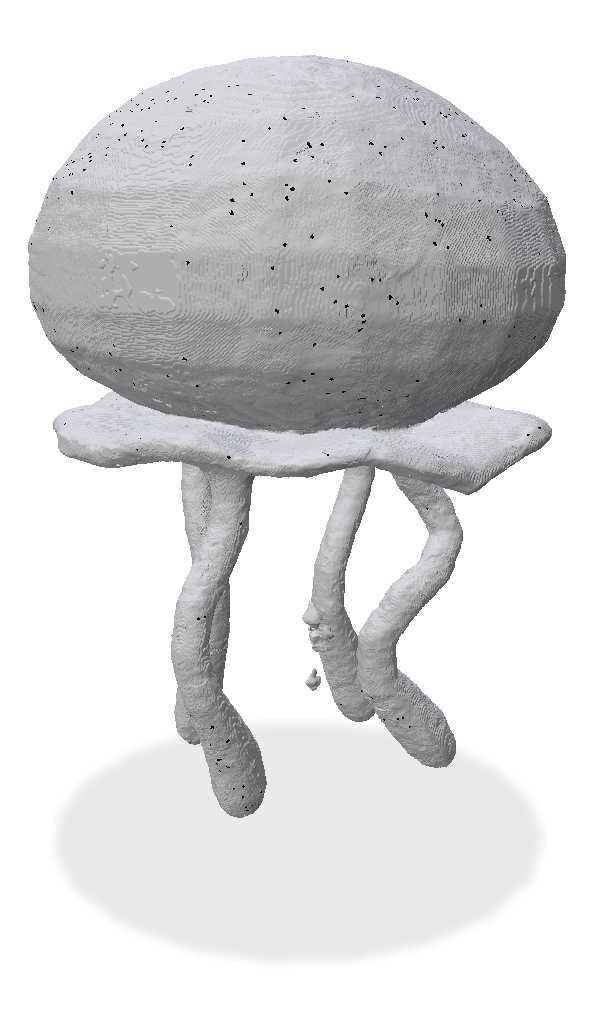} & %
         \includegraphics[width=\jellyimgwidth]{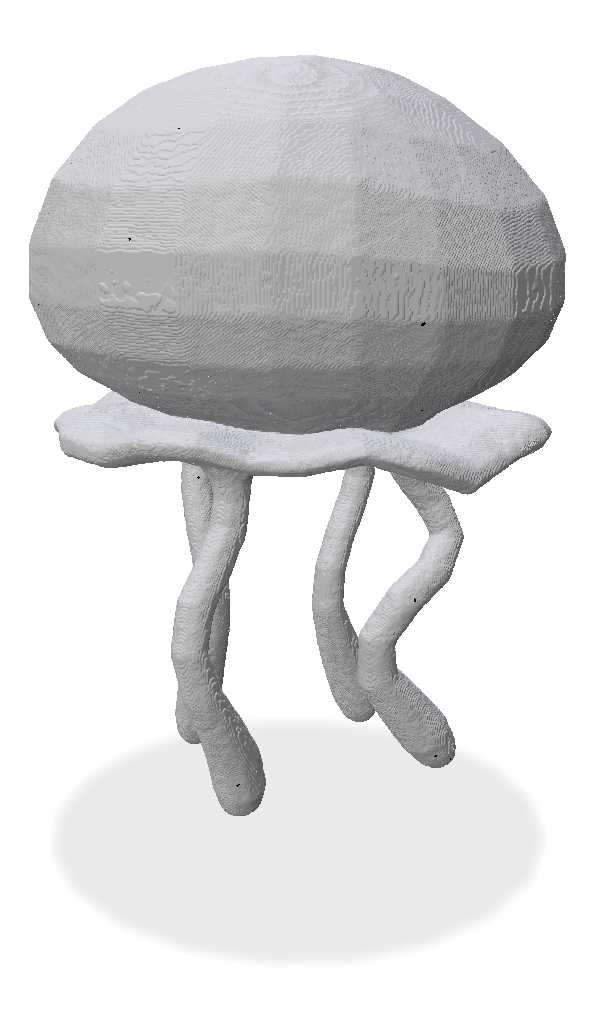} & %
         \includegraphics[width=\jellyimgwidth]{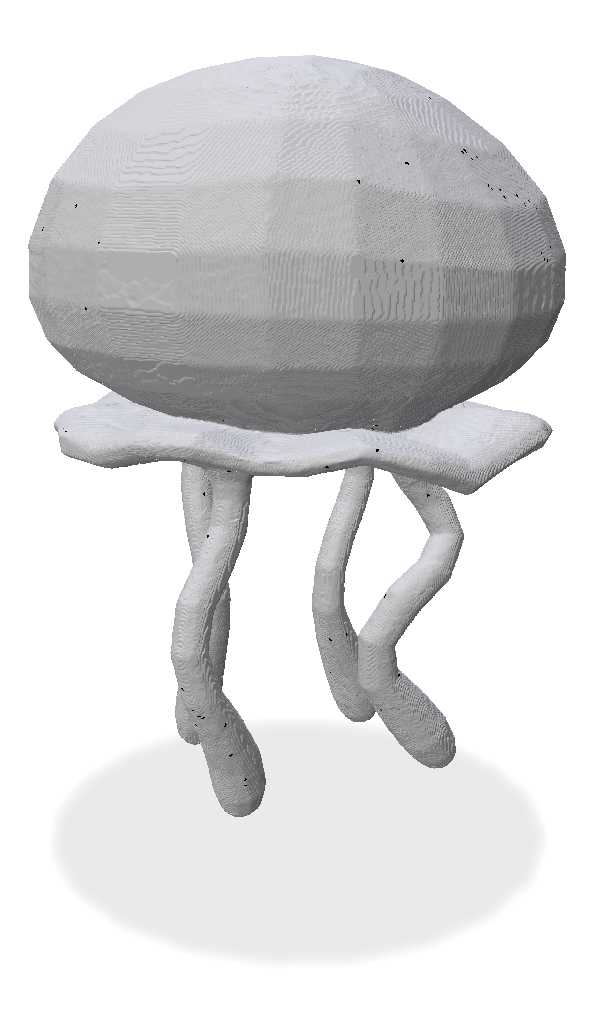} & %
         \includegraphics[width=\jellyimgwidth]{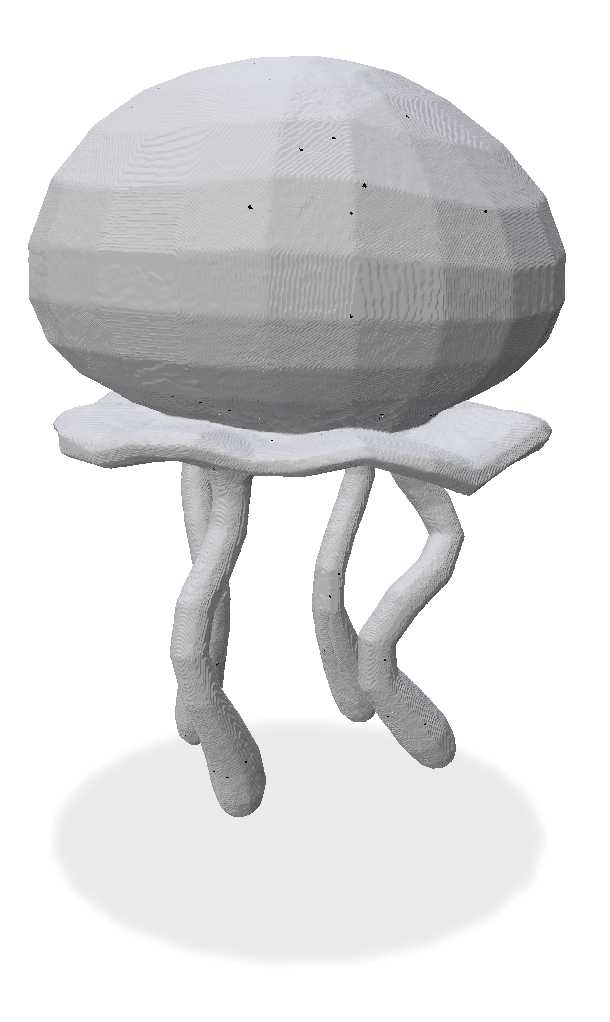} & %
         \includegraphics[width=\jellyimgwidth]{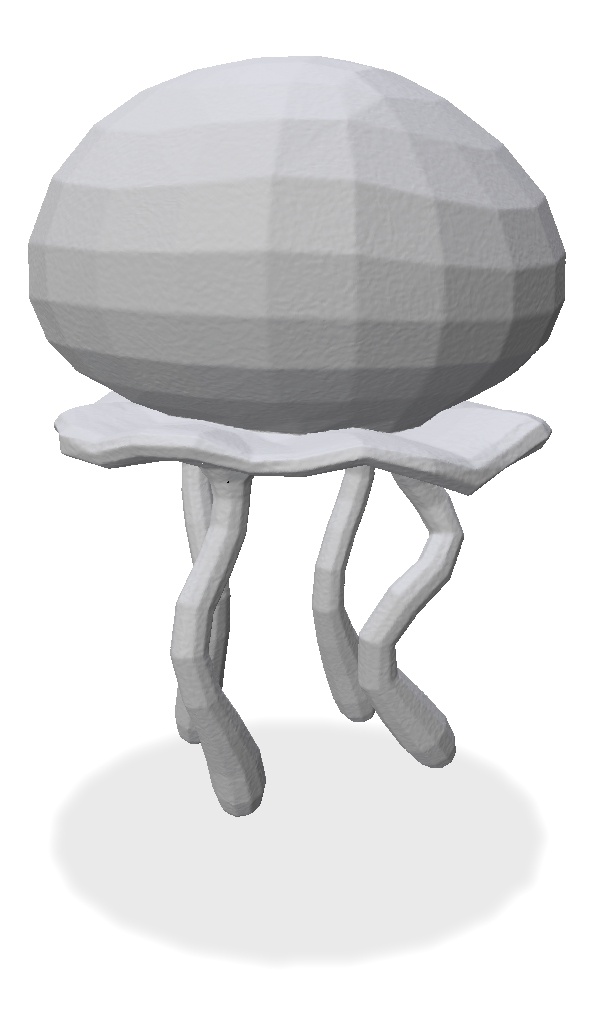} \\ %
         \hline         

     \end{tabular}
     \caption{Surface reconstructions using LoRAs of varying ranks $r$ vs. full fine-tuning (FT). Each row corresponds to a different target surface $\mathcal{D}'$ obtained at increasing timesteps of a physical simulation of $\mathcal{D}$ ~\cite{leticia:2025:elastodynamic}. IoU is shown. \change{See Figure \ref{fig:sdf_ablation_reconstructions} for the parameter count corresponding to each LoRA rank}} 
     \label{fig:jellyfish}
 \end{figure}

\newcommand{\imgwidth}{0.08\textwidth}

\begin{figure*}[ht] 
    \centering
     \begin{tabular}{c | c c  c  c  c  c  c c c}
        Scene & Input ($\mathcal{D}$) & $r=1$ & $r=4$ & $r=8$ & $r=16$ & $r=32$ & $r=64$  & FT & Target ($\mathcal{D'}$)\\                 
         \# Params & & 3.1 k & 12.4 k & 23.8 k  & 46.7 k & 92.4 k & 176.7 k & 340.4 k & \\
         \change {\% Params} & & \change{0.9} & \change{3.6} & \change{7.0}  & \change{13.7} & \change{27.1} & \change{51.9} & \change{100} & \\
         \hline         

         & PSNR & \SI{35.8}{dB} & \SI{38.1}{dB} & \SI{38.9}{dB}  & \SI{39.6}{dB} & \SI{39.9}{dB} & \SI{40.1}{dB} & \SI{40.0}{dB} & \\
         \rotatebox{90}{Sofa} &
         \includegraphics[width=\imgwidth]{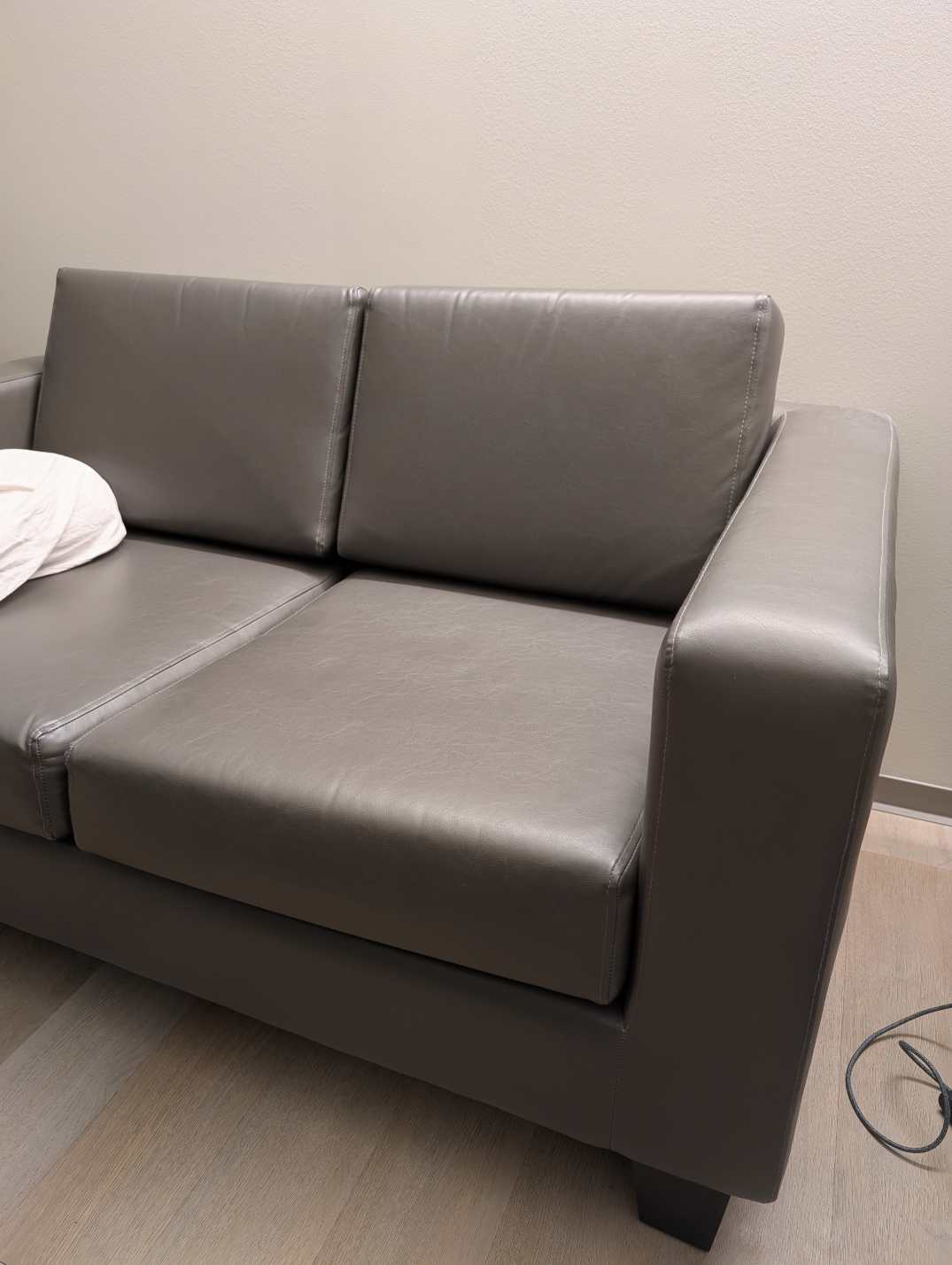} & %
         \includegraphics[width=\imgwidth]{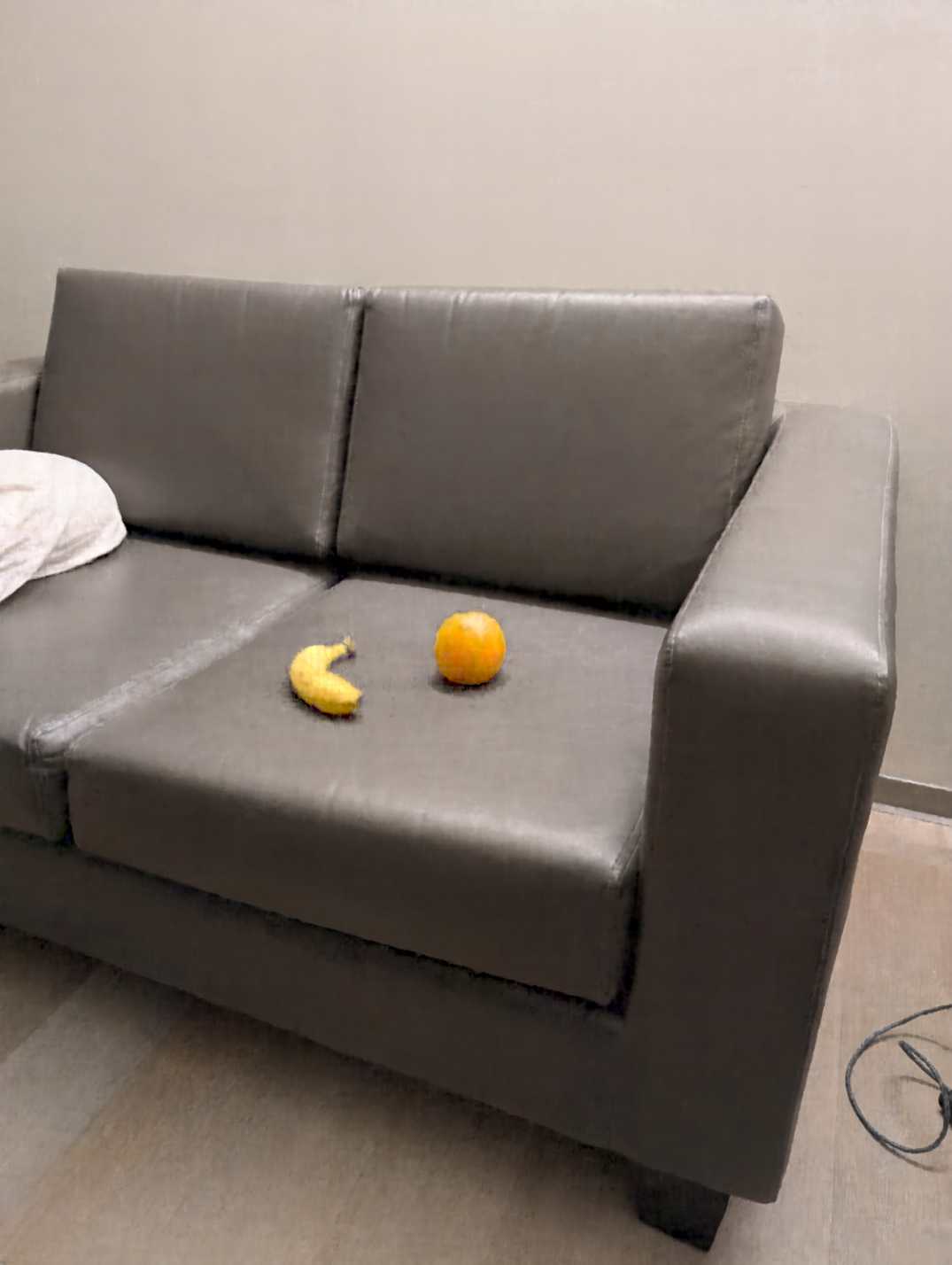} & %
         \includegraphics[width=\imgwidth]{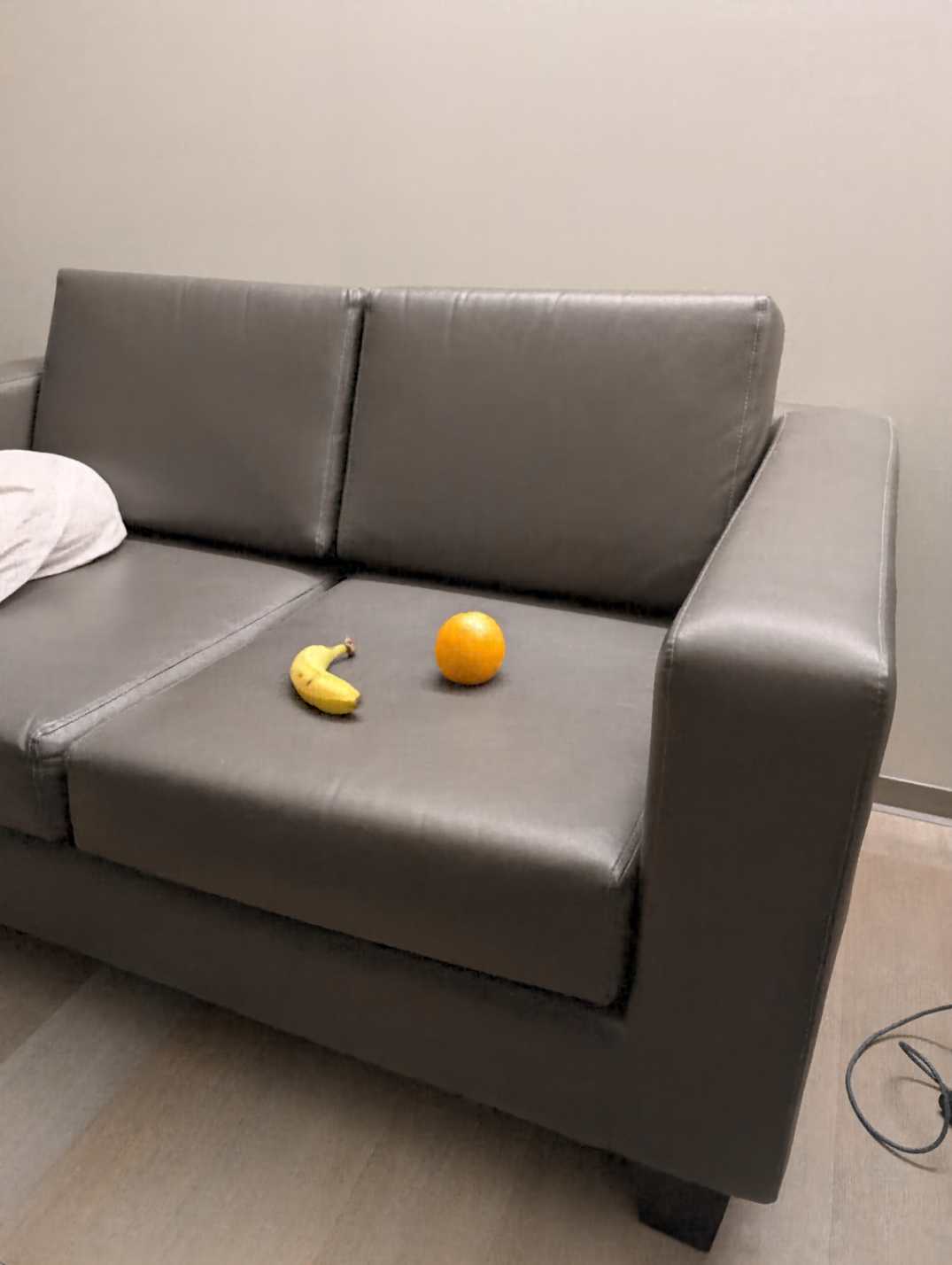} & %
         \includegraphics[width=\imgwidth]{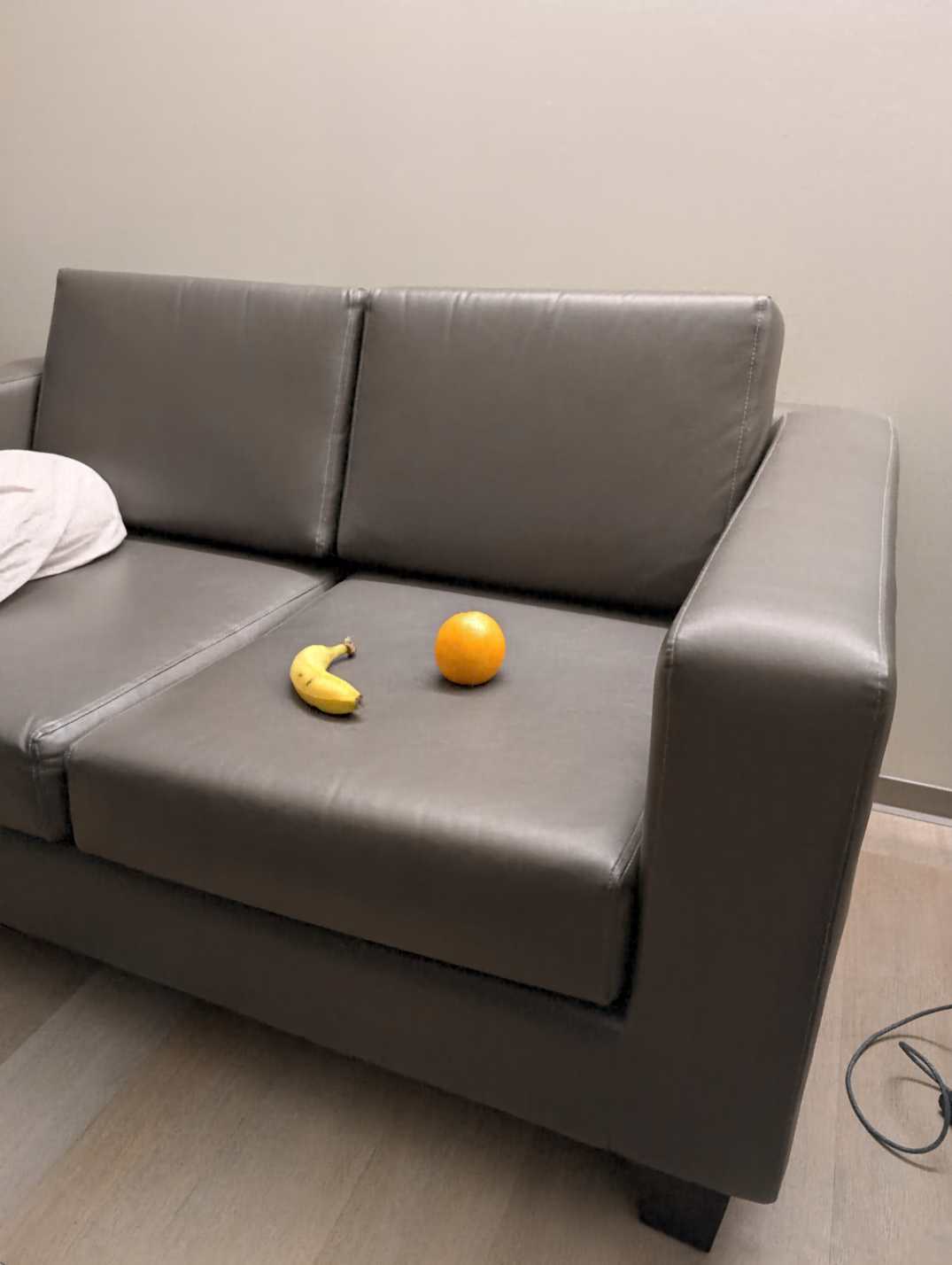} & %
         \includegraphics[width=\imgwidth]{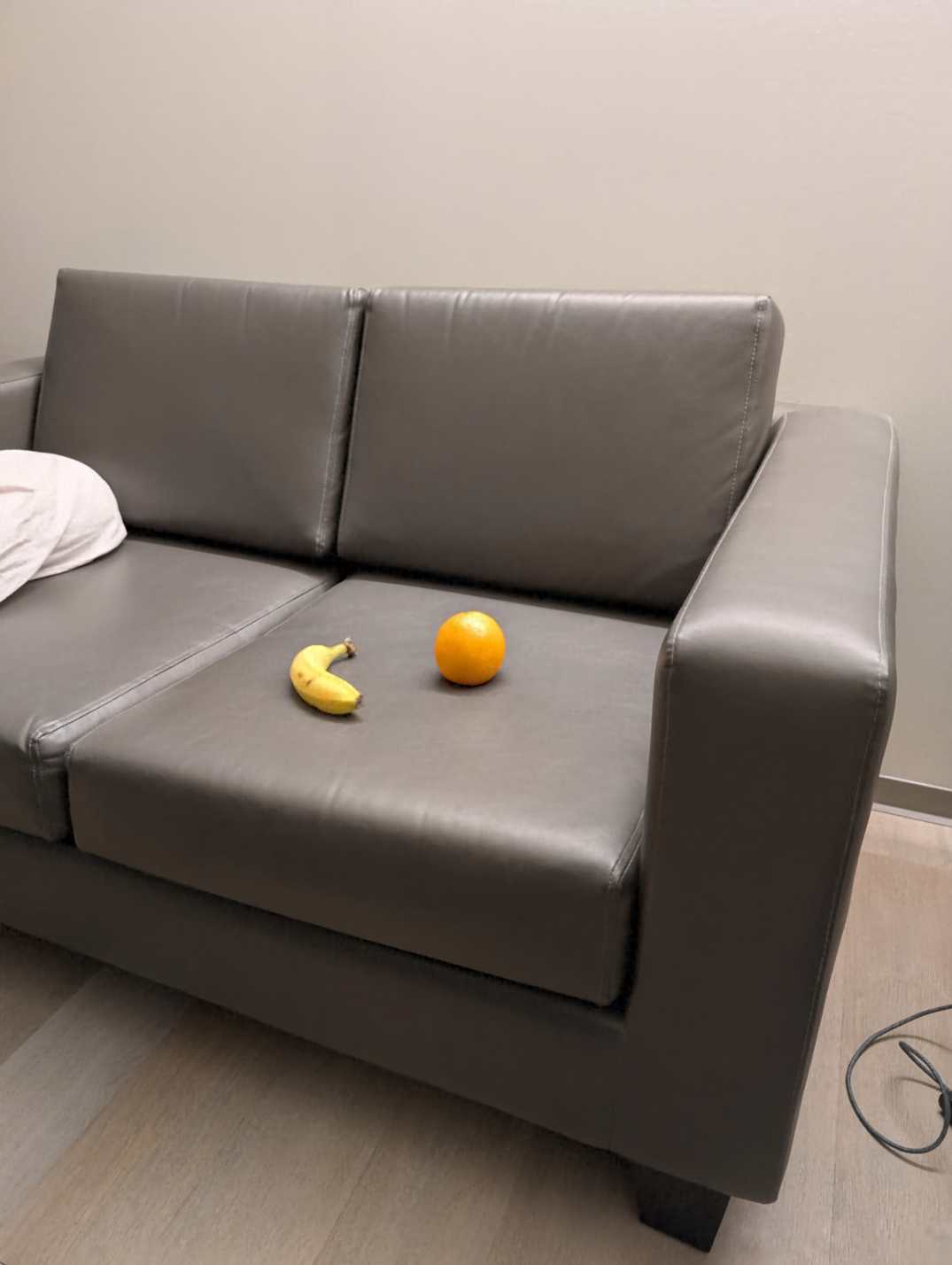} & %
         \includegraphics[width=\imgwidth]{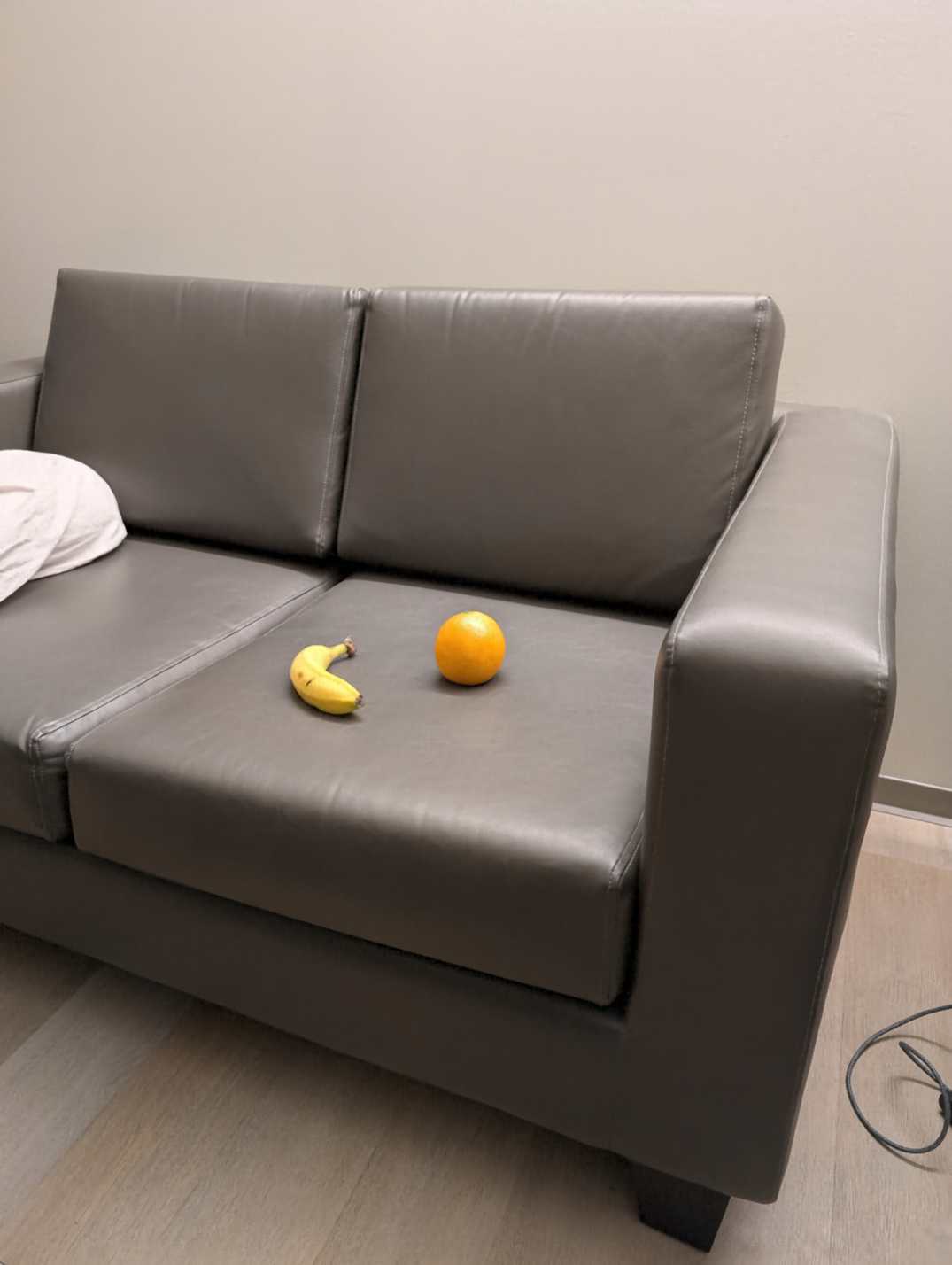} & %
         \includegraphics[width=\imgwidth]{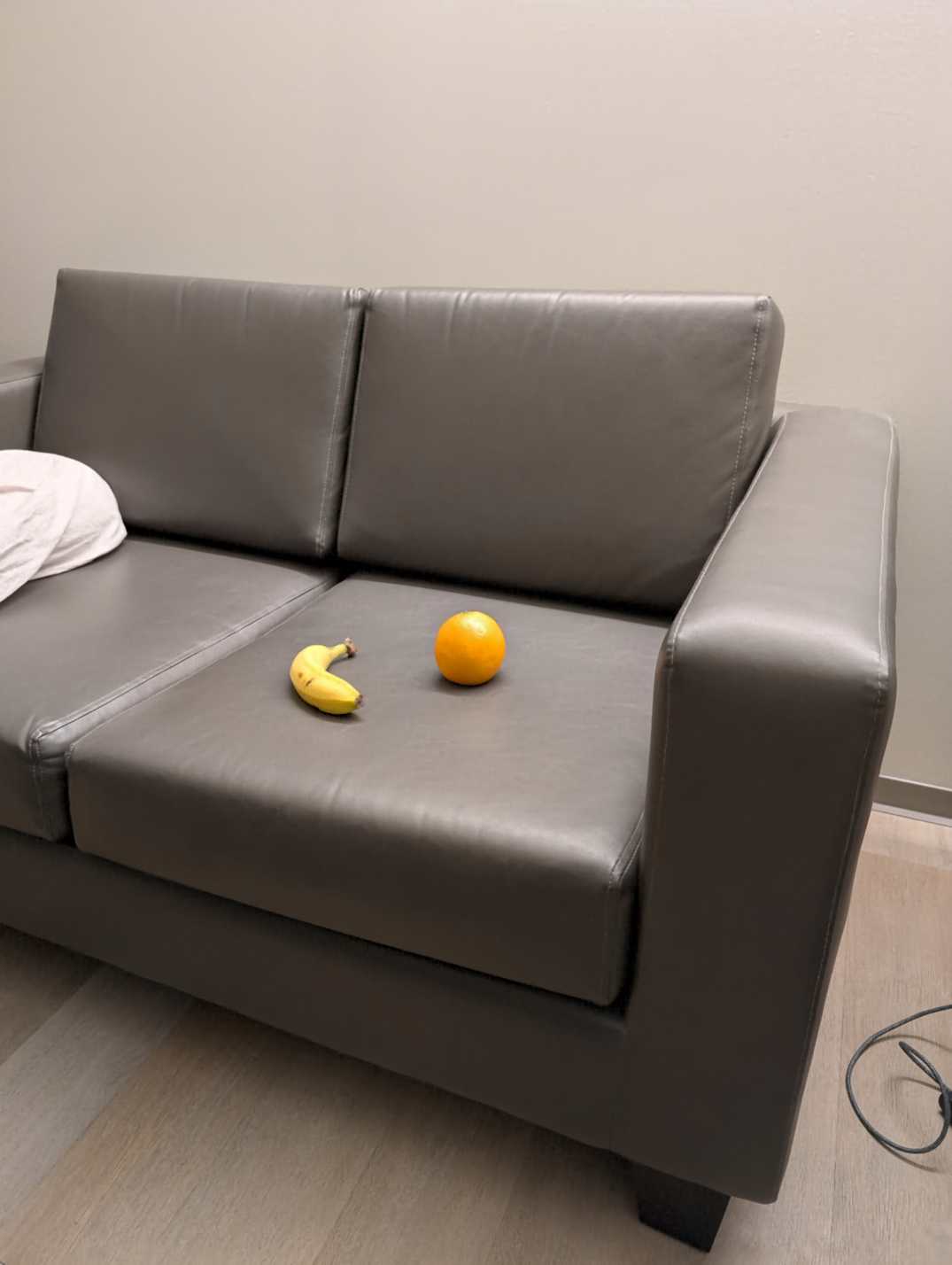} & %
         \includegraphics[width=\imgwidth]{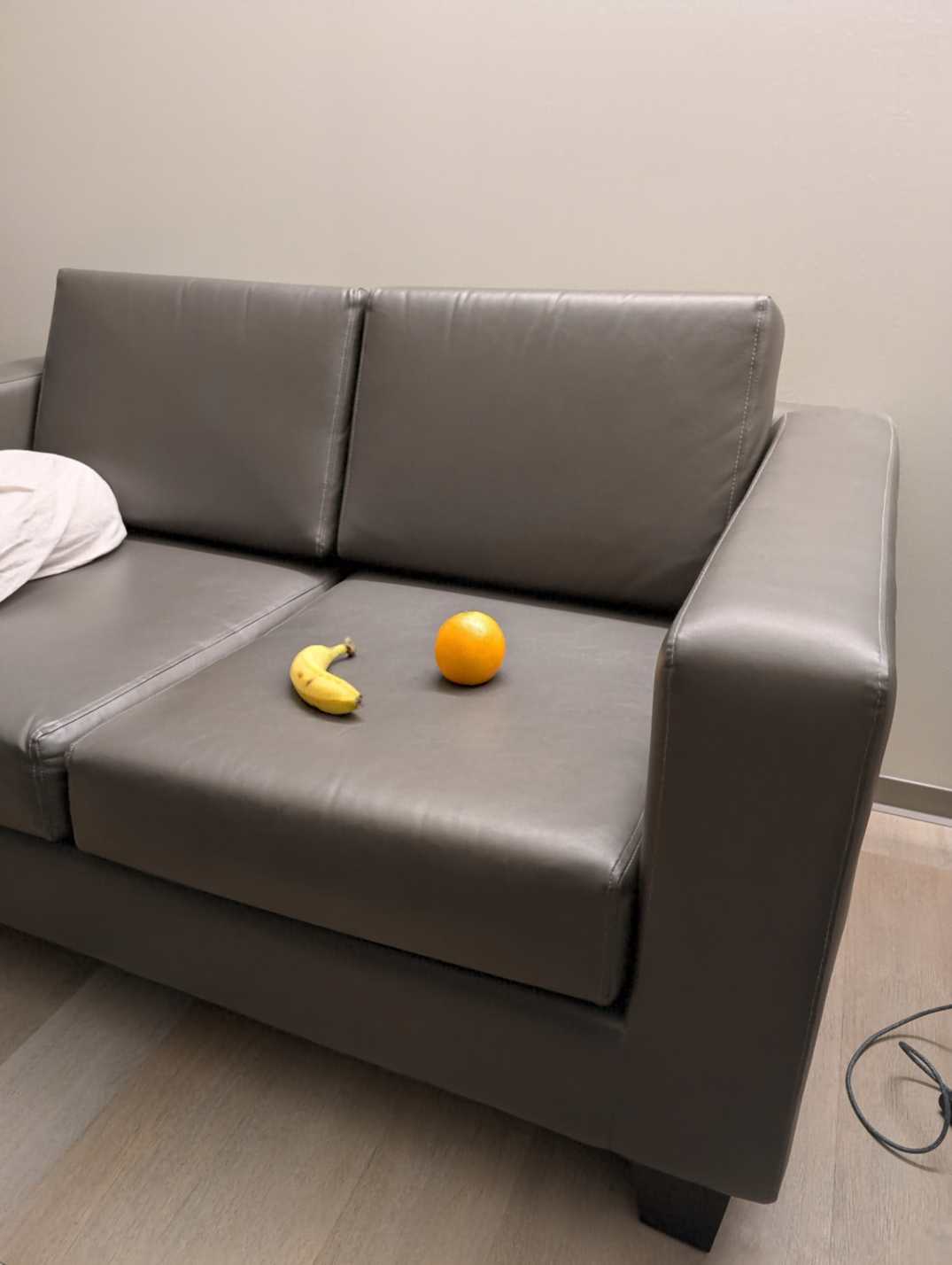} & %
         \includegraphics[width=\imgwidth]{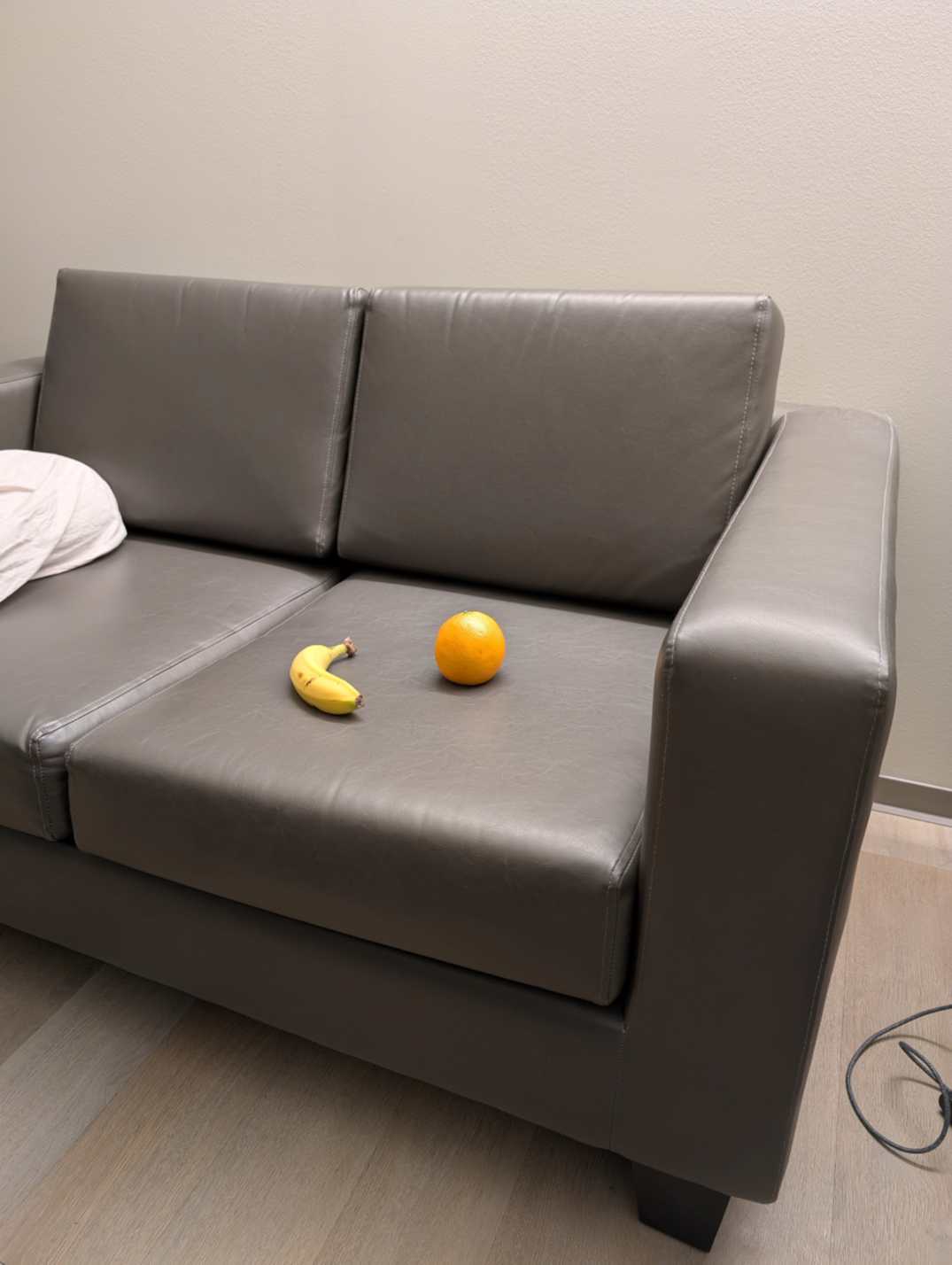} \\ %
         \hline

         & PSNR & \SI{33.4}{dB} & \SI{38.5}{dB} & \SI{41.1}{dB}  & \SI{42.7}{dB} & \SI{43.5}{dB} & \SI{44.4}{dB} & \SI{43.9}{dB} & \\
         \rotatebox{90}{Table} &
         \includegraphics[width=\imgwidth]{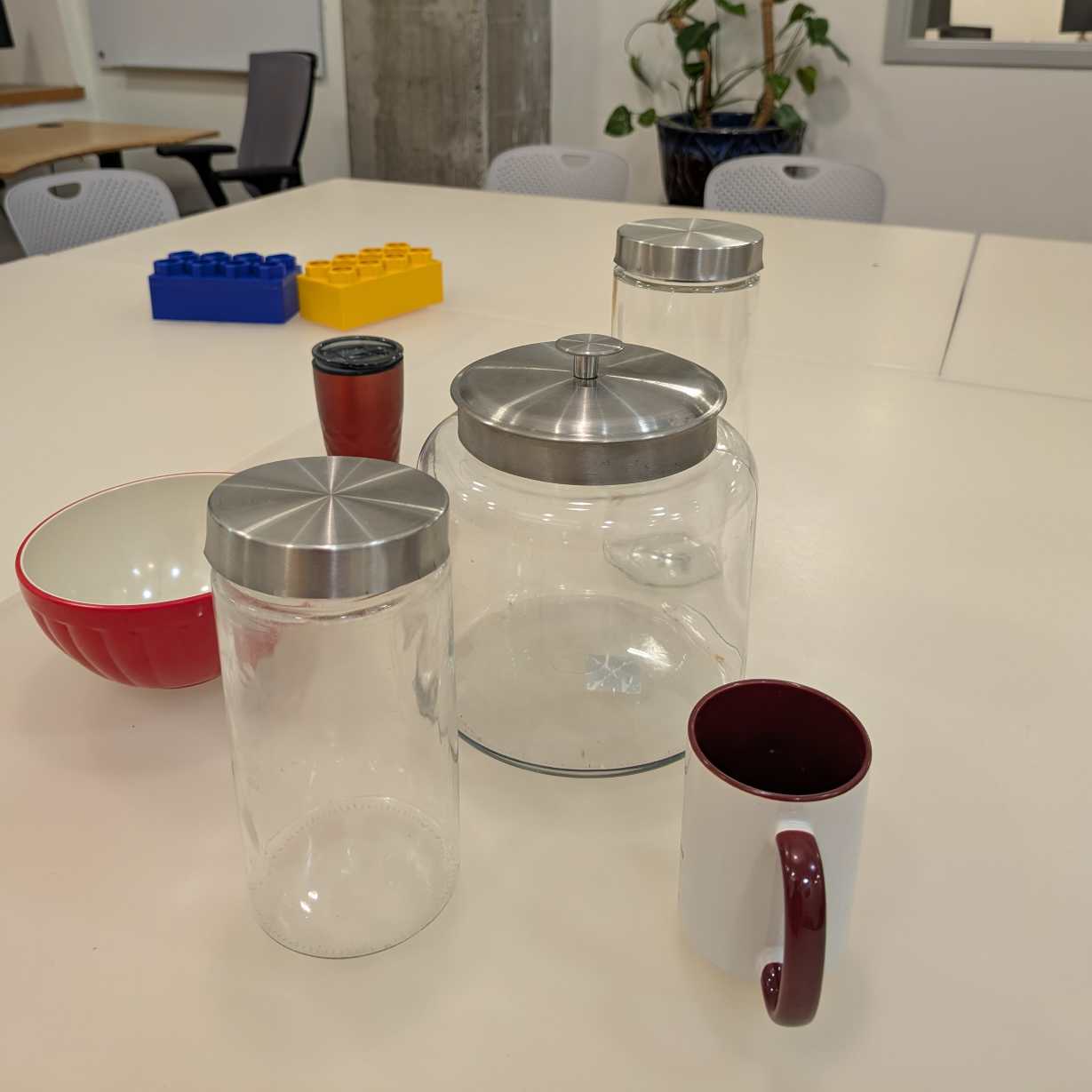} & %
         \includegraphics[width=\imgwidth]{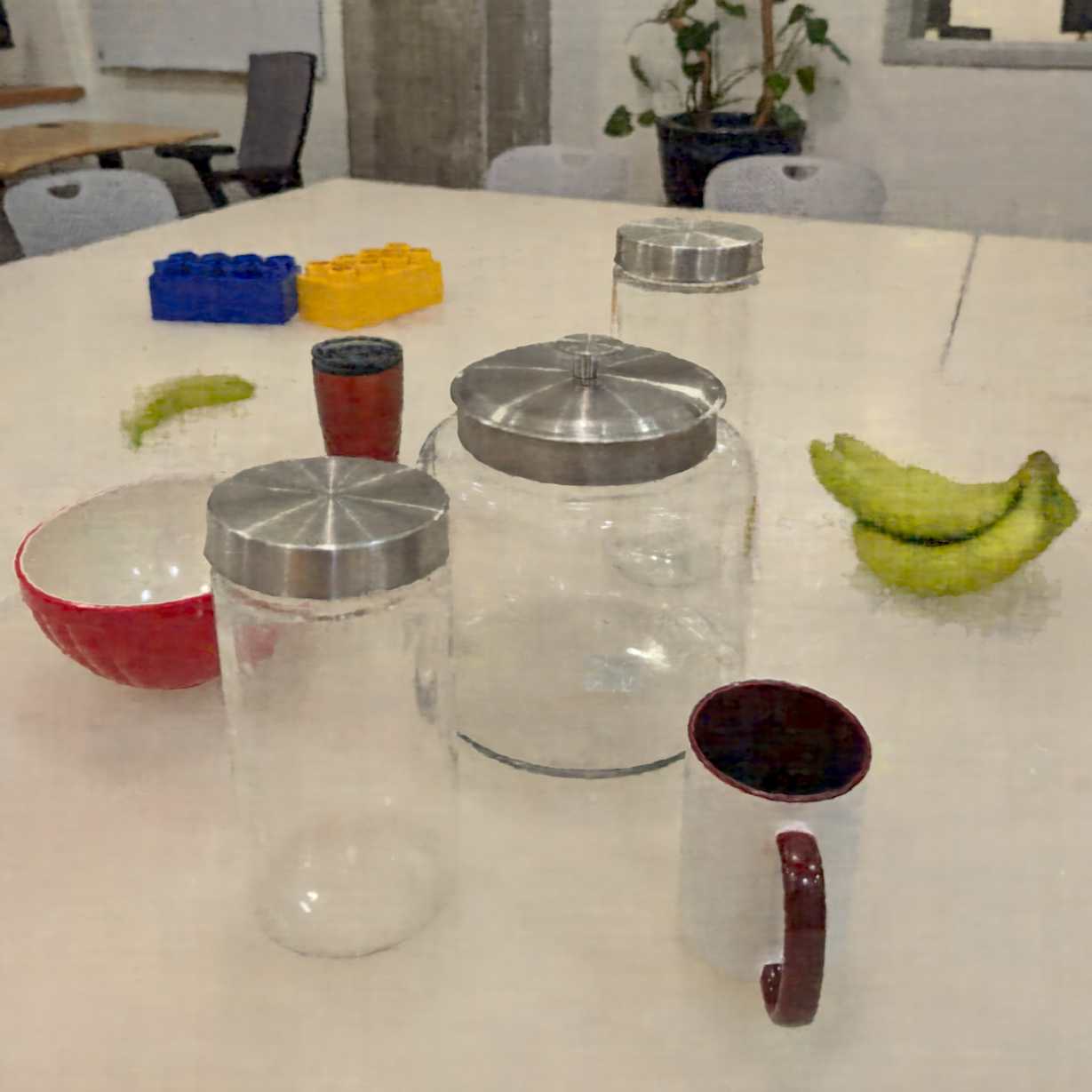} & %
         \includegraphics[width=\imgwidth]{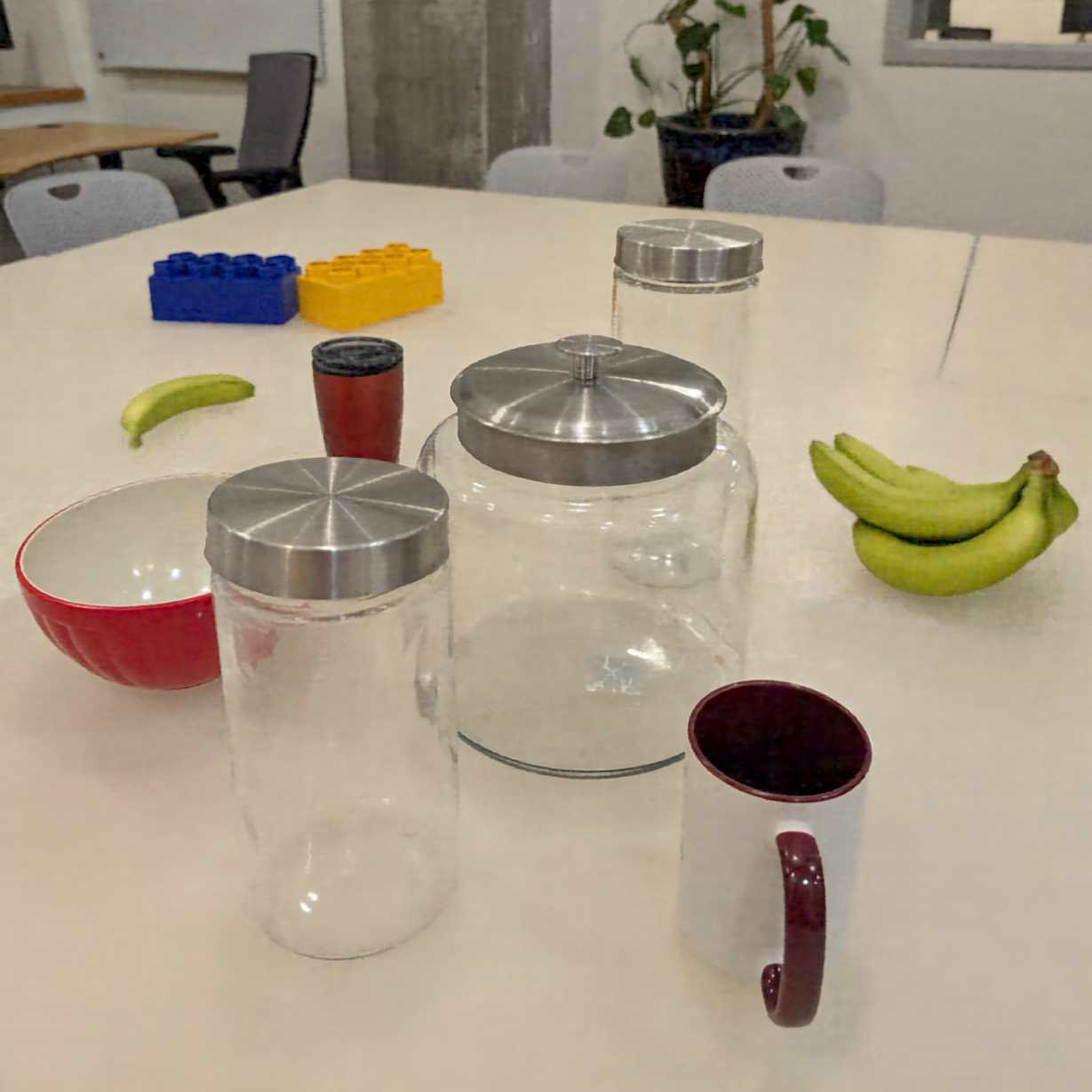} & %
         \includegraphics[width=\imgwidth]{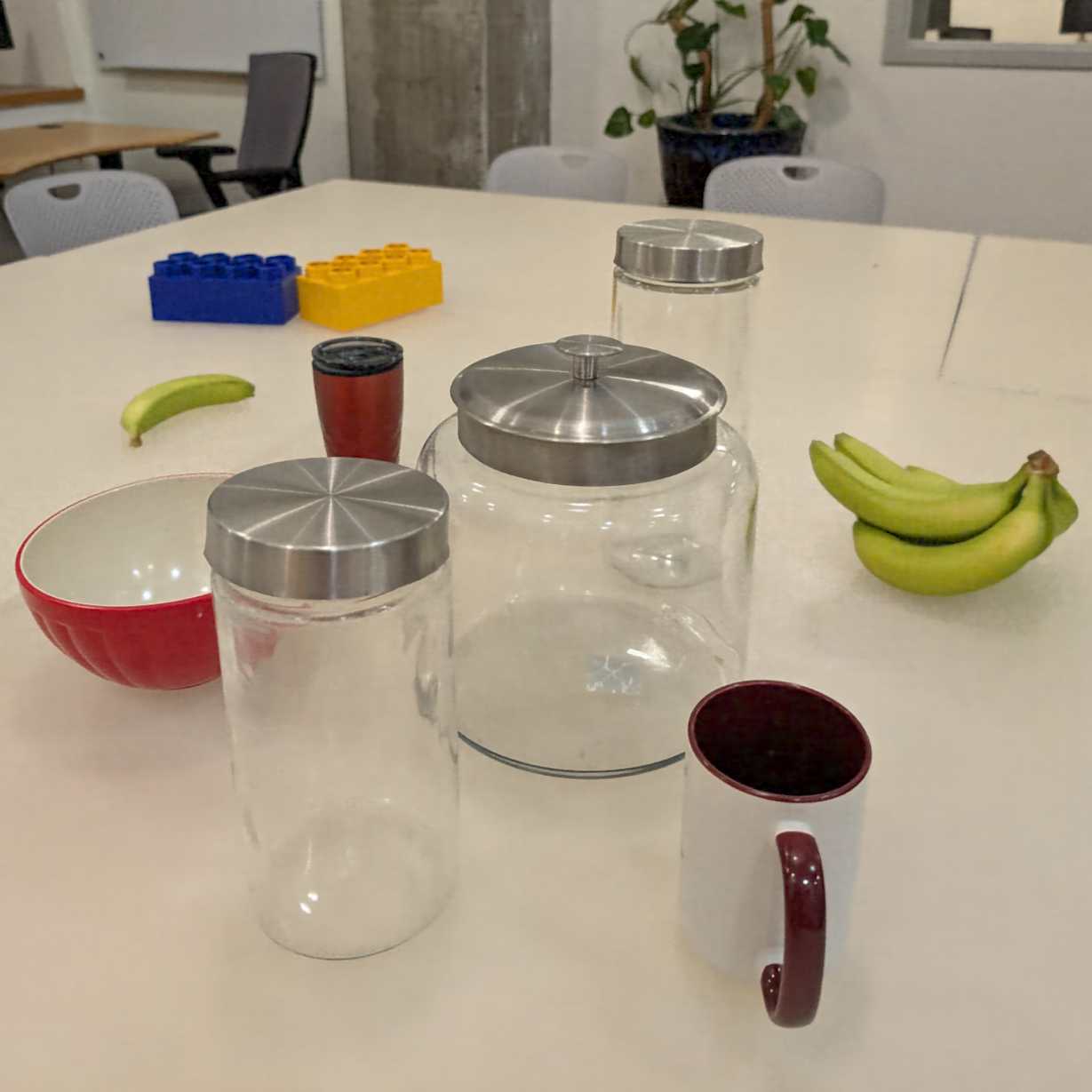} & %
         \includegraphics[width=\imgwidth]{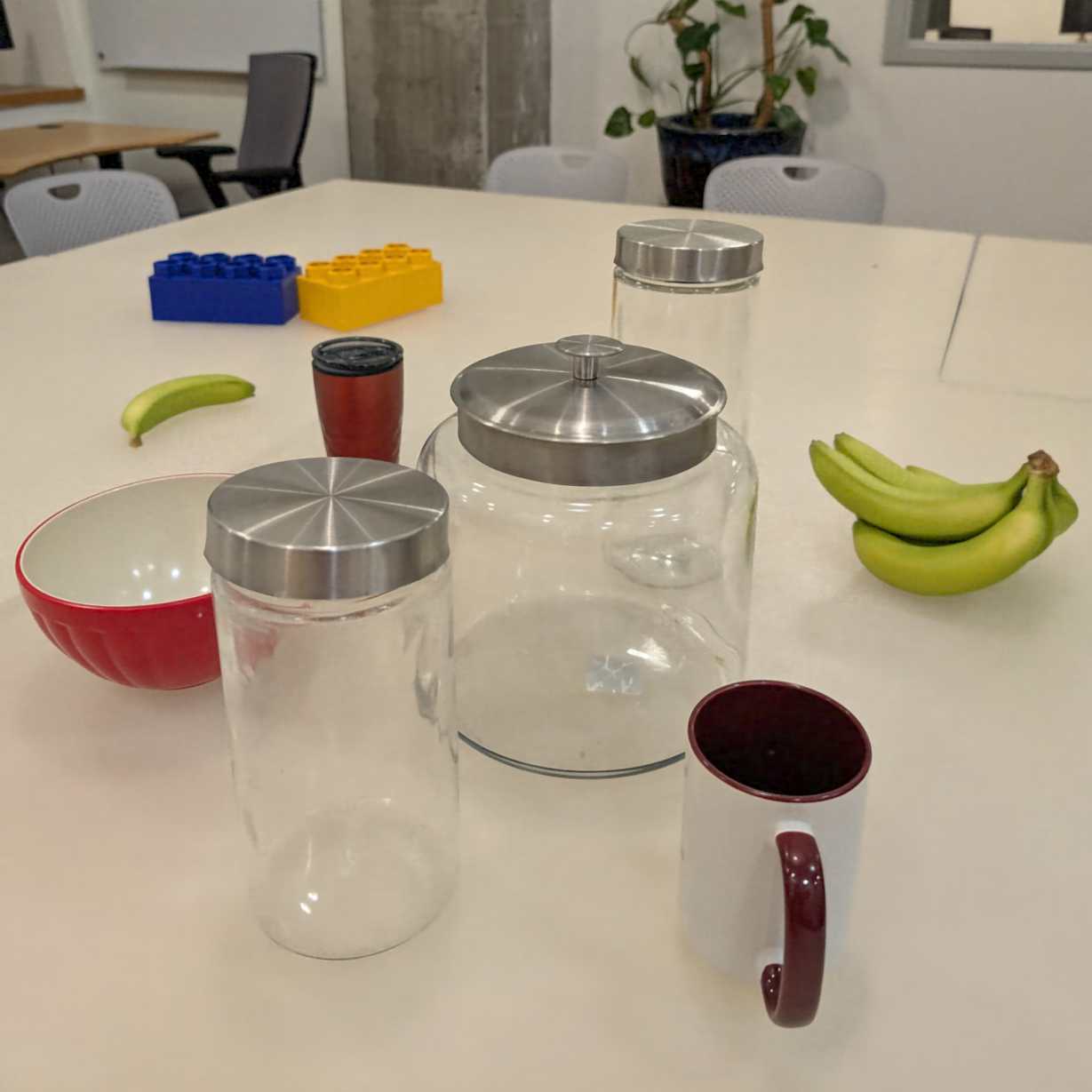} & %
         \includegraphics[width=\imgwidth]{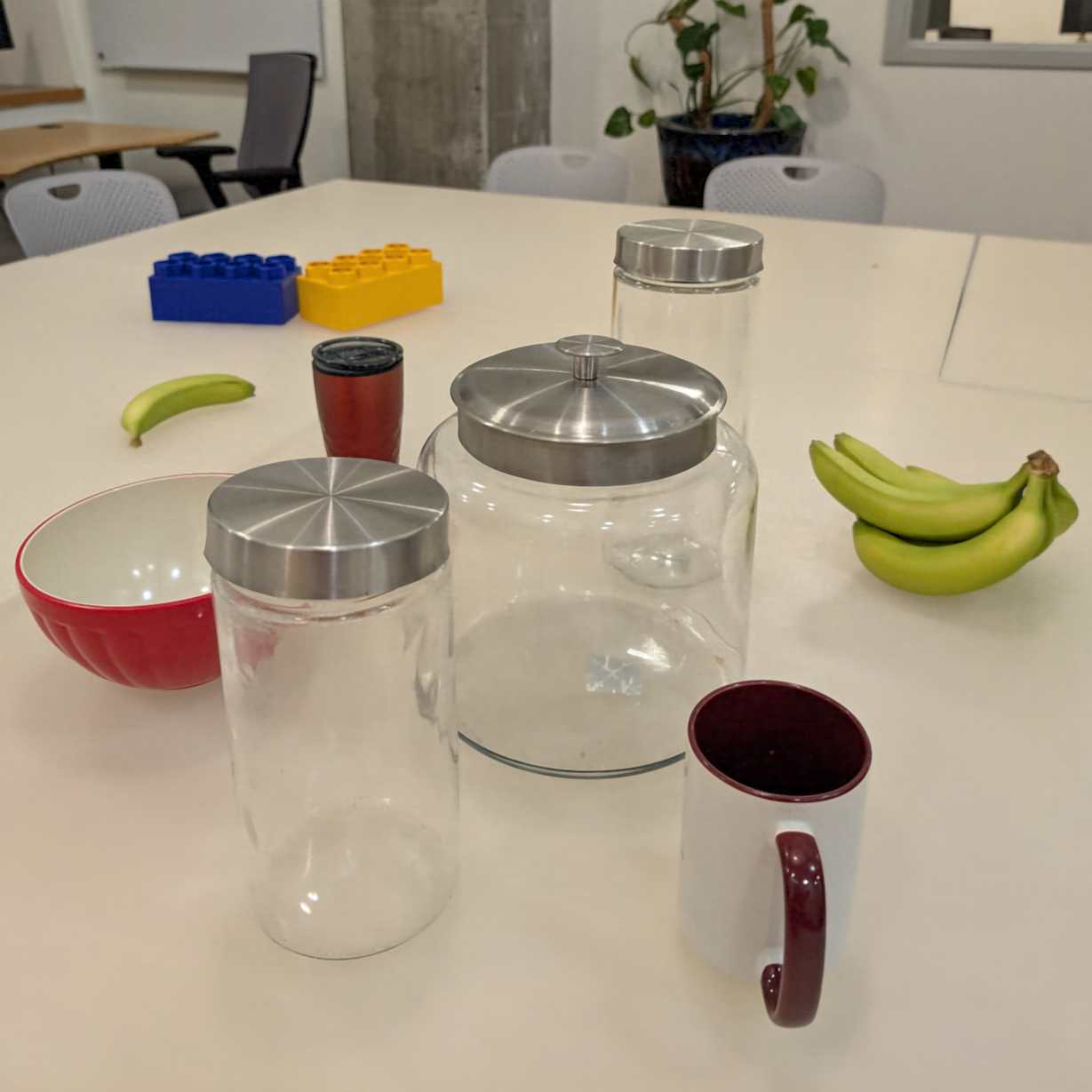} & %
         \includegraphics[width=\imgwidth]{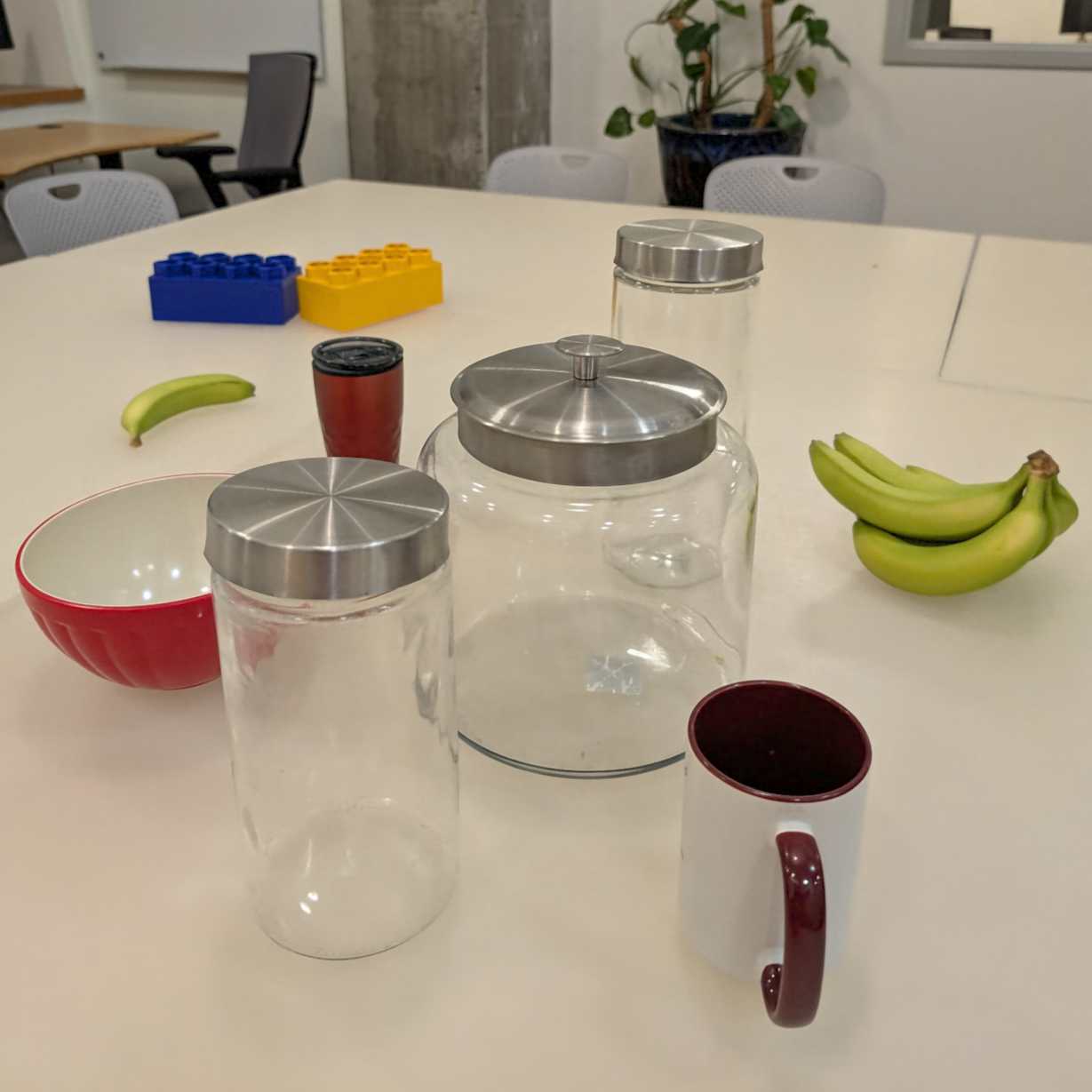} & %
         \includegraphics[width=\imgwidth]{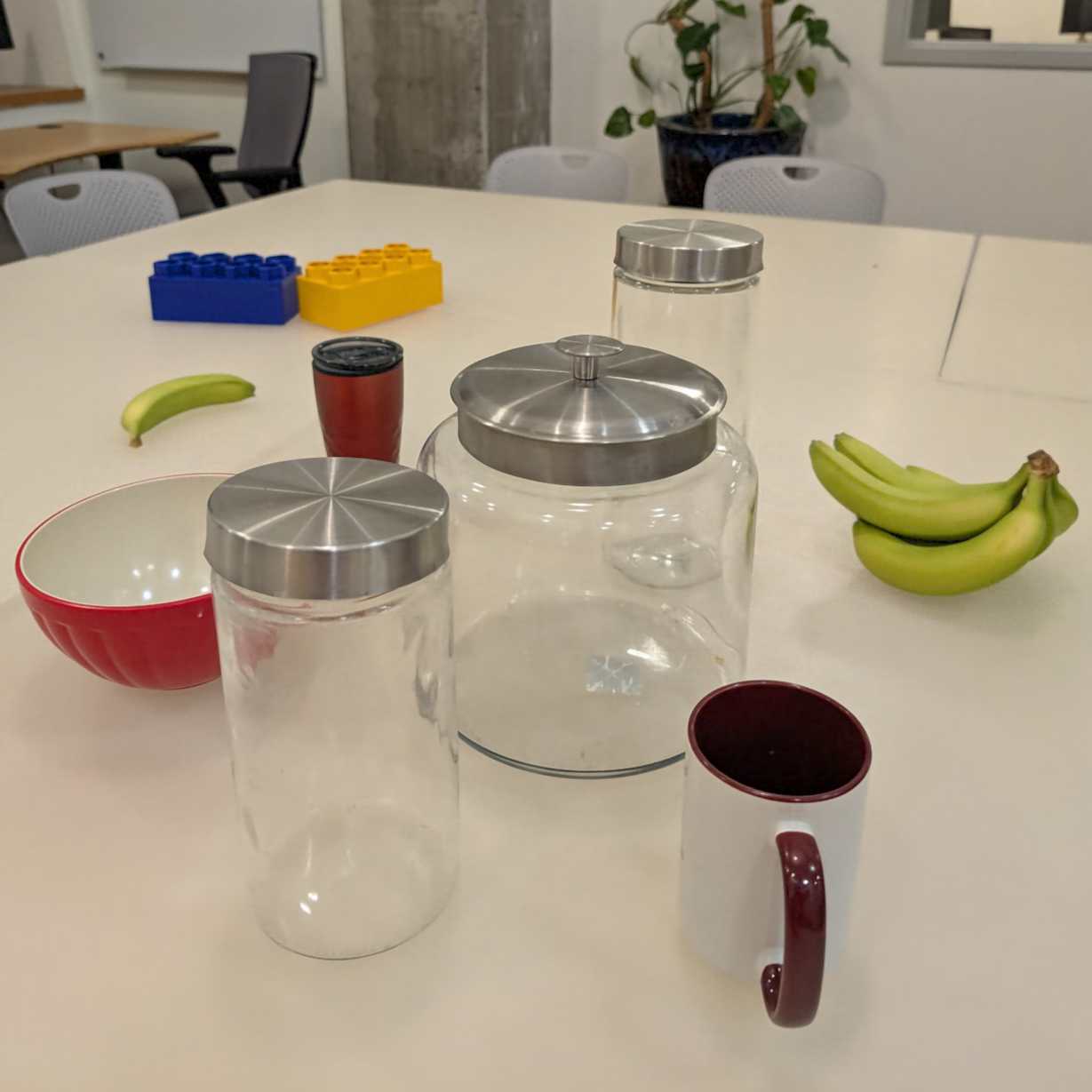} & %
         \includegraphics[width=\imgwidth]{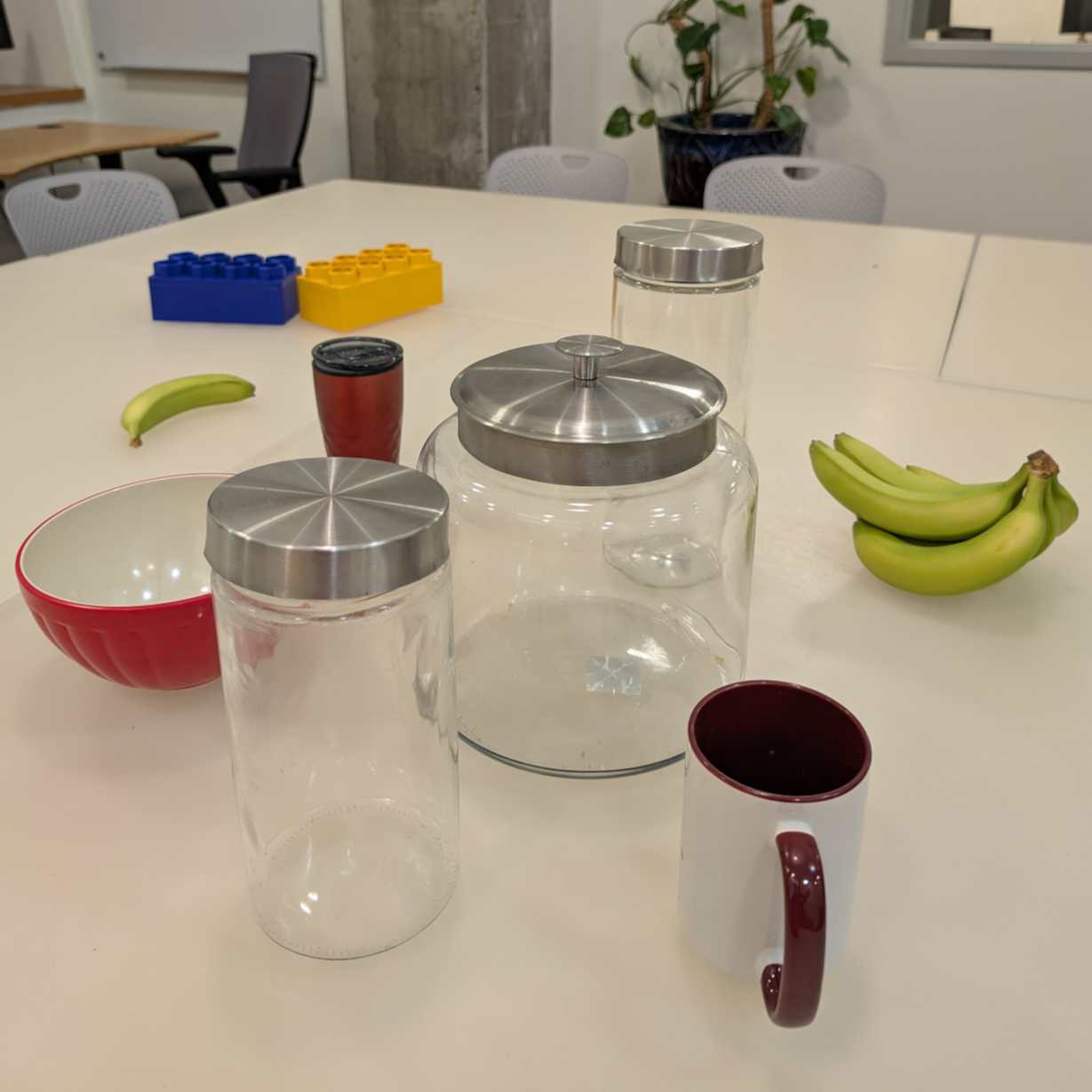} \\ %
         \hline

        & PSNR & \SI{32.8}{dB} & \SI{38.1}{dB} & \SI{39.6}{dB}  & \SI{41.3}{dB} & \SI{42.2}{dB} & \SI{43.1}{dB} & \SI{42.1}{dB} & \\
        \rotatebox{90}{\small{Whiteboard}} &
        \includegraphics[width=\imgwidth]{figs/image_experiments/2025-04-09_01-22-30_whiteboard/whiteboard_before.jpg} & %
        \includegraphics[width=\imgwidth]{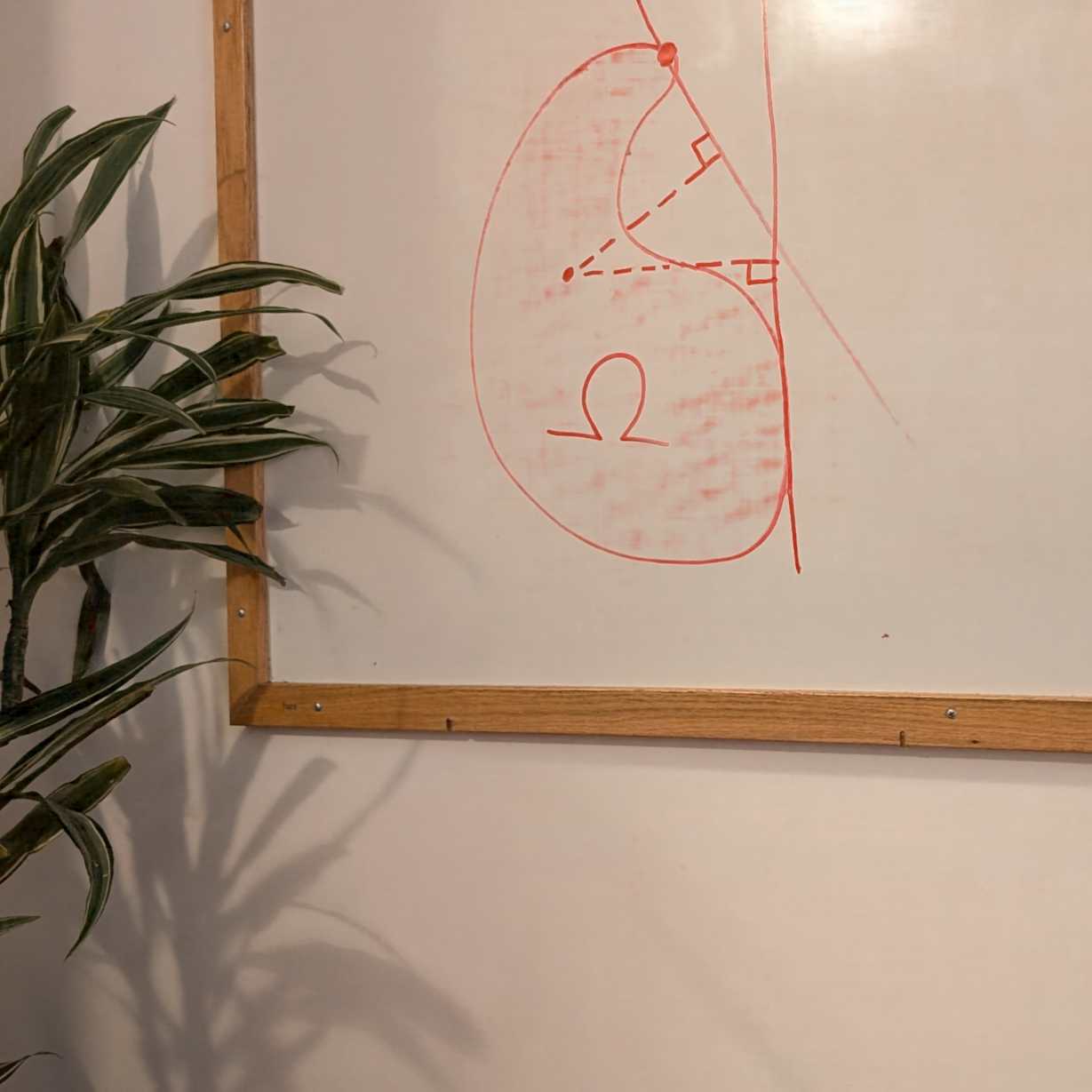} & %
        \includegraphics[width=\imgwidth]{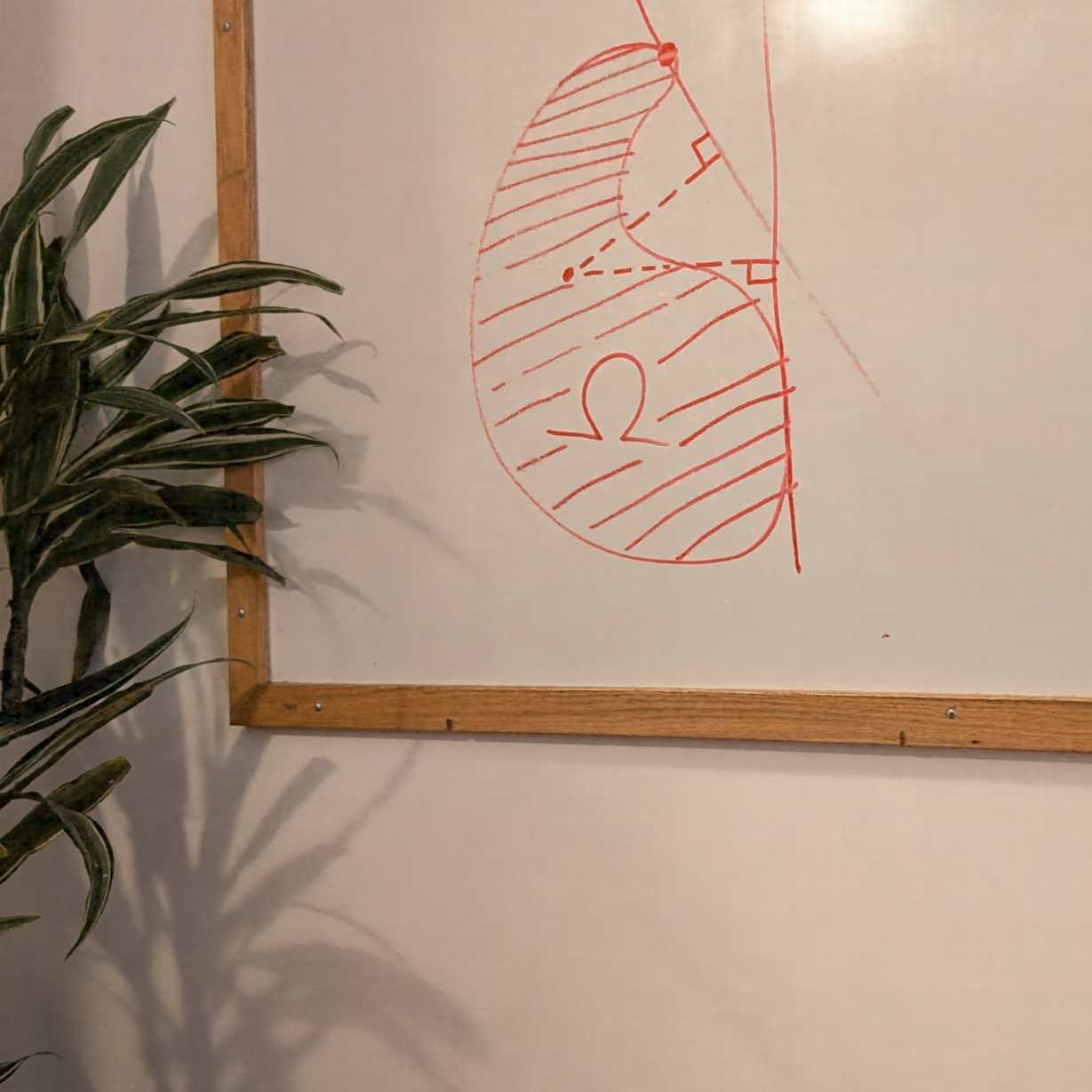} & %
        \includegraphics[width=\imgwidth]{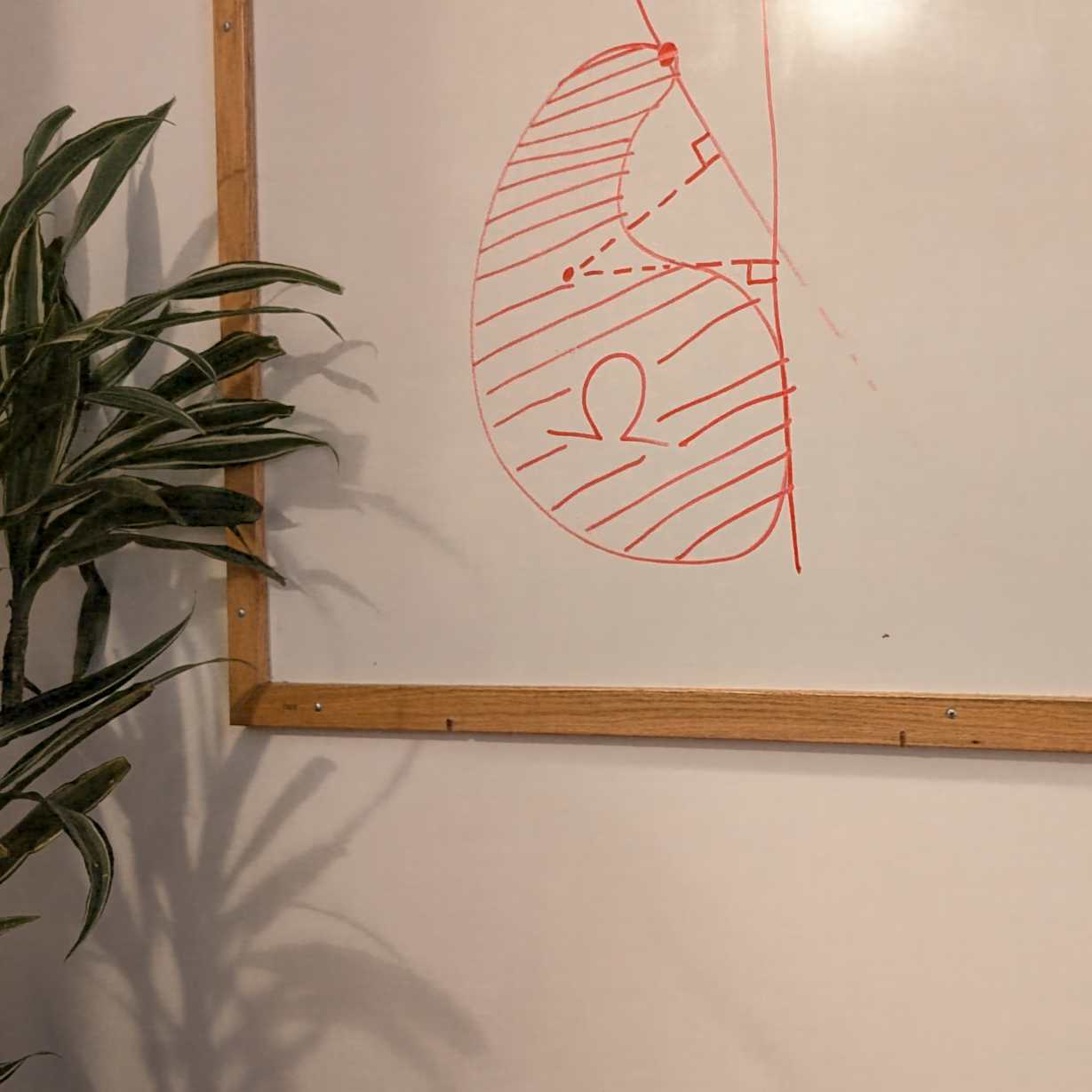} & %
        \includegraphics[width=\imgwidth]{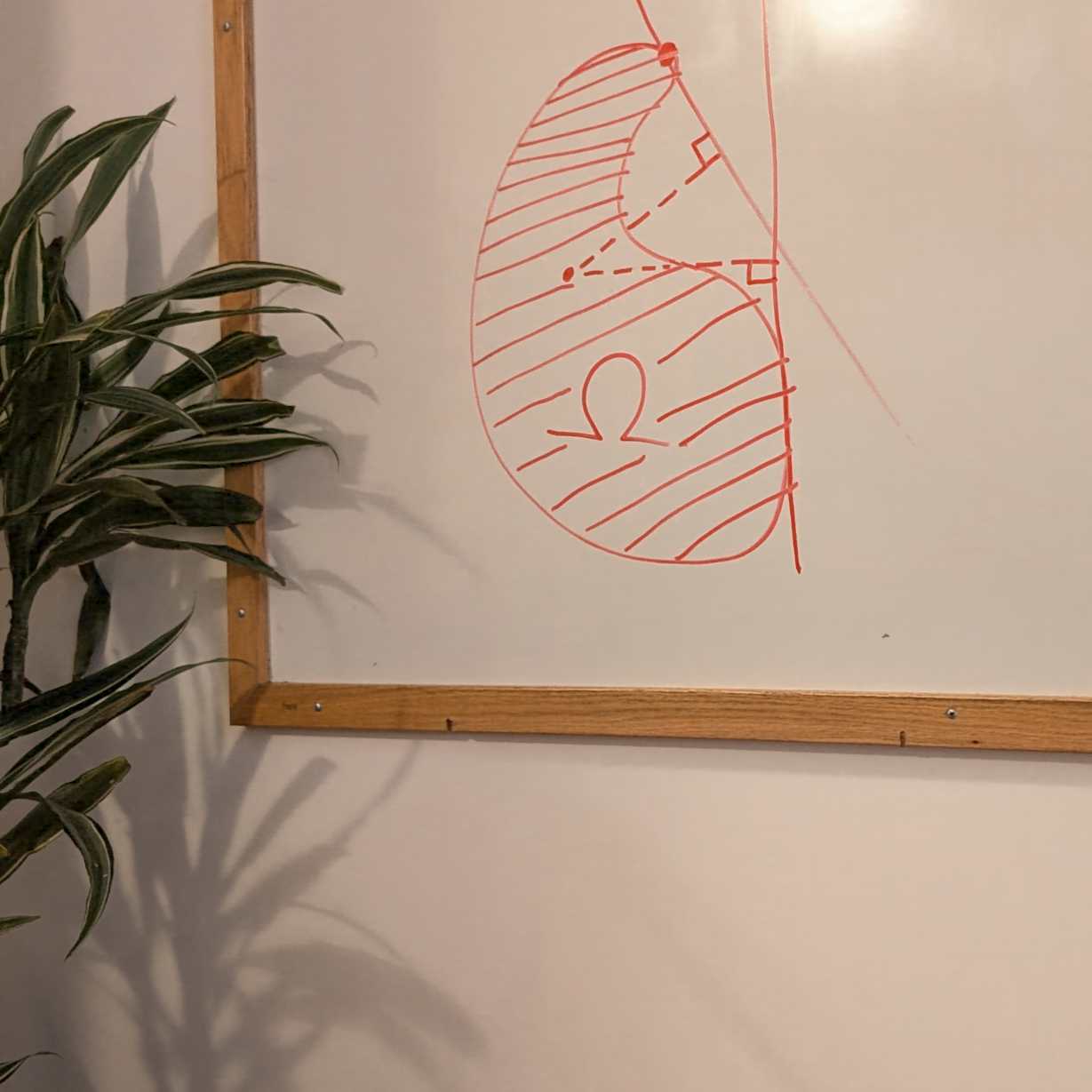} & %
        \includegraphics[width=\imgwidth]{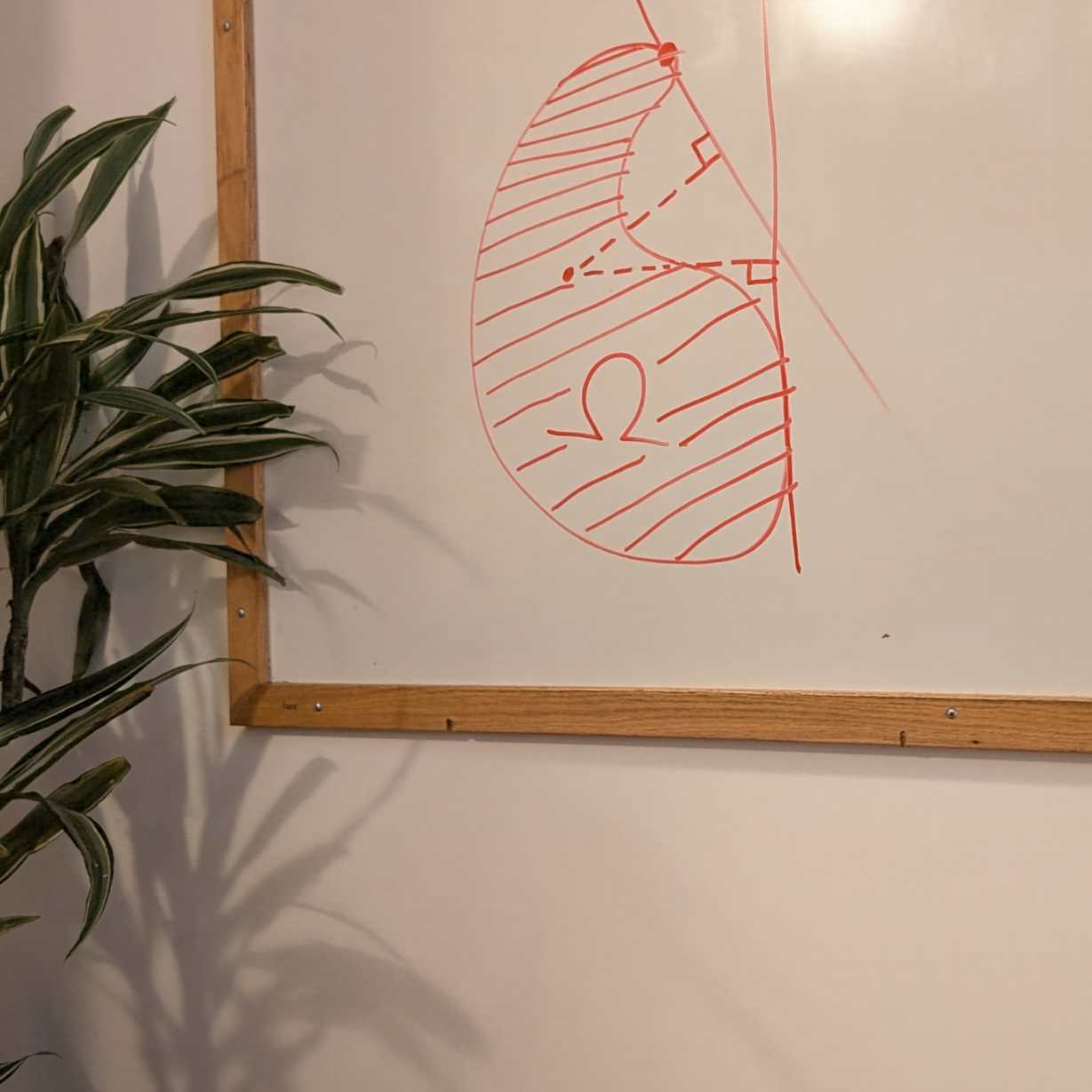} & %
        \includegraphics[width=\imgwidth]{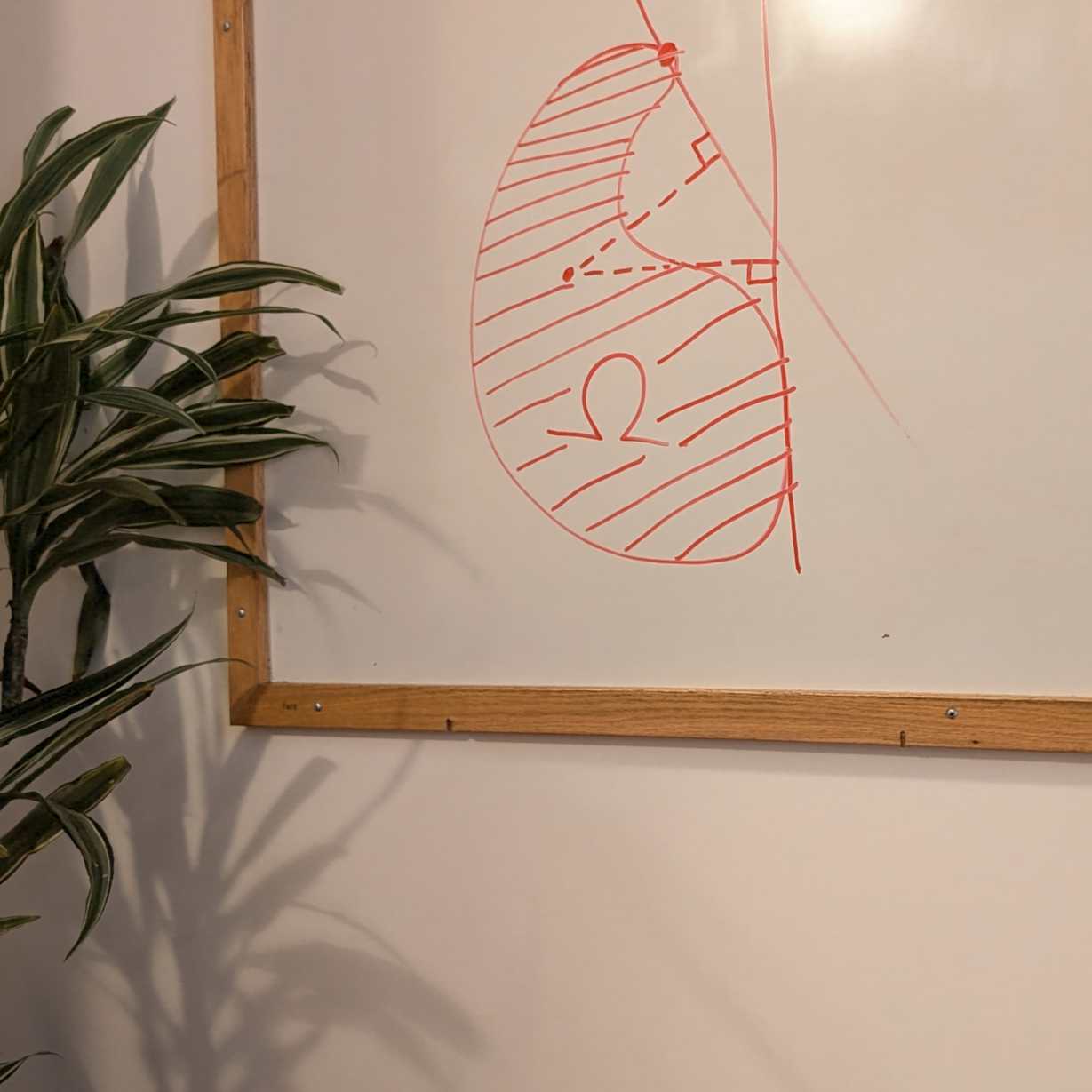} & %
        \includegraphics[width=\imgwidth]{figs/image_experiments/2025-04-09_01-22-30_whiteboard/lora_2025-04-27_01-27-14_finetune/eval_2025-05-03_18-23-20/best_reconstruction.jpg} & %
        \includegraphics[width=\imgwidth]{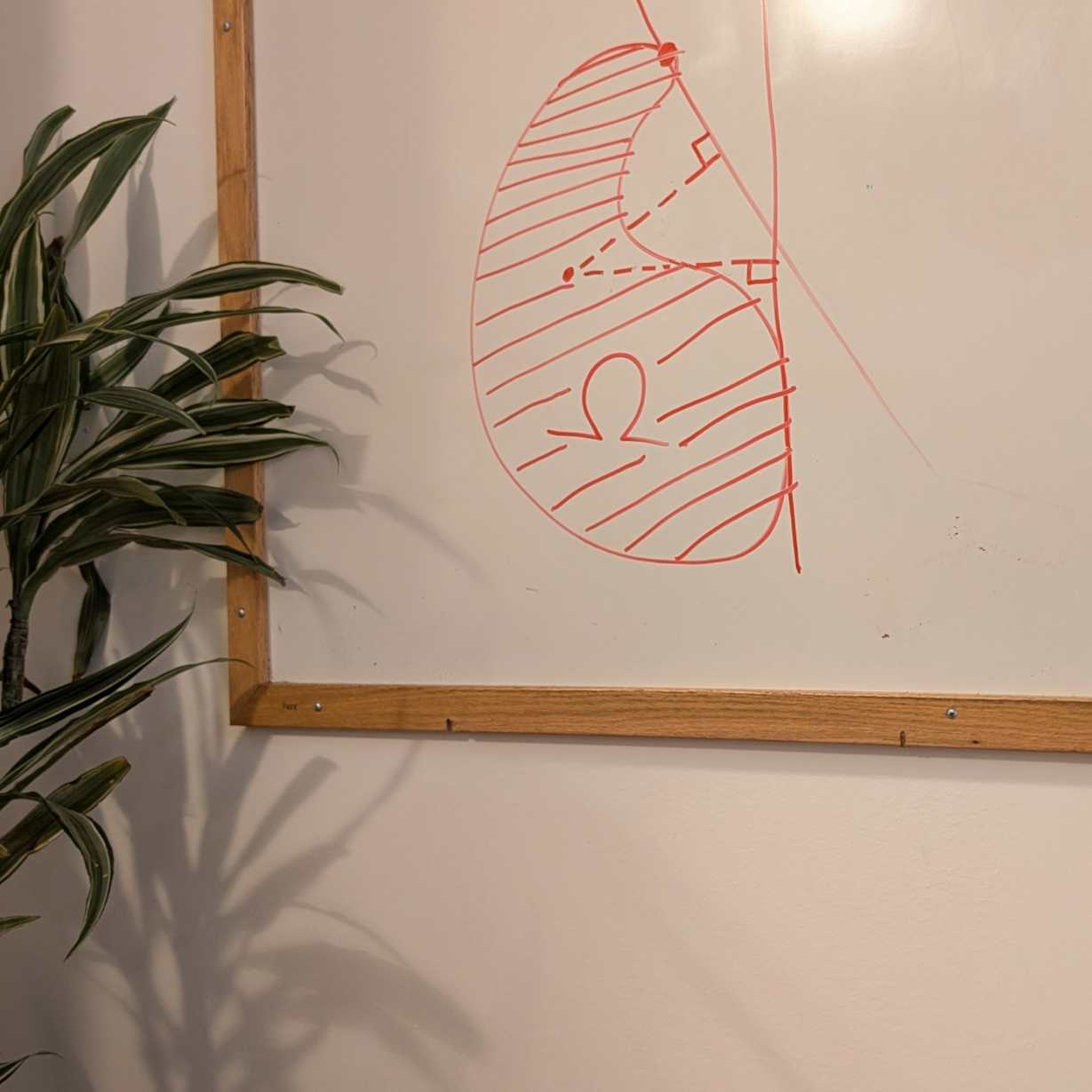} \\ %
        \hline

         & PSNR & \SI{37.4}{dB} & \SI{42.4}{dB} & \SI{44.45}{dB}  & \SI{46.1}{dB} & \SI{47.3}{dB} & \SI{47.8}{dB} & \SI{47.6}{dB} & \\
         \rotatebox{90}{DOF} &
         \includegraphics[width=\imgwidth]{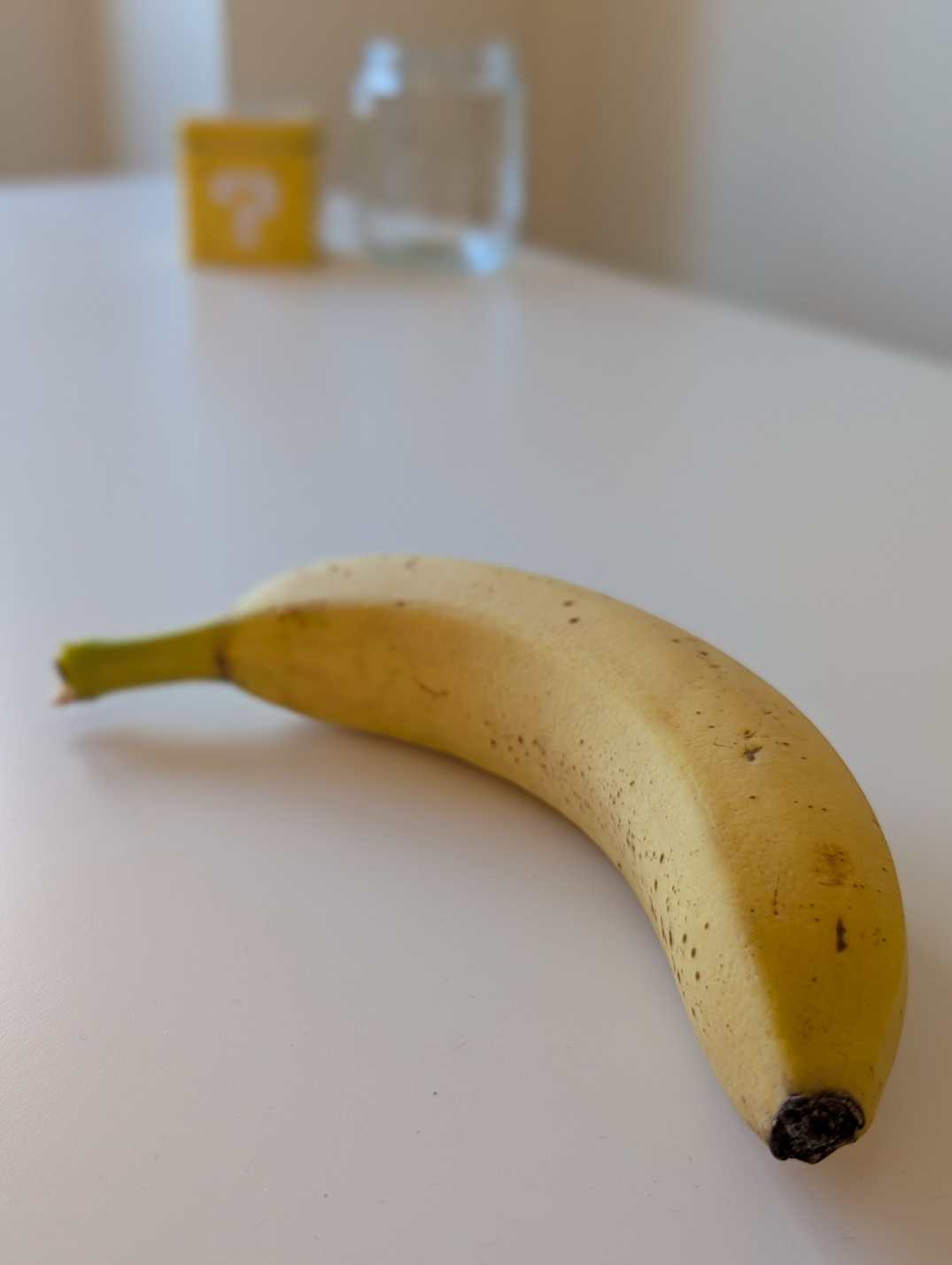} & %
         \includegraphics[width=\imgwidth]{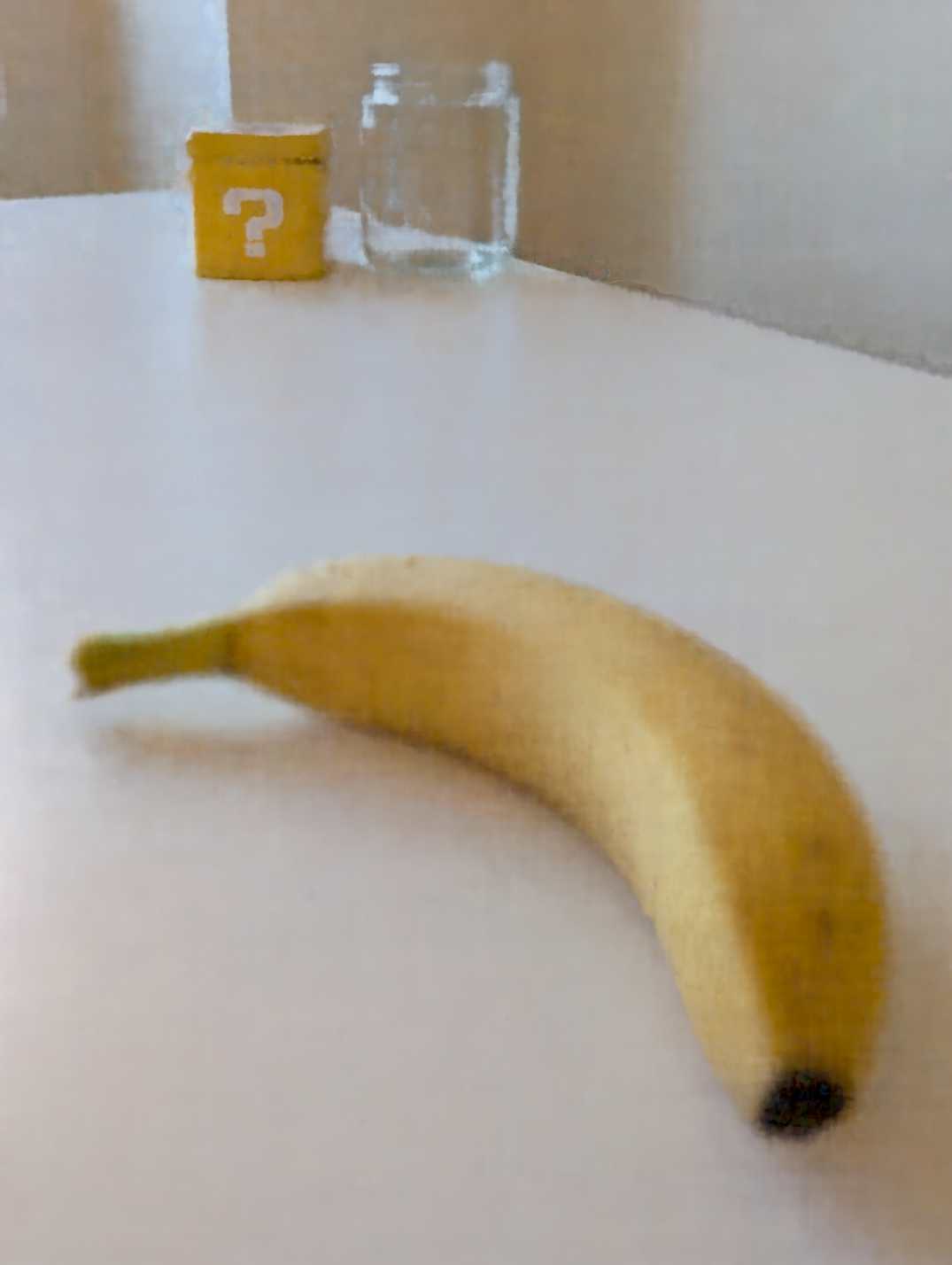} & %
         \includegraphics[width=\imgwidth]{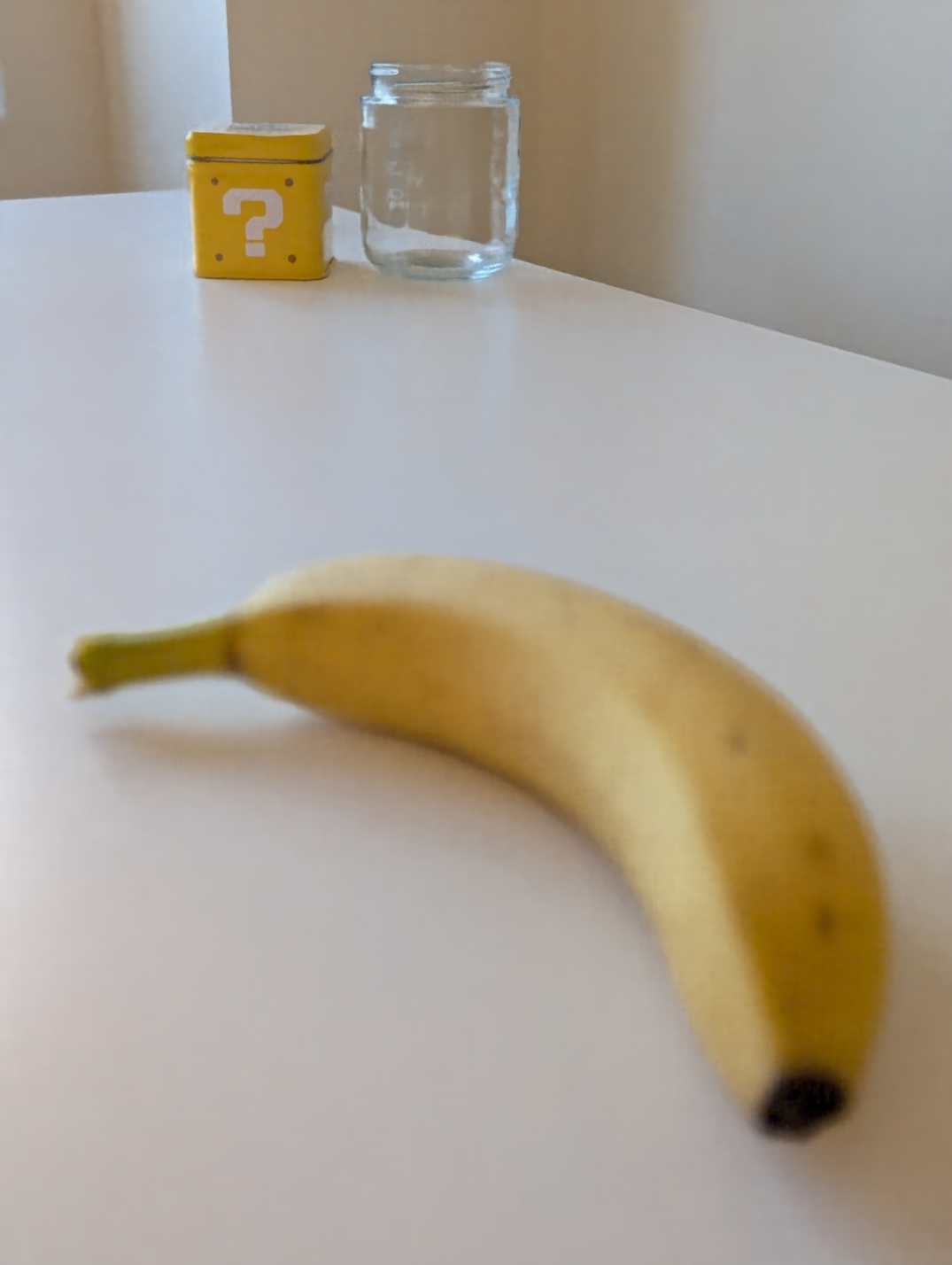} & %
         \includegraphics[width=\imgwidth]{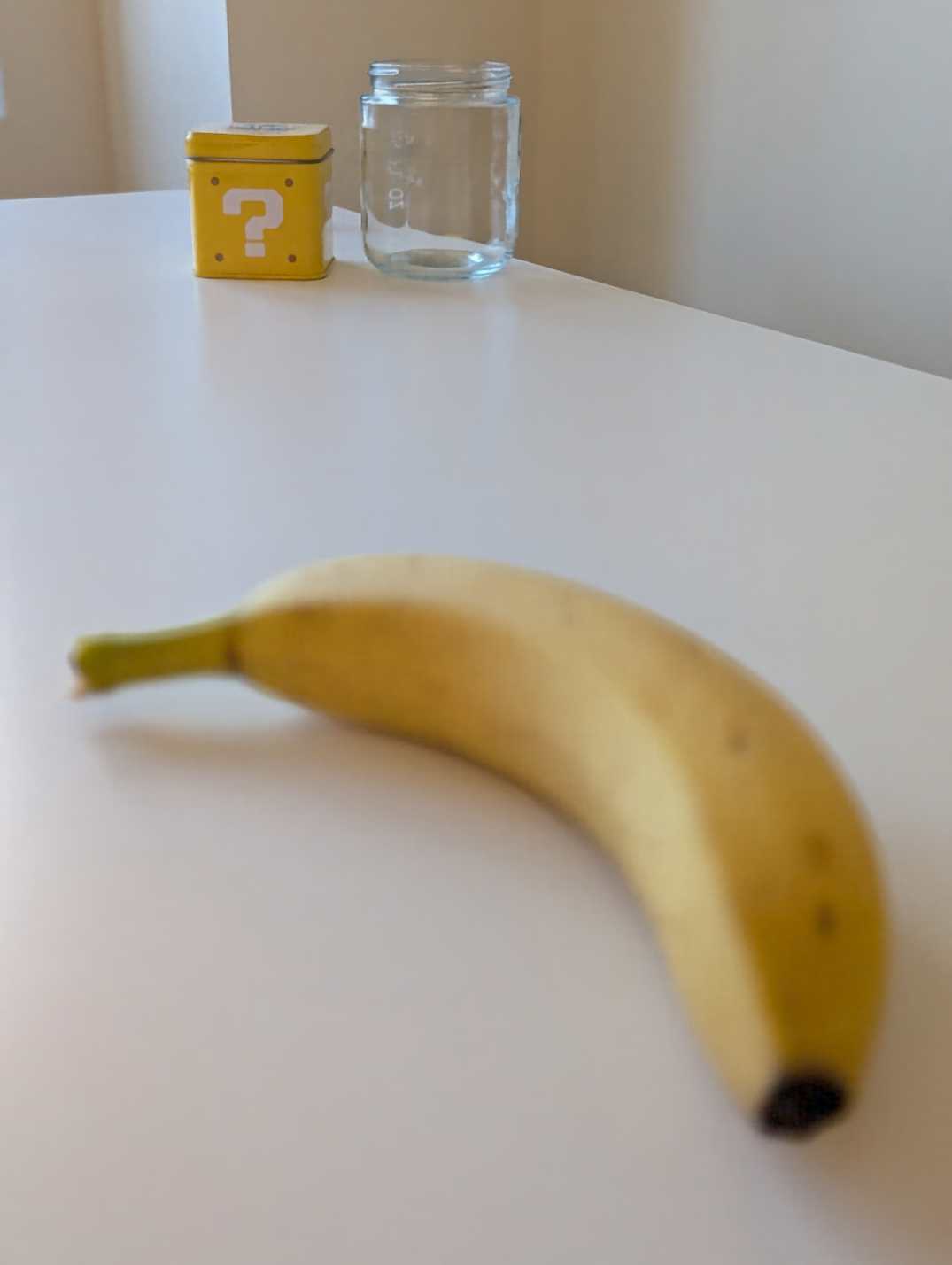} & %
         \includegraphics[width=\imgwidth]{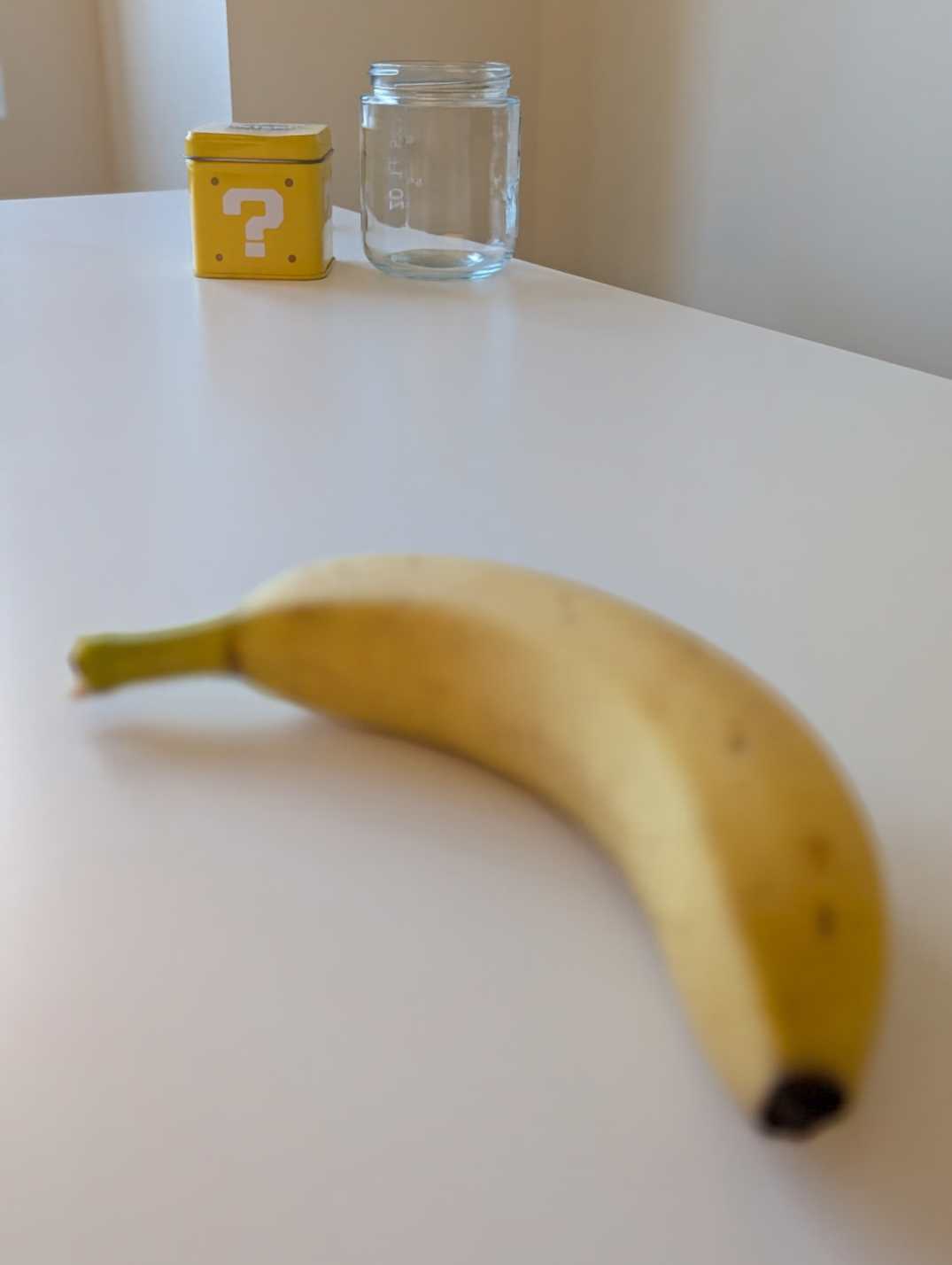} & %
         \includegraphics[width=\imgwidth]{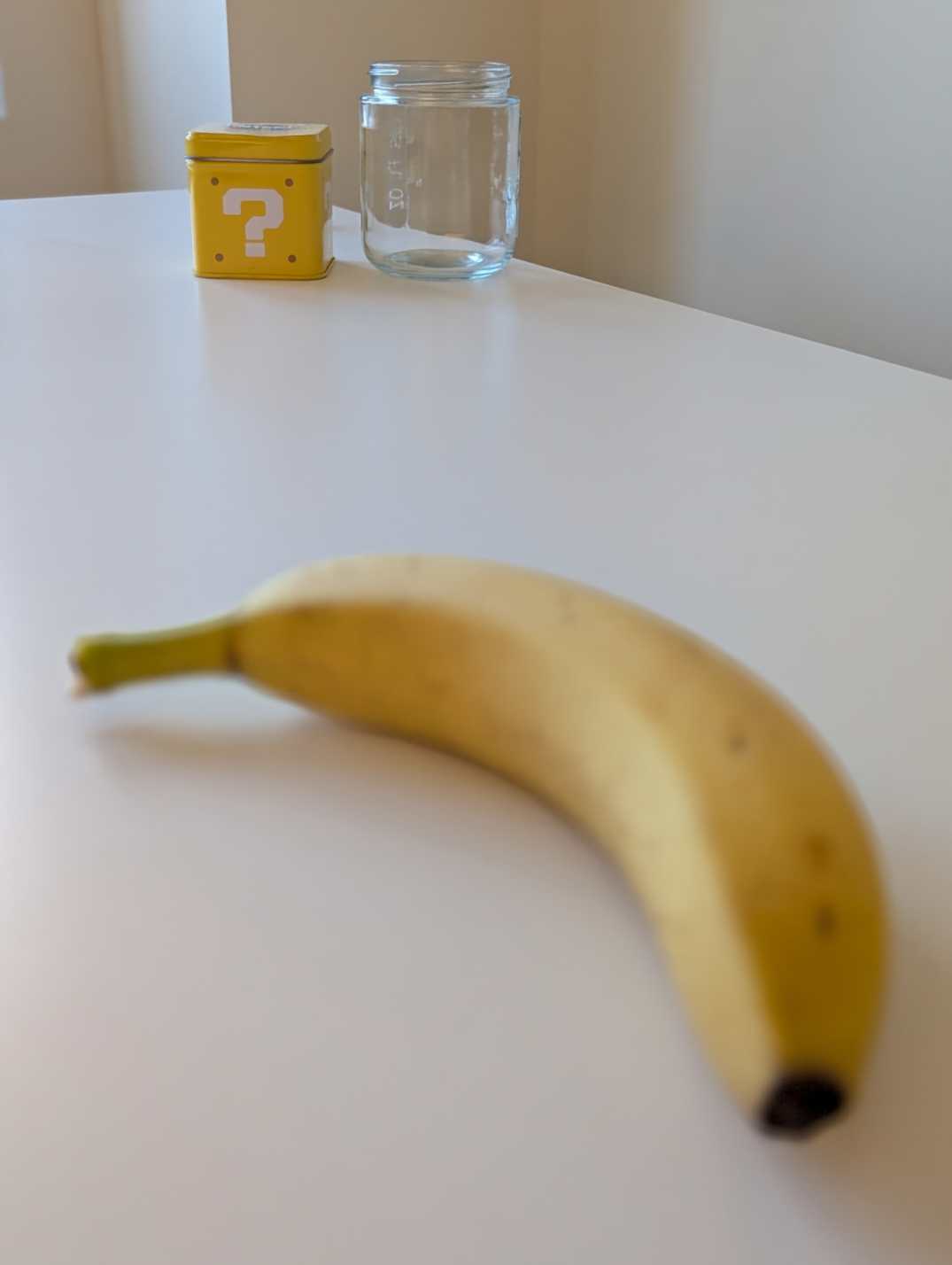} & %
         \includegraphics[width=\imgwidth]{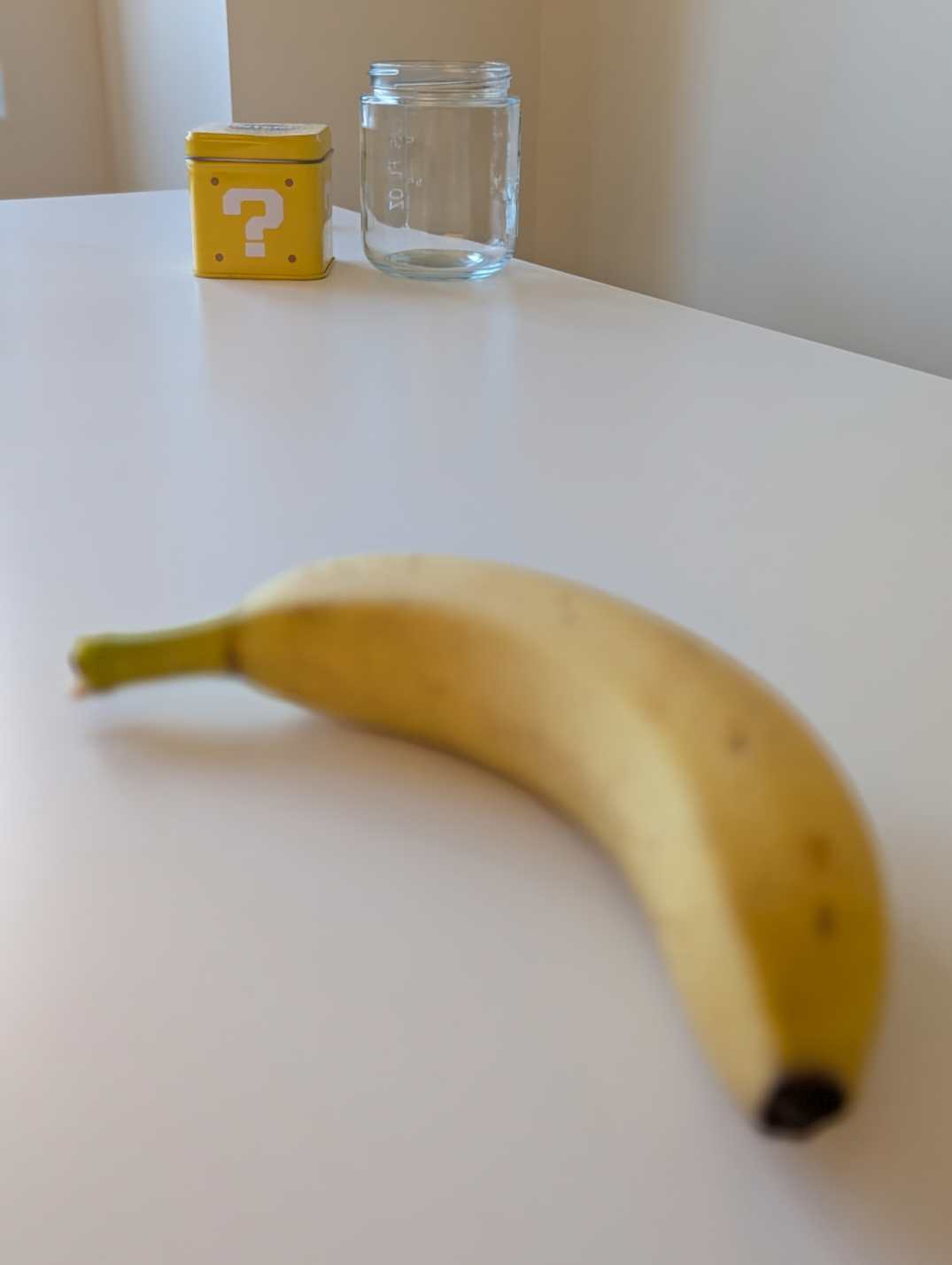} & %
         \includegraphics[width=\imgwidth]{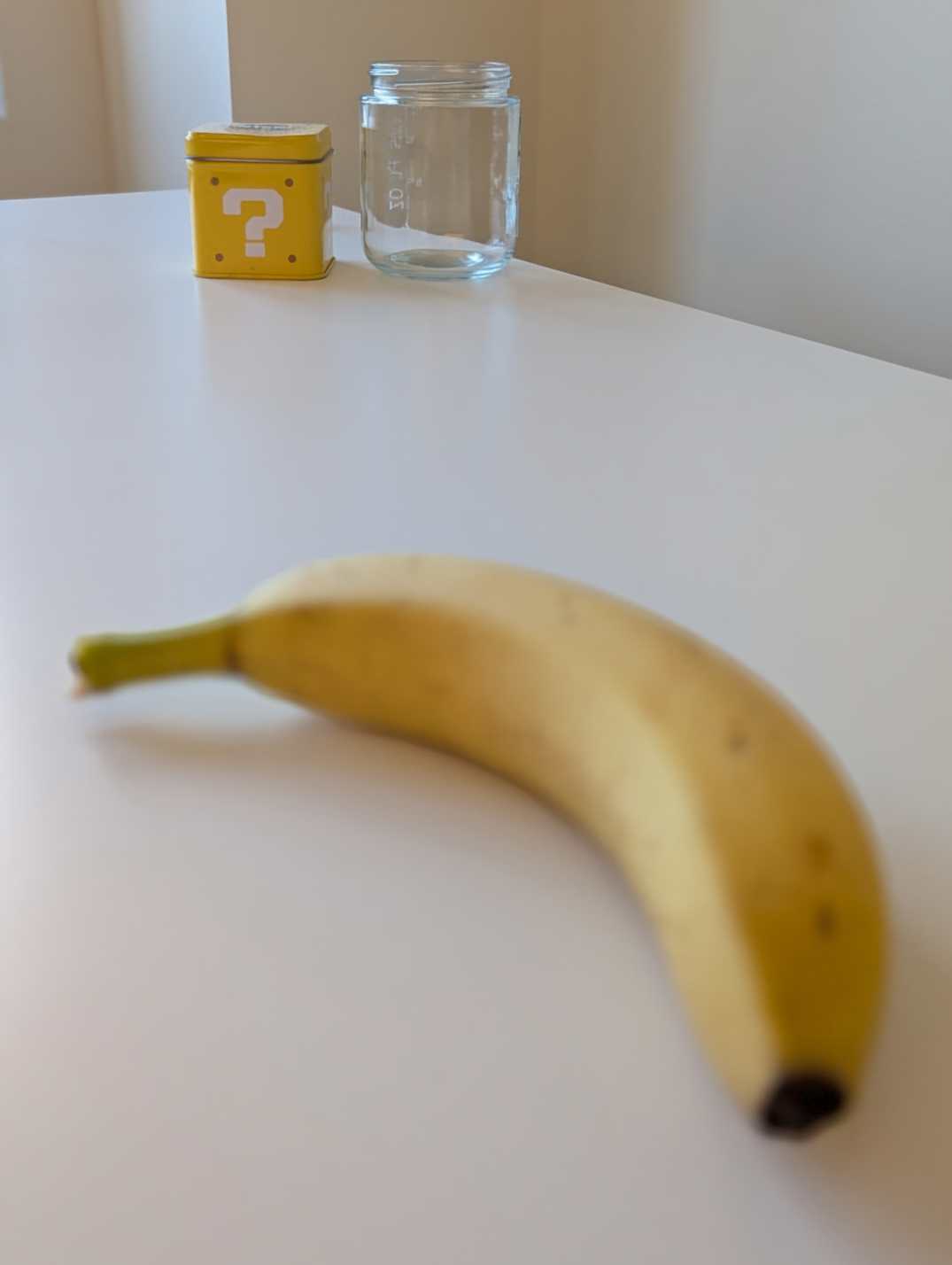} & %
         \includegraphics[width=\imgwidth]{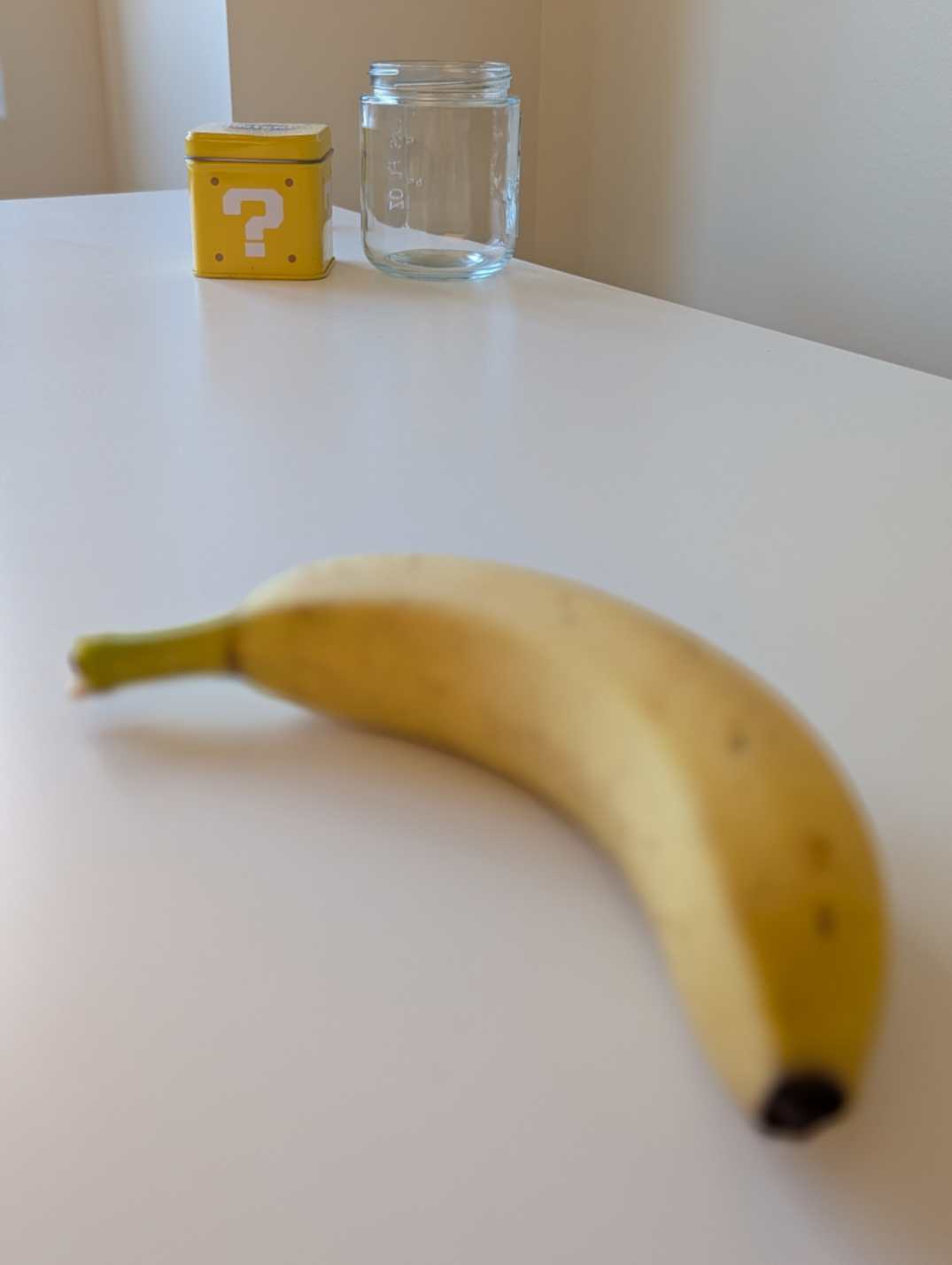} \\ %
         \hline

         & PSNR & \SI{37.9}{dB} & \SI{41.7}{dB} & \SI{43.2}{dB}  & \SI{44.2}{dB} & \SI{45.2}{dB} & \SI{45.9}{dB} & \SI{45.4}{dB} & \\
         \rotatebox{90}{Sunset} &
         \includegraphics[width=\imgwidth]{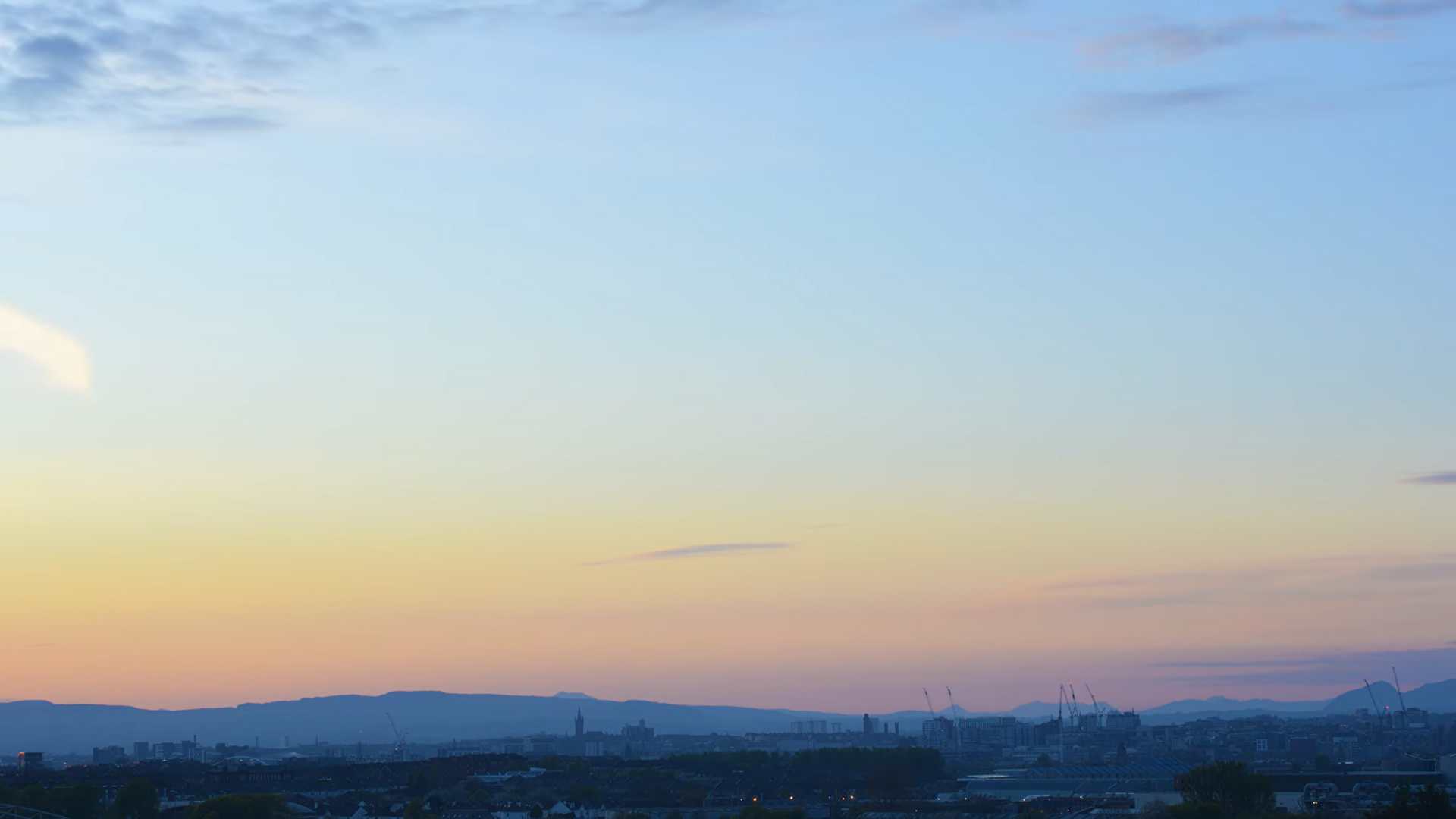} & %
         \includegraphics[width=\imgwidth]{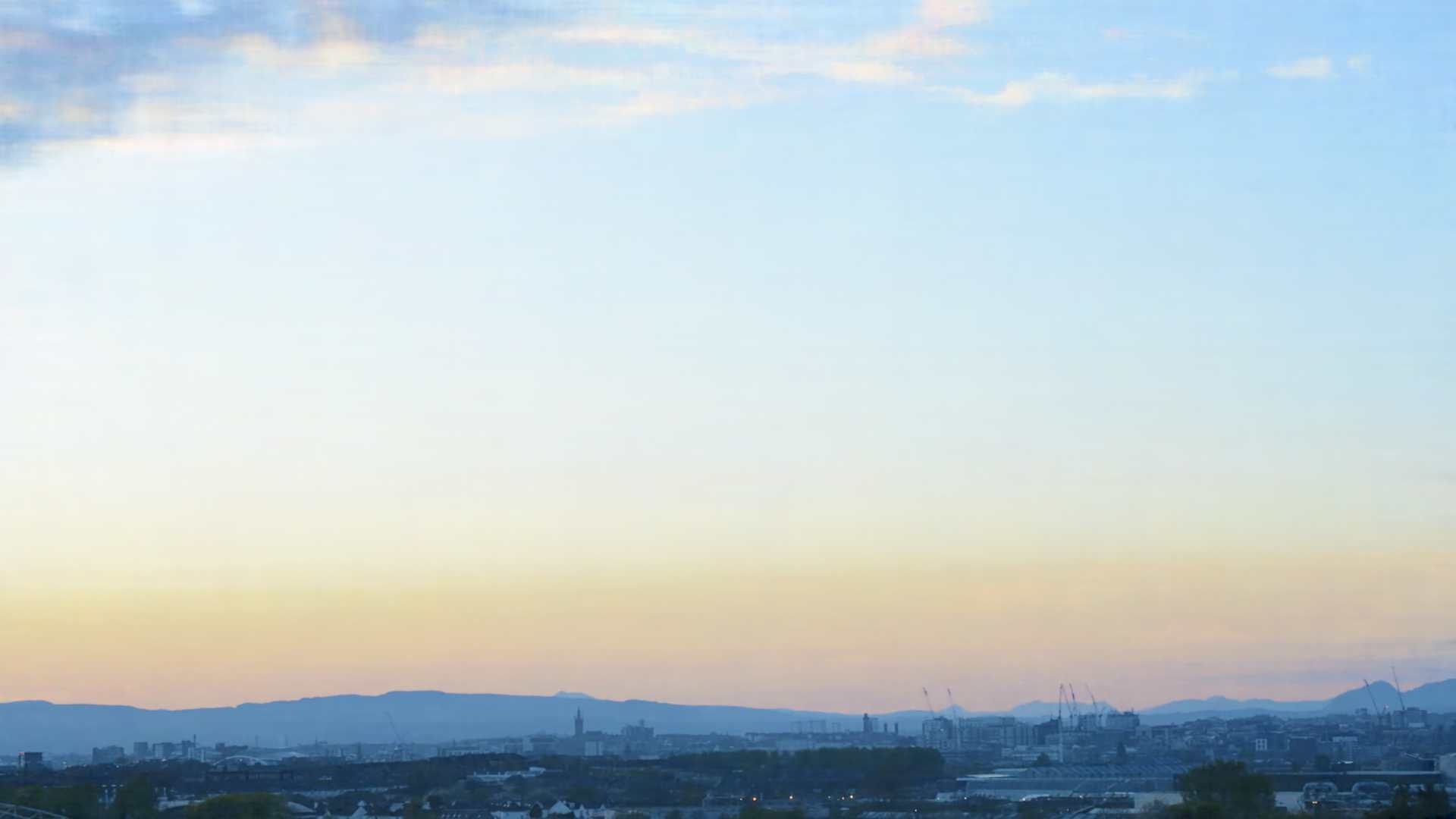} & %
         \includegraphics[width=\imgwidth]{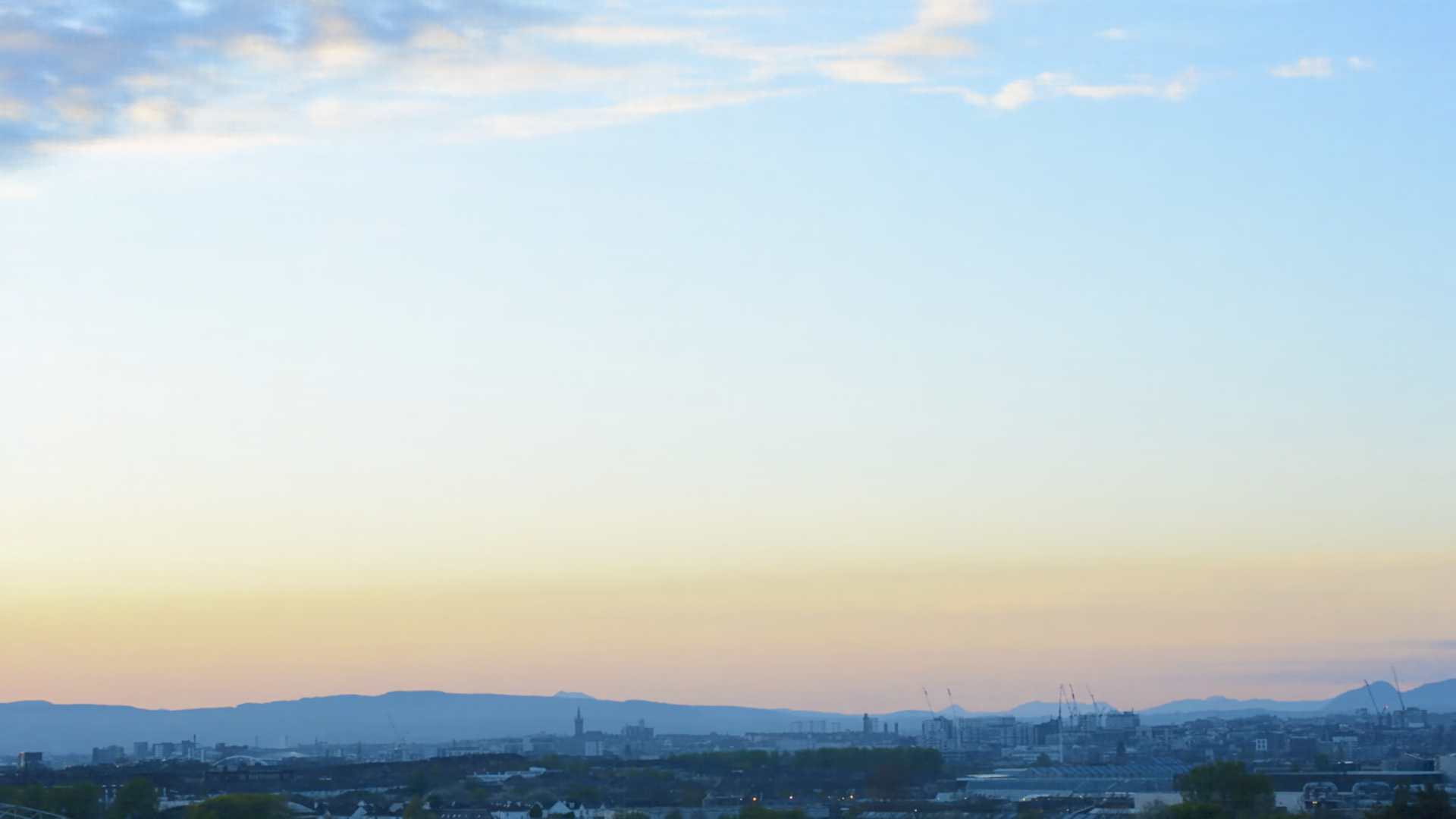} & %
         \includegraphics[width=\imgwidth]{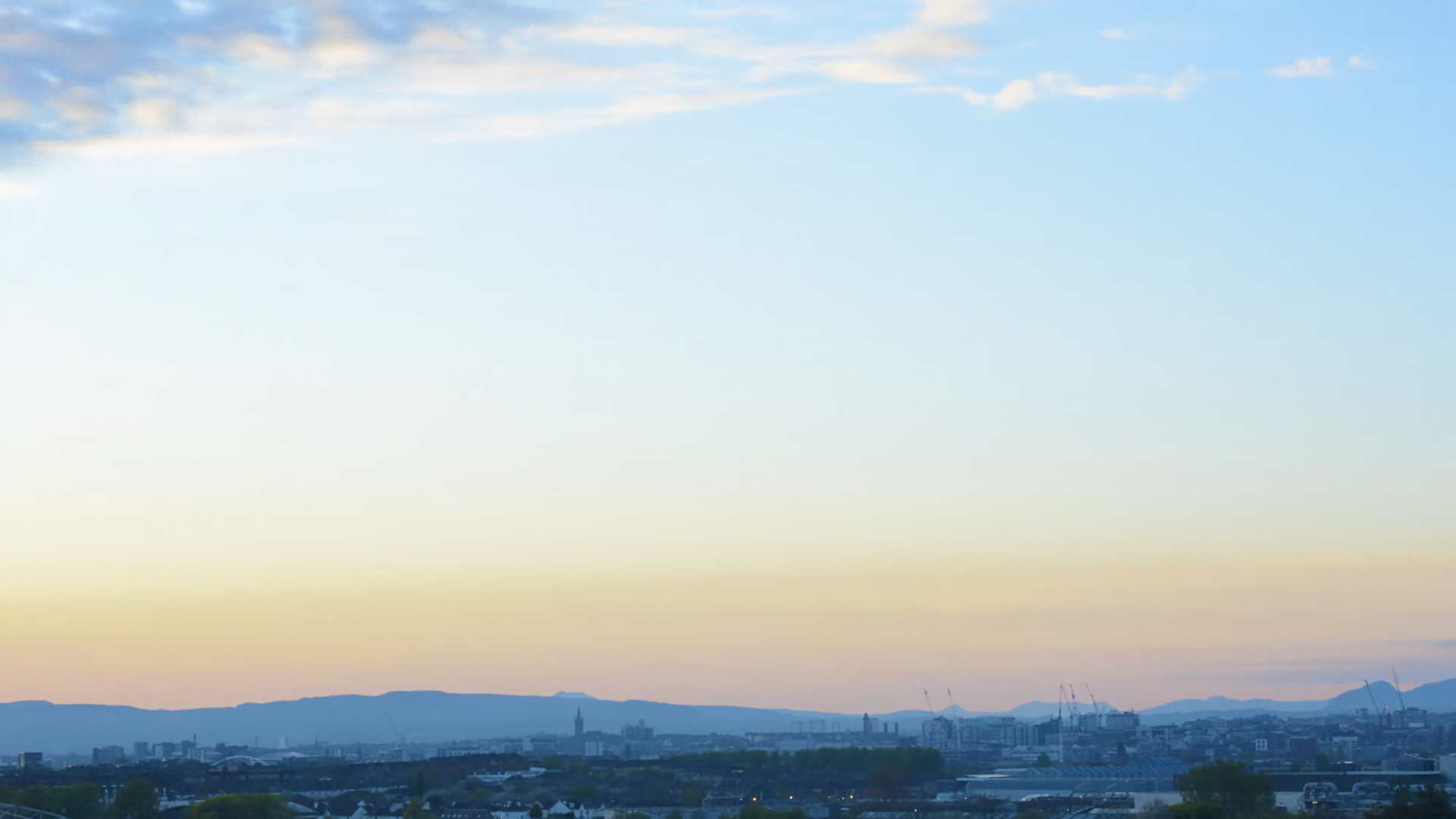} & %
         \includegraphics[width=\imgwidth]{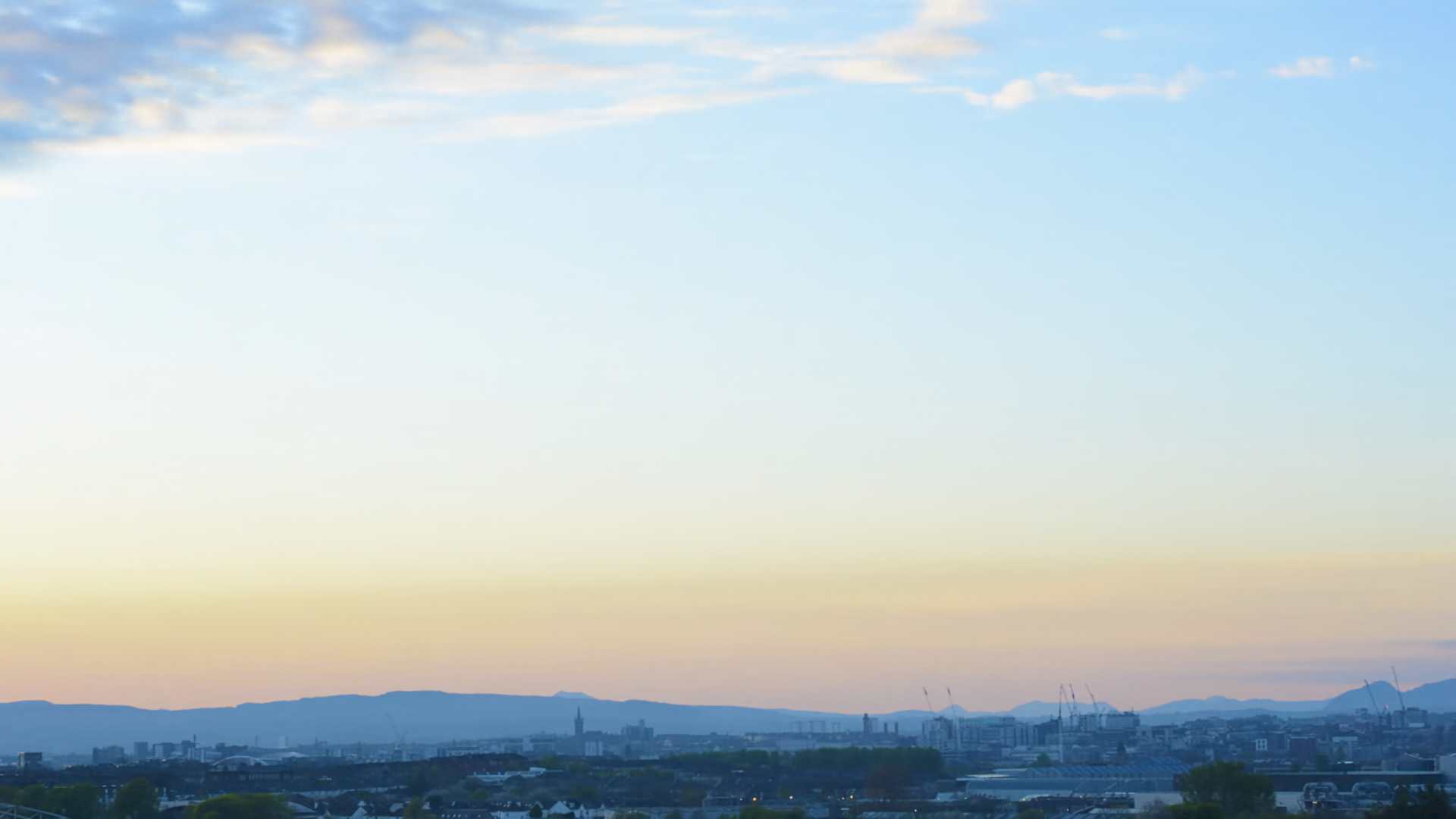} & %
         \includegraphics[width=\imgwidth]{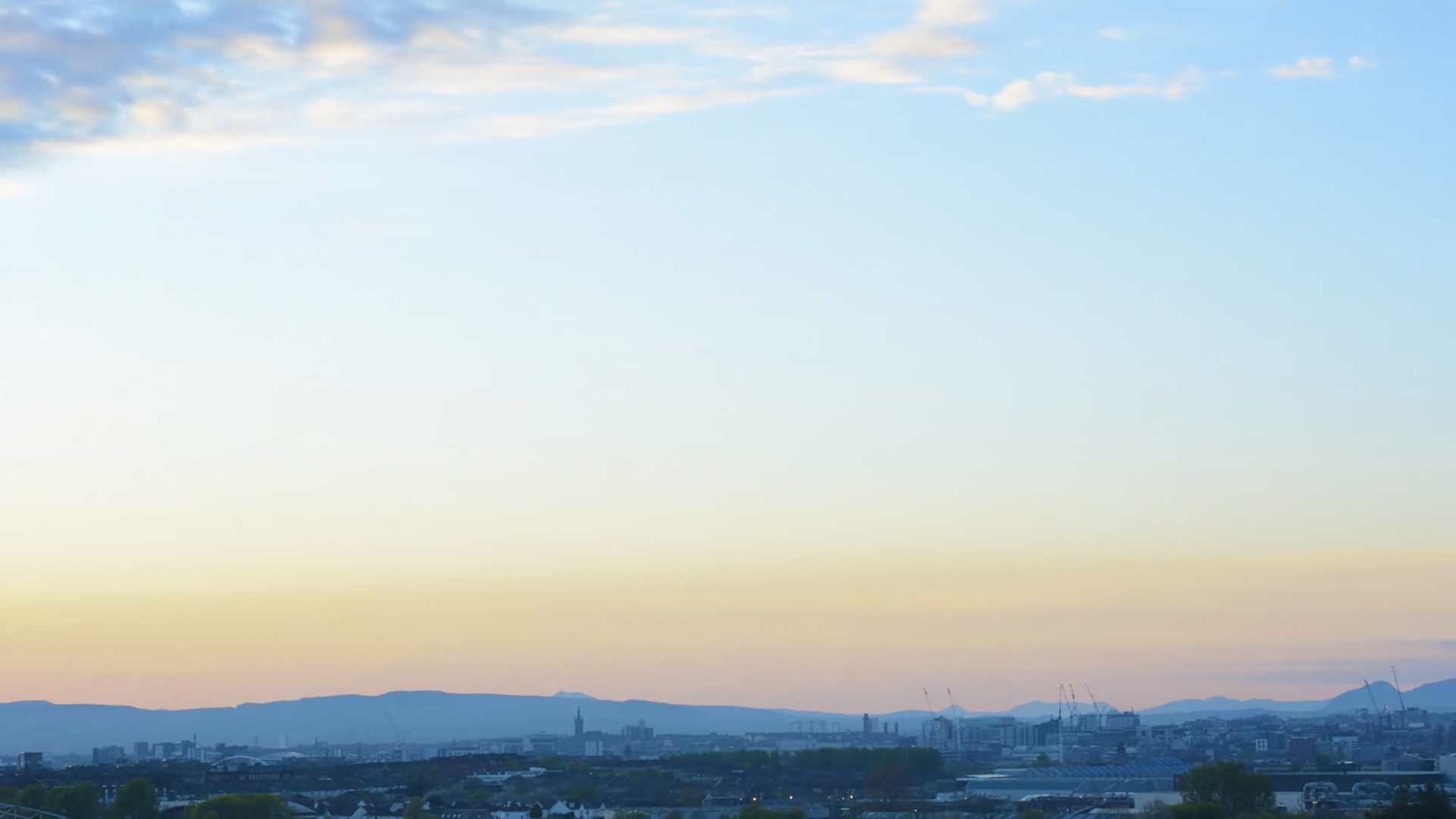} & %
         \includegraphics[width=\imgwidth]{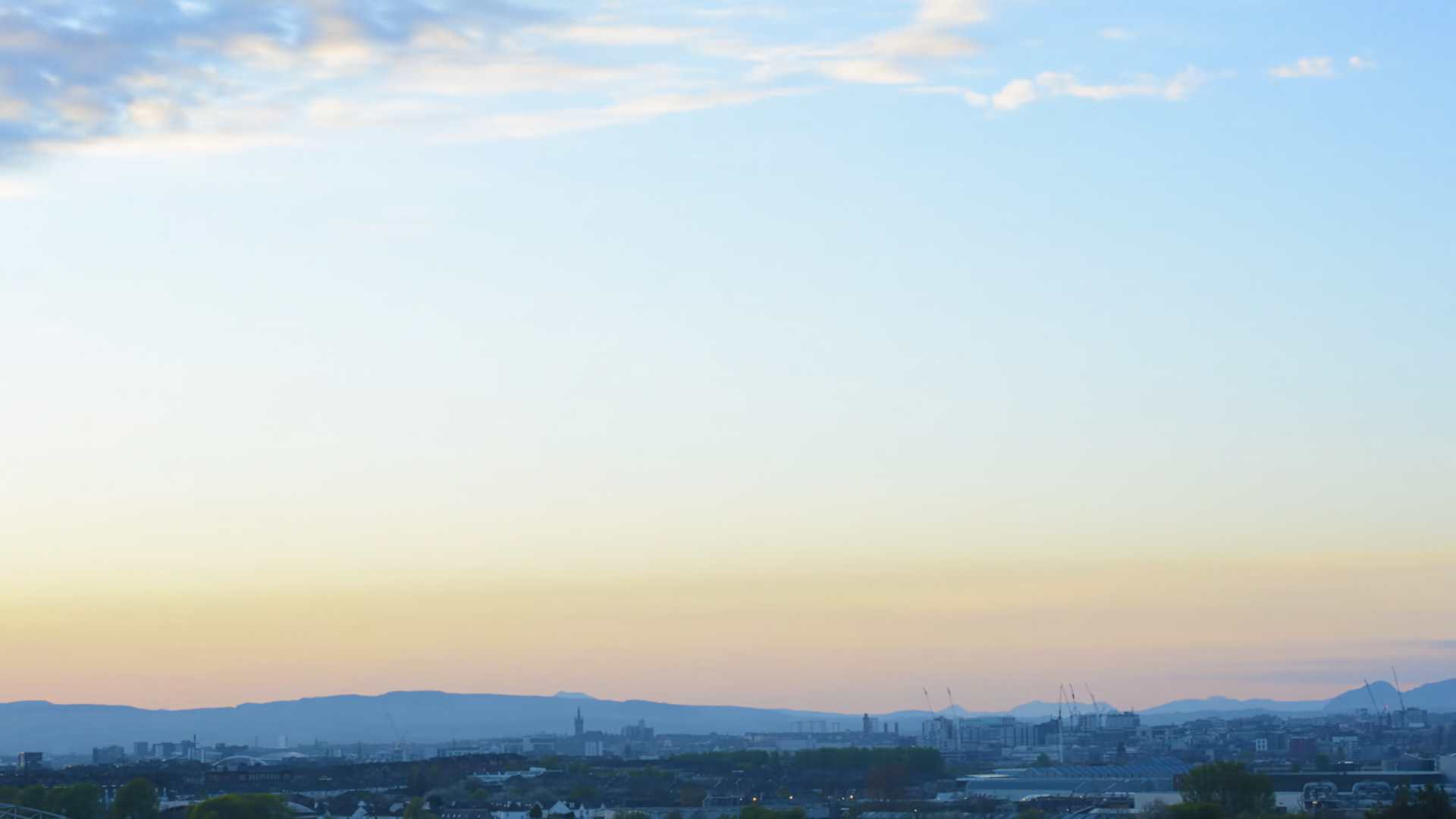} & %
         \includegraphics[width=\imgwidth]{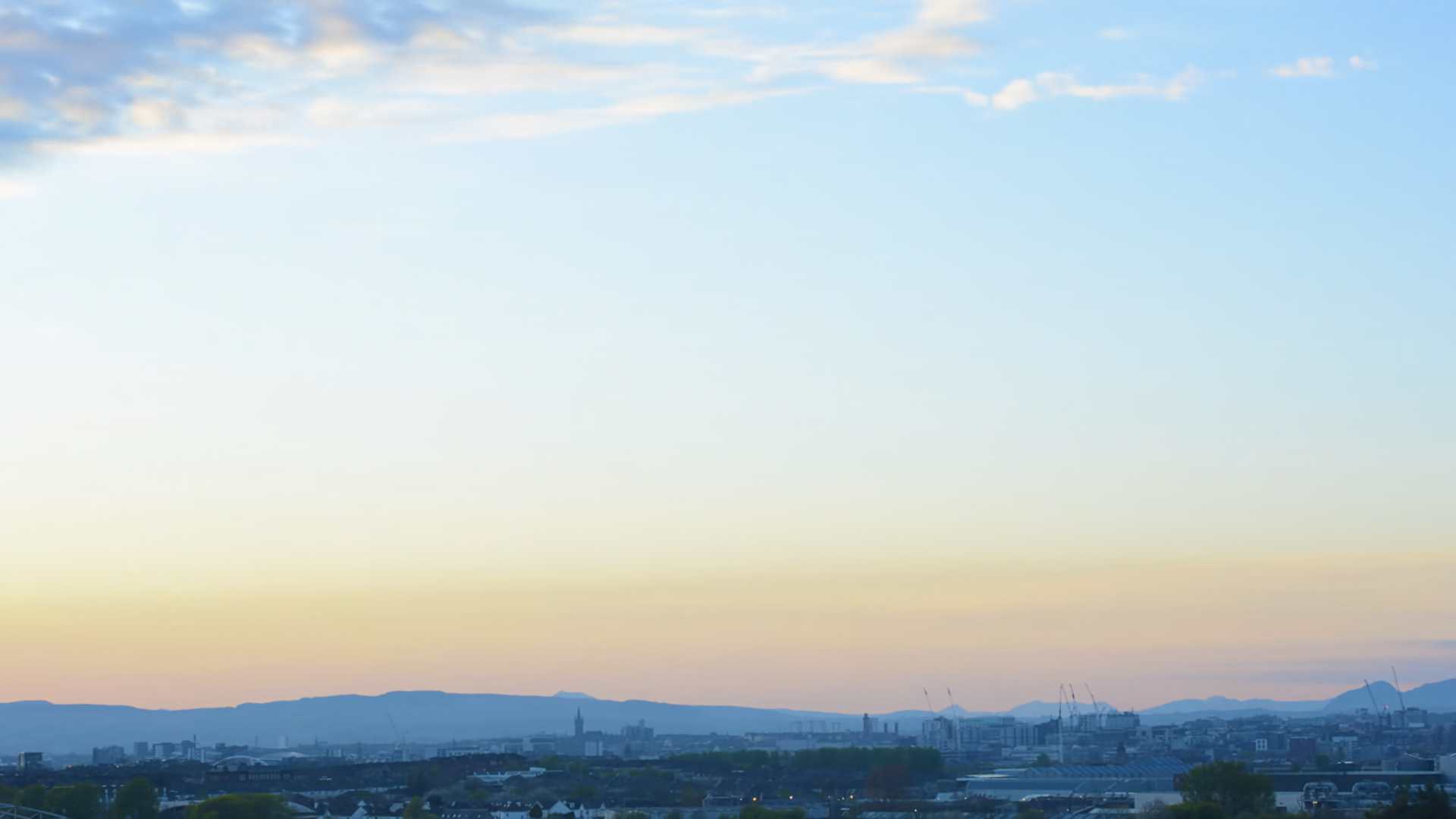} & %
         \includegraphics[width=\imgwidth]{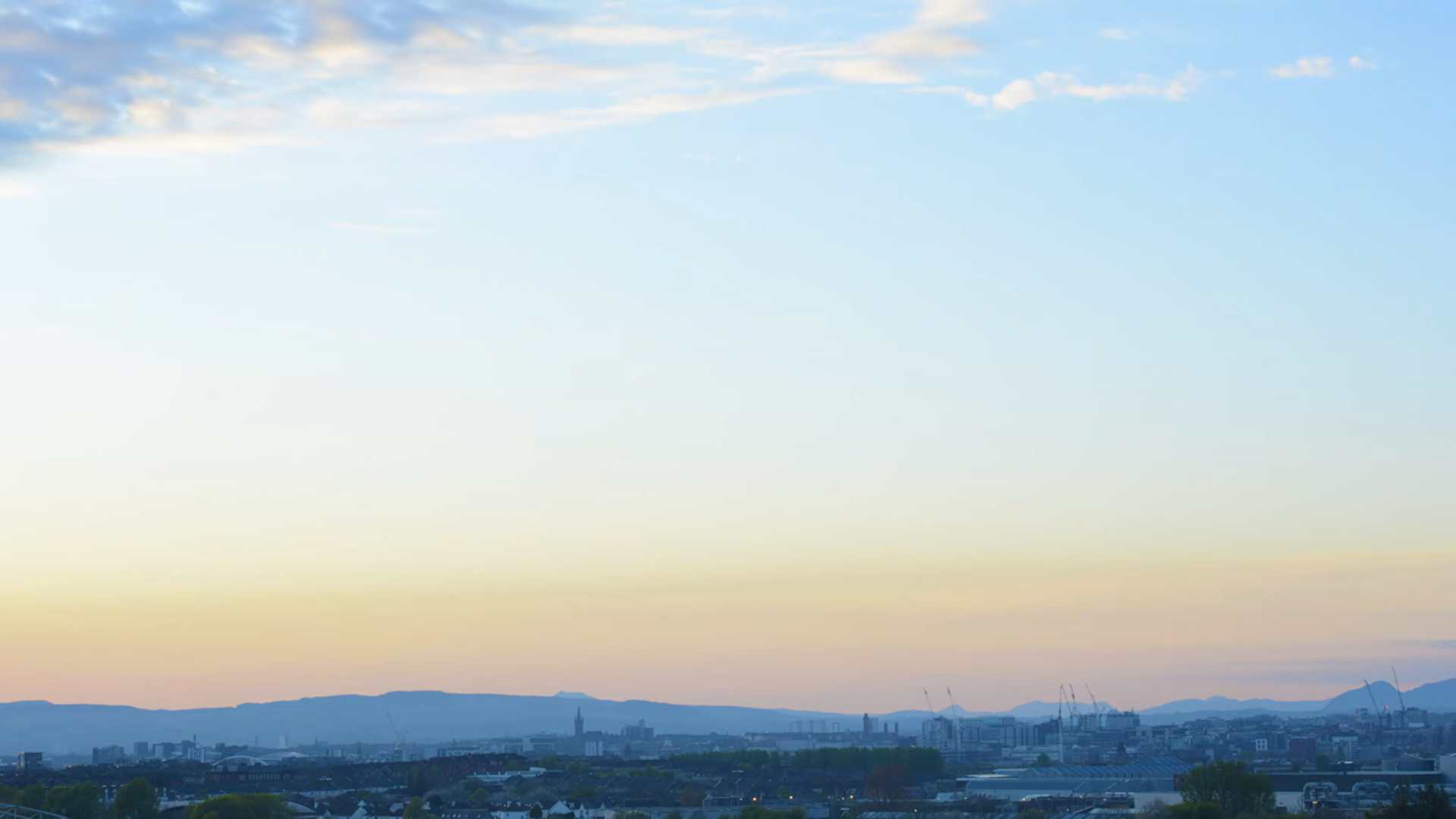} \\ %
         \hline

         & PSNR & \SI{36.2}{dB} & \SI{40.2}{dB} & \SI{41.3}{dB}  & \SI{42.3}{dB} & \SI{43.5}{dB} & \SI{44.2}{dB} & \SI{43.6}{dB} & \\
         \rotatebox{90}{\small{Countertop}} &
         \includegraphics[width=\imgwidth]{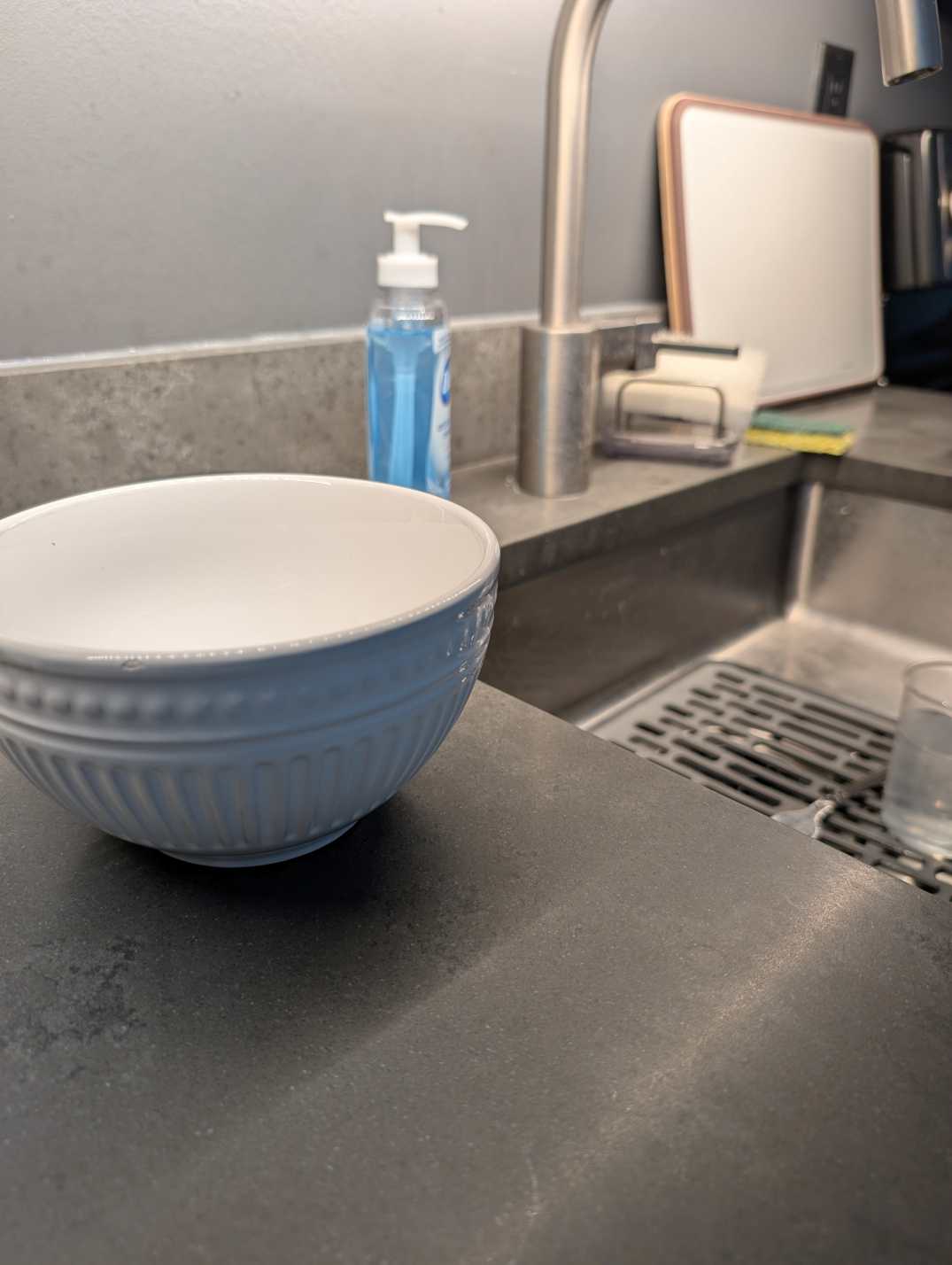} & %
         \includegraphics[width=\imgwidth]{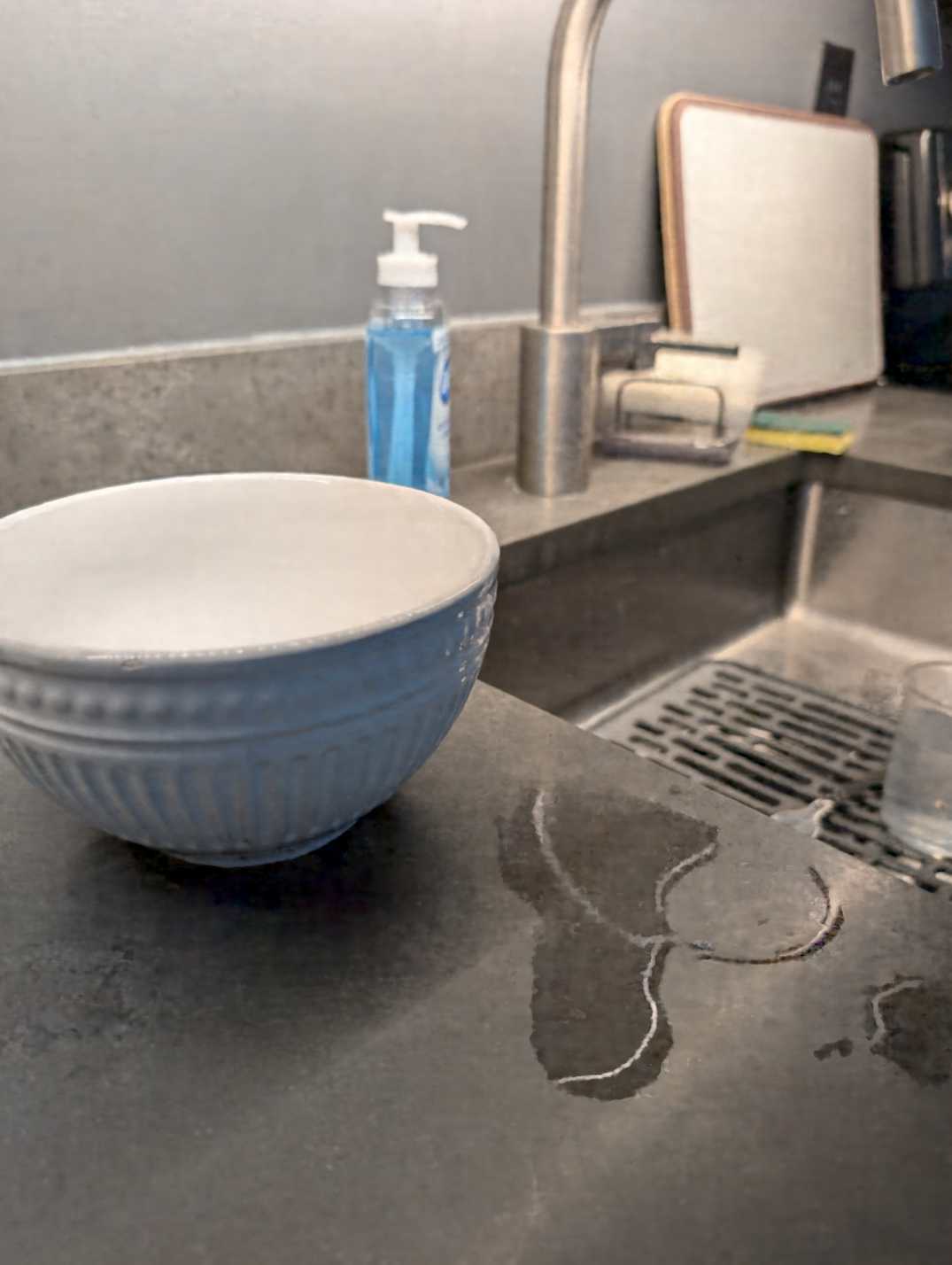} & %
         \includegraphics[width=\imgwidth]{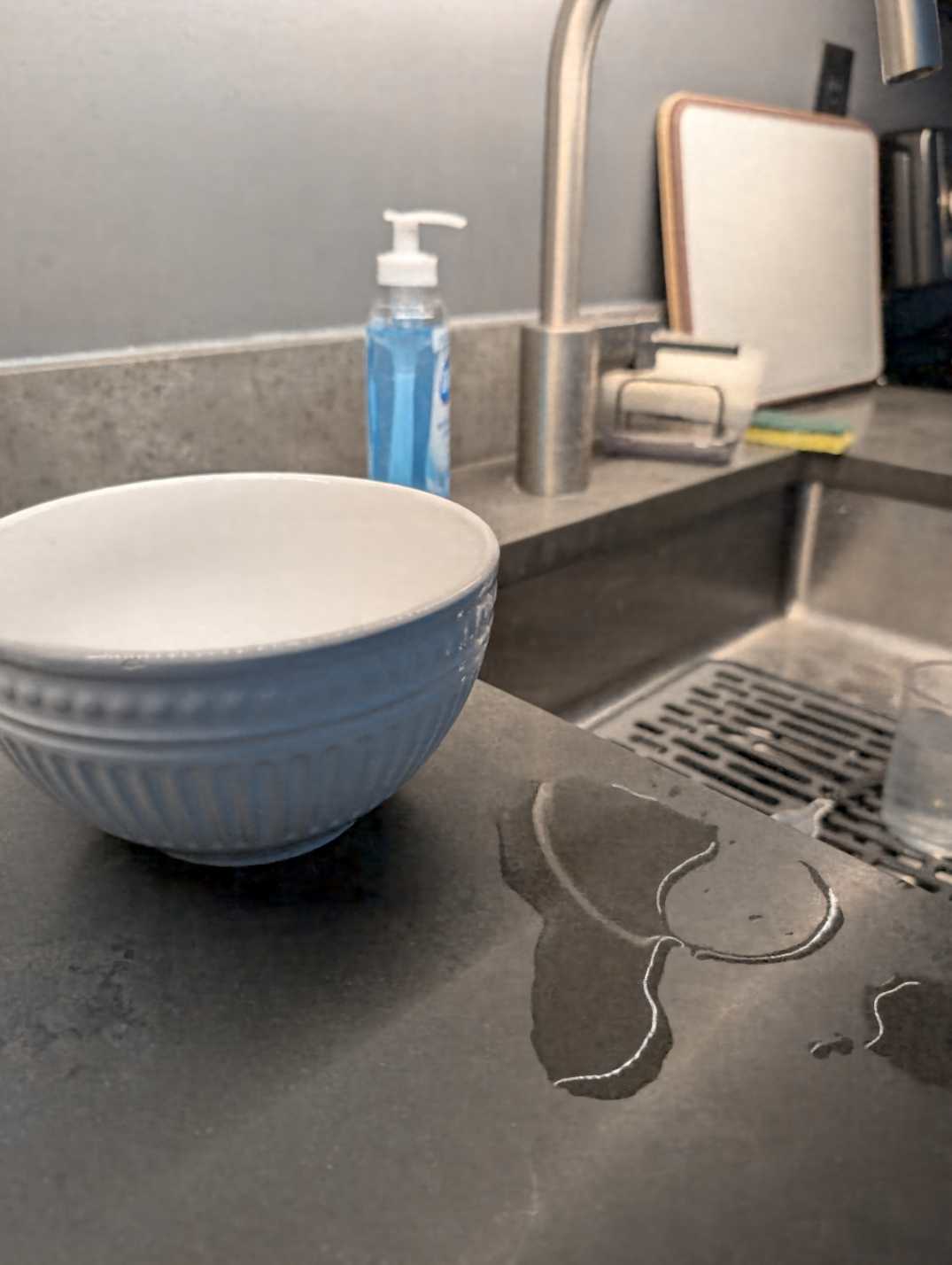} & %
         \includegraphics[width=\imgwidth]{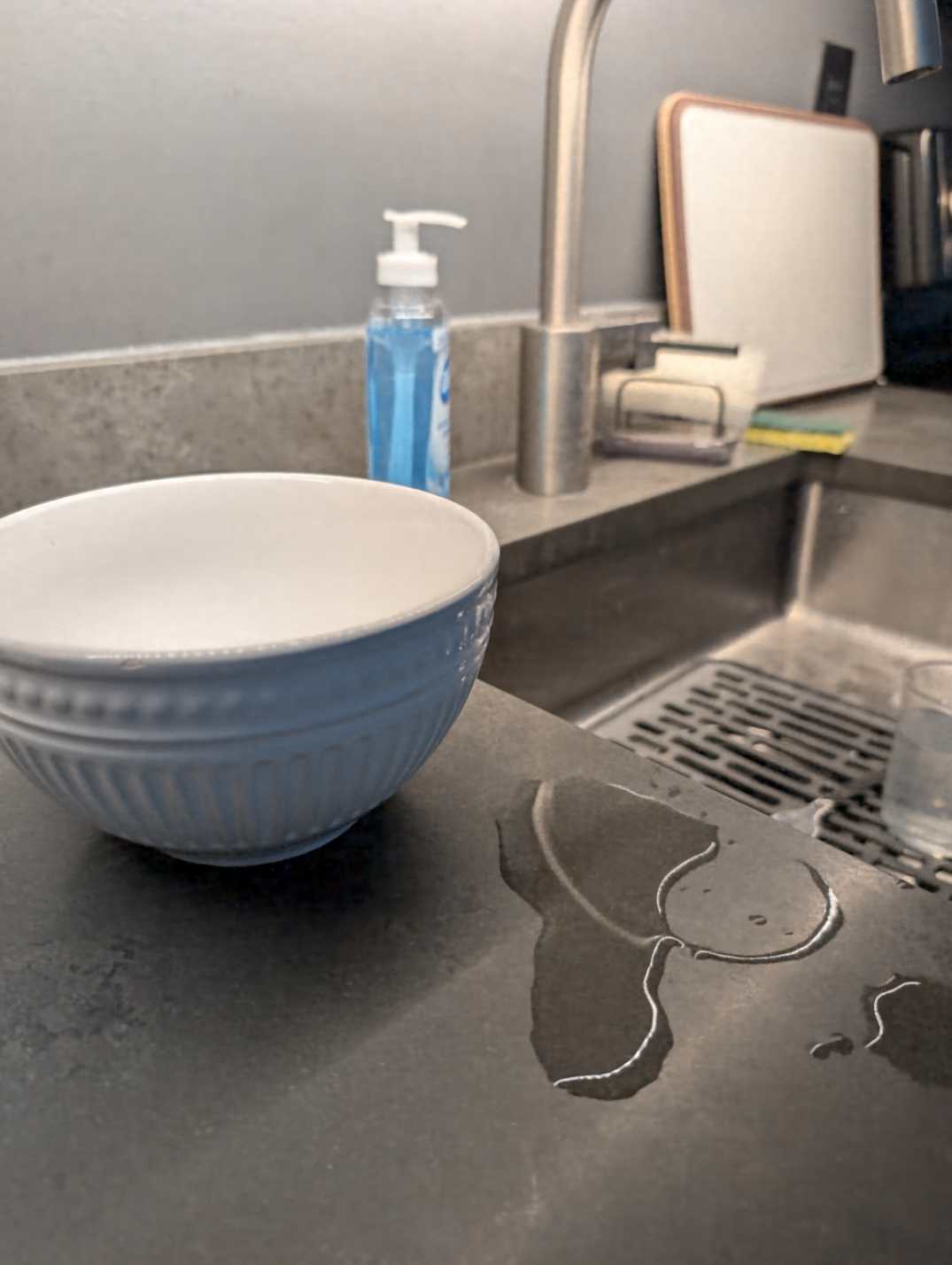} & %
         \includegraphics[width=\imgwidth]{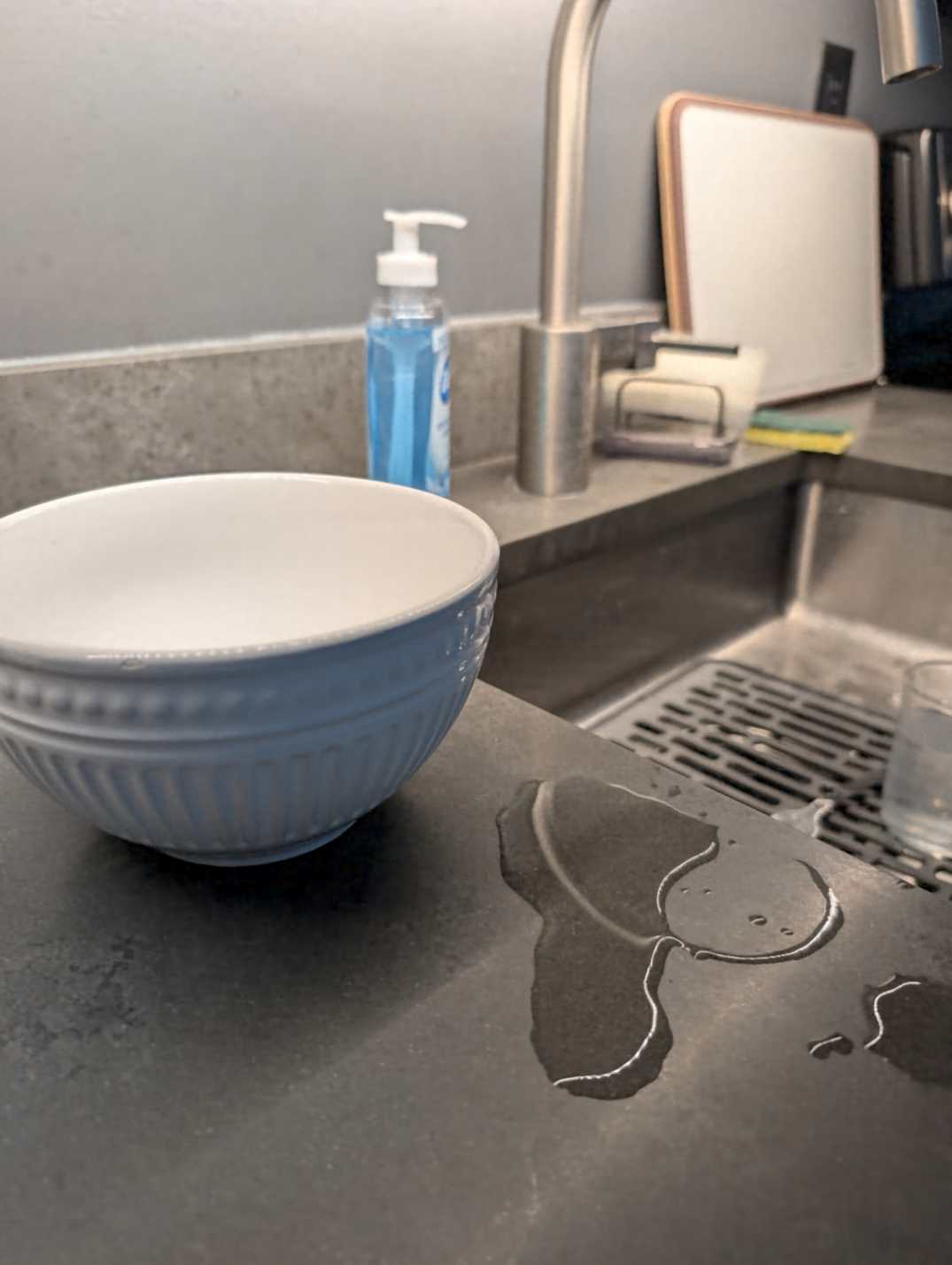} & %
         \includegraphics[width=\imgwidth]{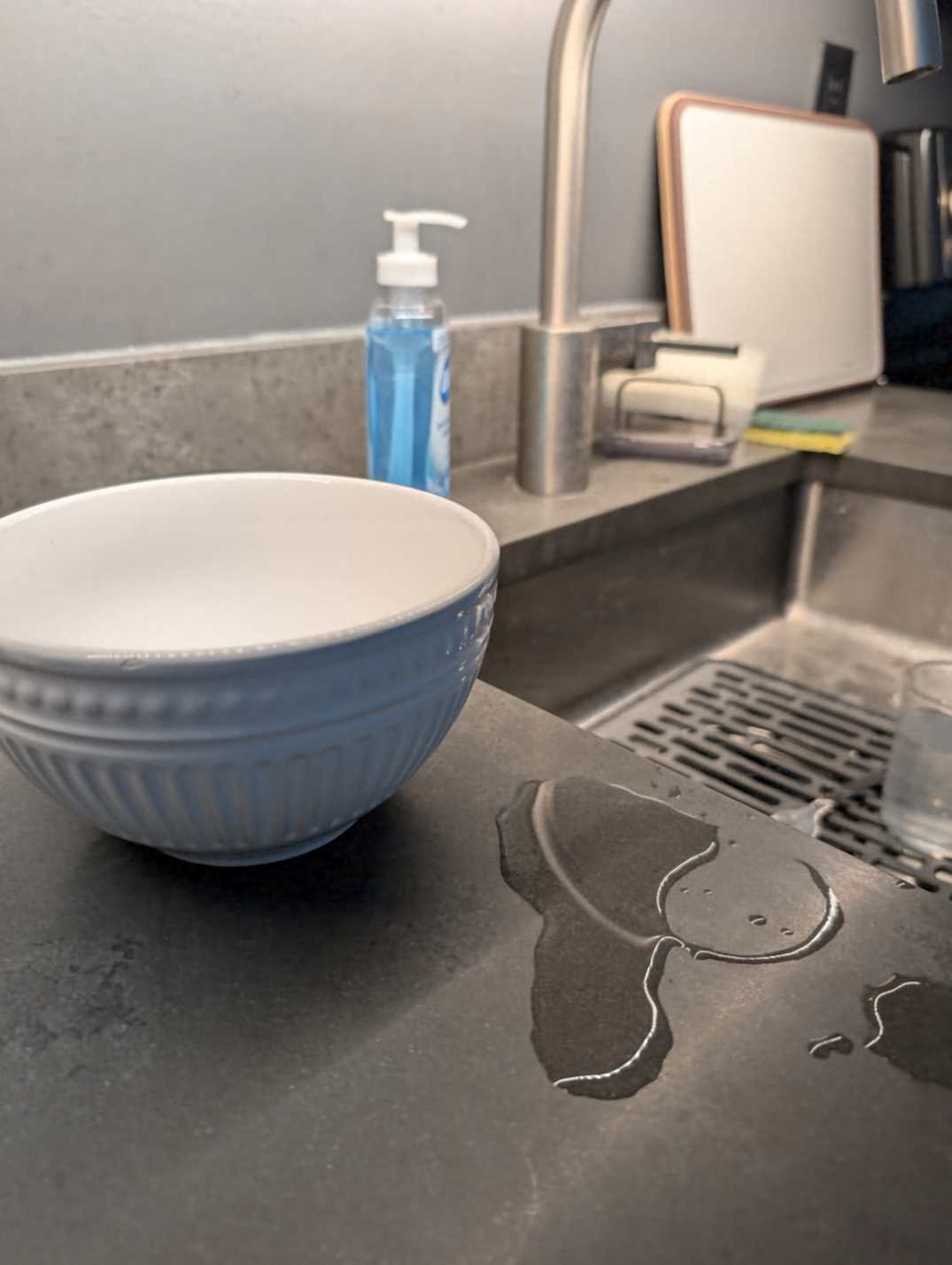} & %
         \includegraphics[width=\imgwidth]{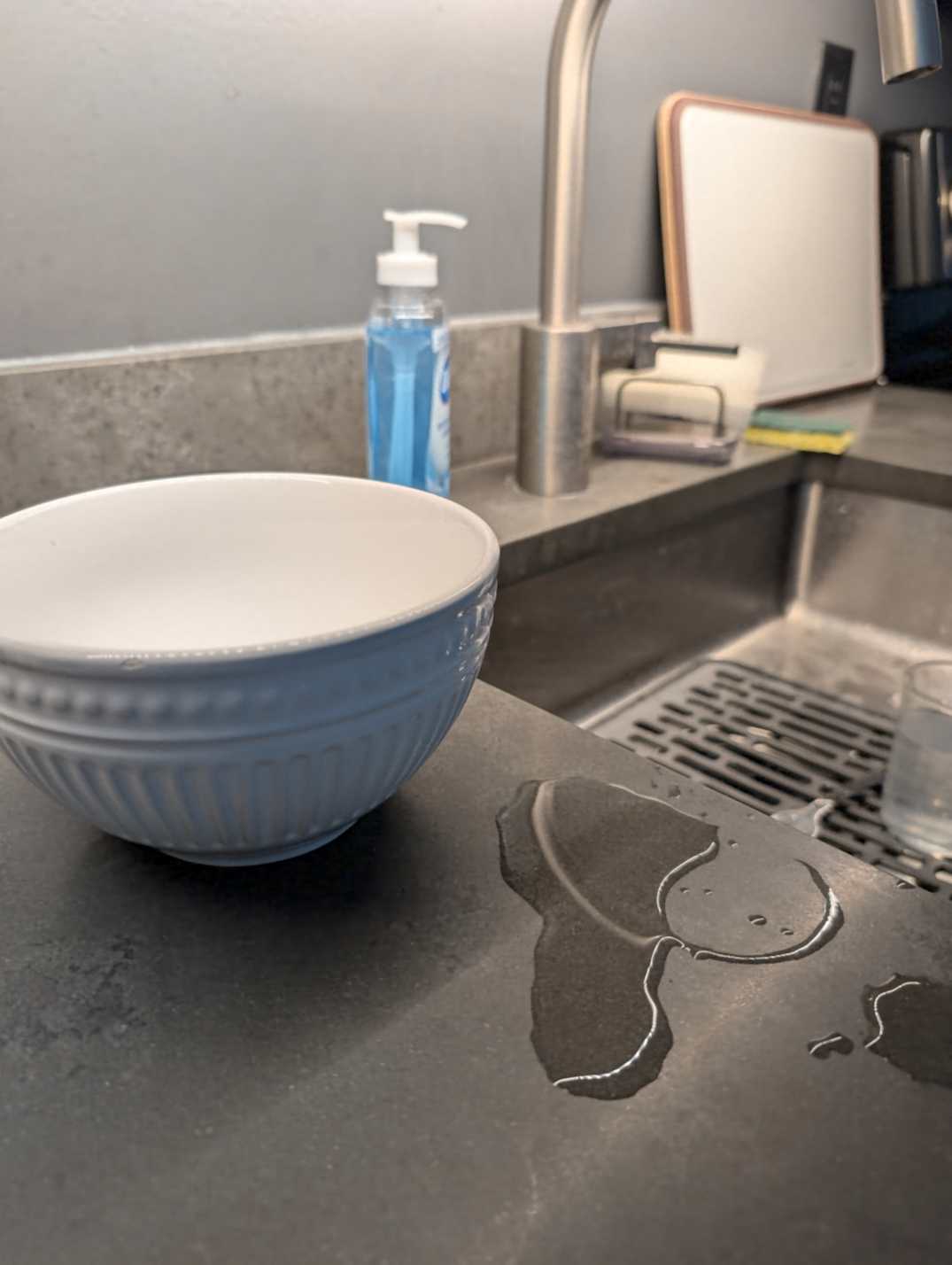} & %
         \includegraphics[width=\imgwidth]{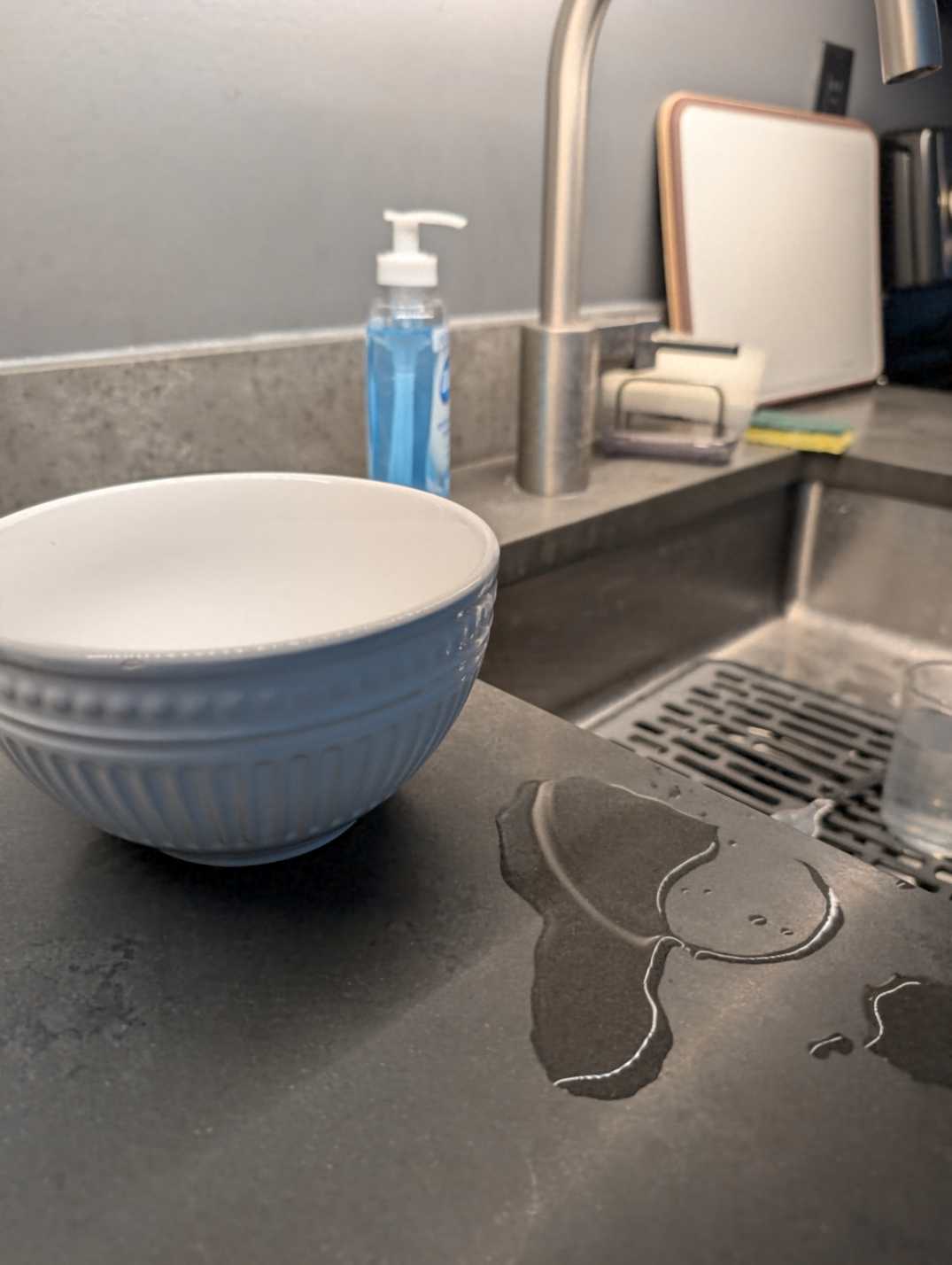} & %
         \includegraphics[width=\imgwidth]{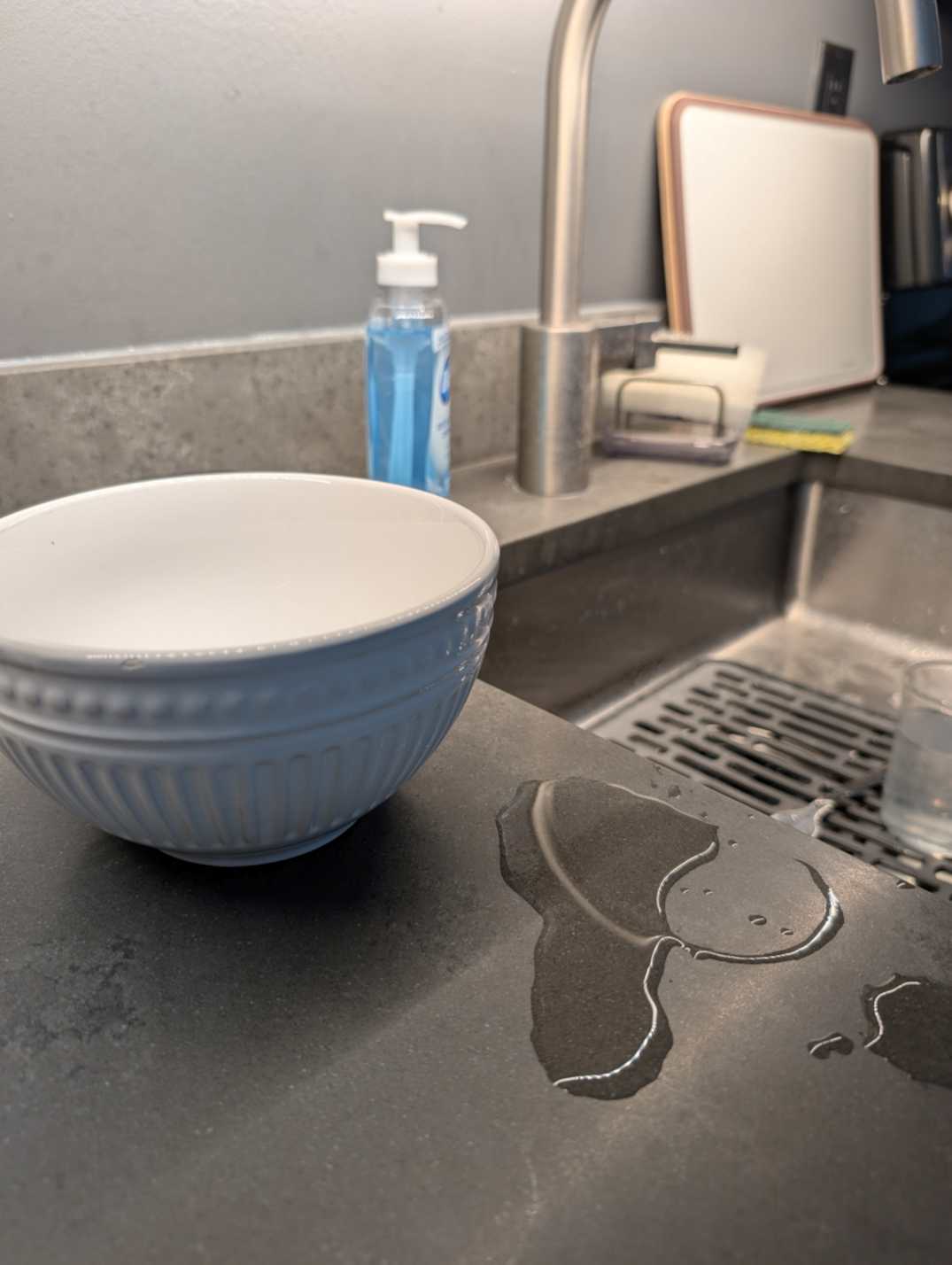} \\ %
         \hline

         & PSNR & \SI{34.4}{dB} & \SI{36.9}{dB} & \SI{38.13}{dB}  & \SI{38.9}{dB} & \SI{39.6}{dB} & \SI{40.2}{dB} & \SI{39.9}{dB} & \\
         \rotatebox{90}{Sink} &
         \includegraphics[width=\imgwidth]{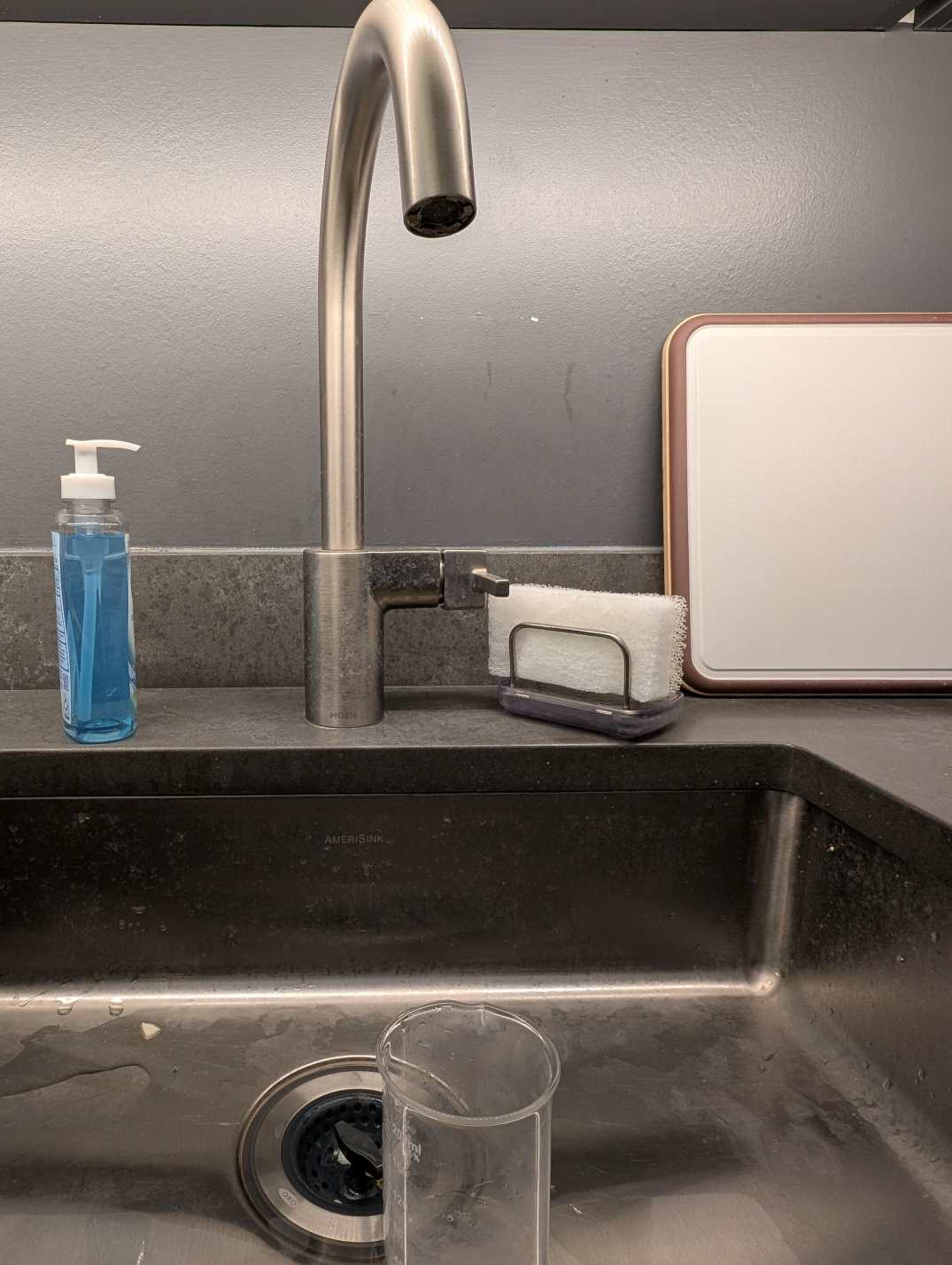} & %
         \includegraphics[width=\imgwidth]{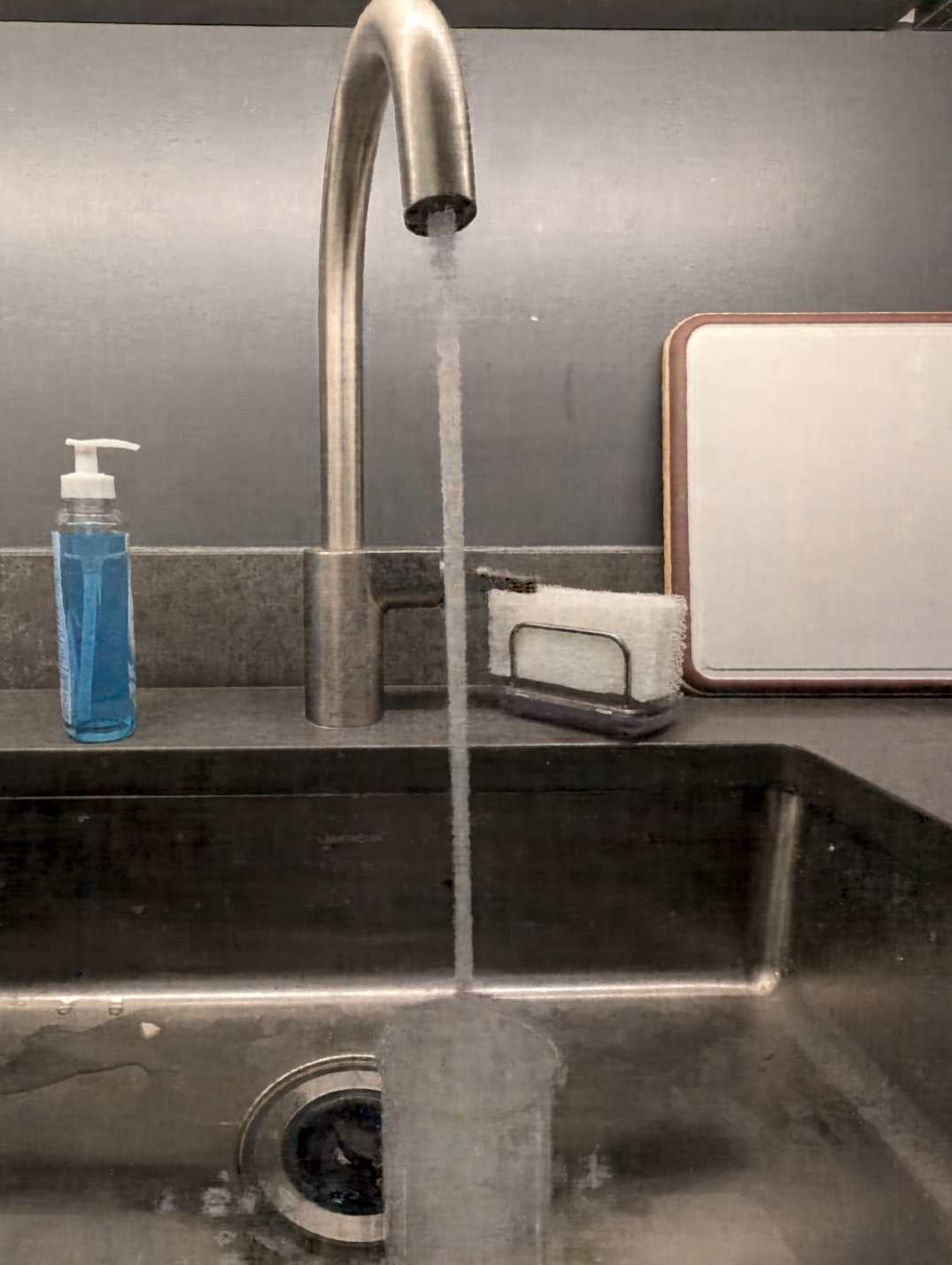} & %
         \includegraphics[width=\imgwidth]{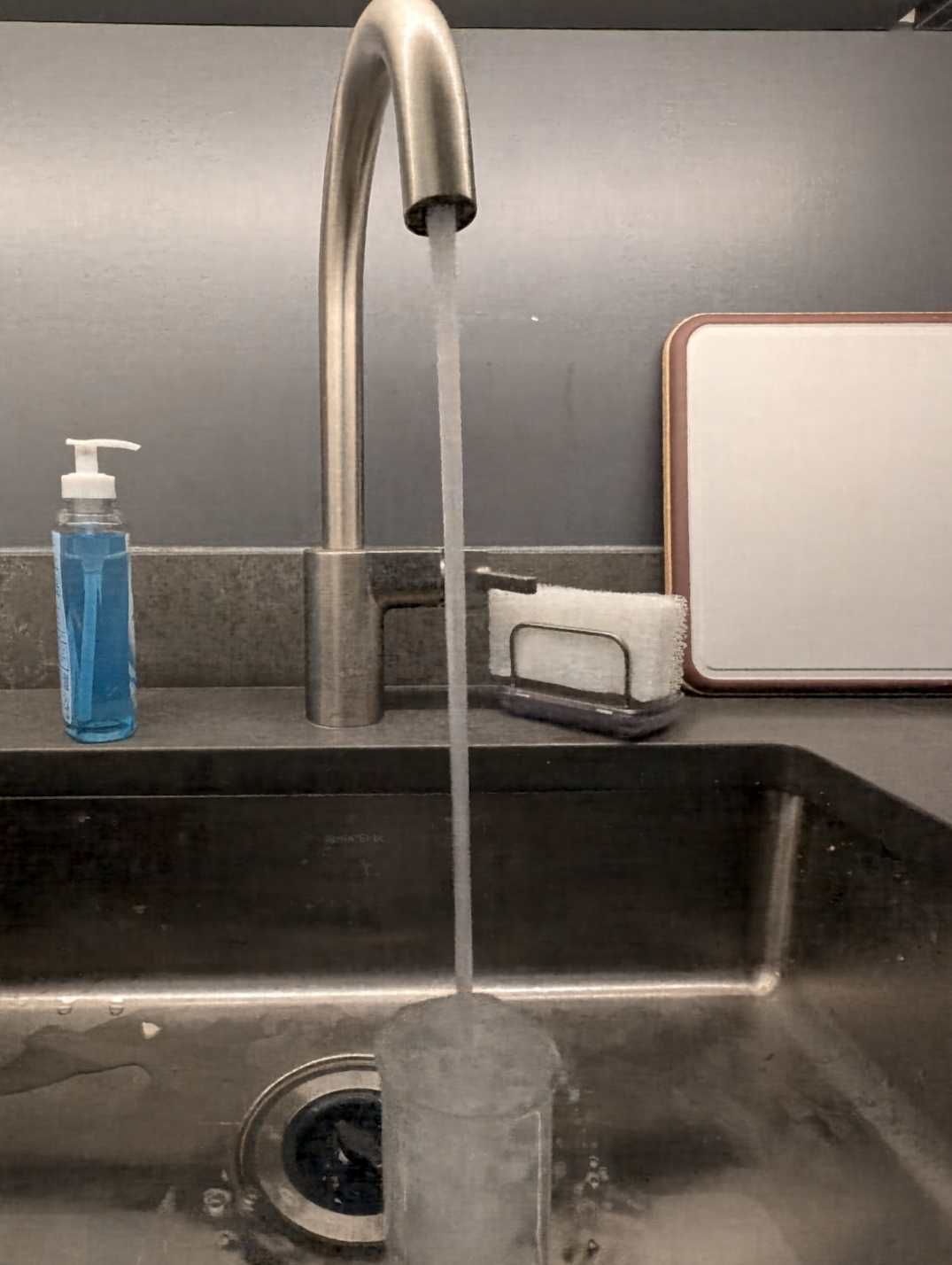} & %
         \includegraphics[width=\imgwidth]{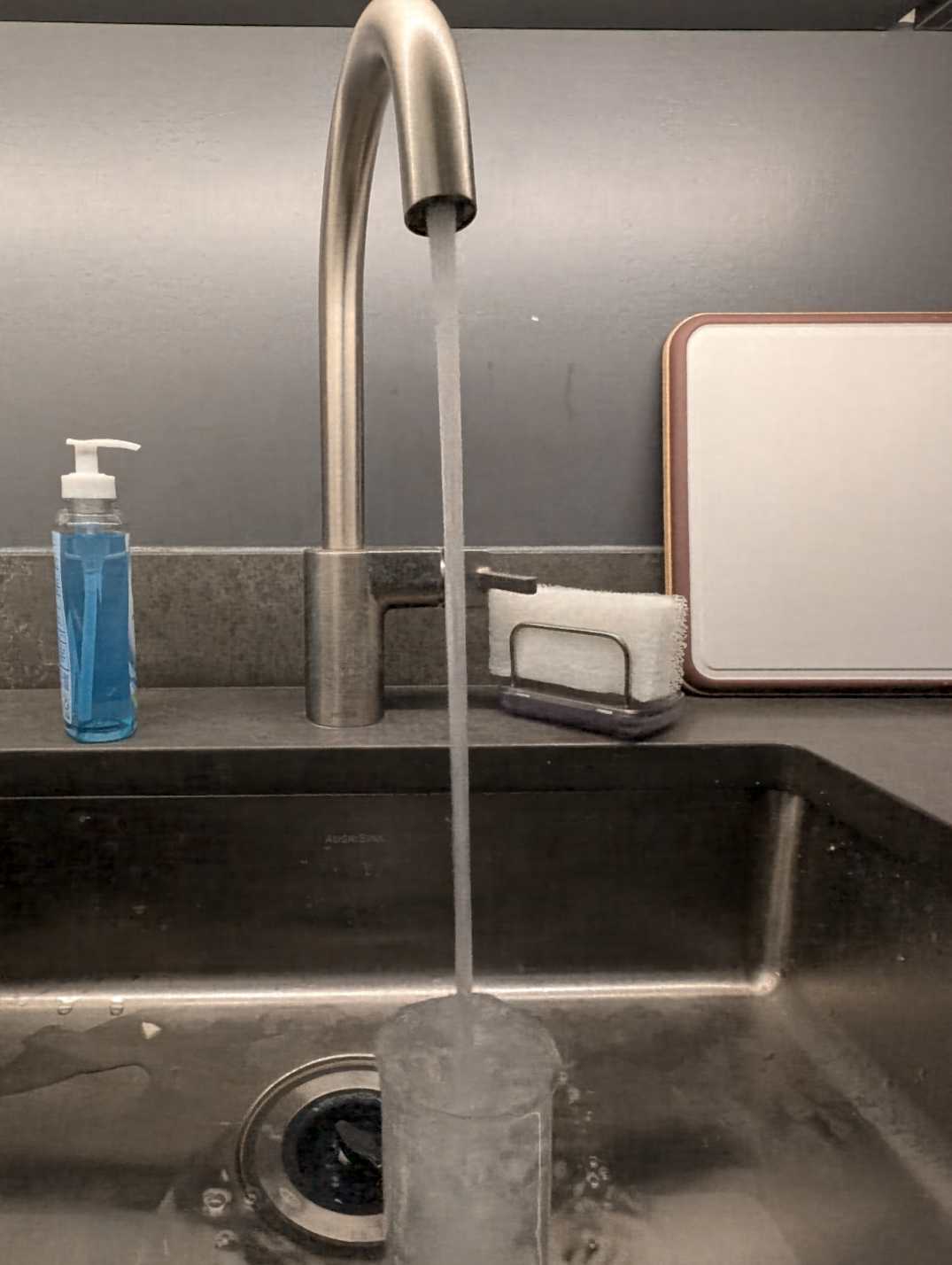} & %
         \includegraphics[width=\imgwidth]{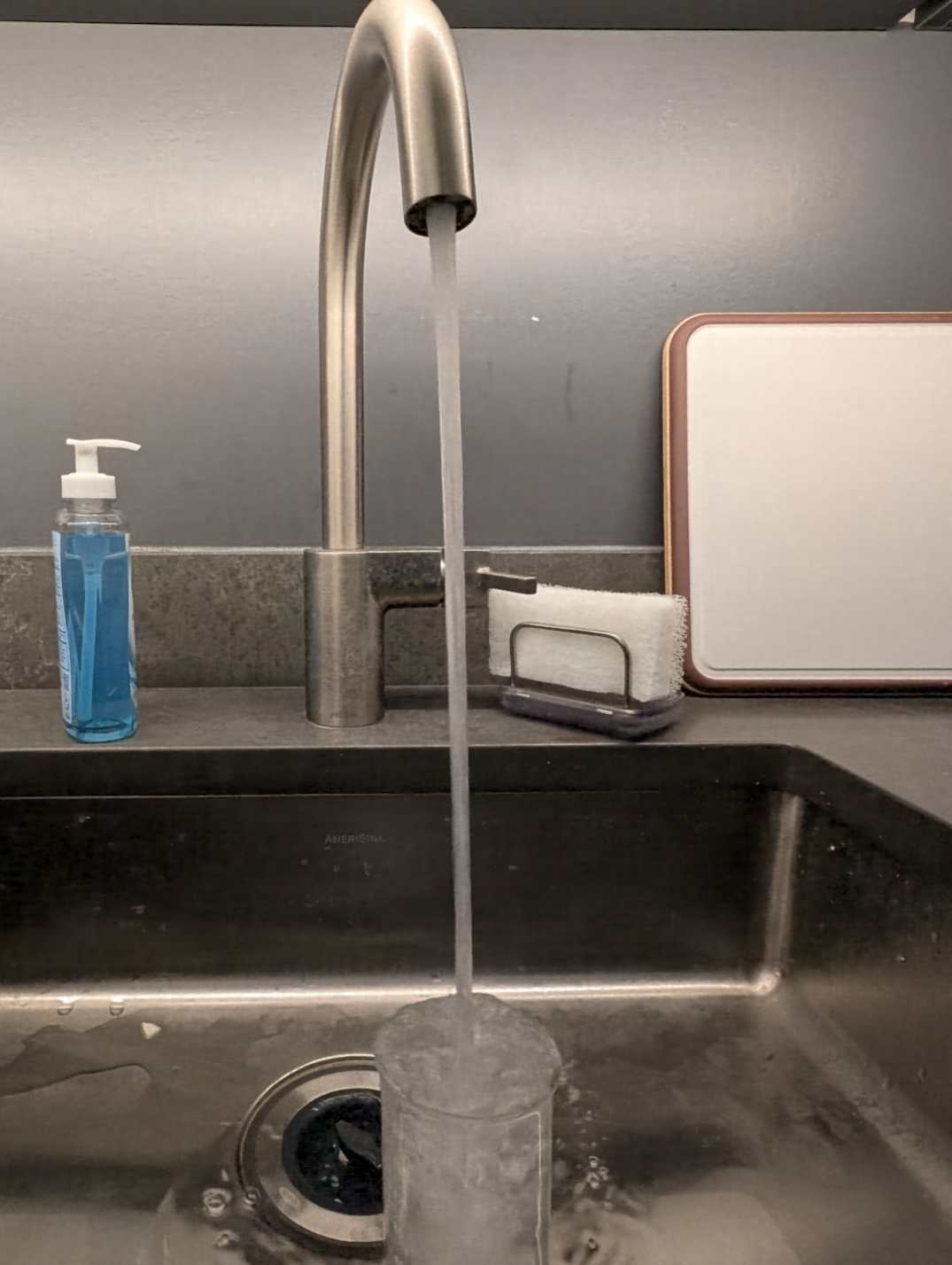} & %
         \includegraphics[width=\imgwidth]{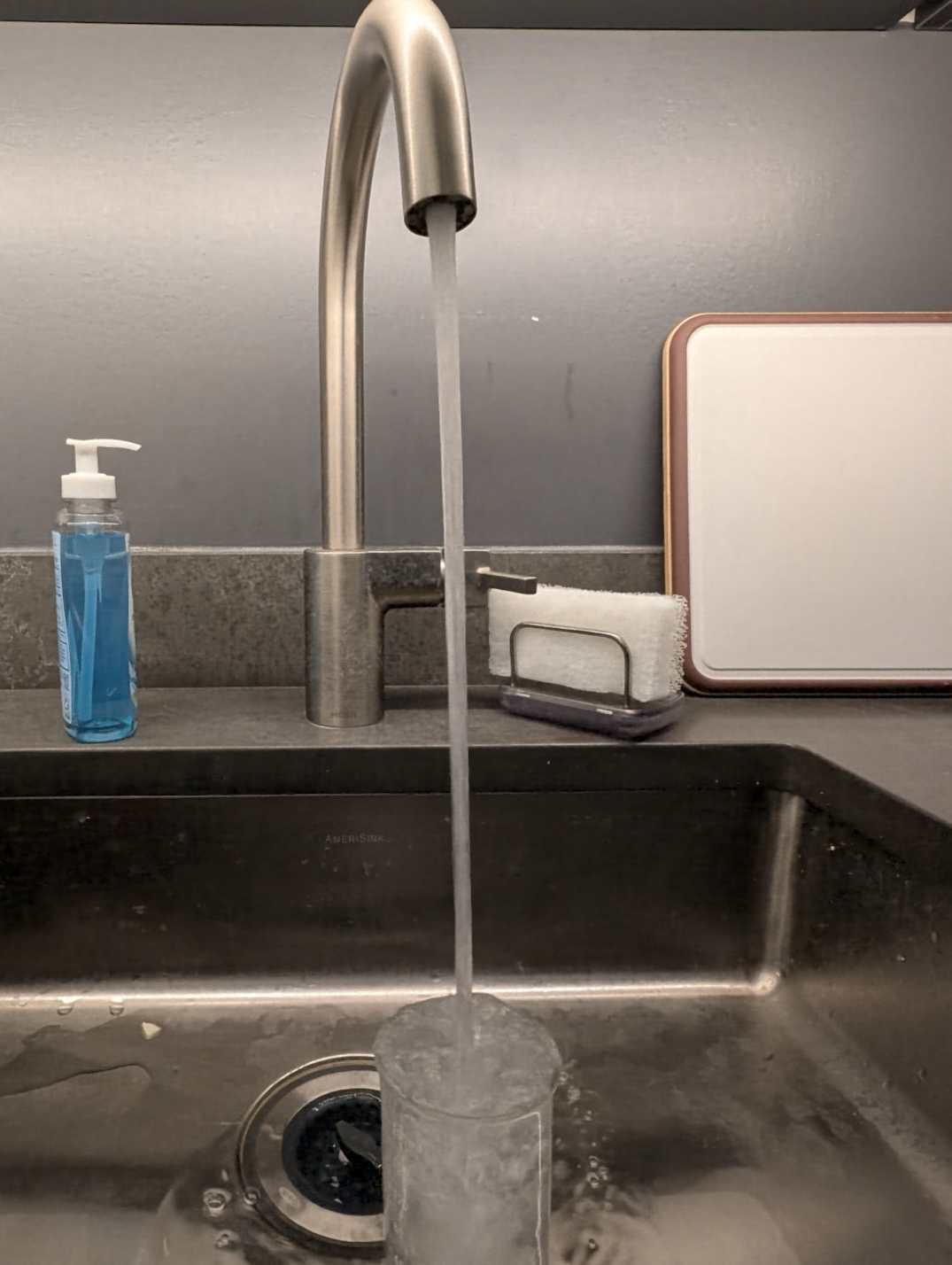} & %
         \includegraphics[width=\imgwidth]{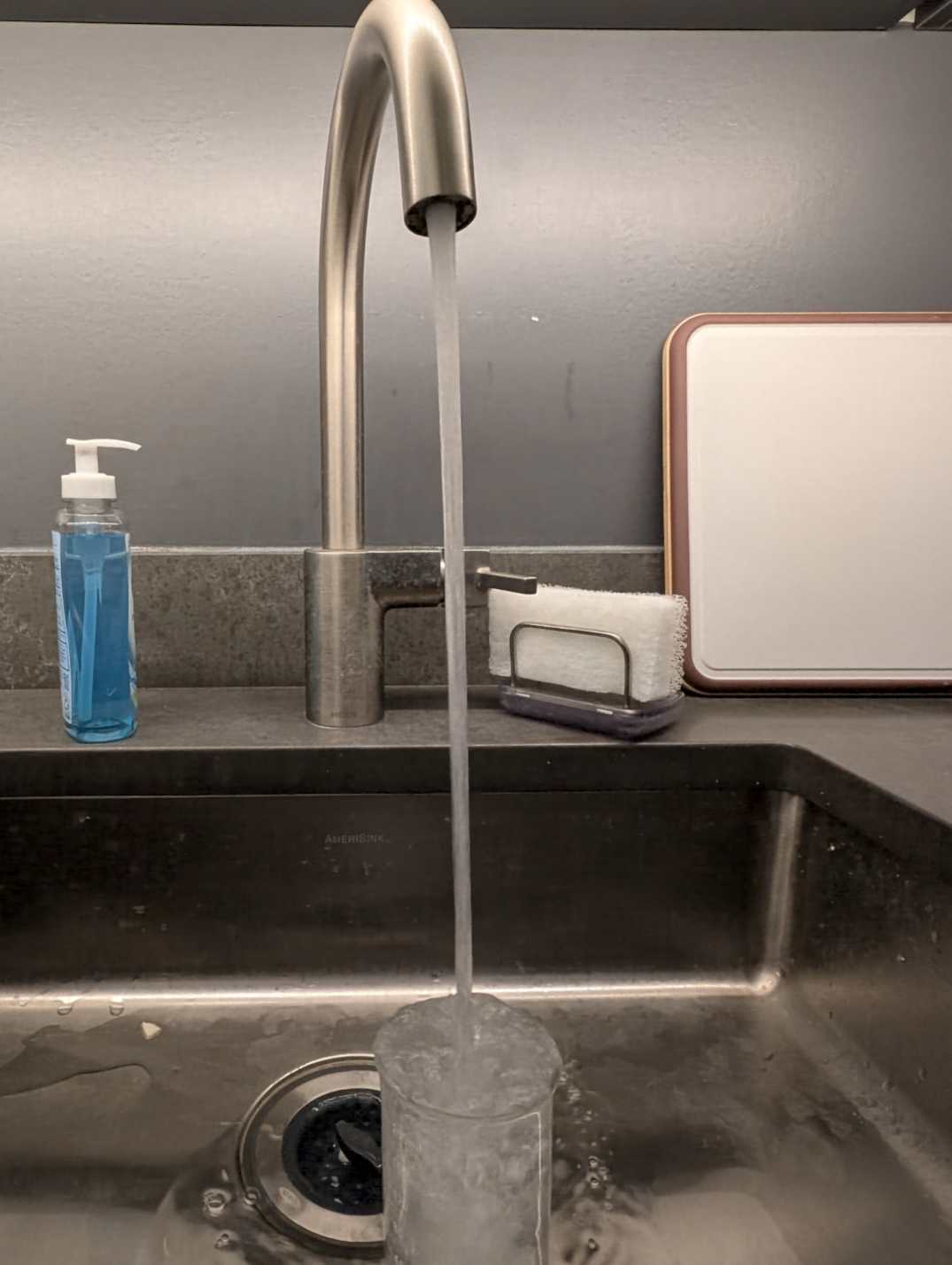} & %
         \includegraphics[width=\imgwidth]{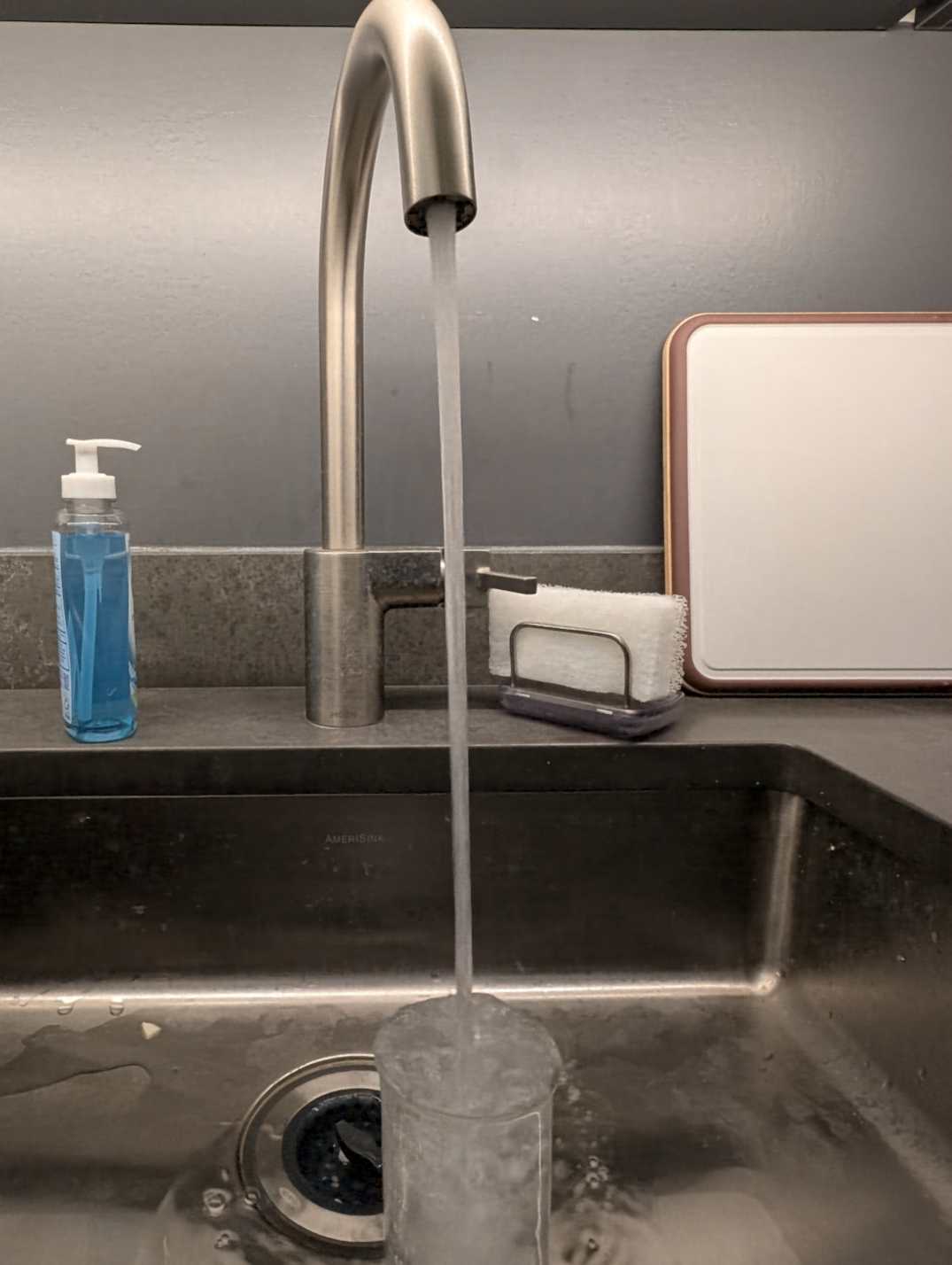} & %
         \includegraphics[width=\imgwidth]{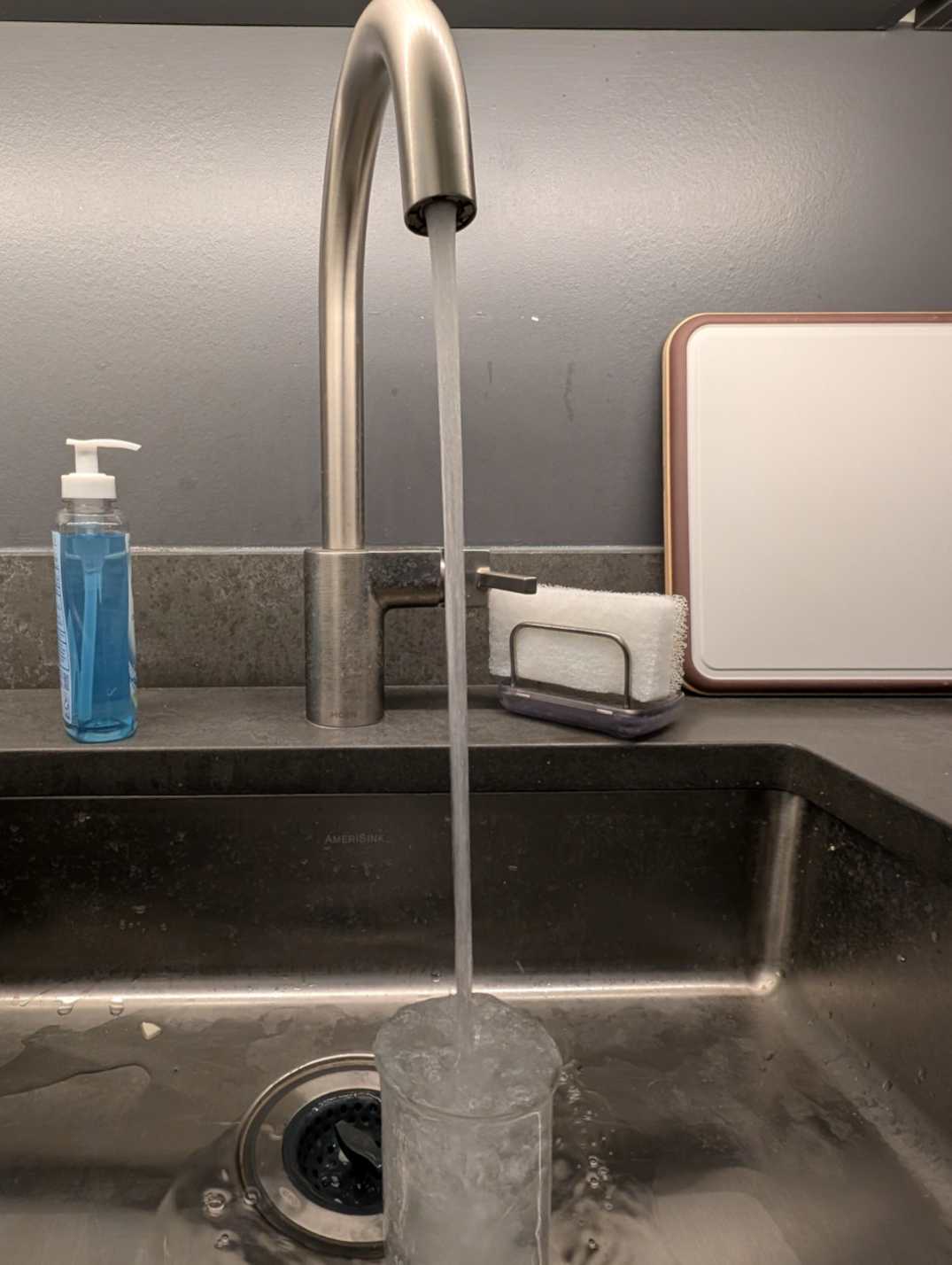} \\ %
         \hline

         & PSNR & \SI{25.5}{dB} & \SI{28.1}{dB} & \SI{30.2}{dB}  & \SI{32.3}{dB} & \SI{34.4}{dB} & \SI{35.7}{dB} & \SI{35.1}{dB} & \\
         \rotatebox{90}{Glass Filter} &
         \includegraphics[width=\imgwidth]{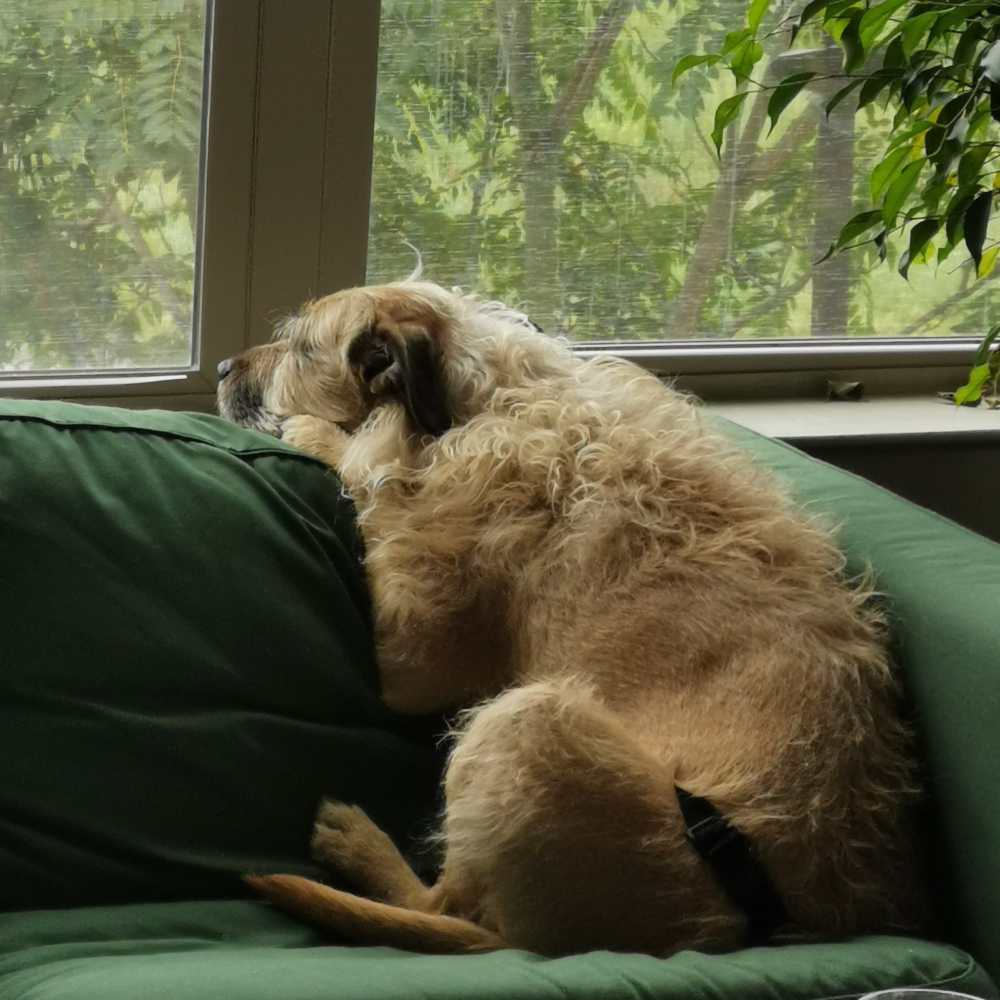} & %
         \includegraphics[width=\imgwidth]{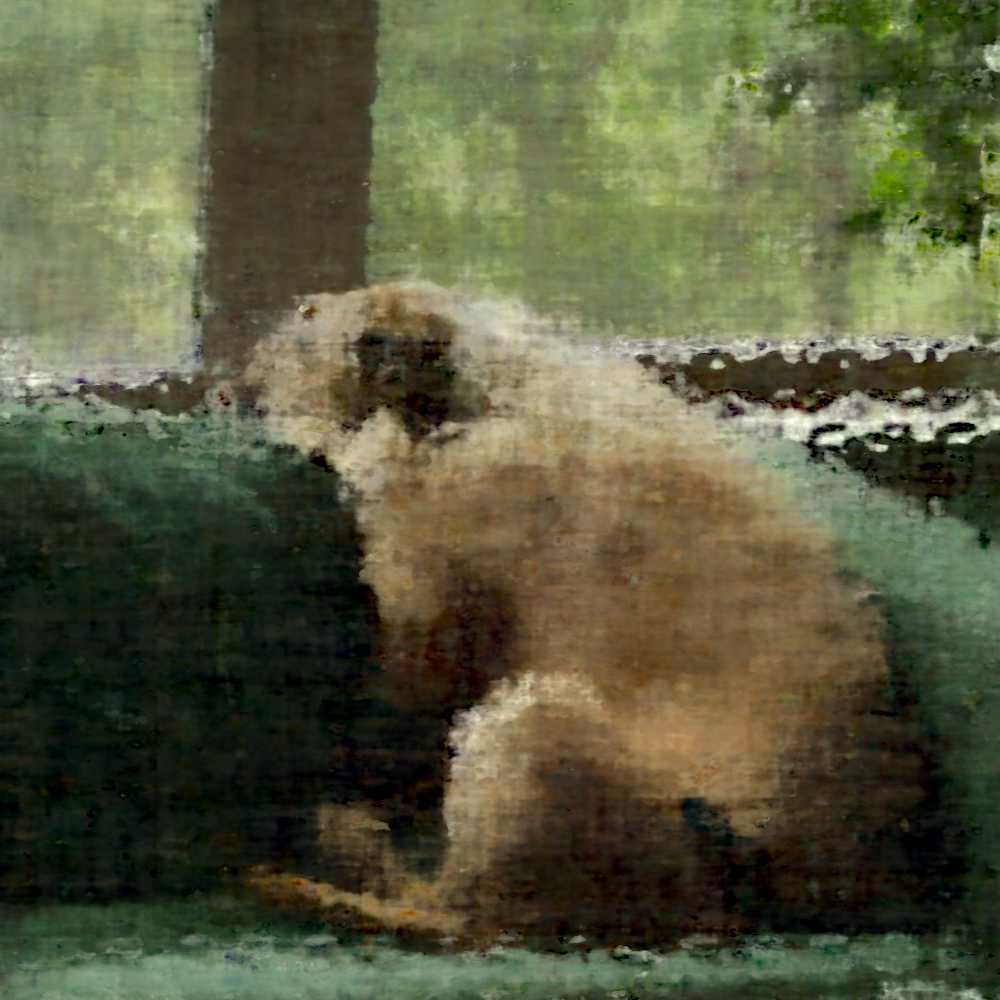} & %
         \includegraphics[width=\imgwidth]{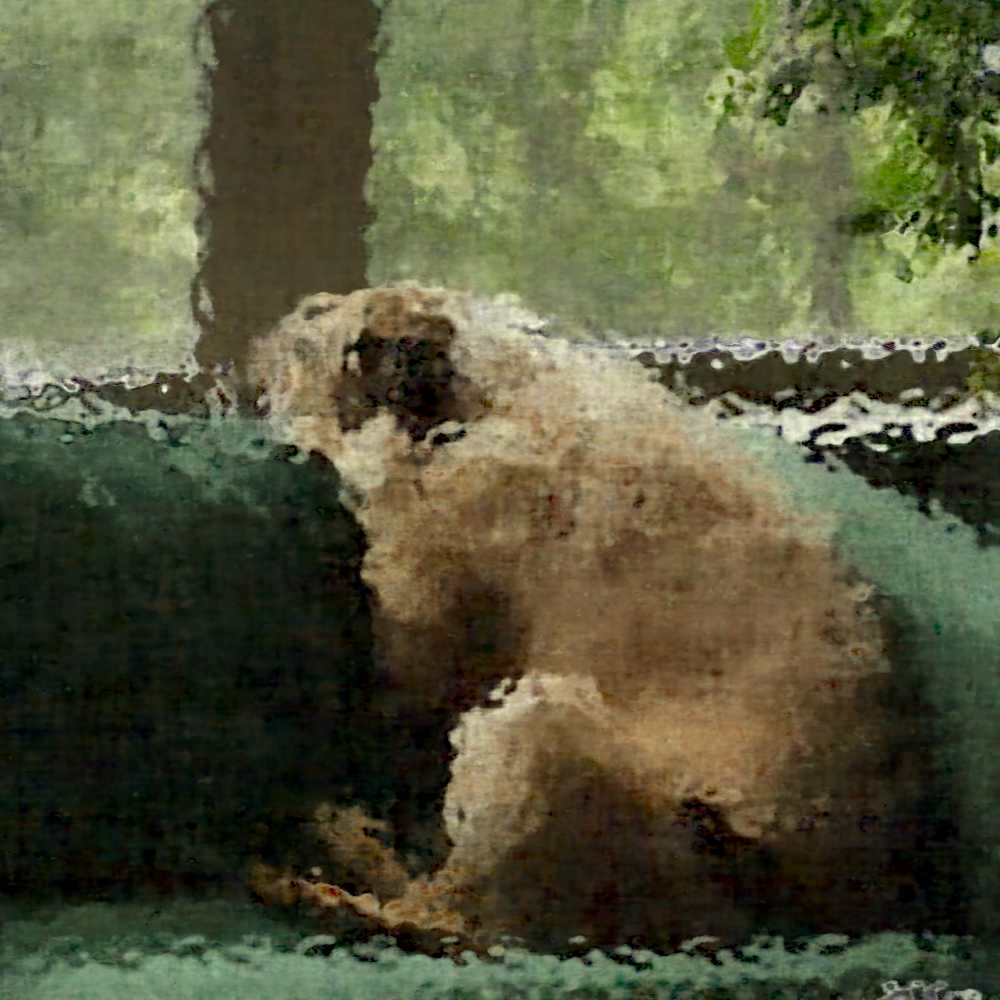} & %
         \includegraphics[width=\imgwidth]{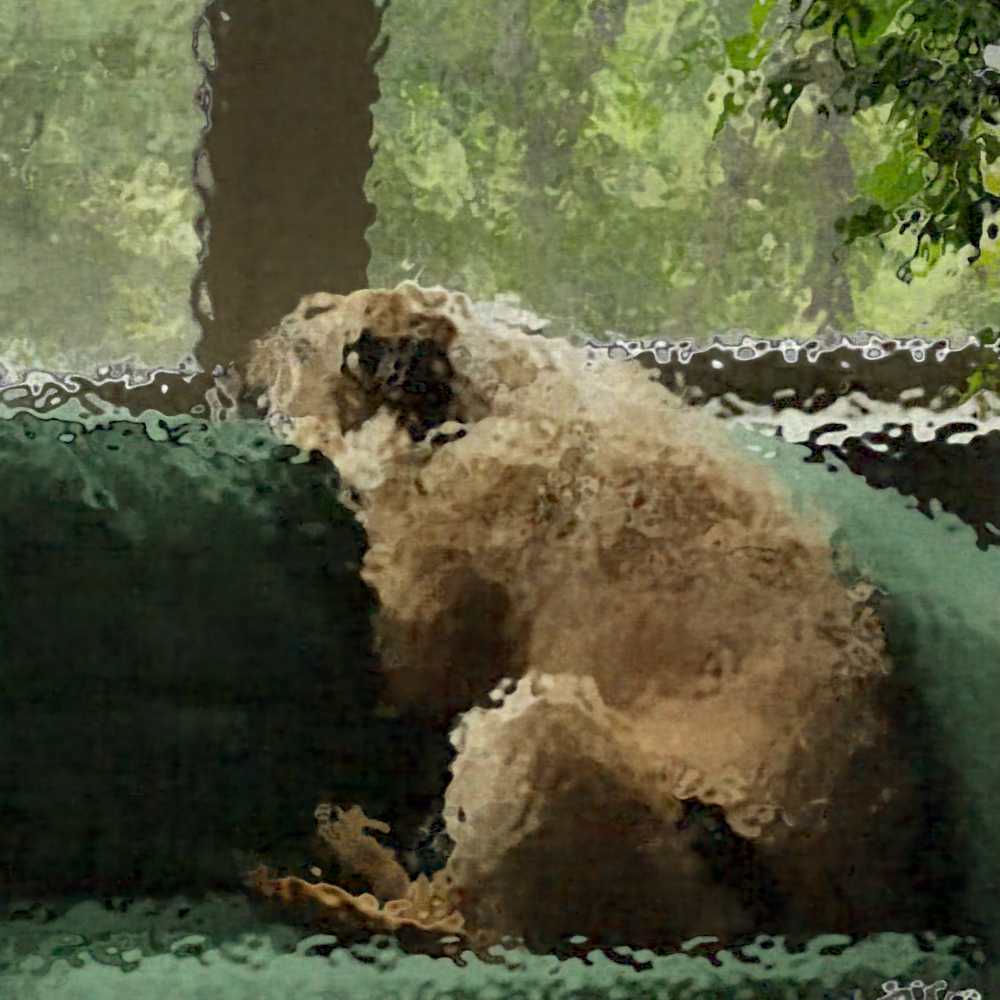} & %
         \includegraphics[width=\imgwidth]{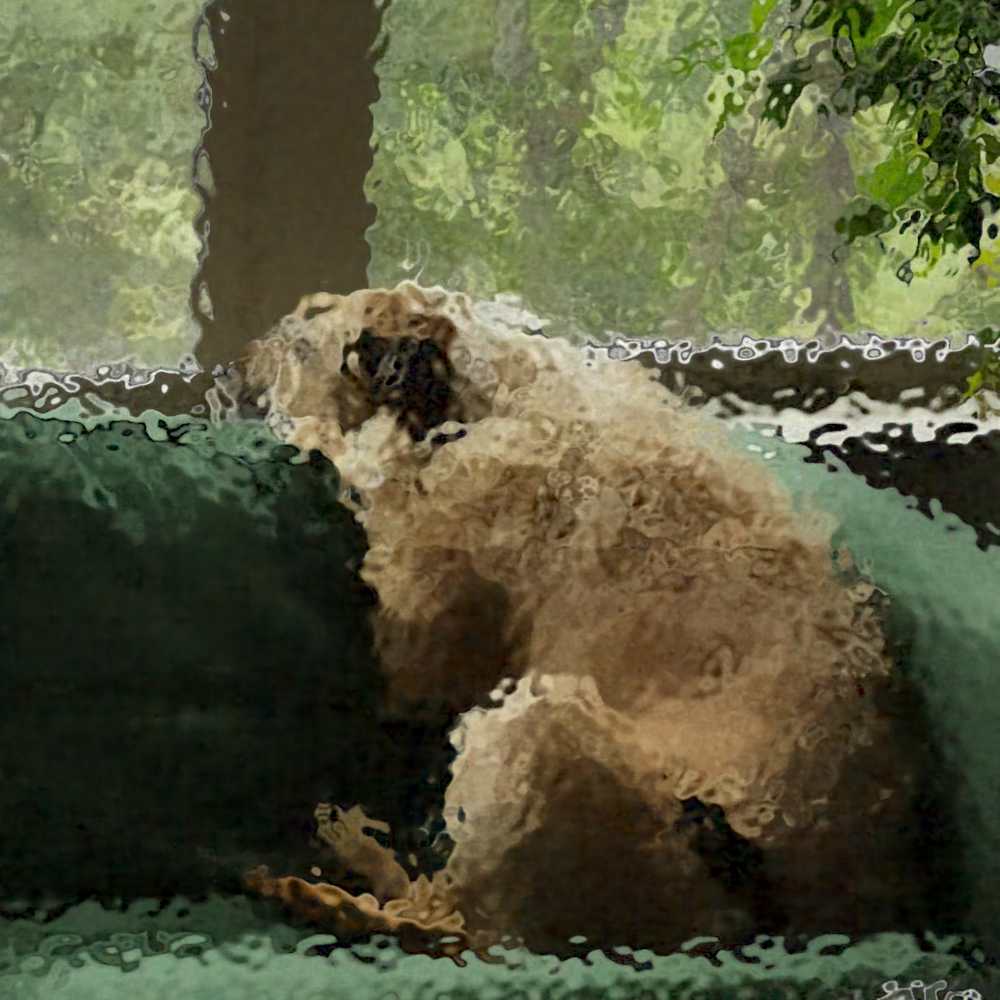} & %
         \includegraphics[width=\imgwidth]{figs/image_experiments/bella_glass_filter/lora_2025-05-07_15-17-08_noise16_rank32/best_reconstruction.jpg} & %
         \includegraphics[width=\imgwidth]{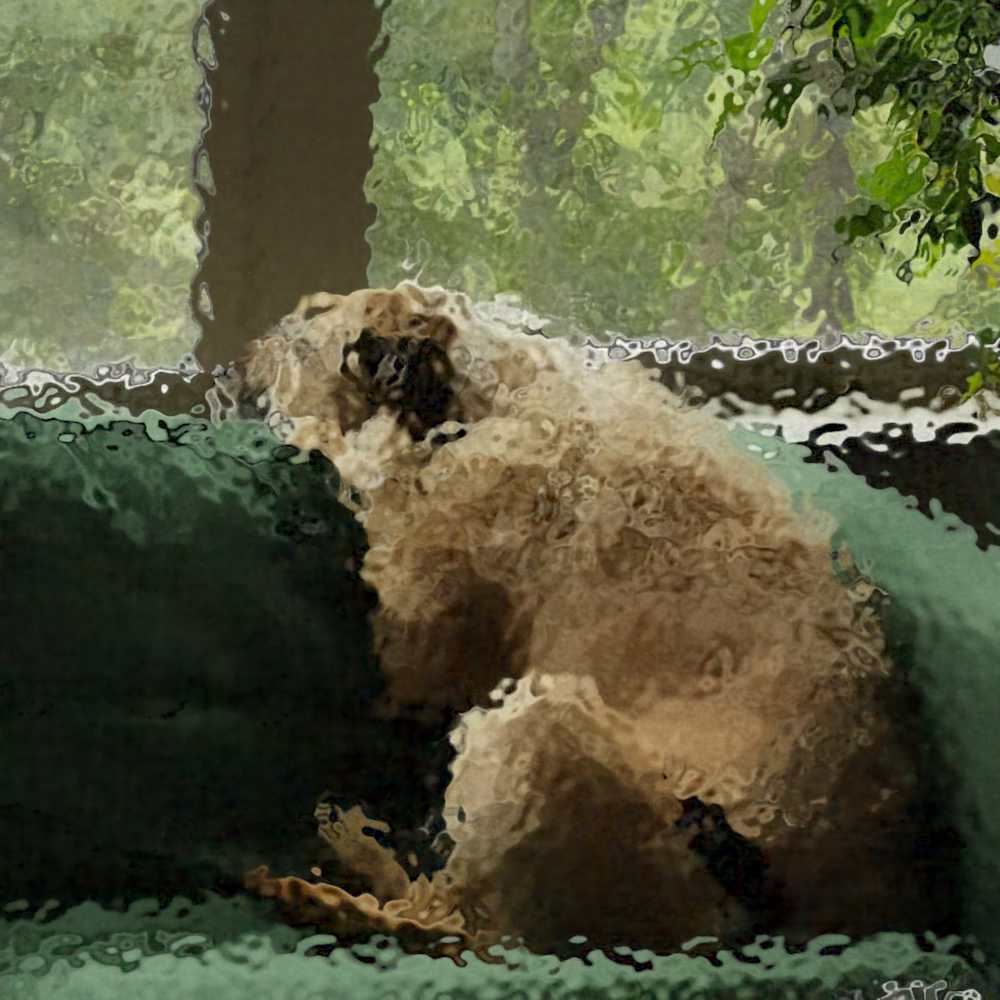} & %
         \includegraphics[width=\imgwidth]{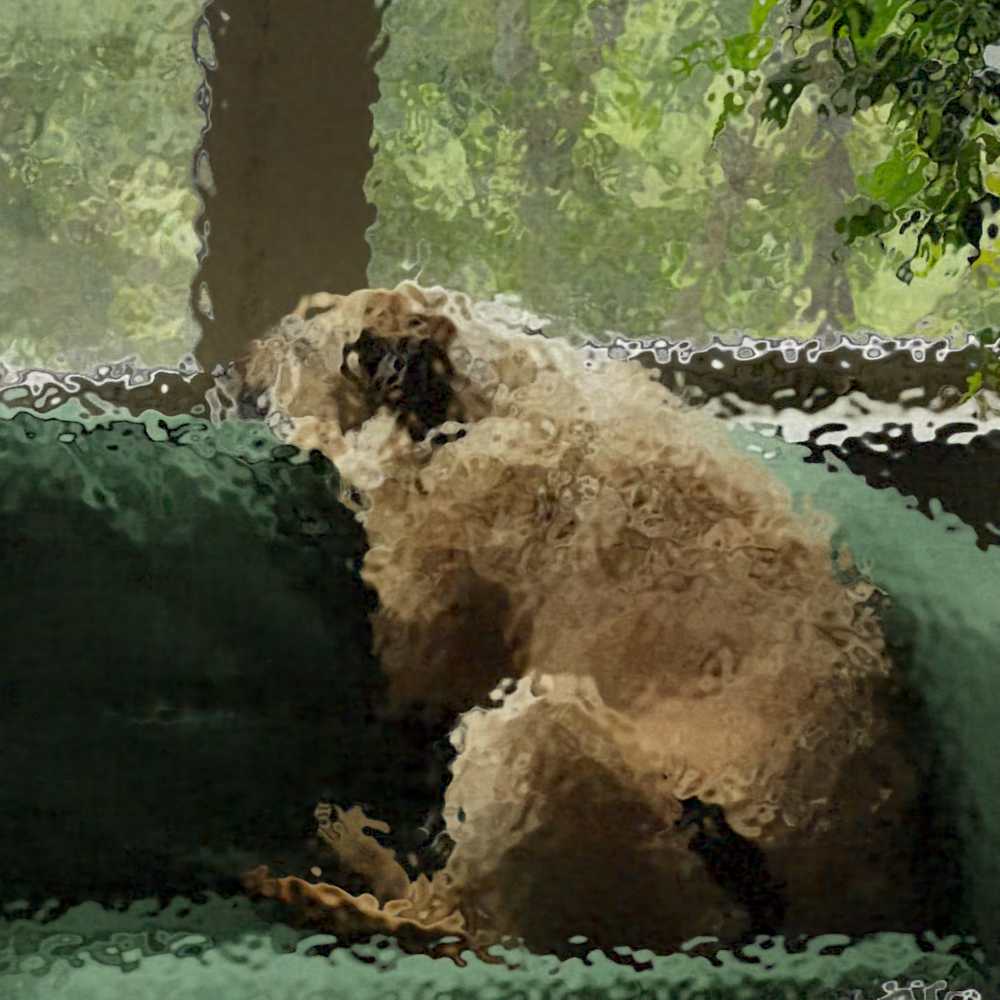} & %
         \includegraphics[width=\imgwidth]{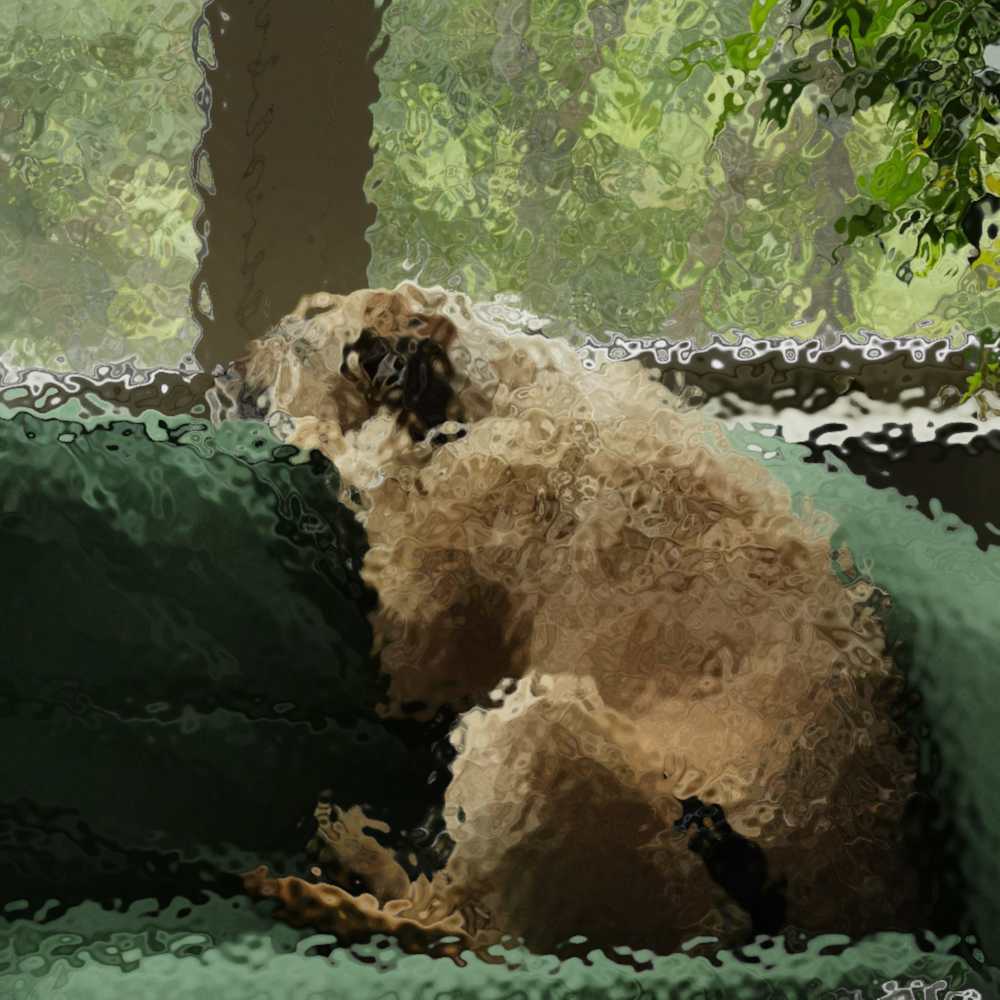} \\ %
         \hline
     \end{tabular}
     \caption{Comparison of LoRA-based image reconstructions using increasing rank $r$ against full fine-tuning (FT) for encoding image variations. \change{The number of fine-tuning parameters as a percentage of the base model size is shown in the top row.}}
     \label{fig:main image result table}
 \end{figure*}

\end{document}